\documentclass[twocolumn]{aa}
\usepackage{graphicx}
\usepackage{latexsym,natbib}
\usepackage{txfonts}
\def\cf{cf.~}
\def\eg{e.g.~}
\def\tab{Table \,}
\def\fig{Fig.\,}
\def\sec{Sect.\,}
\def\hi{$\mathrm{H}$\,{\sc i}\,}
\def\hii{$\mathrm{H}$\,{\sc ii}\,}
\def\ltsim{~\rlap{\lower -0.5ex\hbox{$<$}}{\lower 0.5ex\hbox{$\sim\,$}}}
\def\gtsim{~\rlap{\lower -0.5ex\hbox{$>$}}{\lower 0.5ex\hbox{$\sim\,$}}}
\def\mabs{\hbox{$M_{\rm abs}$}\,}
\def\vvir{\hbox{$v_{\rm vir}$}\,}
\def\vrot{\hbox{$v_{\rm rot}$}\,}
\def\kms{km$\,$s$^{-1}$\,}
\def\deg{\hbox{$^\circ$}\,}

\def\sqdeg{\raisebox{+0.4ex}{\hbox{$\Box^{\circ}$}\,}}
\def\magsqarcsec{mag$/$\raisebox{-0.4ex}{\hbox{$\Box^{\prime\prime}$}\,}}
\def\magsqarcsecfrac{$\frac{\rm mag}{\Box^{\prime\prime}}$}
\def\hin{\hbox{$h_{\rm in}$}\,}
\def\hout{\hbox{$h_{\rm out}$}\,}
\def\hindhout {\hbox{$h_{\rm in}/h_{\rm out}$}\,}
\def\hloc{\hbox{$h_{\rm local}$}\,}
\def\hlocr{\hbox{$h_{\rm local}(R)$}\,}
\def\rbr{\hbox{$R_{\rm br}$}\,}
\def\rbrdhin{\hbox{$R_{\rm br}/h_{\rm in}$}\,}
\def\munin{\hbox{$\mu_{0,\rm in}$}\,}
\def\munout{\hbox{$\mu_{0,\rm out}$}\,}
\def\mun{\hbox{$\mu_0$}\,}
\def\mubr{\hbox{$\mu_{\rm br}$}\,}
\newcommand{\typeo}{Type~I }
\newcommand{\typeoc}{Type~I}
\newcommand{\typet}{Type~II }
\newcommand{\typetc}{Type~II}
\newcommand{\typeti}{Type~II.i }
\newcommand{\typetic}{Type~II.i}
\newcommand{\typetiiii}{Type~II.i + III }
\newcommand{\typeto}{Type~II.o }

\newcommand{\typeiii}{Type~III }
\newcommand{\typeiiic}{Type~III}

\newcommand{\typeiiibc}{Type~III-s}
\newcommand{\typeiiid}{Type~III-d }
\newcommand{\typeiiidc}{Type~III-d}
\newcommand{\typeab}{Type~II-AB }
\newcommand{\typeabc}{Type~II-AB}

\newcommand{\typetoabc}{Type~II.o-AB}
\newcommand{\typeabiii}{Type~II-AB + III }
\newcommand{\typeolr}{Type~II.o-OLR }
\newcommand{\typeolrc}{Type~II.o-OLR}
\newcommand{\typeolrct}{Type~II.o-OLR + II.o-CT }

\newcommand{\typeolriii}{Type~II.o-OLR + III }
\newcommand{\typeolriiic}{Type~II.o-OLR + III}
\newcommand{\typect}{Type~II-CT }
\newcommand{\typectc}{Type~II-CT}
\newcommand{\typetoct}{Type~II.o-CT }
\newcommand{\typetoctc}{Type~II.o-CT}
\newcommand{\typetoctab}{Type~II.o-CT + II.o-AB }

\newcommand{\typectiiic}{Type~II-CT + III}

\begin{document}
%
\title{The structure of galactic disks}
\subtitle{Studying late-type spiral galaxies using SDSS}
\author{M. Pohlen\inst{1,2}\and I. Trujillo \inst{3,4}}
\offprints{M. Pohlen}
\institute{
Kapteyn Astronomical Institute, University of Groningen, P.O. Box 800, 
NL--9700 AV Groningen, The Netherlands \\
\email{pohlen@astro.rug.nl}
\and 
Instituto de Astrof\'{\i}sica de Canarias, C/ Via L\'{a}ctea s/n,
E--38200 La Laguna, Tenerife, Spain
\and
Max--Planck--Institut f\"ur Astronomie, K\"onigstuhl 17, D--69117
Heidelberg, Germany 
\and
School of Physics and Astronomy, University of Nottingham, University Park, Nottingham, NG7 2RD, UK \\
\email{ignacio.trujillo@nottingham.ac.uk}
          }
\date{Received January 23, 2006; accepted April 26, 2006}
\abstract{
Using imaging data from the SDSS survey, we present the g$^{\prime}$ and
r$^{\prime}$ radial stellar light distribution of a complete sample of
$\!\sim\!90$ face-on to intermediate inclined, nearby, late-type (Sb--Sdm) 
spiral galaxies. The surface brightness profiles are reliable ($1 \sigma$
uncertainty less than 0.2 mag) down to $\mu\!\sim\!27\ $\magsqarcsec.  Only 
$\sim\!10$\% of all galaxies have a {\it normal/standard} purely exponential 
disk down to our noise limit. 
The surface brightness distribution of the rest of the galaxies is better
described as a broken exponential. About 60\% of the galaxies have a break in
the exponential profile between $\sim 1.5-4.5$ times the scalelength 
followed by a {\it downbending}, steeper outer region. Another $\sim\!30$\% 
shows also a clear 
break between $\sim 4.0-6.0$ times the scalelength but followed by an 
{\it upbending}, shallower outer region. A few galaxies have even a more 
complex surface brightness distribution.
The shape of the profiles correlates with Hubble type. Downbending breaks
are more frequent in later Hubble types while the fraction of upbending 
breaks rises towards earlier types. No clear relation is found between the 
environment, as characterised by the number of neighbours, and the shape of
the profiles of the galaxies.
%
}          
\maketitle
\section{Introduction}
\label{introduction}
The structural properties of the faintest regions of galactic disks 
must be intimately linked to the mechanisms involved in the growing and 
shaping of galaxies. These {\it outer edges} are easily affected by 
interactions with other galaxies and, consequently, their characteristics 
must be closely connected with the evolutionary path followed by the 
galaxies. Together with their stellar halos, the study of the outer edges 
allows the exploration of the so-called fossil evidence imprinted by the 
galaxy formation process. 
Most of the detailed structural studies of nearby disk galaxies so far 
\citep{courteau1996, dejong1996, graham2001, jansen2001, trujillo2002, 
macarthur2003, moellenhoff2004} have been concentrated on the (most 
easily accessible) inner parts of the galaxies. 
These studies of the brightest region of the stellar disk show that its 
light distribution is almost always well described by a simple exponential 
decline going back to \cite{patterson1940}, \cite{devaucouleurs1959} 
or \cite{free70}. 
This simple description, however, has now been shown to fail at fainter 
surface brightness. In fact, since \cite{vdk1979} we know that this decline 
does not continue to the last measured point, but is {\it truncated} after 
several radial scalelengths.
The main concern with current, larger statistical studies exploring the 
faintest surface brightness region of the galaxies is that they are based 
on edge-on galaxies 
\citep{barteldrees1994,pohlen2000,florido2001,degrijs2001,pohlen2001,kregel2002}. 
This geometry facilitates the
discovery of the truncation in the surface brightness profiles, but 
introduces severe problems caused by the effects of dust and the 
line--of--sight integration \cite[for a recent review]{pohlen2004}, 
such as masking the actual shape of the truncation region, or interfering 
with the identification of other important disk features (\eg bars, rings, 
or spirals).
Only few works have probed the faintest regions of the disk galaxies in low
inclination systems. \cite{pohlen2002} demonstrated for a small sample of 
face-on galaxies that the so-called cut-offs in the surface brightness
profiles discovered by van der Kruit are in fact not complete but better
described by a broken exponential with a shallow inner and a steeper outer
exponential region separated at a relatively well defined break radius.
However, not all the galaxies seem to have a break or a truncation in their
surface brightness profiles. Just recently \cite{ngc300} found  
(using star counts) again a galaxy (NGC\,300) for which the exponential 
decline simply continues down to $\sim 10$ radial scalelength. Together with  
earlier measurements by \cite{barton1997} or \cite{weiner2001} (using 
surface photometry) this provides evidence that indeed there are 
prototypical, model exponential disks. 
Our picture of the faintest regions of spiral galaxies have been even 
broadened with some recent studies. \cite{typeiii}, studying a large sample 
of early-type barred S0--Sb galaxies, discovered that there is more than 
{\it truncated} or {\it untruncated} galaxies. They report the detection 
of {\it antitruncated} galaxies, which show also a broken exponential 
but having the opposite behaviour, an outer upbending profile. Similar 
structures are also found for a large sample of irregular systems (Im) 
and BCDs by \cite{hunter2005}.

The disk structures we describe in this study are the result of 
initial conditions, infalling matter and/or redistribution of gas 
and stars, already settled in the disk, triggered 
by internal (\eg bars) or external (\eg interaction) forces. 
All these processes are linked to the detailed mechanism of forming 
the stars which build up the observed brightness distribution.  
To explain for example the observed truncation in the surface 
brightness profiles radial star-formation thresholds 
\cite[e.g.][]{kennicutt89,schaye2004} have been suggested  
\citep[see][]{pohlen2004}.
However, there is compelling evidence of star-formation in the 
far outer regions of spiral galaxies 
\citep[e.g.][]{ferguson1998,cuillandre2001,thilker2005,gildepaz2005},
well beyond the break of the observed broken exponential 
structure \citep{pohlen2002}.
The outer star formation activity argues against a simple 
threshold scenario. 
Just recently, though, \cite{elmegreen2006} show that their model 
of star-formation is able to produce the variety of observed 
radial profiles. Alternatively, \cite{victor2005} also find 
downbending breaks using purely collisionless $N$-body 
simulations. 

The main goal of the present study is to conduct a complete census of the
outer disk structure of late-type galaxies --together with a complementary 
study by \cite{erwin2006} for early-type (barred) galaxies-- in the local 
universe.
The objective is to provide the frequencies of the different surface
brightness profile types, the structural disk parameters, and search 
for correlations (if any) between them. 
Aside from the observations showing the increasing complexity of 
radial surface brightness profiles, the breaks in the surface 
brightness distribution can be used to directly constrain galaxy 
evolution. \cite{perez2004} showed 
that it is possible to detect stellar truncations even out to higher 
redshift ($z\!\sim\!1$). So using the radial position of the truncation as 
a direct estimator of the size of the stellar disk, \cite{trujillo2005} 
infer a moderate ($\sim25\%$) inside-out growth of the disk galaxies 
since $z\!\sim\!1$, using as a local reference the galaxies studied 
in this paper. 

The remainder of this paper is organised as follows. In \sec\ref{samplep}
we describe the sample selection and our environment parameter. In
\sec\ref{datapp} we give the details about the ellipse fitting and the 
 SDSS imaging used, characterising their quality and addressing 
the crucial issue of sky subtraction. The applied classification schema 
of the observed profiles and  the derived parameters are discussed in 
\sec\ref{analysis}. The results are presented in \sec\ref{results} 
and briefly discussed in \sec\ref{discussion}. 
The colour profiles and the physical implications derived from 
them will be discussed in a subsequent paper. 
In Appendix \ref{atlas} we give detailed comments for all galaxies, 
show their $r^{\prime}$ and $g^{\prime}$-band surface brightness 
profiles and reproduce their SDSS colour pictures.
%
%
\section{The sample}
\label{samplep}
\subsection{Selection}
\label{sample}
We selected our initial galaxy sample from the 
LEDA\footnote{{\ttfamily http://leda.univ-lyon1.fr}} online galaxy 
catalogue (version Nov.~2004), since this is the richest catalogue with
homogeneous parameters of galaxies for the largest available sample.
We restricted the Hubble type ($T$ parameter), the mean heliocentric 
radial velocity relative to the Local Group 
(corrected for virgocentric inflow, \vvir), the axis ratio 
(major axis/minor axis), the absolute B-magnitude (\mabs), 
and the Galactic latitude ($b_{\rm II}$).
The Hubble type is chosen to be between $2.99 < T < 8.49$ 
(corresponding to Sb to Sdm galaxies) building an 
intermediate- to late-type galaxy sample.
This allows a complementary study to the work by \cite{erwin2006} on 
the disk structure of early type (barred) galaxies. To have some, 
but not too much, overlap with their sample we did not include all 
Sb galaxies by excluding the range $2.49<T<2.99$.
The axis ratio is selected to be $\log r_{25} < 0.301$ (equal to $a/b<2$ 
or $e<0.5$, which corresponds to an inclination of $\ltsim 61\deg$ 
assuming an intrinsic flattening of $q_0=0.14$) bringing a face-on to 
intermediate inclined galaxy sample. 
This is necessary to avoid the influence of the dust and is convenient 
to provide reliable information on the morphological features such as 
bars, rings, or spiral arm structure. 
The recession velocity is chosen to be $\vvir < 3250$ km/s and the galaxies 
brighter than $-18.4$ \mabs in B band (the total B-magnitude, 
provided by LEDA, is corrected for galactic and internal extinction, k 
corrected, and the distance modulus is derived from the recession velocity 
corrected for Virgo infall with a Hubble constant of 
$H_{0}\!=\!70$ km s$^{-1}$Mpc$^{-1}$).
We did not apply a diameter limit. 
Since SDSS, our data source, is a survey of the high Galactic latitude 
sky, visible from the northern hemisphere, there is no worry about 
incompleteness there. In any case, we restricted the initial sample to 
$|b_{\rm II}| > 20\deg$.

\begin{figure*}
\includegraphics[width=6.1cm,angle=270]{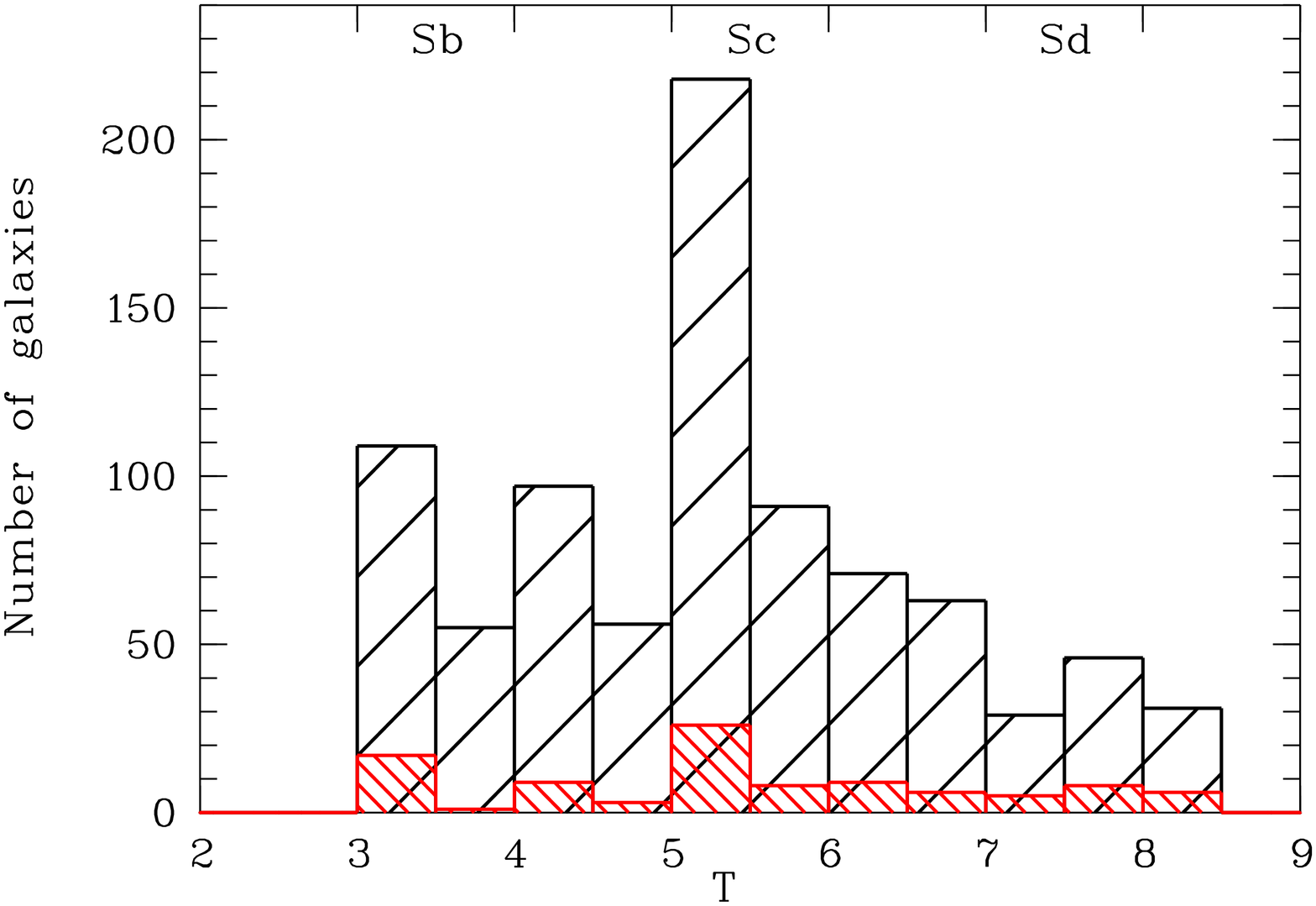}
\includegraphics[width=6.1cm,angle=270]{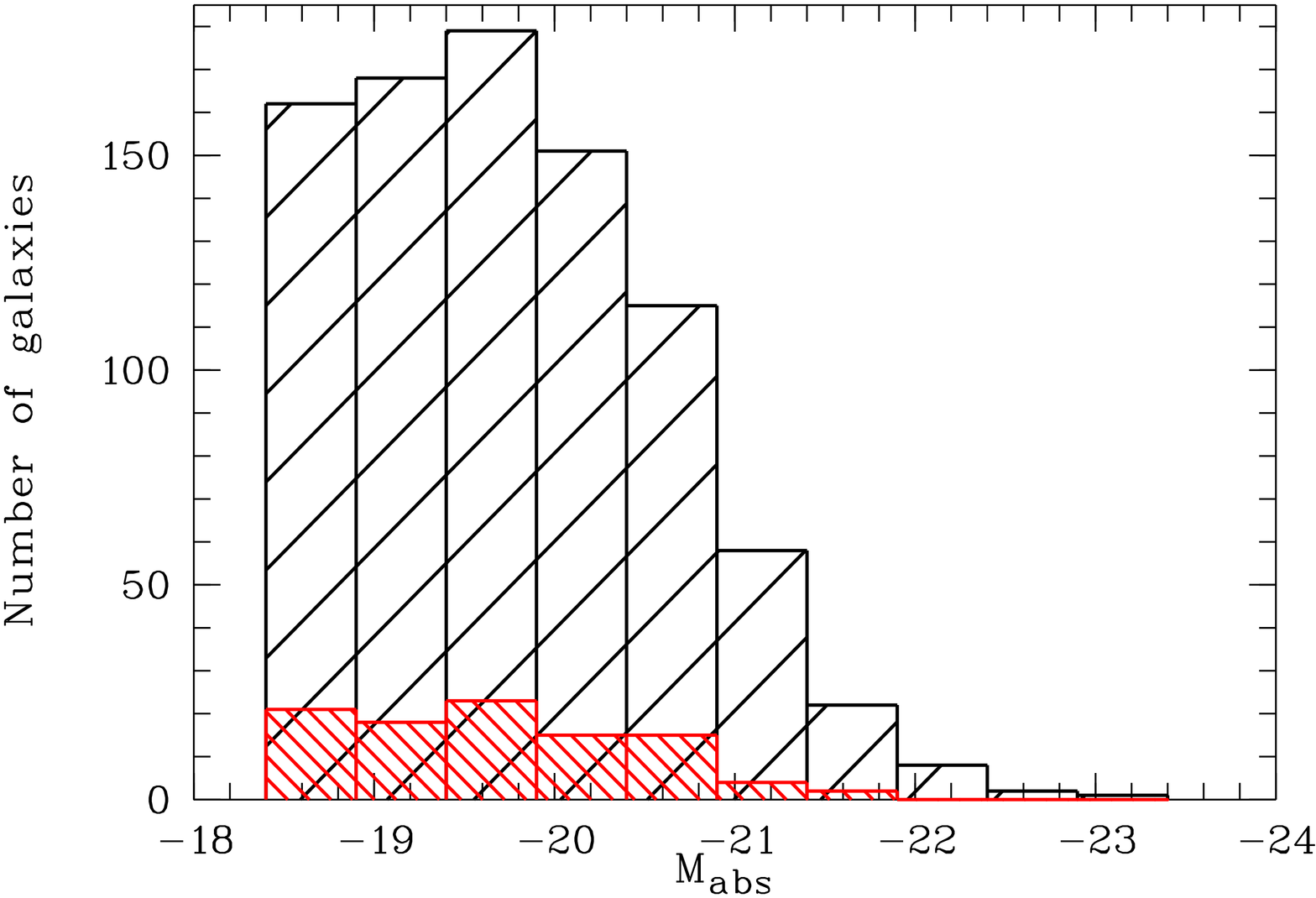}
\caption{Histogram showing the distribution of galaxies covering 
the morphological types ($T$) and absolute magnitude (\mabs) for the 
full LEDA sample {\it (rising hashed lines)} and the studied SDSS DR2 
subsample {\it (falling hashed lines)}.}
\label{HtpHmabsmabsVSt}
\end{figure*}

To assess the completeness of the LEDA catalogue \cite{ledacom} draw 
several cumulative curves $\log N$ versus $\log r$ (N is the total 
number of observed galaxies within the radius $r$ in Mpc) for 
different absolute magnitudes and compare these with an expected 
galaxy distribution. 
Using the same approach we have confirmed that LEDA 
(version Nov.~2004) is almost complete for galaxies of $\mabs<-18.4$ 
within our $\sim 46\,$Mpc survey distance (estimated 
following the Hubble relation with $H_{0}\!=\!70$ km s$^{-1}$Mpc$^{-1}$). 
Our selection criteria results in a total number of 
655 galaxies, of which 98 (15\%) are part of the SDSS Second 
Data Release (DR2), surveying the north Galactic cap and three 
stripes in the southern Galactic cap \cite[]{sloan,sloandr2}. 
The global parameters\footnote{Note: Some of the bar classifications 
from LEDA have slightly changed after our study. Table \ref{LEDAsample} 
still shows the 2004 version. For comparison we always listed the 
RC3 classification.} for this subsample of 
98 galaxies are presented in Table \ref{LEDAsample}. 
Histograms showing the distribution of Hubble type and absolute 
magnitude (\cf\fig\ref{HtpHmabsmabsVSt}) reveal that the subsample 
restricted by the SDSS DR2 coverage is apparently unbiased in this 
sense. There is a lack of very bright galaxies 
($\mabs<-21.9$ B-mag) due to the small number statistics. 
Plotting absolute magnitude \mabs with Hubble Type $T$ reveals 
--although with a lot of scatter-- a mild trend of earlier Hubble 
type galaxies being on average brighter.
Only three galaxies (NGC\,4273, NGC\,4480, NGC\,4496A) are possible 
members of the Virgo Cluster according to their VCC numbers 
by \cite{vcc}. 
In this sense our sample is a faithful representation of 
the local spiral galaxy population providing a volume limited 
sample ($D\ltsim 46\,$Mpc) of late-type, intermediate inclined to
 face-on nearby 
galaxies brighter than $-18.4$ B-mag. 
The Appendix \ref{atlas} and \ref{rejected} show colour images 
for all galaxies.
%
\begin{table*}
\begin{center}
{\normalsize
\begin{tabular}{ l  c c c l  c  c  c  c r r }
\hline
\rule[+0.4cm]{0mm}{0.0cm}
Galaxy
&\multicolumn{1}{c}{RA}
&\multicolumn{1}{c}{DEC}
&\multicolumn{1}{c}{RC3}
&\multicolumn{1}{c}{LEDA}
&\multicolumn{1}{c}{$T$}
&\multicolumn{1}{c}{\mabs}
&\multicolumn{1}{c}{Diam.}
&\multicolumn{1}{c}{$v_{\rm vir}$}
&\multicolumn{1}{c}{Dist.} 
&\multicolumn{1}{c}{$v_{\rm rot}$}
\\[+0.1cm]
&\multicolumn{2}{c}{(J2000.0)}
&\multicolumn{1}{c}{type}
&\multicolumn{1}{c}{type} 
&
&\multicolumn{1}{c}{[\ B-mag\ ]}
&\multicolumn{1}{c}{[\ \arcmin\ ]}
&\multicolumn{1}{c}{[\ \kms]}
&\multicolumn{1}{c}{[Mpc]} 
&\multicolumn{1}{c}{[\ \kms]} \\ 
\rule[-3mm]{0mm}{5mm}{\scriptsize{\raisebox{-0.7ex}{\it (1)}}}
&{\scriptsize{\raisebox{-0.7ex}{\it (2)}}}
&{\scriptsize{\raisebox{-0.7ex}{\it (3)}}}
&\hspace*{0.1cm}{\scriptsize{\raisebox{-0.7ex}{\it (4)}}}
&{\scriptsize{\raisebox{-0.7ex}{\it (5)}}}
&{\scriptsize{\raisebox{-0.7ex}{\it (6)}}} 
&{\scriptsize{\raisebox{-0.7ex}{\it (7)}}}
&{\scriptsize{\raisebox{-0.7ex}{\it (8)}}}
&{\scriptsize{\raisebox{-0.7ex}{\it (9)}}} 
&{\scriptsize{\raisebox{-0.7ex}{\it (10)}}} 
&{\scriptsize{\raisebox{-0.7ex}{\it (11)}}} 
\\
\hline\hline \\[-0.2cm]
NGC\,0428$^\star$    &01\,12\,55.7&$+$00\,58\,54&\texttt{.SXS9..}& SBd &   8.2 &$-$19.39  &3.6&1118  &16.0  &  90\\
NGC\,0450            &01\,15\,30.8&$-$00\,51\,38&\texttt{.SXS6*.}& SBc &   6.0 &$-$19.63  &3.0&1712  &24.5  & 118\\
PGC\,006667$^{\dagger}$&01\,49\,10.2&$-$10\,03\,45&\texttt{.SBS7..}& SBcd&   7.1 &$-$18.77  &2.9&1887  &27.0  & 129\\
NGC\,0701            &01\,51\,03.8&$-$09\,42\,09&\texttt{.SBT5..}& SBc &   5.1 &$-$19.74  &2.5&1729  &24.7  & 140\\
NGC\,0853            &02\,11\,41.5&$-$09\,18\,17&\texttt{.S..9P?}& Sd  &   8.3 &$-$18.42  &1.5&1405  &20.1  & \ldots\\
NGC\,0941            &02\,28\,27.9&$-$01\,09\,06&\texttt{.SXT5..}& SBc &   5.4 &$-$19.02  &2.5&1535  &21.9  &  91\\
NGC\,0988$^\star$    &02\,35\,27.4&$-$09\,21\,17&\texttt{.SBS6*.}& SBc &   5.8 &$-$20.72  &4.5&1392  &19.9  & 135\\
UGC\,02081           &02\,36\,00.9&$+$00\,25\,12&\texttt{.SXS6..}& Sc  &   5.8 &$-$18.56  &1.8&2549  &36.4  & 100\\
NGC\,1042            &02\,40\,23.9&$-$08\,25\,58&\texttt{.SXT6..}& SBc &   6.1 &$-$19.83  &4.4&1264  &18.1  &  50\\
NGC\,1068$^{\dagger}$  &02\,42\,40.8&$-$00\,00\,48&\texttt{RSAT3..}& Sb  &   3.0 &$-$21.69  &7.6&1068  &15.3  & 310\\
NGC\,1084            &02\,45\,59.7&$-$07\,34\,37&\texttt{.SAS5..}& Sc  &   5.1 &$-$20.20  &3.2&1299  &18.6  & 173\\
NGC\,1087            &02\,46\,25.2&$-$00\,29\,55&\texttt{.SXT5..}& SBc &   5.3 &$-$20.46  &3.7&1443  &20.6  & 125\\
NGC\,1299            &03\,20\,09.4&$-$06\,15\,50&\texttt{.SBT3?.}& Sb  &   3.2 &$-$19.25  &1.2&2197  &31.4  & \ldots\\
NGC\,2543            &08\,12\,58.0&$+$36\,15\,16&\texttt{.SBS3..}& SBb &   3.1 &$-$20.66  &2.5&2590  &37.0  & 158\\
NGC\,2541            &08\,14\,40.1&$+$49\,03\,41&\texttt{.SAS6..}& SBc &   6.2 &$-$18.50  &5.8& 734  &10.5  &  97\\
UGC\,04393           &08\,26\,04.3&$+$45\,58\,03&\texttt{.SB?...}& SBc &   5.5 &$-$19.49  &2.2&2290  &32.7  &  70\\
NGC\,2684            &08\,54\,54.1&$+$49\,09\,37&\texttt{.S?....}& Sd  &   7.8 &$-$19.76  &1.0&3043  &43.5  &  99\\
UGC\,04684$^\star$   &08\,56\,40.7&$+$00\,22\,30&\texttt{.SAT8*.}& Sd  &   7.9 &$-$18.86  &1.4&2477  &35.4  & 108\\
NGC\,2701          &08\,59\,05.9&$+$53\,46\,16&\texttt{.SXT5*.}& SBc &   5.1 &$-$20.36  &2.1&2528  &36.1  & 168\\
NGC\,2712$^\star$  &08\,59\,30.5&$+$44\,54\,50&\texttt{.SBR3*.}& SBb &   3.2 &$-$20.02  &2.9&1987  &28.4  & 171\\
NGC\,2776            &09\,12\,14.3&$+$44\,57\,17&\texttt{.SXT5..}& SBc &   5.1 &$-$21.04  &2.9&2796  &39.9  & 118\\
NGC\,2967            &09\,42\,03.3&$+$00\,20\,11&\texttt{.SAS5..}& Sc  &   4.8 &$-$20.26  &2.6&1858  &26.5  & 172\\
NGC\,3023$^\star$    &09\,49\,52.6&$+$00\,37\,05&\texttt{.SXS5P*}& SBc &   5.4 &$-$19.20  &2.6&1852  &26.5  &  67\\
NGC\,3055            &09\,55\,17.9&$+$04\,16\,11&\texttt{.SXS5..}& SBc &   5.0 &$-$19.90  &2.1&1816  &25.9  & 155\\
NGC\,3246            &10\,26\,41.8&$+$03\,51\,43&\texttt{.SX.8..}& SBd &   8.0 &$-$18.91  &2.2&2138  &30.5  & 117\\
NGC\,3259            &10\,32\,34.8&$+$65\,02\,28&\texttt{.SXT4*.}& SBbc&   4.0 &$-$19.84  &2.1&1929  &27.6  & 119\\
NGC\,3310            &10\,38\,45.8&$+$53\,30\,12&\texttt{.SXR4P.}& SBbc&   4.0 &$-$20.25  &2.8&1208  &17.3  & 280\\
NGC\,3359            &10\,46\,36.3&$+$63\,13\,28&\texttt{.SBT5..}& SBc &   5.0 &$-$20.42  &7.2&1262  &18.0  & 139\\
PGC\,032356$^{\dagger\star}$&10\,49\,04.6&$+$52\,19\,57&\texttt{.S?....}& Sc  &   6.0 &$-$18.46  &0.8&2679  &38.3  & \ldots\\
NGC\,3423            &10\,51\,14.3&$+$05\,50\,24&\texttt{.SAS6..}& Sc  &   6.0 &$-$19.54  &3.9&1032  &14.7  & 122\\
NGC\,3488            &11\,01\,23.6&$+$57\,40\,39&\texttt{.SBS5*.}& SBc &   5.1 &$-$18.80  &1.6&3226  &46.1  & 132\\
UGC\,06162$^\star$   &11\,06\,54.7&$+$51\,12\,12&\texttt{.S..7..}& SBcd&   6.7 &$-$19.08  &2.3&2426  &34.7  & 105\\
NGC\,3583            &11\,14\,11.0&$+$48\,19\,06&\texttt{.SBS3..}& SBb &   3.2 &$-$19.57  &2.3&2347  &33.5  & 194\\
NGC\,3589            &11\,15\,13.2&$+$60\,42\,01&\texttt{.S..7*.}& SBcd&   7.0 &$-$18.49  &1.5&2217  &31.7  &  75\\
UGC\,06309           &11\,17\,46.4&$+$51\,28\,36&\texttt{.SB?...}& SBc &   5.0 &$-$19.63  &1.3&3097  &44.2  & 124\\
NGC\,3631            &11\,21\,02.4&$+$53\,10\,08&\texttt{.SAS5..}& Sc  &   5.1 &$-$20.62  &4.9&1388  &19.8  &  82\\
NGC\,3642            &11\,22\,18.0&$+$59\,04\,27&\texttt{.SAR4*.}& Sbc &   4.0 &$-$20.60  &5.5&1831  &26.2  &  45\\
UGC\,06518           &11\,32\,20.4&$+$53\,54\,17&\texttt{.......}& Sbc &   4.3 &$-$18.94  &1.0&3044  &43.5  &  84\\
NGC\,3756            &11\,36\,48.3&$+$54\,17\,37&\texttt{.SXT4..}& SBbc&   4.3 &$-$19.89  &3.8&1525  &21.8  & 153\\
NGC\,3888            &11\,47\,34.7&$+$55\,58\,00&\texttt{.SXT5..}& SBc &   5.3 &$-$20.43  &1.8&2648  &37.8  & 191\\
NGC\,3893            &11\,48\,38.4&$+$48\,42\,34&\texttt{.SXT5*.}& SBc &   4.8 &$-$20.33  &4.1&1193  &17.0  & 175\\
UGC\,06903           &11\,55\,36.9&$+$01\,14\,15&\texttt{.SBS6..}& SBc &   5.8 &$-$18.80  &2.5&1916  &27.4  & 158\\
NGC\,3982            &11\,56\,28.3&$+$55\,07\,29&\texttt{.SXR3*.}& SBb &   3.2 &$-$19.54  &2.3&1351  &19.3  & 193\\
NGC\,3992$^{\dagger}$  &11\,57\,35.9&$+$53\,22\,35&\texttt{.SBT4..}& SBbc&   3.8 &$-$20.98  &6.9&1286  &18.4  & 285\\
NGC\,4030            &12\,00\,23.4&$-$01\,06\,03&\texttt{.SAS4..}& Sbc &   4.1 &$-$20.27  &3.9&1476  &21.1  & 231\\
NGC\,4041            &12\,02\,12.2&$+$62\,08\,14&\texttt{.SAT4*.}& Sbc &   4.3 &$-$19.92  &2.7&1486  &21.2  & 284\\
NGC\,4102            &12\,06\,23.1&$+$52\,42\,39&\texttt{.SXS3\$.}& SBb &   3.0 &$-$19.38  &3.1&1082  &15.5  & 169\\
NGC\,4108            &12\,06\,44.0&$+$67\,09\,44&\texttt{PSA.5*.}& Sc  &   5.0 &$-$20.13  &1.7&2828  &40.4  & 242\\
\hline
\end{tabular}
}
\caption[]{Global parameters of the LEDA--SDSS\,DR2 subsample: 
{\scriptsize{\it (1)}} Principal name in LEDA, 
{\scriptsize{\it (2)}} right ascension, and 
{\scriptsize{\it (3)}} declination,  
{\scriptsize{\it (4)}} RC3 \cite{rc3} Hubble-type, and 
{\scriptsize{\it (5)}} LEDA Hubble-type,
{\scriptsize{\it (6)}} coded LEDA Hubble parameter $T$, 
{\scriptsize{\it (7)}} absolute B band magnitude, corrected for 
galactic plus internal extinction, and k-corrected,
{\scriptsize{\it (8)}} apparent diameter, defined by the isophote at the 
brightness of 25 B-\magsqarcsec, 
{\scriptsize{\it (9)}} heliocentric radial velocities corrected for the 
Local Group infall onto Virgo, 
{\scriptsize{\it (10)}} estimated distance according to the Hubble relation 
with the Hubble constant of $H_{0}\!=\!70$ km s$^{-1}$Mpc$^{-1}$, 
{\scriptsize{\it (11)}} weighted average of the measurements maximum rotation
velocity from radio (\hi) and optical rotation curves (H$_{\alpha}$).
\label{LEDAsample} }
\end{center}
\end{table*}
\addtocounter{table}{-1}
\begin{table*}
\begin{center}
{\normalsize
\begin{tabular}{ l  c c c l  c  c  c  c r r }
\hline
\rule[+0.4cm]{0mm}{0.0cm}
Galaxy
&\multicolumn{1}{c}{RA}
&\multicolumn{1}{c}{DEC}
&\multicolumn{1}{c}{RC3}
&\multicolumn{1}{c}{LEDA}
&\multicolumn{1}{c}{$T$}
&\multicolumn{1}{c}{\mabs}
&\multicolumn{1}{c}{Diam.}
&\multicolumn{1}{c}{$v_{\rm vir}$}
&\multicolumn{1}{c}{Dist.} 
&\multicolumn{1}{c}{$v_{\rm rot}$}
\\[+0.1cm]
&\multicolumn{2}{c}{(J2000.0)}
&\multicolumn{1}{c}{type}
&\multicolumn{1}{c}{type} 
&
&\multicolumn{1}{c}{[\ B-mag\ ]}
&\multicolumn{1}{c}{[\ \arcmin\ ]}
&\multicolumn{1}{c}{[\ \kms]}
&\multicolumn{1}{c}{[Mpc]} 
&\multicolumn{1}{c}{[\ \kms]} \\ 
\rule[-3mm]{0mm}{5mm}{\scriptsize{\raisebox{-0.7ex}{\it (1)}}}
&{\scriptsize{\raisebox{-0.7ex}{\it (2)}}}
&{\scriptsize{\raisebox{-0.7ex}{\it (3)}}}
&\hspace*{0.1cm}{\scriptsize{\raisebox{-0.7ex}{\it (4)}}}
&{\scriptsize{\raisebox{-0.7ex}{\it (5)}}}
&{\scriptsize{\raisebox{-0.7ex}{\it (6)}}} 
&{\scriptsize{\raisebox{-0.7ex}{\it (7)}}}
&{\scriptsize{\raisebox{-0.7ex}{\it (8)}}}
&{\scriptsize{\raisebox{-0.7ex}{\it (9)}}} 
&{\scriptsize{\raisebox{-0.7ex}{\it (10)}}} 
&{\scriptsize{\raisebox{-0.7ex}{\it (11)}}} 
\\
\hline\hline \\[-0.2cm]
NGC\,4108B           &12\,07\,11.3&$+$67\,14\,10&\texttt{.SXS7P? }& SBcd&   7.0 &$-$18.46  &1.3&2840  &40.6  & 212\\
NGC\,4116$^\star$    &12\,07\,37.0&$+$02\,41\,29&\texttt{.SBT8.. }&SBd &   7.5 &$-$19.32  &3.5&1346  &19.2   &104\\
NGC\,4123            &12\,08\,11.1&$+$02\,52\,42&\texttt{.SBR5.. }&SBc &   4.8 &$-$19.55  &4.1&1364  &19.5   &134\\
NGC\,4210            &12\,15\,15.9&$+$65\,59\,08&\texttt{.SBR3.. }&SBb &   3.0 &$-$19.89  &2.0&3000  &42.9   &202\\
NGC\,4273            &12\,19\,56.2&$+$05\,20\,36&\texttt{.SBS5.. }&SBc &   5.1 &$-$20.41  &2.1&2435  &34.8   &187\\
NGC\,4480            &12\,30\,26.8&$+$04\,14\,47&\texttt{.SXS5.. }&SBc &   5.1 &$-$20.14  &2.1&2494  &35.6   &169\\
NGC\,4496A$^\star$   &12\,31\,39.5&$+$03\,56\,22&\texttt{.SBT9.. }&SBd &   8.2 &$-$20.25  &3.8&1780  &25.4   & 95\\
NGC\,4517A           &12\,32\,28.1&$+$00\,23\,25&\texttt{.SBT8*. }&SBd &   7.9 &$-$19.18  &3.8&1562  &22.3   & 73\\
UGC\,07700           &12\,32\,32.8&$+$63\,52\,38&\texttt{.SBS8.. }&SBd &   7.8 &$-$18.50  &1.8&3239  &46.3   & 94\\
NGC\,4545            &12\,34\,34.2&$+$63\,31\,30&\texttt{.SBS6*. }&SBc &   5.9 &$-$20.49  &2.5&3000  &42.9   &136\\
NGC\,4653            &12\,43\,50.9&$-$00\,33\,41&\texttt{.SXT6.. }&SBc &   6.1 &$-$20.37  &3.0&2658  &38.0   &181\\
NGC\,4668            &12\,45\,32.0&$-$00\,32\,09&\texttt{.SBS7*. }&SBcd&   7.1 &$-$18.88  &1.5&1654  &23.6   & 63\\
UGC\,08041           &12\,55\,12.7&$+$00\,07\,00&\texttt{.SBS7.. }&SBcd&   6.8 &$-$18.62  &3.2&1376  &19.7   & 93\\
UGC\,08084           &12\,58\,22.4&$+$02\,47\,33&\texttt{.SBS8.. }&SBd &   8.1 &$-$18.61  &1.3&2824  &40.3   & 90\\
NGC\,4900$^\star$    &13\,00\,39.2&$+$02\,30\,02&\texttt{.SBT5.. }&SBc &   5.2 &$-$19.01  &2.2&1020  &14.6   &185\\
NGC\,4904            &13\,00\,58.6&$-$00\,01\,38&\texttt{.SBS6.. }&SBc &   5.7 &$-$18.69  &2.1&1213  &17.3   &122\\
UGC\,08237           &13\,08\,54.5&$+$62\,18\,23&\texttt{PSB.3*. }&SBb &   3.2 &$-$19.61  &0.9&3120  &44.6   &\ldots\\
NGC\,5147            &13\,26\,19.2&$+$02\,06\,04&\texttt{.SBS8.. }&SBd &   7.9 &$-$18.77  &1.8&1154  &16.5   &118\\
NGC\,5218$^\star$    &13\,32\,10.4&$+$62\,46\,04&\texttt{.SBS3\$P}&SBb &   3.1 &$-$20.46  &1.9&3154  &45.1   &243\\
UGC\,08658$^{\dagger}$  &13\,40\,39.9&$+$54\,19\,59&\texttt{.SXT5.. }&SBc &   5.1 &$-$19.91  &2.6&2285  &32.6   &126\\
NGC\,5300            &13\,48\,16.1&$+$03\,57\,03&\texttt{.SXR5.. }&SBc &   5.0 &$-$18.93  &3.5&1246  &17.8   &124\\
NGC\,5334            &13\,52\,54.4&$-$01\,06\,51&\texttt{.SBT5*. }&SBc &   5.0 &$-$19.22  &3.9&1433  &20.5   &145\\
NGC\,5376            &13\,55\,16.1&$+$59\,30\,24&\texttt{.SXR3\$.}&SBb &   3.0 &$-$20.00  &2.1&2293  &32.8   &216\\
NGC\,5364$^\star$    &13\,56\,11.9&$+$05\,00\,55&\texttt{.SAT4P. }&Sbc &   4.4 &$-$20.53  &6.0&1322  &18.9   &157\\
NGC\,5430            &14\,00\,45.8&$+$59\,19\,43&\texttt{.SBS3.. }&SBb &   3.1 &$-$20.86  &2.2&3238  &46.3   &195\\
NGC\,5480            &14\,06\,21.5&$+$50\,43\,32&\texttt{.SAS5*. }&Sc  &   5.0 &$-$19.72  &1.7&2119  &30.3   &138\\
NGC\,5584            &14\,22\,23.8&$-$00\,23\,15&\texttt{.SXT6.. }&SBc &   6.1 &$-$19.80  &3.2&1702  &24.3   &136\\
UGC\,09215$^\star$   &14\,23\,27.3&$+$01\,43\,32&\texttt{.SBS7.. }&SBcd&   6.6 &$-$18.81  &2.2&1462  &20.9   &102\\
NGC\,5624            &14\,26\,35.3&$+$51\,35\,03&\texttt{.S?.... }&Sc  &   5.0 &$-$18.69  &1.1&2186  &31.2   & 67\\
NGC\,5660            &14\,29\,49.8&$+$49\,37\,22&\texttt{.SXT5.. }&SBc &   5.1 &$-$20.62  &2.8&2585  &36.9   &138\\
NGC\,5667            &14\,30\,22.9&$+$59\,28\,11&\texttt{.S..6*P }&SBc &   5.8 &$-$19.69  &1.7&2222  &31.7   &119\\
NGC\,5668            &14\,33\,24.4&$+$04\,27\,02&\texttt{.SAS7.. }&Scd &   6.7 &$-$19.65  &2.8&1672  &23.9   & 87\\
NGC\,5693            &14\,36\,11.2&$+$48\,35\,04&\texttt{.SBT7.. }&SBcd&   6.8 &$-$18.64  &1.9&2537  &36.2   & 39\\
NGC\,5713            &14\,40\,11.4&$-$00\,17\,25&\texttt{.SXT4P. }&SBbc&   4.1 &$-$20.50  &2.8&1958  &28.0   &135\\
NGC\,5768            &14\,52\,07.9&$-$02\,31\,47&\texttt{.SAT5*. }&Sc  &   5.0 &$-$19.24  &1.6&2018  &28.8   &122\\
 IC\,1067       &14\,53\,05.2&$+$03\,19\,54&\texttt{.SBS3.. }&SBb &   3.0 &$-$18.97  &2.0&1665  &23.8   &158\\
NGC\,5774            &14\,53\,42.6&$+$03\,35\,00&\texttt{.SXT7.. }&SBcd&   6.9 &$-$19.09  &2.8&1655  &23.6   & 91\\
NGC\,5806            &15\,00\,00.4&$+$01\,53\,29&\texttt{.SXS3.. }&SBb &   3.3 &$-$19.74  &3.0&1440  &20.6   &182\\
NGC\,5850            &15\,07\,07.7&$+$01\,32\,40&\texttt{.SBR3.. }&SBb &   3.0 &$-$21.43  &4.3&2637  &37.7   &133\\
UGC\,09741$^{\dagger}$  &15\,08\,33.5&$+$52\,17\,46&\texttt{.S?.... }&Sc  &   6.0 &$-$18.54  &0.9&2735  &39.1   &\ldots\\
UGC\,09837           &15\,23\,51.7&$+$58\,03\,11&\texttt{.SXS5.. }&Sc  &   5.1 &$-$19.45  &1.8&2938  &42.0   &221\\
NGC\,5937            &15\,30\,46.1&$-$02\,49\,46&\texttt{PSXT3P. }&Sb  &   3.2 &$-$20.90  &1.9&2870  &41.0   &202\\
 IC\,1125       &15\,33\,05.6&$-$01\,37\,42&\texttt{.S..8*. }&SBd &   7.7 &$-$20.05  &1.7&2868  &41.0   &106\\
 IC\,1158       &16\,01\,34.1&$+$01\,42\,28&\texttt{.SXR5*. }&SBc &   5.3 &$-$19.36  &2.2&2018  &28.8   &125\\
NGC\,6070            &16\,09\,58.9&$+$00\,42\,34&\texttt{.SAS6.. }&Sc  &   6.0 &$-$21.04  &3.4&2085  &29.8   &215\\
NGC\,6155            &16\,26\,08.3&$+$48\,22\,01&\texttt{.S?.... }&Sd  &   7.8 &$-$19.45  &1.3&2684  &38.3   &105\\
UGC\,10721           &17\,08\,25.6&$+$25\,31\,03&\texttt{.S..6?. }&Sc  &   5.8 &$-$19.68  &1.2&3118  &44.5   &144\\
NGC\,7437            &22\,58\,10.0&$+$14\,18\,32&\texttt{.SXT7.. }&Scd &   7.1 &$-$18.88  &1.8&2190  &31.3   &175\\
NGC\,7606            &23\,19\,04.8&$-$08\,29\,06&\texttt{.SAS3.. }&Sb  &   3.2 &$-$21.38  &4.3&2187  &31.2   &291\\
UGC\,12709           &23\,37\,24.0&$+$00\,23\,27&\texttt{.SXS9.. }&Sd  &   8.3 &$-$19.05  &3.0&2672  &38.2   & 70\\
\hline
\end{tabular}
}
\caption[]{(continued): Global parameters of the LEDA--SDSS\,DR2 subsample 
\newline 
($^\star$): Galaxy removed from further analysis (see text). 
\newline ($\dagger$): Galaxy with relevant alternative name: 
NGC\,1068 $\equiv$ M\,77; 
NGC\,3992 $\equiv$ M\,109; 
PGC\,006667 $\equiv$ UGCA\,021; 
PGC\,032356 $\equiv$ UGCA\,219;
UGC\,08658 $\equiv$ Holmberg\,V; 
UGC\,09741 $\equiv$ NGC\,5875A
}
\end{center}
\end{table*}
%
%
\subsection{Environment}
\label{environment}
To characterise the environment we counted neighbours around our 
sample galaxies using the SDSS database from a more recent 
Data Release \citep[DR3,][]{sloandr3}.
According to the distance, derived from the infall corrected velocity, 
we sum all the DR3 galaxies within a projected radius of 1\,Mpc that 
satisfy the following criteria: the difference in velocity with respect 
to the targeted galaxy is less than $350\,$\kms \cite[]{nogsample} 
and their absolute magnitude is brighter than $\mabs<-16$ $r^{\prime}$-mag. 
It is important to note, however, that the density of galaxies could 
be affected by the position of the galaxy within the survey. 
If the targeted galaxy is close to the edge of the survey or to a 
region where the redshift completeness (fraction of photometrically 
detected galaxies with redshifts) is not high, the number of 
neighbours could be underestimated. 
To eliminate such a problem we have estimated whether the distance 
of our galaxy to the edge of the survey or to a redshift incomplete 
region is less than 1\,Mpc. 
If this is the case this galaxy is not used in the analysis 
of the effect of the density on the galaxy profile type.
%
\section{Data and profile extraction}
\label{datapp}
\subsection{The data}
\label{data}
The Sloan Digital Sky Survey (SDSS) \cite[]{sloan} 
will map one-quarter ($\sim 10^4\sqdeg$) of the entire sky mainly 
around the north galactic cap (above Galactic latitude $\sim30\deg$) 
in five bands, $u^{\prime}$, $g^{\prime}$, $r^{\prime}$, $i^{\prime}$ 
and $z^{\prime}$ \citep{phot1,phot2}. SDSS imaging are obtained
using a drift-scanning mosaic CCD camera \citep{sloancam} with a pixel 
size of $0.396\arcsec$. 
We downloaded the $g^{\prime}$ and $r^{\prime}$ band images (which have a median PSF 
of $\sim\!1.4$\arcsec in $r^{\prime}$) of the night sky ("corrected frames") 
in fits format using the Data Archive Server\footnote{
\ttfamily http://www.sdss.org/dr2/access/index.html\#DAS}. 
The $u^{\prime}$, $i^{\prime}$ and $z^{\prime}$ bands
images are less sensitive 
and, consequently, less useful to study the profile towards the outer disk. 
The corrected frames, having been bias subtracted, flat-fielded, 
and purged of bright stars are stored at SDSS in integer format 
 to save disk space. According to the 
SDSS helpdesk the pixel values get randomised appropriately before 
being rounded to make sure that the statistics of the background 
counts are what they should be. 
An additional offset (SOFTBIAS) of 1000 counts is added to each 
pixel to avoid negative pixel values and should be subtracted together 
with the sky value (as described in \sec\ref{skysub}).
After inspecting all images for the 98 galaxies we had to remove
13 galaxies (marked with a $\star$ in Table \ref{LEDAsample}) from 
further analysis (\cf Appendix \ref{rejected}). 
For four galaxies (NGC\,0988, NGC\,4900, UGC\,04684, and UGC\,06162) 
there are one (or more) very bright foreground stars too close to the 
galaxy preventing any useful surface brightness measurement. 
Another six galaxies (NGC\,0428, NGC\,2712, NGC\,4116, NGC\,4496A, 
NGC\,5364, and UGC\,09215) are very close to the border of the SDSS 
field (more than $1/3$ of the galaxy is off the image). To avoid any 
problems while trying to recover the galaxies with a mosaic out of 
two or even three images we discarded them. 
Two peculiar galaxies (NGC\,3023 and NGC\,5218) are removed since 
they are clearly interacting systems. They are gravitationally 
highly distorted and in the case of NGC\,5218 connected by an intergalactic 
bridge to its companion.
One galaxy (PGC\,032356 $\equiv$ UGCA\,219) turns out to be a BCD 
galaxy and probably even a double system. 
This leave 85 galaxies for the final analysis. 
%
\subsection{Photometric calibration}
\label{phot}
The photometric calibration is done using the $aa$, $kk$, and $airmass$ 
coefficients (the photometric zeropoint, the extinction term
and the airmass) out of the associated TsField table file for 
each image.
From this we calculated\footnote{\ttfamily 
http://www.sdss.org/dr2/algorithms/fluxcal.html} 
our surface brightness zeropoints as:
$-2.5*\left(0.4*\left[ aa + kk * airmass \right]\right) + 
2.5*\log\left(53.907456*0.396^2\right)$
using the pixel scale of $0.396\arcsec$/pixel and the exposure 
time of $53.907456\,s$ for each SDSS pixel. 
Our magnitudes are the conventional Pogson astronomical 
magnitudes in the SDSS $g^{\prime}$ and $r^{\prime}$ AB system in contrast to
the asinh magnitudes used in the SDSS database. 
The applied zero points are given in \tab\ref{ellskyresults}. 
To transform the SDSS $g^{\prime}$ and $r^{\prime}$ standard-star system 
magnitudes 
into the commonly used Johnson-Cousins $B$, $R$ system, we applied
the transformation equations given in \cite{phot2}: 
$B     = g^{\prime} + 0.47(g^{\prime}-r^{\prime}) + 0.17$
and 
$R     = g^{\prime} - 1.14(g^{\prime}-r^{\prime}) - 0.14$.
For eight galaxies of the final SDSS sample we found aperture 
measurements in the literature. Synthetic aperture measurements
on the SDSS images agree with these previous published values in 
the majority of cases to better than $0.1\,$mag. 
All profiles and surface brightness results, given in Table 
\ref{resultstab}, are the measured values from the uncorrected SDSS 
data. Only at calculating mean values and plotting histograms we 
applied the galactic extinction correction according to 
\cite{schlegel} using 
$\mu_{\rm corrected}\!=\mu_{\rm measured}\!-\!A_{r^{\prime}/g^{\prime}}$. 
The values we used for $A_{r^{\prime}}$ and $A_{g^{\prime}}$ are also shown 
in Table \ref{resultstab}.
No attempt was made to correct the surface brightness measurements for 
internal extinction, since no unique recipe is available to do this. 
In addition, the galaxies studied here are all fairly face-on 
systems ($\!<\!e\!>=\!0.3 \Leftrightarrow i\!\sim\!46\deg$) so the expected 
corrections are only small. Since we are working with a local 
sample ($z\ltsim0.01$) we also do not correct for the cosmological surface 
brightness dimming.
%
%
\subsection{Sky subtraction}
\label{skysub}
\begin{figure}
\hspace*{1.7cm}
\includegraphics[width=5.6cm,angle=0]{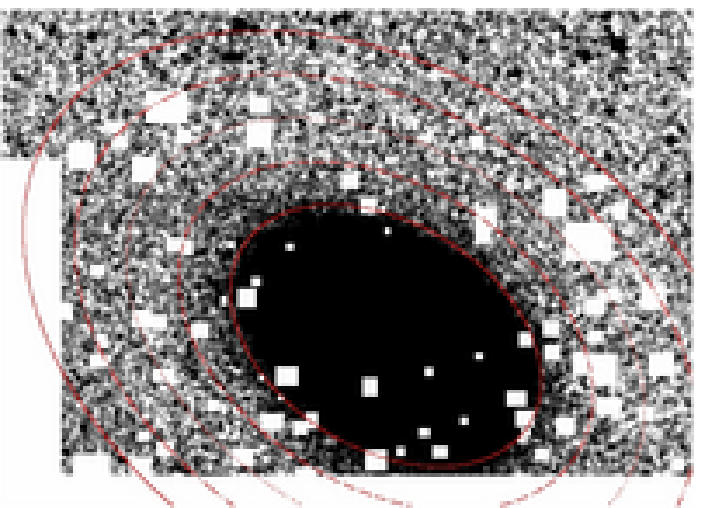} \\
\includegraphics[width=6.1cm,angle=270]{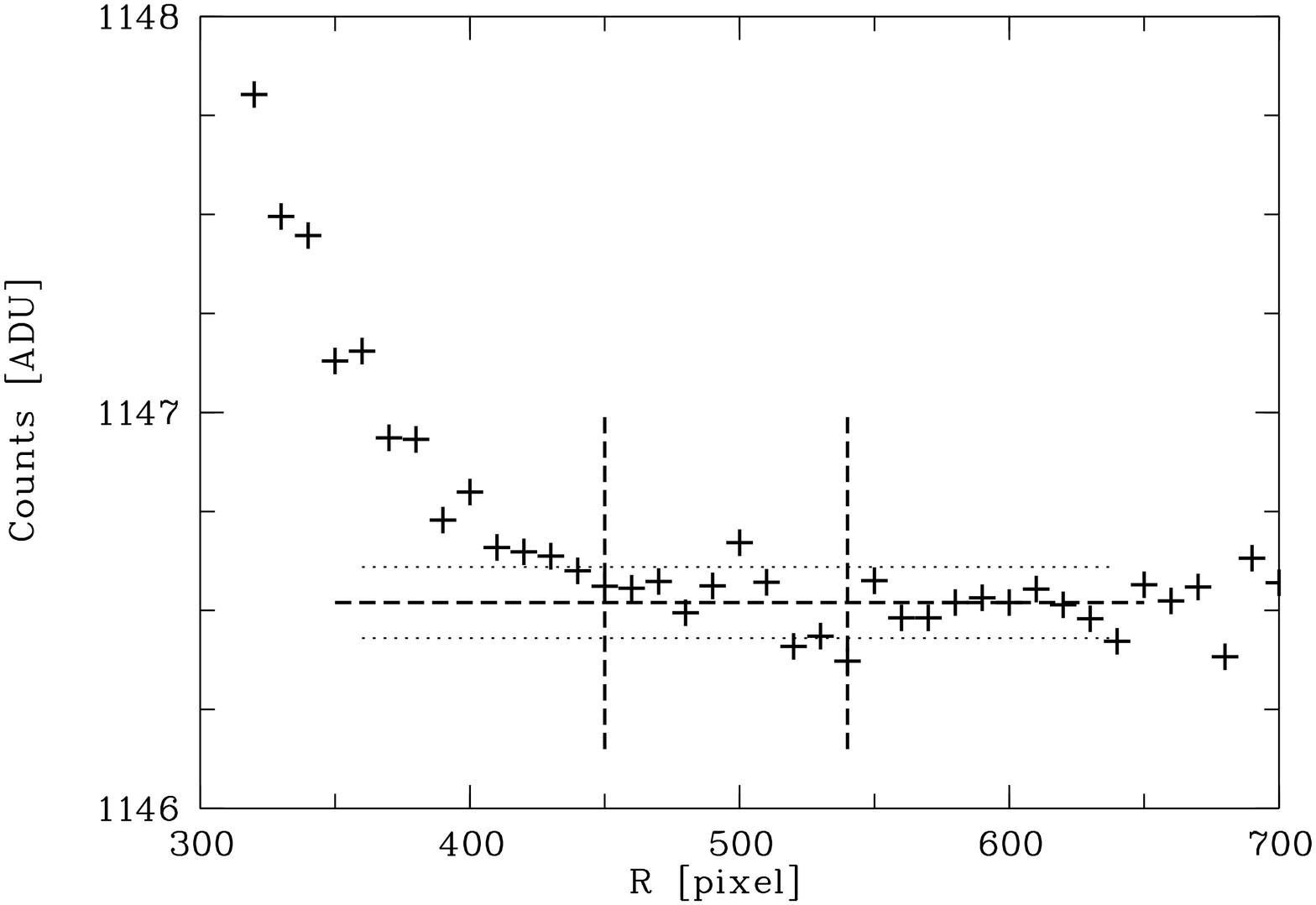}
\includegraphics[width=6.1cm,angle=270]{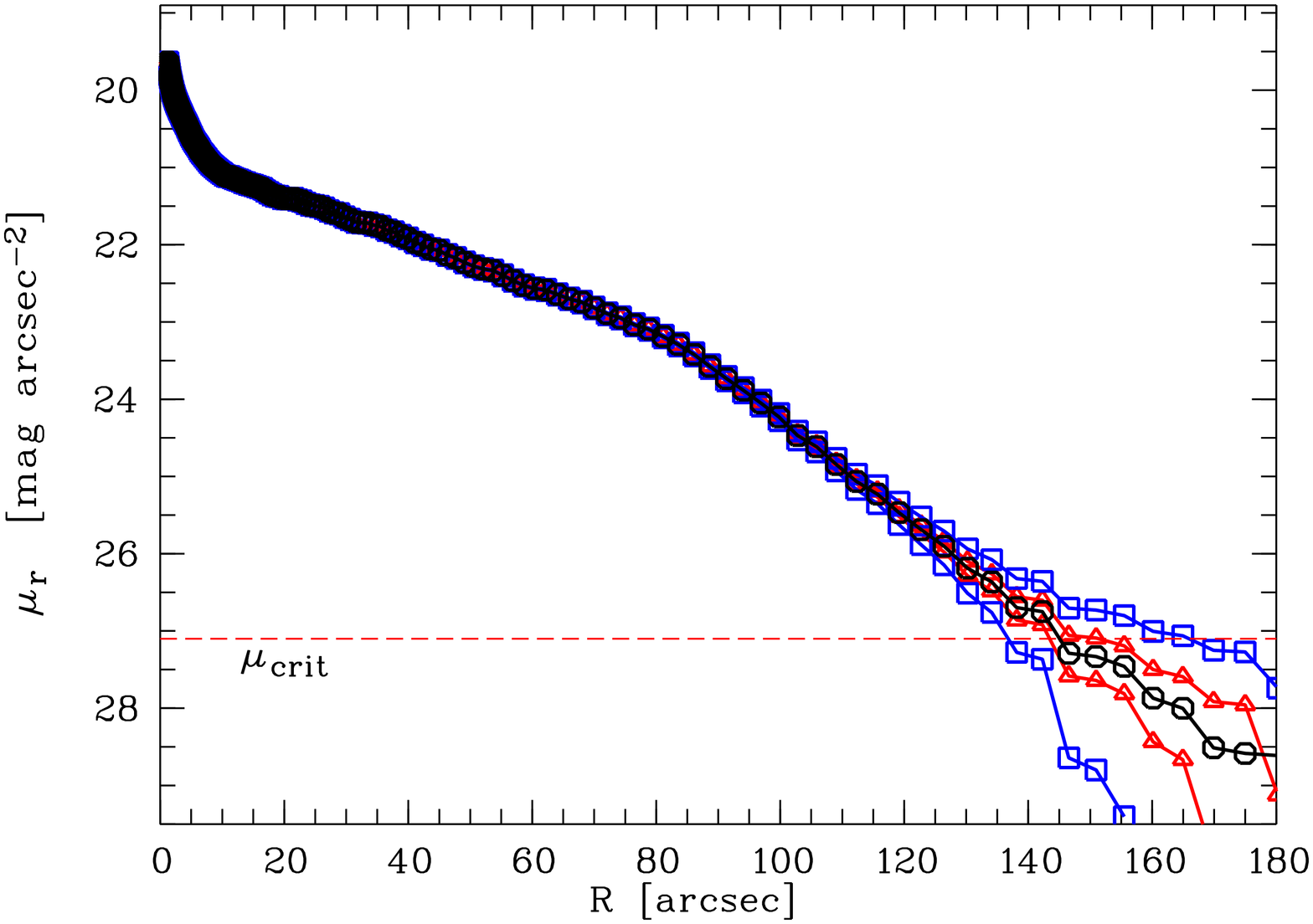}
\caption{Sky obtained by ellipse fitting:  
{\it Upper panel:} $r^{\prime}$-band image of NGC\,5300 exemplary 
overlayed by eight fixed ellipses at every 100 pixels in the range 
300-700 pixels. The white rectangular regions are masked areas. 
{\it Middle panel:} mean isophotal intensity (in counts) of 
the fixed ellipses fitted every 10 pixels. The region between 
the vertical dashed lines is used to determine the sky value 
($1146.52 \pm 0.09$), which is indicated as a horizontal 
dashed line together with the two dotted lines at $\pm 1\sigma$. 
{\it Lower panel:} Final surface brightness profile {\it (circles)} 
overlayed by the four profiles obtained by either over- or 
undersubtracting the sky by $\pm 1\sigma$ {\it (triangles)} or
by $\pm 3\sigma$ {\it (squares)}.
The horizontal dashed line gives the critical surface brightness 
($\mu_{\rm crit}$) up to where we trust the profile i.e. where 
the $\pm 1\sigma$ profiles deviate by more than 0.2\,mag. 
}
\label{N5300skysub}
\end{figure}

The crucial point in using SDSS data (and in general) to study surface 
brightness profiles at very faint levels is the measurement of an 
appropriate sky value. SDSS provides in the header of each image a 
first estimate of such a value. However, this is just a global value
which is in more than half of the cases off by more than $\pm0.2$\,counts 
(\gtsim 0.2\% level) and therefore for our purposes not accurate 
enough. 
We need a more elaborated value depending on the size of the 
galaxy, thus the area of sky covered by the galaxy, and the 
exact position on the chip. This means that for different galaxies 
on the same chip we may derive different sky values. 
In deep surface photometry the quality (flatness and noise) 
of the background around (and 'below') the galaxy determines 
the surface brightness level down to which one 
can trust the profiles.
A common way of measuring the sky is, after the removal of a possible large scale 
gradient (typically of only first order), to manually place several 
small boxes homogeneously around {\it --but outside--} the galaxy, 
avoiding foreground stars or obvious background galaxies, and 
measure the median of the pixel distribution in each box. The best 
sky value is then the mean value of all median measurements.
However, the fact of storing SDSS data as integer not real numbers 
(\cf\sec\ref{data}) hampers the use of median values in small boxes. 
Only by using a large number of pixels (or an inappropriate high 
number of boxes dealing with about 100 galaxies) one is able to recover 
an accurate estimate of the mean value of the background pixel 
distribution. 
Therefore we decided to use two independent methods to determine 
the sky value. First, we strategically placed 2-5 large, rectangular
sky boxes (of roughly 100k-200k pixels) as close as possible to the 
galaxy avoiding bright foreground stars or obvious structure in 
the background (\eg from halos of very bright foreground stars) 
while trying to match the brightness of the sky around the galaxy. 
Within each box we derived the mean sky after three, $3\sigma$, clipping 
iterations to remove the unavoidable contamination by faint foreground 
stars. The mean value of all boxes serves as the first estimate of
the sky. 
For the second method we applied a free ellipse fit to the original 
image. Using the ellipse task in IRAF\footnote{Image Reduction and 
Analysis Facility (IRAF) \newline \hspace*{0.5cm} http://iraf.noao.edu/} 
we calculated the flux in linear steps of 10 pixels between successive 
ellipses forcing the program to extend the fit well beyond the galaxy
with the fixed ellipticity and PA of the outer disk (see 
\fig\ref{N5300skysub}). By plotting the flux at these ellipses as 
a function of radius it becomes clear at which radial distance the ellipses 
leave the galaxy and enters the background, flat flux (noise), region. After 
visual confirmation that these ellipses are really outside the galaxy, 
by overplotting the ellipses on the image, the mean value and standard 
deviation of the fluxes within the above radius and the radius extended by
 $\sim\!20\%$ are used as the final sky values. 
In most cases the two methods agree well within $\pm 0.15$ counts 
($\sim 0.15\%$ of the sky). Since it is possible to automatise it,
we choose the ellipse method to fix our final sky values. They are 
listed in \tab\ref{ellskyresults} together with a limiting surface 
brightness ($\mu_{\rm lim}$) due to a $3\sigma$-sky error, which 
is used to constrain the outer boundary for the exponential fits. 
Our sky values can be easily inspected by downloading the public 
SDSS images.  
To specify the error on the profiles from our sky estimate 
(which is much larger than the statistical errors produced 
by \eg photon noise) in more detail we defined a more conservative 
critical surface brightness ($\equiv \mu_{\rm crit}$) up to where 
we really trust the final profile. 
This is placed where the profiles obtained by either over- or 
undersubtracting the sky by $\pm 1\sigma$ start to deviate by more 
than 0.2\,mag, which is at $0.54\,$mag above the limiting surface 
brightness defined above.  
So the typical value for the full sample of this critical surface 
brightness is $\mu_{\rm crit} \!\sim\!27.0\ r^{\prime}$-\magsqarcsec.
This is the limit down to which the slope of the outer profile could 
be traced confidently. 
The pure background noise in the $r^{\prime}$ band images is typically slightly 
higher compared to the $g^{\prime}$ band image, which is expected for images 
taken at new moon. However, the $g^{\prime}$ band images suffer often from an 
electronic noise pattern in addition to the normal patchy background. 
Typically 6-8 more or less pronounced stripes are visible where all 
rows show a slightly different background level, alternating between
being higher or lower. Unfortunately, this structure is not common on 
all images so there is no way to properly remove it by combining 
several images as usually done for fringe patterns. In many cases the galaxies 
are small compared to the strip size and the above pattern is not an issue.
In other cases only small parts of the galaxies  reach  another 
strip and can be masked. However, there are some galaxies extended over
 two or more  stripes. For these systems the sky estimate is less 
reliable because the contribution of each region changes with 
changing size of the ellipse.           
For five galaxies (NGC\,1042, NGC\,1084, NGC\,5480, NGC\,5624, and 
PGC\,006667) we had to remove at first a large scale gradient (from 
top to bottom) in the background of the $r^{\prime}$ band images (for NGC\,1042 
also in the $g^{\prime}$ band) with a linear fit (using IRAFs {\sl imsurfit}).   
%
\subsection{Data quality}
%
\begin{figure}
\includegraphics[width=6.1cm,angle=270]{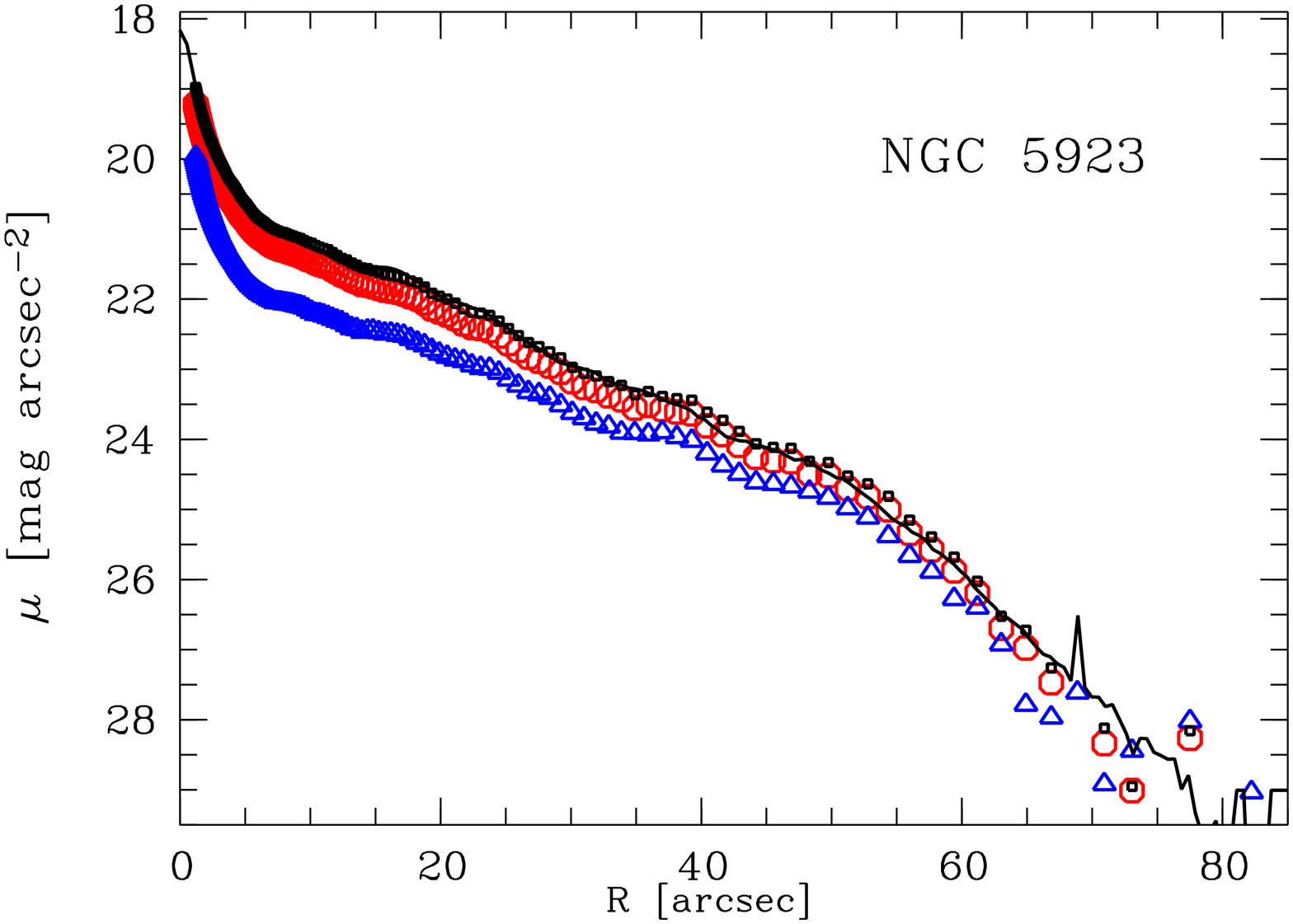}
\includegraphics[width=6.1cm,angle=270]{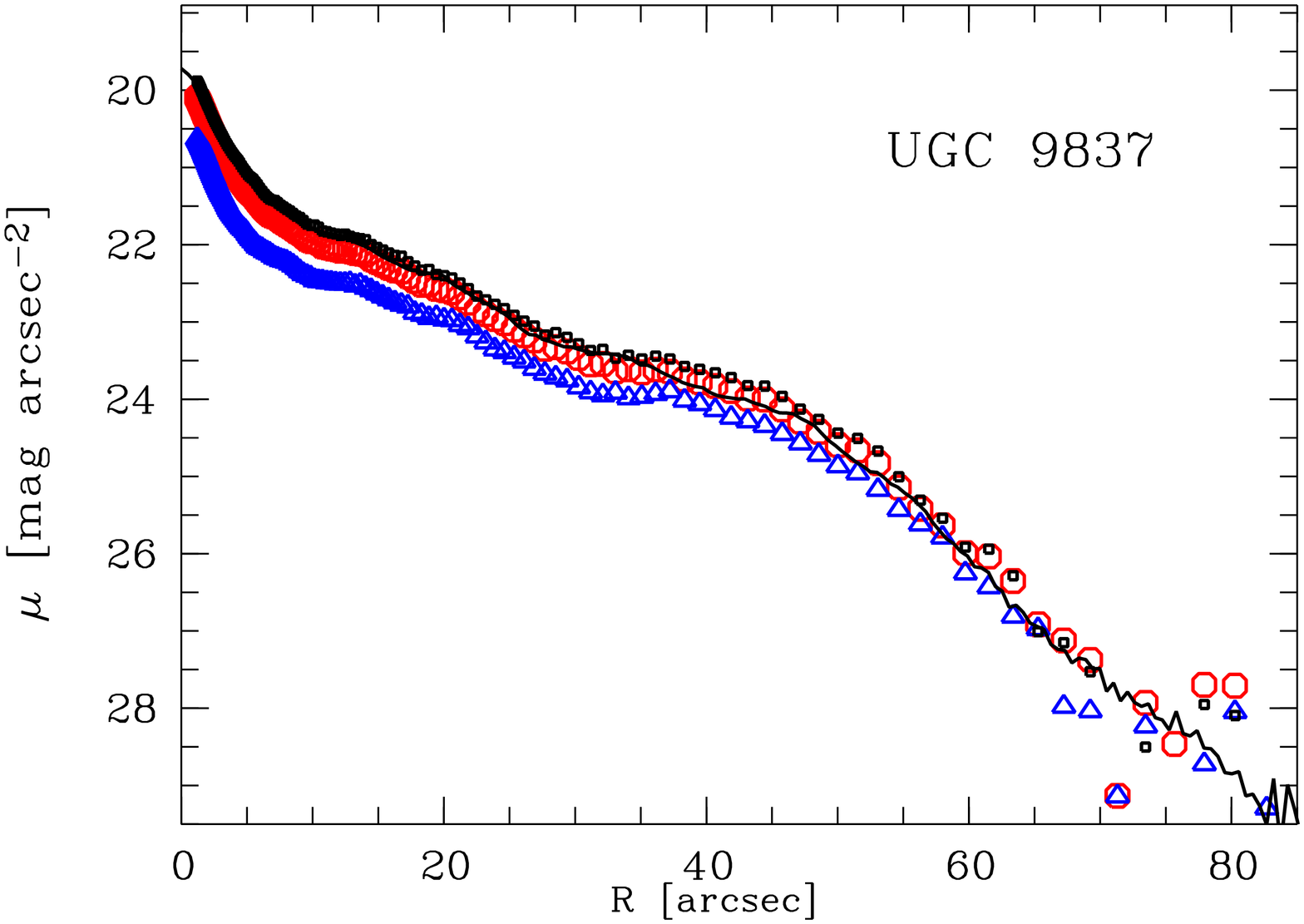}
\caption{SDSS quality check: Azimuthally averaged, radial surface 
brightness profiles obtained from SDSS images {\it ($r^{\prime}$ band with 
big circles; $g^{\prime}$ band with big triangles))} are compared with much deeper imaging 
{\it (thin, continuous line)} presented in \cite{pohlen2002}. 
The {\it small squares} are obtained by using the transformation 
to convert SDSS $g^{\prime}$ and $r^{\prime}$ into standard Johnson $R$ band 
\cite[used by][]{pohlen2002} following \cite{phot2}.}
\label{SDSSvsCCD}
\end{figure}
%
We have checked our sky estimation method comparing the profiles obtained 
from the SDSS images with deep surface photometry. We have used the 
profiles of three face-on Sbc--Sc galaxies from \cite{pohlen2002}, from 
which UGC\,9837 is also part of our sample. 
The photometrically transformed SDSS profiles match the deep R-band surface
brightness profiles in all three cases quite well (two of the galaxies are 
shown in \fig \ref{SDSSvsCCD}). The remaining differences visible are easily 
explained by the two very different methods applied to extract these profiles. 
Whereas the SDSS profiles are standard fixed ellipse fits to the images, the 
deep R-band comparison profiles are obtained by vertically averaging after 
polar transformation \citep[see][]{pohlen2002}. This test confirms that SDSS 
profiles can be obtained reliably down to 
$\sim 27.0\ r^{\prime}$-\magsqarcsec.
%
\subsection{Ellipse fitting and Masking}
\label{ellipse}
We used the IRAF task {\sl ellipse} (STSDAS package) on the sky 
subtracted images to produce our final surface brightness profiles. 
We apply an extensive manual masking for each galaxy image. This is 
necessary to remove contamination from different sources. 
We masked all non-galaxy components like foreground stars, companion 
galaxies, or faint background galaxies. Not masked are extended spiral 
arms, or outer asymmetries clearly belonging to the galaxy itself.
The masking is primarily done in a 9x9 pixel ($3.6\arcsec$ x $3.6\arcsec$)
median smoothed version of the image, ideally suited to visualise 
structure in the background. This step also ensures to reliably 
trace the halos of the foreground stars to large radii. 
Whereas this is necessary for bright stars it is a conservative measure 
for fainter ones. In a second round any remaining stars in the unsmoothed 
image are also masked.
The common center for each concentric ellipse is fixed by a Gaussian 
fit to the bright nucleus of the galaxy. For 21 galaxies (marked in 
\tab\ref{ellskyresults} with a $\dagger$) the disk is lopsided or 
there is no bright center visible so we first allowed the ellipse 
centers to vary, choosing the best center representative for the 
outer disk. 
We used a logarithmic radial sampling with steps of $0.03$ to increase 
the S/N especially in the outer parts at cost of larger radial bins
compared to linear sampling. Iterative $3\sigma$ rejection along the 
ellipse is applied to minimise the influence of cosmic rays or 
any remaining faint foreground stars. 
We fitted two different sets of ellipses (three in case of a 
necessary free center fit) to each galaxy image \citep[for general 
information on ellipse fitting see][]{erwin2006}. The first, 
free ellipse fit tends to follow morphological features like 
bars, spirals, or asymmetries so it is not ideal to characterise 
the outer, underlying disk component, which we want to address in 
this work. Therefore we used a fixed ellipse fit to produce our 
final surface brightness profiles which will be the only ones 
discussed in the following sections. Fixing a single ellipticity and 
position angle for a galaxy is based on the assumption that the
disk is axisymmetric and round and the ellipticity and PA values 
are representative for the orientation (inclination angle) of the 
galaxy.
The initial free ellipse fit (fixed center, free ellipticity and PA) 
is used to determine the best set of ellipticity and PA describing 
the outer disk. These values are typically taken at the radius where 
the mean flux of the best fitted free ellipse reaches the value of 
the standard deviation of our background measurement in large boxes 
($1\sigma$ criterion). 
This limit ensures enough S/N to fit a free ellipse but is small enough 
to be in the radial region dominated by the outer disk. 
Ideally one would like to determine the ellipticity and PA 
of the outer disk not photometrically, but by means of kinematical 
informations. Since this is not possible for the whole sample 
we used only the photometric way to get a set of homogeneous values 
for all our galaxies. However, for some galaxies the $1\sigma$ criterion 
clearly marked a radius in a region with rapidly changing ellipticity 
and PA values, or a region which is not representative to the outer disk. 
In this cases we chose, after visual inspection, a different radius to 
represent the disk, which is typically the outermost successfully 
fitted free ellipse. 
The $r^{\prime}$ band and $g^{\prime}$ band ellipticity and PA values 
are determined independently. Since a galaxy should have only a single 
pair of PA and ellipticity (depending on the inclination and intrinsic 
circularity) independent of the wavelength used, we averaged the 
two values in case of no major problem in one of the bands. 
The fixed ellipse fits to the $r^{\prime}$ and $g^{\prime}$ band is 
therefore repeated using these mean values (see \tab\ref{ellskyresults})
and the resulting profiles are shown in Appendix \ref{atlas}.
%
\begin{table*}
\begin{center}
{\normalsize
\begin{tabular}{ l  c r c c  c  c r c c  c  c c c}
\hline
\rule[+0.4cm]{0mm}{0.0cm}
&
&\multicolumn{4}{l}{$r^{\prime}$ band}
& \hspace*{-2.1cm}
&\multicolumn{4}{l}{$g^{\prime}$ band}
&
&
&
\\[+0.05cm]
\cline{2-5}\cline{7-10} \\[-0.15cm]
Galaxy
&$ZP$ 
&$sky$
&$\sigma$  
&$\mu_{\rm lim}$ 
&
&$ZP$ 
&$sky$
&$\sigma$
&$\mu_{\rm lim}$ 
&
&$R_{\rm ell}$
&$PA$
&$ME$
\\[+0.1cm]
& [\magsqarcsecfrac]
& [ADU]
& [ADU]
& [\magsqarcsecfrac]
&
& [\magsqarcsecfrac]
& [ADU]
& [ADU]
& [\magsqarcsecfrac]
&
& [\arcsec]
& [\deg]
& \\
\rule[-3mm]{0mm}{5mm}{\scriptsize{\raisebox{-0.7ex}{\it (1)}}}
&{\scriptsize{\raisebox{-0.7ex}{\it (2)}}}
&{\scriptsize{\raisebox{-0.7ex}{\it (3)}}}
&{\scriptsize{\raisebox{-0.7ex}{\it (4)}}}
&{\scriptsize{\raisebox{-0.7ex}{\it (5)}}}
&
&{\scriptsize{\raisebox{-0.7ex}{\it (6)}}}
&{\scriptsize{\raisebox{-0.7ex}{\it (7)}}}
&{\scriptsize{\raisebox{-0.7ex}{\it (8)}}}
&{\scriptsize{\raisebox{-0.7ex}{\it (9)}}}
&
&{\scriptsize{\raisebox{-0.7ex}{\it (10)}}} 
&{\scriptsize{\raisebox{-0.7ex}{\it (11)}}} 
&{\scriptsize{\raisebox{-0.7ex}{\it (12)}}} 
\\
\hline\hline \\[-0.2cm]
NGC\,0450                    &26.267 &165.11 &0.14 &27.2&&26.686 &  99.19 &0.09 &28.1 &&  84&  -8.8 &0.36    \\
PGC\,006667                  &26.213 &175.59 &0.06 &28.1&&26.451 &  74.16 &0.12 &27.6 &&  67&  35.8 &0.19    \\
NGC\,0701${^{^\dagger}}$     &26.136 &157.72 &0.08 &27.7&&26.343 &  69.07 &0.09 &27.8 &&  74& -41.9 &0.49    \\
NGC\,0853${^{^\dagger}}$     &26.293 &144.89 &0.11 &27.5&&26.542 &  75.84 &0.07 &28.2 &&  59& -11.7 &0.36    \\
NGC\,0941                    &26.114 & 99.42 &0.02 &29.2&&26.507 &  68.58 &0.11 &27.7 &&  69&  78.6 &0.30    \\
UGC\,02081                   &26.106 & 99.91 &0.03 &28.7&&26.282 &  56.34 &0.08 &27.8 &&  51& -18.8 &0.42    \\
NGC\,1042                    &26.266 &175.77 &0.09 &27.7&&26.508 &  75.11 &0.07 &28.2 && 136& -69.0 &0.20    \\
NGC\,1068                    &26.248 &186.31 &0.12 &27.4&&26.551 &  94.08 &0.09 &28.0 && 312&   9.8 &0.09    \\
NGC\,1084                    &26.259 &170.55 &0.18 &26.9&&26.416 &  70.41 &   * &   * &&  80& -46.9 &0.39    \\
NGC\,1087${^{^\dagger}}$     &26.246 &175.97 &0.09 &27.7&&26.533 &  92.17 &0.08 &28.1 && 114& -88.4 &0.39    \\
NGC\,1299                    &26.275 &190.56 &0.13 &27.3&&26.480 &  82.11 &0.14 &27.4 &&  43& -35.0 &0.45    \\
NGC\,2543                    &26.271 &129.29 &0.17 &27.0&&26.563 &  79.22 &0.09 &28.0 && 108&  -3.1 &0.50    \\
NGC\,2541${^{^\dagger}}$     &26.259 &181.06 &0.10 &27.6&&26.532 &  91.73 &0.04 &28.8 && 135& -53.5 &0.52    \\
UGC\,04393${^{^\dagger}}$    &26.345 &116.47 &0.08 &27.9&&26.564 &  66.68 &0.07 &28.3 &&  61&  11.9 &0.45    \\
NGC\,2684                    &26.314 &122.01 &0.08 &27.9&&26.591 &  70.12 &0.19 &27.2 &&  35&   8.7 &0.18    \\
NGC\,2701${^{^\dagger}}$     &26.360 &162.44 &0.09 &27.8&&26.653 &  86.65 &0.08 &28.2 &&  60& -19.3 &0.32    \\
NGC\,2776                    &26.291 &156.28 &0.05 &28.4&&26.508 &  67.37 &0.05 &28.6 &&  83&  30.8 &0.05    \\
NGC\,2967                    &26.278 &150.33 &0.06 &28.1&&26.605 & 106.62 &0.12 &27.7 &&  77& -70.7 &0.07    \\
NGC\,3055                    &26.255 &117.73 &0.11 &27.5&&26.566 &  73.52 &0.11 &27.8 &&  65& -24.1 &0.42    \\
NGC\,3246${^{^\dagger}}$     &26.281 &125.07 &0.11 &27.5&&26.539 &  72.34 &0.09 &28.0 &&  65&   8.0 &0.48    \\
NGC\,3259                    &26.260 &158.22 &0.14 &27.2&&26.530 &  70.55 &0.07 &28.2 &&  53& -47.0 &0.43    \\
NGC\,3310                    &26.229 &134.53 &0.07 &27.9&&26.527 &  70.73 &0.24 &26.9 &&  83& -50.2 &0.05    \\
NGC\,3359                    &26.274 &150.77 &0.08 &27.8&&26.623 &  84.83 &0.10 &27.9 && 236& -80.9 &0.48    \\
NGC\,3423                    &26.286 &103.65 &0.07 &28.0&&26.720 &  75.84 &0.10 &28.0 && 112& -53.5 &0.19    \\
NGC\,3488                    &26.266 &119.86 &0.10 &27.6&&26.548 &  70.64 &0.08 &28.1 &&  52& -75.8 &0.34    \\
NGC\,3583                    &26.270 &104.52 &0.11 &27.5&&26.550 &  59.18 &0.08 &28.1 &&  81&  46.1 &0.32    \\
NGC\,3589${^{^\dagger}}$     &26.312 &139.12 &0.10 &27.6&&26.567 &  60.49 &0.03 &29.2 &&  46& -21.9 &0.50    \\
UGC\,06309                   &26.241 &109.30 &0.07 &27.9&&26.569 &  58.96 &0.05 &28.6 &&  50&  45.8 &0.31    \\
NGC\,3631                    &26.261 &140.70 &0.07 &28.0&&26.661 &  77.36 &0.06 &28.5 && 129&  32.1 &0.16    \\
NGC\,3642                    &26.266 & 94.01 &0.05 &28.3&&26.569 &  54.00 &0.04 &28.9 &&  77&   5.1 &0.05    \\
UGC\,06518                   &26.263 &137.25 &0.13 &27.3&&26.514 &  65.27 &0.08 &28.1 &&  31& -59.0 &0.39    \\
NGC\,3756                    &26.205 &135.84 &0.17 &26.9&&26.507 &  67.02 &0.14 &27.4 && 118& -82.2 &0.50    \\
NGC\,3888                    &26.223 &133.12 &0.07 &27.9&&26.512 &  65.09 &0.05 &28.6 &&  52&  30.9 &0.26    \\
NGC\,3893                    &26.279 &107.16 &0.09 &27.7&&26.691 &  66.61 &0.10 &28.0 && 133&  83.6 &0.41    \\
UGC\,06903                   &26.264 &136.09 &0.04 &28.6&&26.571 &  76.94 &0.08 &28.1 &&  69&  48.1 &0.12    \\
NGC\,3982                    &26.204 &133.59 &0.11 &27.4&&26.540 &  68.43 &0.04 &28.8 &&  58& -87.4 &0.11    \\
NGC\,3992                    &26.235 &113.06 &0.09 &27.7&&26.517 &  59.44 &0.18 &27.2 && 264& -15.4 &0.42    \\
NGC\,4030                    &26.261 &137.95 &0.06 &28.1&&26.707 &  96.55 &0.09 &28.1 && 124& -59.6 &0.20    \\
NGC\,4041                    &26.353 &152.10 &0.05 &28.4&&26.596 &  62.19 &0.07 &28.3 &&  84&  -8.5 &0.05    \\
NGC\,4102                    &26.267 &123.57 &0.06 &28.1&&26.594 &  68.12 &0.08 &28.1 &&  94& -47.9 &0.43    \\
NGC\,4108${^{^\dagger}}$     &26.317 &174.08 &0.11 &27.5&&26.603 &  92.76 &0.06 &28.5 &&  55&  13.2 &0.22    \\
NGC\,4108B${^{^\dagger}}$    &26.317 &173.86 &0.06 &28.2&&26.603 &  92.71 &0.03 &29.2 &&  37&  39.8 &0.23    \\
NGC\,4123                    &26.195 &161.48 &0.09 &27.6&&26.524 &  90.50 &0.11 &27.7 && 111&  36.1 &0.33    \\
NGC\,4210                    &26.290 &119.97 &0.04 &28.6&&26.607 &  62.28 &0.05 &28.7 &&  55&   9.3 &0.24    \\
NGC\,4273${^{^\dagger}}$     &26.252 &130.84 &0.10 &27.6&&26.566 &  73.89 &0.11 &27.8 &&  72& -83.3 &0.39    \\
NGC\,4480                    &26.228 &143.92 &0.08 &27.8&&26.489 &  77.53 &0.08 &28.0 &&  65&  87.2 &0.50    \\
\hline
\end{tabular}
}
\caption[]{Background (sky) and ellipse parameters: 
{\scriptsize{\it (1)}} Principal name in LEDA, 
{\scriptsize{\it (2+6)}} photometric $r^{\prime}$ and $g^{\prime}$ band zero points, 
{\scriptsize{\it (3+7)}} background value estimated around each galaxy,  
{\scriptsize{\it (4+8)}} $\sigma$ of background from the ellipse method 
(\cf\sec\ref{skysub}), 
{\scriptsize{\it (5+9)}} limiting surface brightness due to this background, used to
constrain the outer boundary of the exponential fits, 
{\scriptsize{\it (10)}} radius used for fixing the ellipse parameters 
(\cf\sec\ref{ellipse}),   
{\scriptsize{\it (11)}} mean position angle $PA$ (for $r^{\prime}$ and $g^{\prime}$ band) of 
the fixed ellipse (as measured on the SDSS image),
{\scriptsize{\it (12)}} mean ellipticity $ME$ ($r^{\prime}$ and $g^{\prime}$ band) of the fixed 
ellipse
\label{ellskyresults} }
\end{center}
\end{table*}
\addtocounter{table}{-1}
\begin{table*}
\begin{center}
{\normalsize
\begin{tabular}{ l  c r c c  c  c r c c  c  c c c}
\hline
\rule[+0.4cm]{0mm}{0.0cm}
&
&\multicolumn{4}{l}{$r^{\prime}$ band}
& \hspace*{-2.1cm}
&\multicolumn{4}{l}{$g^{\prime}$ band}
&
&
&
\\[+0.05cm]
\cline{2-5}\cline{7-10} \\[-0.15cm]
Galaxy
&$ZP$ 
&$sky$
&$\sigma$  
&$\mu_{\rm lim}$ 
&
&$ZP$ 
&$sky$
&$\sigma$
&$\mu_{\rm lim}$ 
&
&$R_{\rm ell}$
&$PA$
&$ME$
\\[+0.1cm]
& [\magsqarcsecfrac]
& [ADU]
& [ADU]
& [\magsqarcsecfrac]
&
& [\magsqarcsecfrac]
& [ADU]
& [ADU]
& [\magsqarcsecfrac]
&
& [\arcsec]
& [\deg]
& \\
\rule[-3mm]{0mm}{5mm}{\scriptsize{\raisebox{-0.7ex}{\it (1)}}}
&{\scriptsize{\raisebox{-0.7ex}{\it (2)}}}
&{\scriptsize{\raisebox{-0.7ex}{\it (3)}}}
&{\scriptsize{\raisebox{-0.7ex}{\it (4)}}}
&{\scriptsize{\raisebox{-0.7ex}{\it (5)}}}
&
&{\scriptsize{\raisebox{-0.7ex}{\it (6)}}}
&{\scriptsize{\raisebox{-0.7ex}{\it (7)}}}
&{\scriptsize{\raisebox{-0.7ex}{\it (8)}}}
&{\scriptsize{\raisebox{-0.7ex}{\it (9)}}}
&
&{\scriptsize{\raisebox{-0.7ex}{\it (10)}}} 
&{\scriptsize{\raisebox{-0.7ex}{\it (11)}}} 
&{\scriptsize{\raisebox{-0.7ex}{\it (12)}}} 
\\
\hline\hline \\[-0.2cm]
NGC\,4517A                   &26.278 &136.44 &0.07 &28.0&&26.605 &  78.57 &0.08 &28.2 && 113& -61.4 &0.43    \\
UGC\,07700${^{^\dagger}}$    &26.346 &124.40 &0.02 &29.4&&26.766 &  71.96 &0.13 &27.8 &&  66& -23.6 &0.24    \\
NGC\,4545                    &26.225 &115.06 &0.04 &28.5&&26.541 &  54.71 &0.07 &28.2 &&  66& -87.9 &0.42    \\
NGC\,4653                    &26.292 &135.45 &0.05 &28.4&&26.584 &  75.43 &0.08 &28.1 &&  79&  78.5 &0.17    \\
NGC\,4668                    &26.292 &135.17 &0.07 &28.0&&26.584 &  75.49 &0.09 &28.0 &&  53& -85.5 &0.48    \\
UGC\,08041                   &26.253 &127.60 &0.11 &27.5&&26.566 &  78.23 &0.14 &27.5 && 101&  70.0 &0.48    \\
UGC\,08084${^{^\dagger}}$    &26.195 &157.33 &0.05 &28.3&&26.524 &  86.02 &0.01 &30.3 &&  43& -30.9 &0.18    \\
NGC\,4904                    &26.322 &142.49 &0.08 &27.9&&26.588 &  75.36 &0.06 &28.4 &&  70& -63.4 &0.29    \\
UGC\,08237                   &26.228 &103.85 &0.11 &27.4&&26.574 &  56.11 &0.03 &29.2 &&  35&  30.7 &0.22    \\
NGC\,5147                    &26.242 &179.24 &0.04 &28.5&&26.494 &  92.26 &0.06 &28.4 &&  60&  28.8 &0.21    \\
UGC\,08658                   &26.174 &122.81 &0.13 &27.2&&26.465 &  66.52 &0.07 &28.2 &&  65&   3.2 &0.39    \\
NGC\,5300                    &26.235 &146.52 &0.09 &27.7&&26.495 &  81.87 &0.09 &27.9 && 100&  58.5 &0.33    \\
NGC\,5334                    &26.261 &125.54 &0.07 &28.0&&26.707 &  86.37 &0.05 &28.8 && 107& -76.0 &0.23    \\
NGC\,5376                    &26.237 &108.94 &0.11 &27.4&&26.528 &  56.56 &0.03 &29.1 &&  63& -49.8 &0.39    \\
NGC\,5430                    &26.237 &108.70 &0.07 &27.9&&26.528 &  55.89 &0.05 &28.6 &&  69&  70.5 &0.35    \\
NGC\,5480                    &26.226 &149.31 &0.08 &27.8&&26.670 &  69.86 &0.08 &28.2 &&  60& -73.3 &0.15    \\
NGC\,5584                    &26.312 &135.42 &0.08 &27.9&&26.567 &  74.06 &0.07 &28.3 && 106&  67.9 &0.26    \\
NGC\,5624${^{^\dagger}}$     &26.163 &147.69 &0.09 &27.6&&26.487 &  63.15 &0.09 &27.9 &&  38&  82.8 &0.34    \\
NGC\,5660                    &26.199 &123.03 &0.06 &28.1&&26.488 &  64.66 &0.08 &28.0 &&  69&  24.3 &0.05    \\
NGC\,5667                    &26.216 & 99.47 &0.09 &27.6&&26.566 &  55.67 &0.04 &28.9 &&  55&  51.3 &0.47    \\
NGC\,5668                    &26.257 &117.18 &0.05 &28.3&&26.578 &  70.51 &0.09 &28.0 &&  95&  43.2 &0.15    \\
NGC\,5693${^{^\dagger}}$     &26.225 &151.52 &0.08 &27.8&&26.669 &  74.68 &0.09 &28.1 &&  61&  13.2 &0.05    \\
NGC\,5713${^{^\dagger}}$     &26.312 &134.21 &0.04 &28.6&&26.567 &  73.88 &0.06 &28.4 &&  88& -81.6 &0.13    \\
NGC\,5768${^{^\dagger}}$     &26.212 &128.52 &0.07 &27.9&&26.493 &  76.07 &0.03 &29.1 &&  70&  24.1 &0.11    \\
 IC\,1067                    &26.261 &123.11 &0.09 &27.7&&26.537 &  69.83 &0.17 &27.3 &&  60&  25.8 &0.26    \\
NGC\,5774${^{^\dagger}}$     &26.254 &173.84 &0.06 &28.1&&26.522 &  85.40 &0.06 &28.4 &&  87&  41.8 &0.22    \\
NGC\,5806${^{^\dagger}}$     &26.246 &179.83 &0.09 &27.7&&26.518 &  82.76 &0.06 &28.4 &&  79&  73.7 &0.46    \\
NGC\,5850                    &26.241 &195.14 &0.09 &27.7&&26.492 &  94.97 &0.06 &28.4 && 141&  74.9 &0.22    \\
UGC\,09741                   &26.192 &126.43 &0.04 &28.5&&26.548 &  56.25 &0.05 &28.6 &&  45&  42.2 &0.11    \\
UGC\,09837                   &26.246 &115.43 &0.08 &27.8&&26.731 &  62.31 &0.07 &28.4 &&  53&  20.9 &0.05    \\
NGC\,5937                    &26.232 &121.74 &0.07 &27.9&&26.650 &  84.21 &0.12 &27.8 &&  61& -69.1 &0.44    \\
 IC\,1125                    &26.189 &113.30 &0.12 &27.3&&26.525 &  74.76 &0.09 &27.9 &&  46&  56.0 &0.44    \\
 IC\,1158                    &26.233 &164.49 &0.10 &27.5&&26.689 & 112.46 &0.12 &27.8 &&  69&  39.1 &0.44    \\
NGC\,6070                    &26.296 &149.32 &0.15 &27.2&&26.519 &  63.01 &0.13 &27.5 && 114& -30.8 &0.57    \\
NGC\,6155${^{^\dagger}}$     &26.304 &125.81 &0.07 &28.0&&26.584 &  63.54 &0.11 &27.8 &&  47&  -7.7 &0.28    \\
UGC\,10721                   &26.198 &118.98 &0.10 &27.5&&26.524 &  55.93 &0.05 &28.6 &&  38& -31.0 &0.33    \\
NGC\,7437                    &26.191 &153.77 &0.12 &27.3&&26.495 &  87.08 &0.09 &27.9 &&  57& -65.7 &0.10    \\
NGC\,7606                    &26.278 &225.46 &0.12 &27.4&&26.576 &  89.15 &0.17 &27.3 && 146&  51.7 &0.62    \\
UGC\,12709${^{^\dagger}}$    &26.236 &154.79 &0.11 &27.4&&26.565 &  80.44 &0.07 &28.3 &&  71&  50.4 &0.39    \\
\hline
\end{tabular}
}
\caption[]{(continued): Background and ellipse parameters
\newline
$\dagger$: The centers for these galaxies are not obtained by a Gaussian 
fit to the central region but by a free ellipse fit to determine the best 
center representative for the outer disk.
}
\end{center}
\end{table*}
%
%
\section{Analysis}
\label{analysis}
\subsection{Classification}
\label{classi}
\begin{figure}
\includegraphics[width=8.0cm,angle=0]{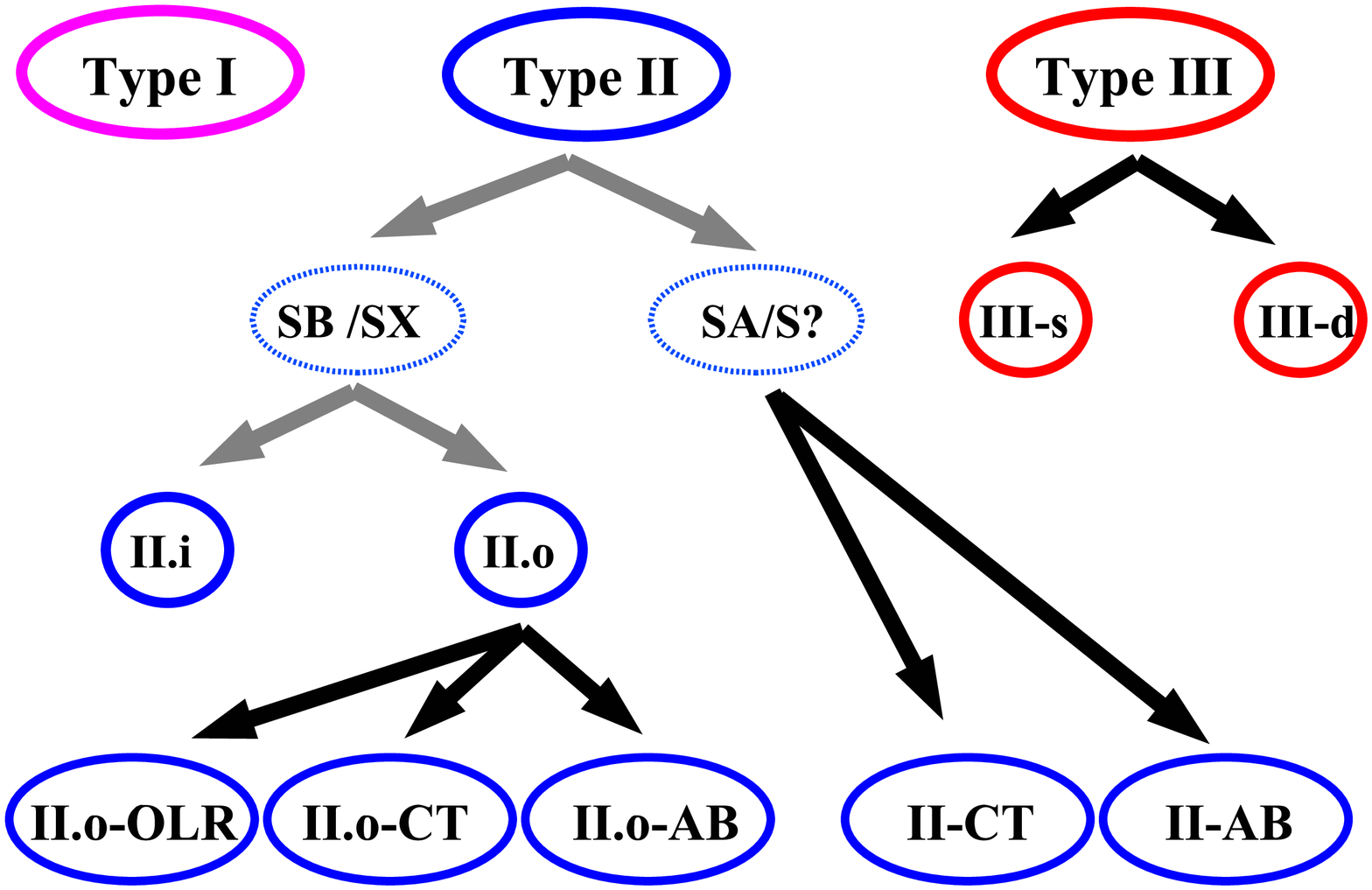}
\caption{Classification schema \citep[following][]{erwin2006}: The 
three main disk types according to break features in their surface 
brightness profiles are 
\typeo (no break), \typet ({\it downbending} break), 
and \typeiii ({\it upbending} break). 
In case the galaxy is barred (SB/SX) the {\it downbending} 
breaks (\typetc) are subdivided according to the position of 
the break in respect to the bar: 
\typeti (inside) or \typeto (outside). 
An additional level of subclassification is applied for this
class trying to relate the observed break to a physical origin.  
These are \typetoabc, \typeolrc, and \typetoct (see text). 
For apparently unbarred galaxies (SA/S?) we use only \typect or 
\typeabc. 
The {\it upbending} breaks (\typeiiic) are subdivided according 
to the possible nature of the outer region: {\it (d)}isk- or 
{\it (s)}pheroid-like. 
Some galaxies are better described with two breaks, each of 
which could be associated to one of the individual types. 
They are assigned a mixed classification (\eg \typectiiic). 
}
\label{classischema}
\end{figure}
%
%
We have classified \citep[following][]{erwin2006} each profile according 
to observed break feature using the following nomenclature: \typeo 
(no break), \typet (downbending break), and \typeiii (upbending break). 
A schematic view of our classification is shown in \fig\ref{classischema}. 
Images and profiles for some typical examples of each class are given 
in \fig\ref{typeexamples}. 
The {\bf \typeoc} classification follows the nomenclature established by 
\cite{free70}. Aside from a varying bulge component these galaxies exhibit 
a region well described by a single exponential allowing for some wiggles 
associated with substructure in the disk such as spiral arms or prominent 
star forming regions.
The galaxies displaying a break with a downbending steeper outer 
region could be labelled 
{\bf \typet} as \cite[]{free70} originally proposed. Although he states at 
some point the inner deficit is {\it not far from the center of the system} 
there is no further restriction on this distance. In addition, for the three 
most prominent \typet galaxies in his sample he states that the outer 
exponential disk {\it begins outside the main region of spiral-arm activity}, 
so clearly outside a central region. 
Following \cite{erwin2006} the \typet class is subdivided 
into two groups in case the galaxy is barred. A {\bf \typeti} (inside) 
and a {\bf \typeto} (outside) depending on the position of the break in 
respect to the bar length (inside or outside the bar region).
Moreover, we tried to subclassify the \typet and \typeto profiles 
even further into additional three groups, which now tend to 
categorise the observed profiles according to their potential 
physical origin.  
First, profiles where the apparent break can be associated with 
large scale asymmetries of the galaxy in itself (only present 
in Sc--Sd galaxies). These are either associated to a lopsided disk 
having a well defined center, which is different from the applied 
ellipse centered on the outer disk, or to some large scale asymmetries 
in (or beyond) the main body of the outer disk (\cf extended spiral arms). 
These profiles are called {\bf \typeabc}, for {\it A}\/pparent 
(or {\it A}\/symmetric) {\it B}reak.
Another group, showing the break radius at around $2-2.5$ times the 
bar radius, is probably related to a resonance of the bar as described 
in \cite{erwin2006}. The bar itself is most often nicely marked by the 
presence of an additional inner ring. 
These profiles are classified as {\bf \typeolrc}, for {\it O}\/uter 
{\it L}\/indblad {\it R}\/esonance which happens to be at the 
position of $\sim 2$ times the bar radius.
The remaining profiles revealing a broken exponential behaviour 
are called {\bf \typectc}, since they are best associated to what we 
call now {\it C}\/lassical {\it T}\/runcations discovered by 
\cite{vdk1979}. Three nice examples of unbarred Sbc--Sc face-on 
galaxies are presented by \cite{pohlen2002}. 
For \typect galaxies with bars we made sure that the break is 
significantly further out than typical for the \typeolr breaks.  
The galaxies with a break followed by an upbending (shallower outer)
 profile are 
called {\bf \typeiii} according to \cite{typeiii} who find two 
sub-classes. One ({\bf \typeiiibc}) showing a fairly gradual 
transition and outer isophotes that are progressively rounder 
than the isophotes in the main disk, suggesting a disk embedded 
within a more spheroidal outer zone associated to a halo or an 
extended bulge component. The other type ({\bf \typeiiidc}) 
exhibits a rather sharp transition with the outer isophotes being 
not significantly rounder.     
Whereas measuring the ellipticity is an appropriate way to 
disentangle for early-type galaxies, devoid of prominent 
spiral arm structure misleading the free ellipse fits, it 
will most probably fail for late-types. So we decided here 
to classify them all as \typeiii only subclassifying those 
with visible spiral arm structure beyond the break radius, 
obviously associated with an outer disk-like region, 
as \typeiiid (\cf Appendix \ref{atlas}).
In our subsequent paper we will study if colour variations
at and beyond the break could potentially be used to decide 
between these subclasses. The colour profile of a transition
between a star-forming inner disk and a halo/envelope 
({\bf \typeiiibc}) may look different from a transition 
between an intermediate-age inner disk to a predominantly 
younger outer disk ({\bf \typeiiidc}).

Some galaxies have a more complex surface brightness 
distribution. They are better described with two breaks 
each of which could be associated to an individual 
type (as described above). These are called {\it mixed} 
classification (see \fig\ref{typeexamples}). 
For all the galaxies with a break in the profile we verified 
that changes in the sky value (\eg subtracting $\pm3\sigma$ 
to the sky level) will not produce an untruncated \typeo 
profile (see \fig\ref{N5300skysub}).
%
%
\begin{figure*}
\begin{center}
\includegraphics[width=5.8cm,angle=270]{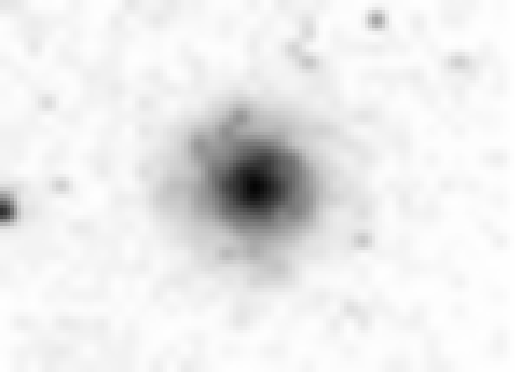}
\includegraphics[width=5.8cm,angle=270]{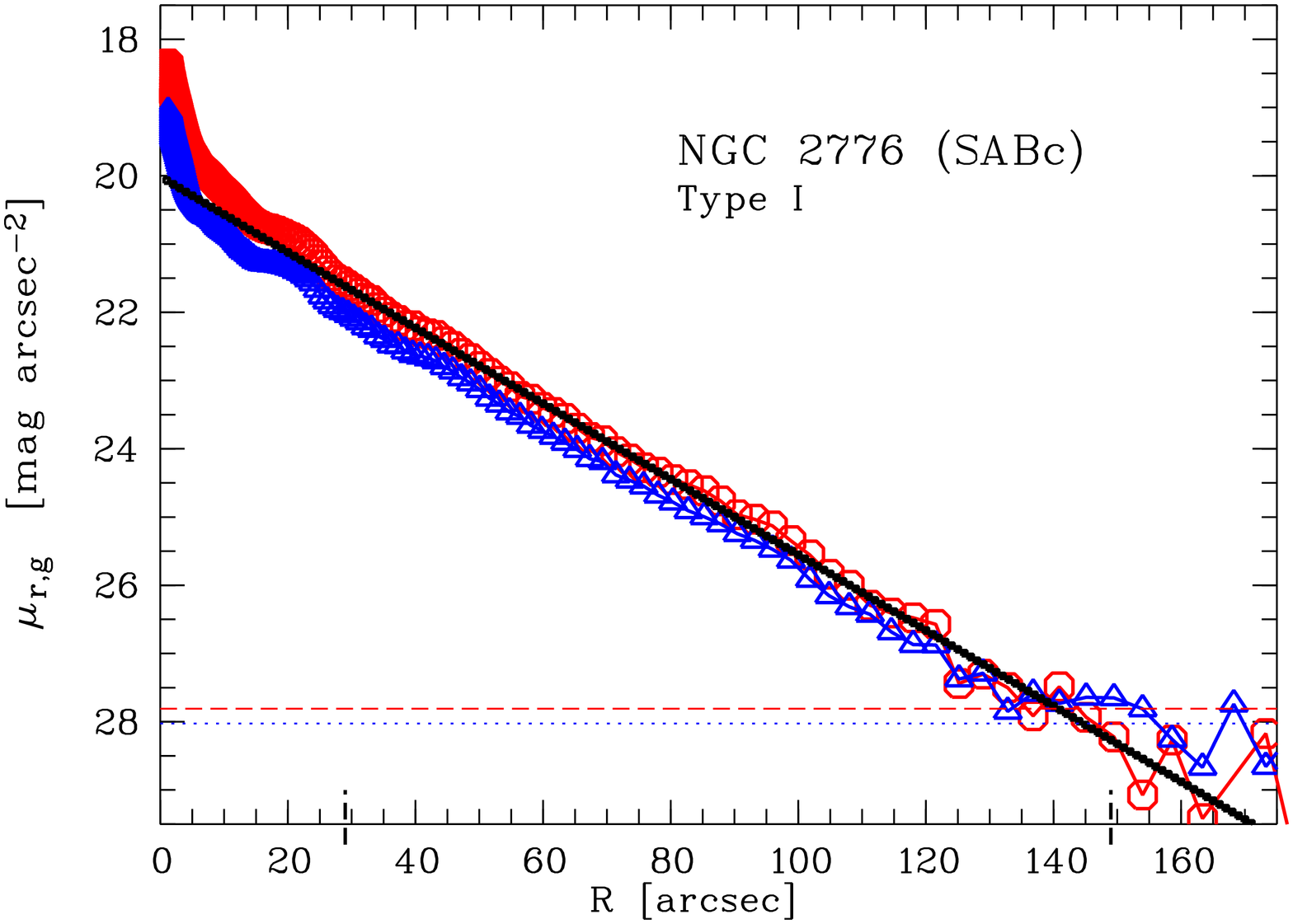}
\includegraphics[width=5.8cm,angle=270]{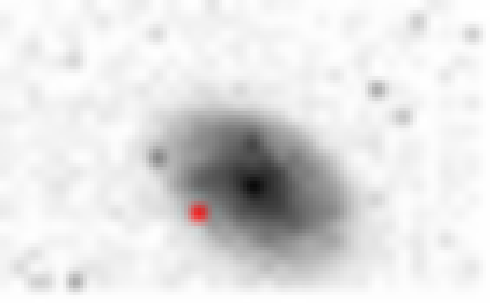}
\includegraphics[width=5.8cm,angle=270]{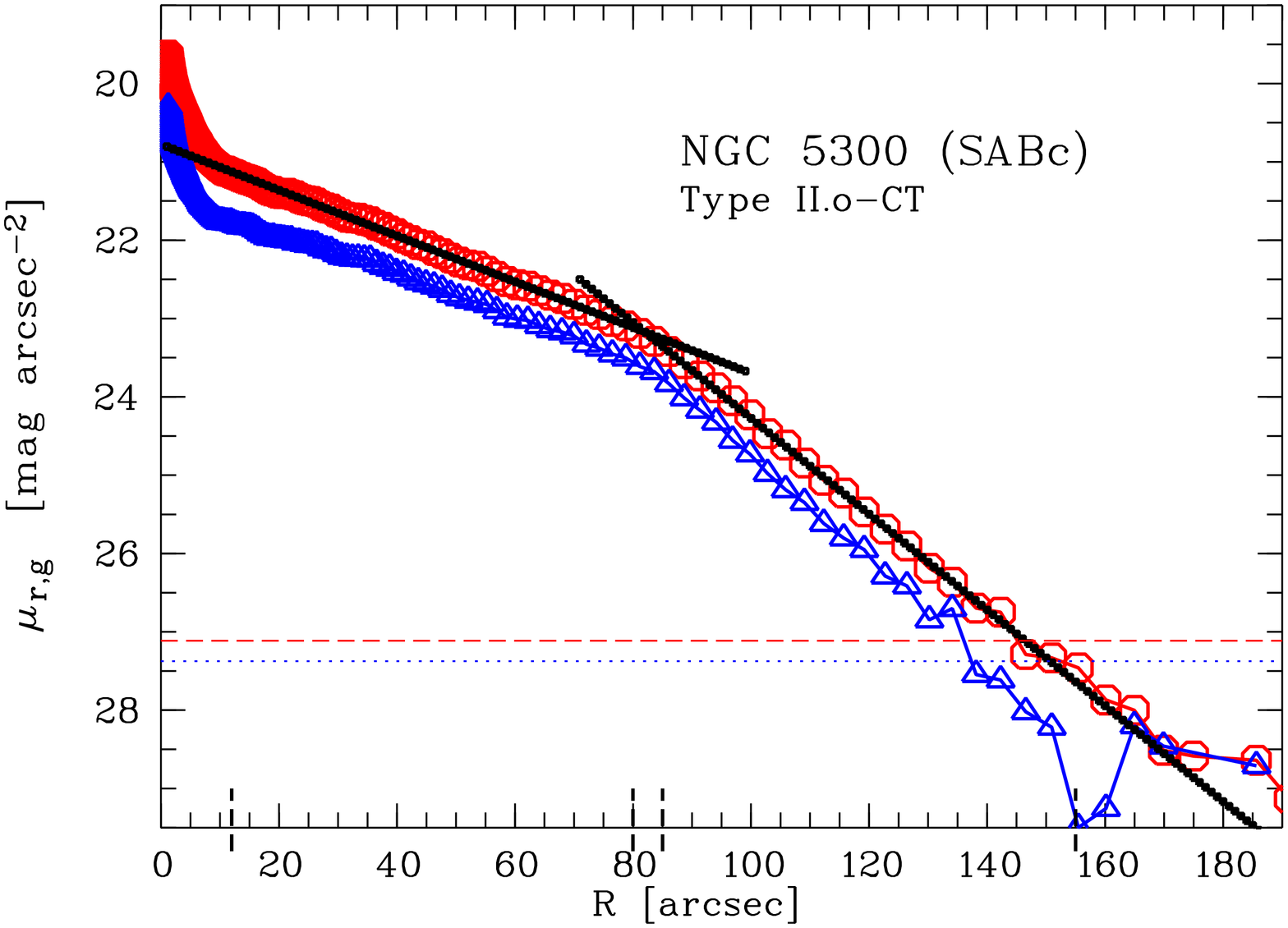}
\includegraphics[width=5.8cm,angle=270]{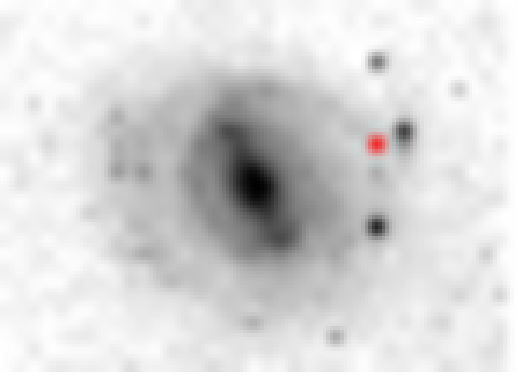}
\includegraphics[width=5.8cm,angle=270]{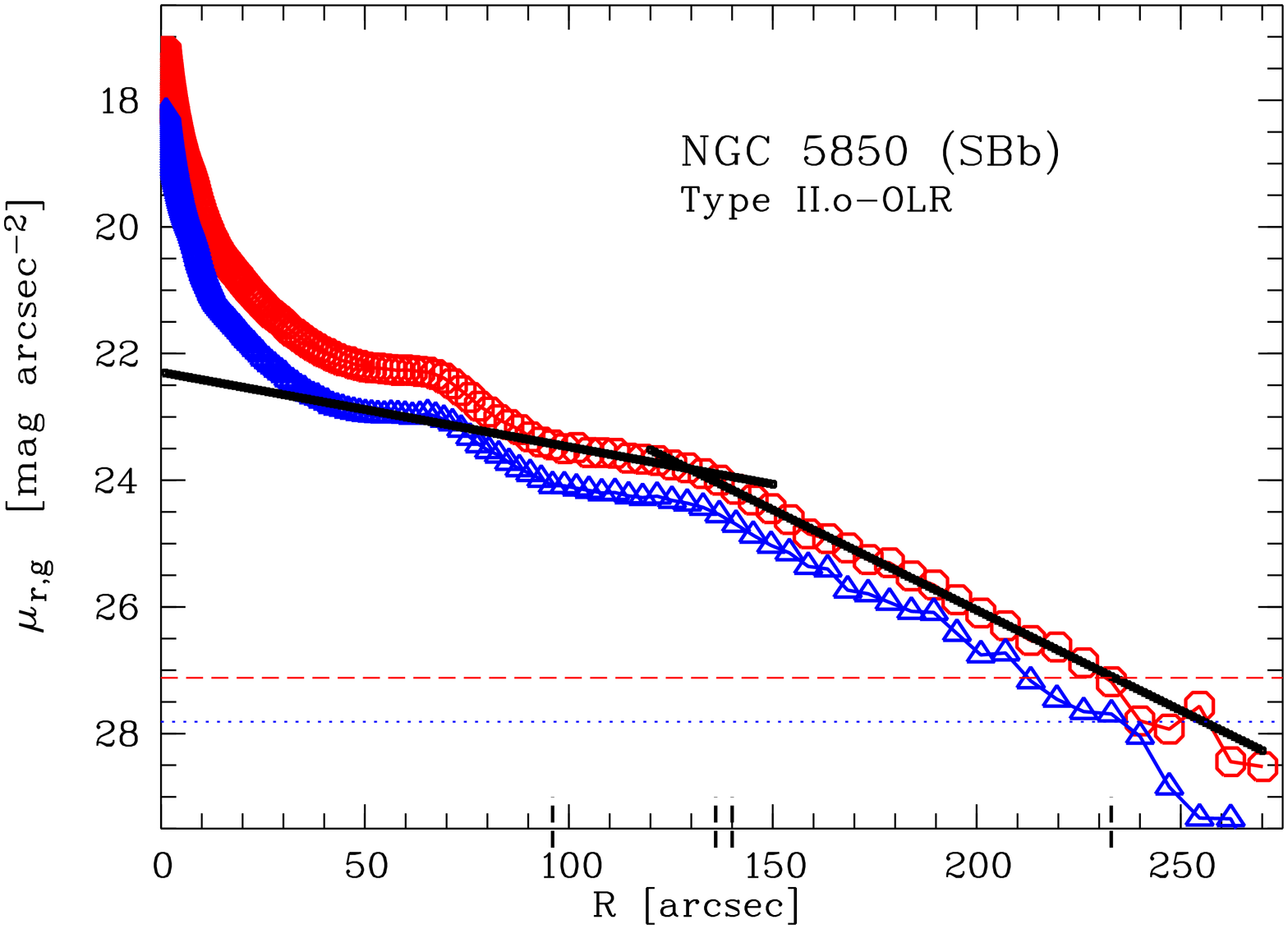}
\includegraphics[width=5.8cm,angle=270]{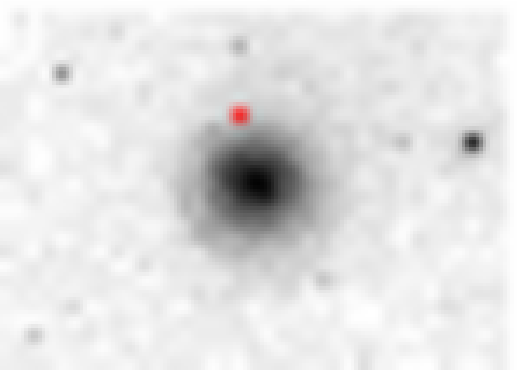}
\includegraphics[width=5.8cm,angle=270]{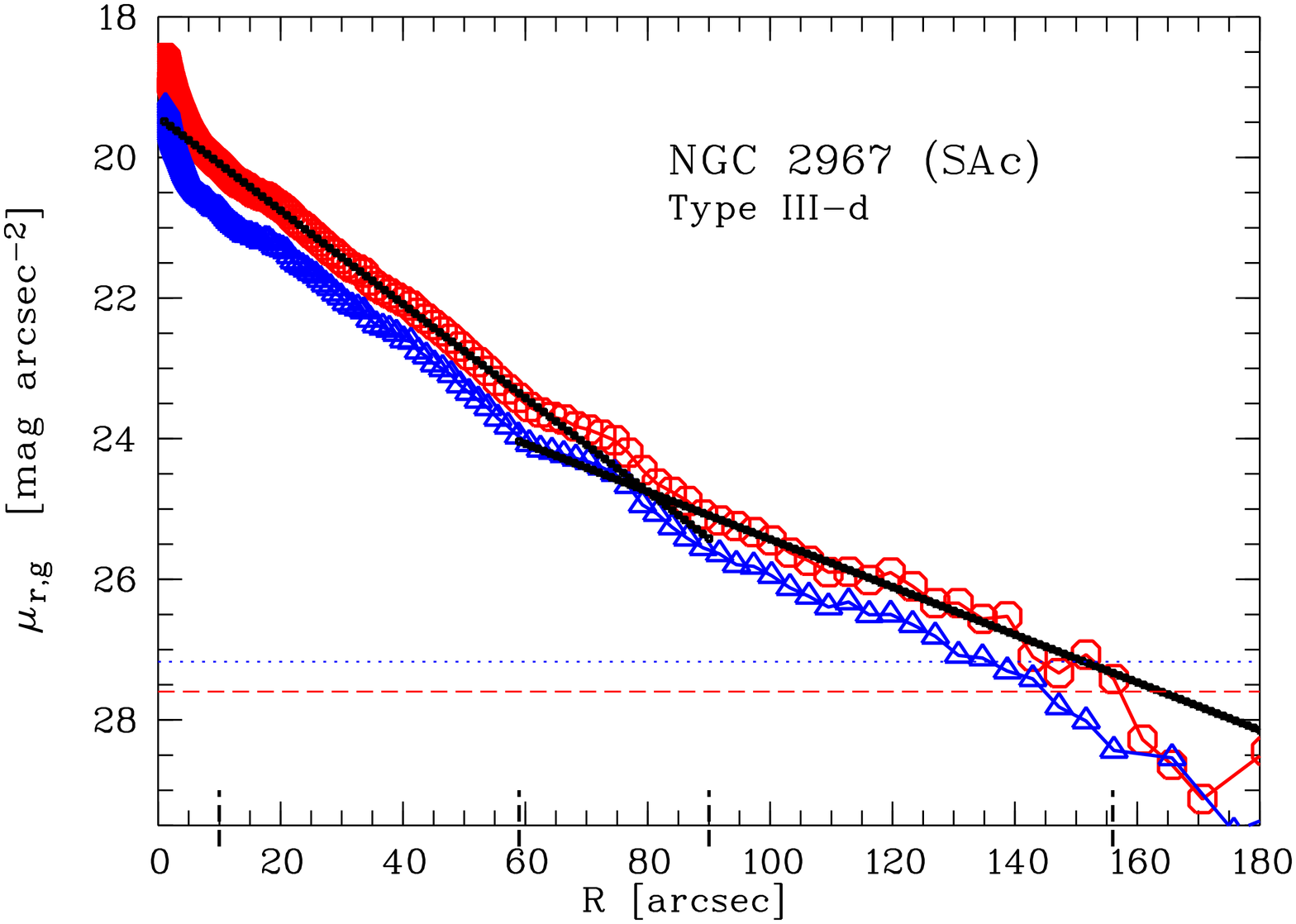}
\caption{Prototypical examples for each class of profiles:
\typeoc, \typectc, \typeolrc, and \typeiii {\it (top to bottom)}. 
{(\it Left panels):} $r^{\prime}$-band images (unrotated cut-outs 
from the SDSS fields) with the break radius marked with an ellipse. 
The ellipse for the first \typeo 
galaxy corresponds to the noise limit at $\sim\,140''$.
{(\it Right panels):} Azimuthally averaged, radial SDSS surface 
brightness profiles in the 
$g^{\prime}$ {\it (triangles)} and $r^{\prime}$ {\it (circles)} 
band overlayed by $r^{\prime}$ band exponential fits to the 
individual regions: single disk or inner and outer disk.
In addition we show the critical surface brightness 
($\mu_{\rm crit}$) for each band (dotted and dashed 
horizontal lines), down to which the profile is reliable 
(see Appendix \ref{atlas} for a detailed legend). 
}
\label{typeexamples}
\end{center}
\end{figure*}
%
\addtocounter{figure}{-1}
%
\begin{figure*}
\begin{center}
\includegraphics[width=5.8cm,angle=270]{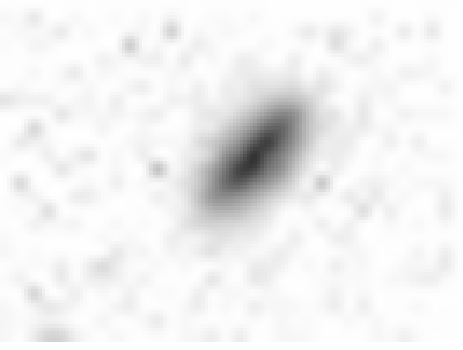}
\includegraphics[width=5.8cm,angle=270]{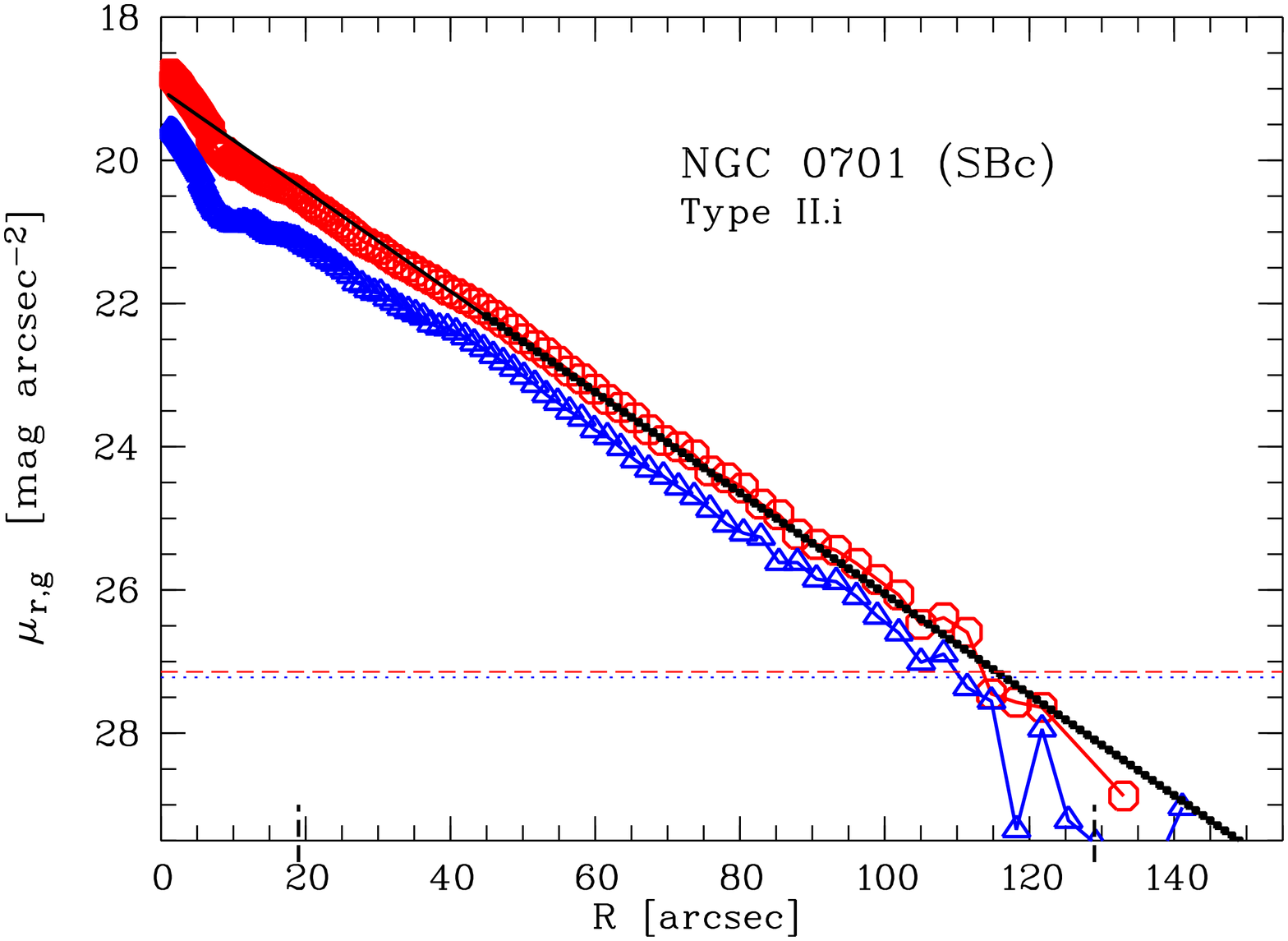}
\includegraphics[width=5.8cm,angle=270]{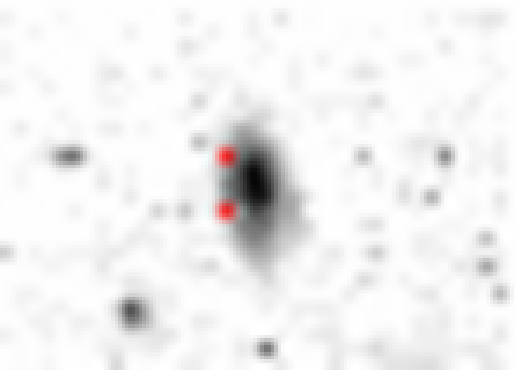}
\includegraphics[width=5.8cm,angle=270]{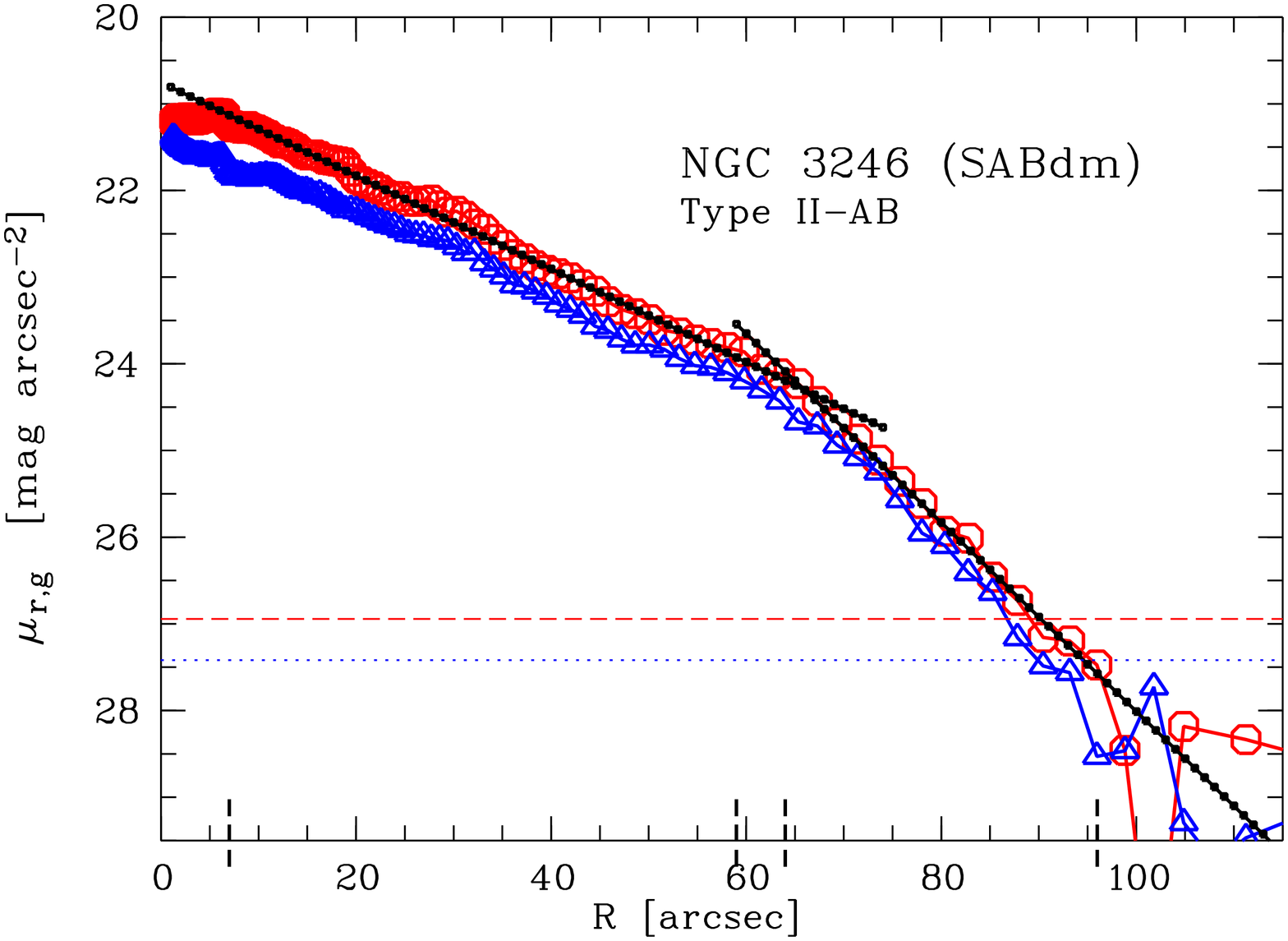}
\includegraphics[width=5.8cm,angle=270]{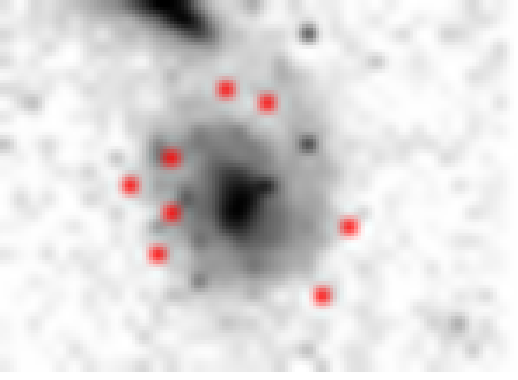}
\includegraphics[width=5.8cm,angle=270]{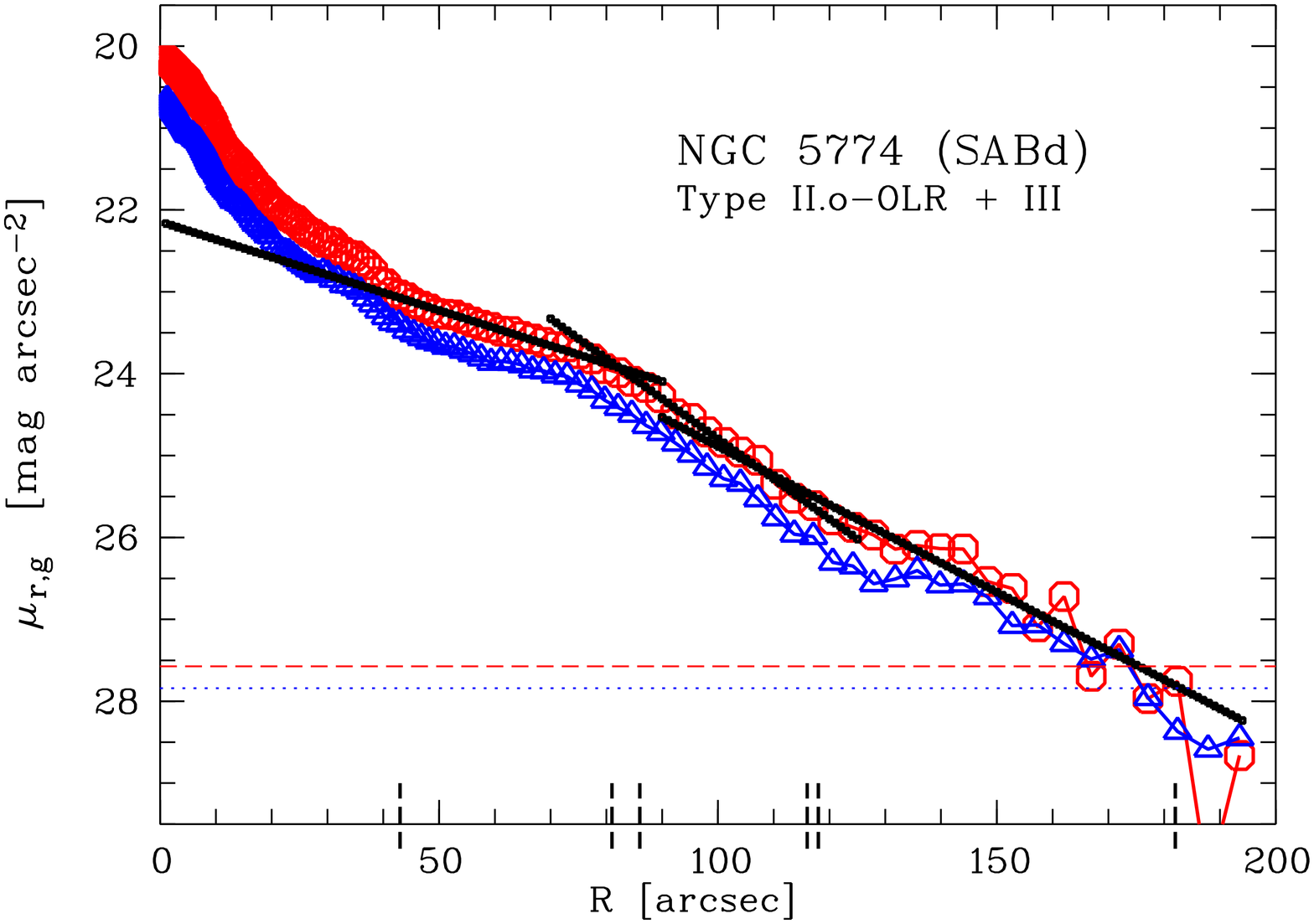}
\includegraphics[width=5.8cm,angle=270]{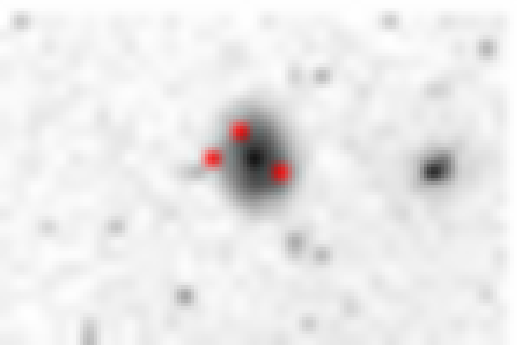}
\includegraphics[width=5.8cm,angle=270]{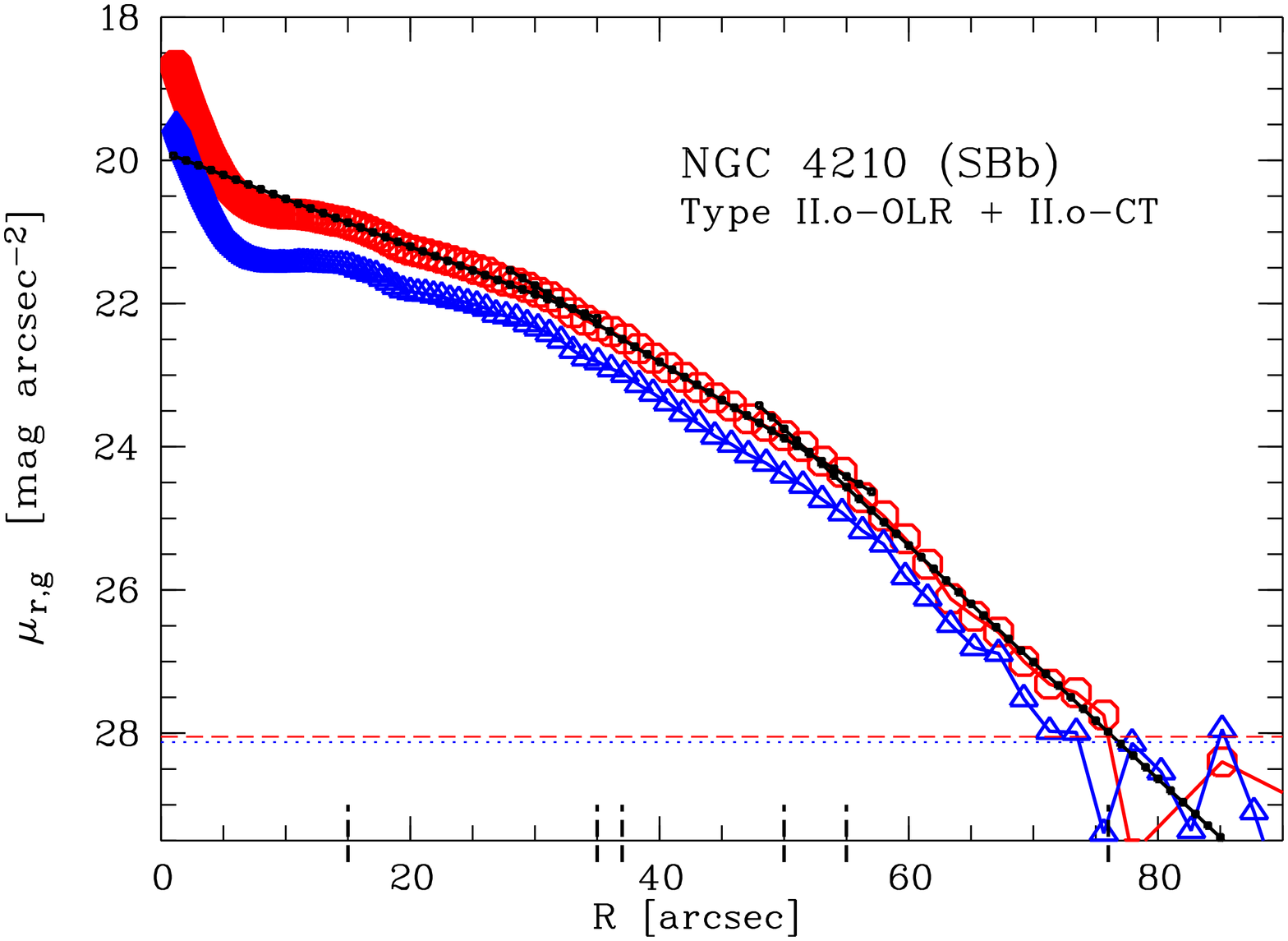}
\caption{(continued): Prototypical examples: \typetic, \typeabc, 
\typeolriiic, and \typeolrct {\it (top to bottom)}. 
The ellipse for the first \typeti galaxy corresponds to 
the inner boundary at $\sim\,20''$. For the mixed classifications 
two ellipses are shown corresponding to the associated breaks. 
} 
\end{center}
\end{figure*}
%

Please note that our classification \typeo and \typet are used 
slightly different from \cite{macarthur2003}. Our \typect class
is called \typeo with truncation in their study and their \typet
is a mix of our \typeti and all other \typeto classes. Since we 
do not have infrared images we cannot comment on their 'transition' 
class. However, we note that for IC\,1067 (see Appendix \ref{atlas} 
for a detailed discussion) the $r^{\prime}$-band profile is barely 
consistent with being \typeo whereas in the $g^{\prime}$-band 
it is clearly a \typeolr which is consistent with the profile 
shown by \cite{erwin2006}. This might imply that in very few
cases dust is playing a role in shaping the observed profile 
close to the center. So this profile could be a \typeo 
in the K-band, which has to be investigated in further 
studies. 
%
\subsection{Deriving break radius, break region and scalelengths}
\label{expfit}
\begin{figure}
\begin{center}
\includegraphics[width=6.1cm,angle=270]{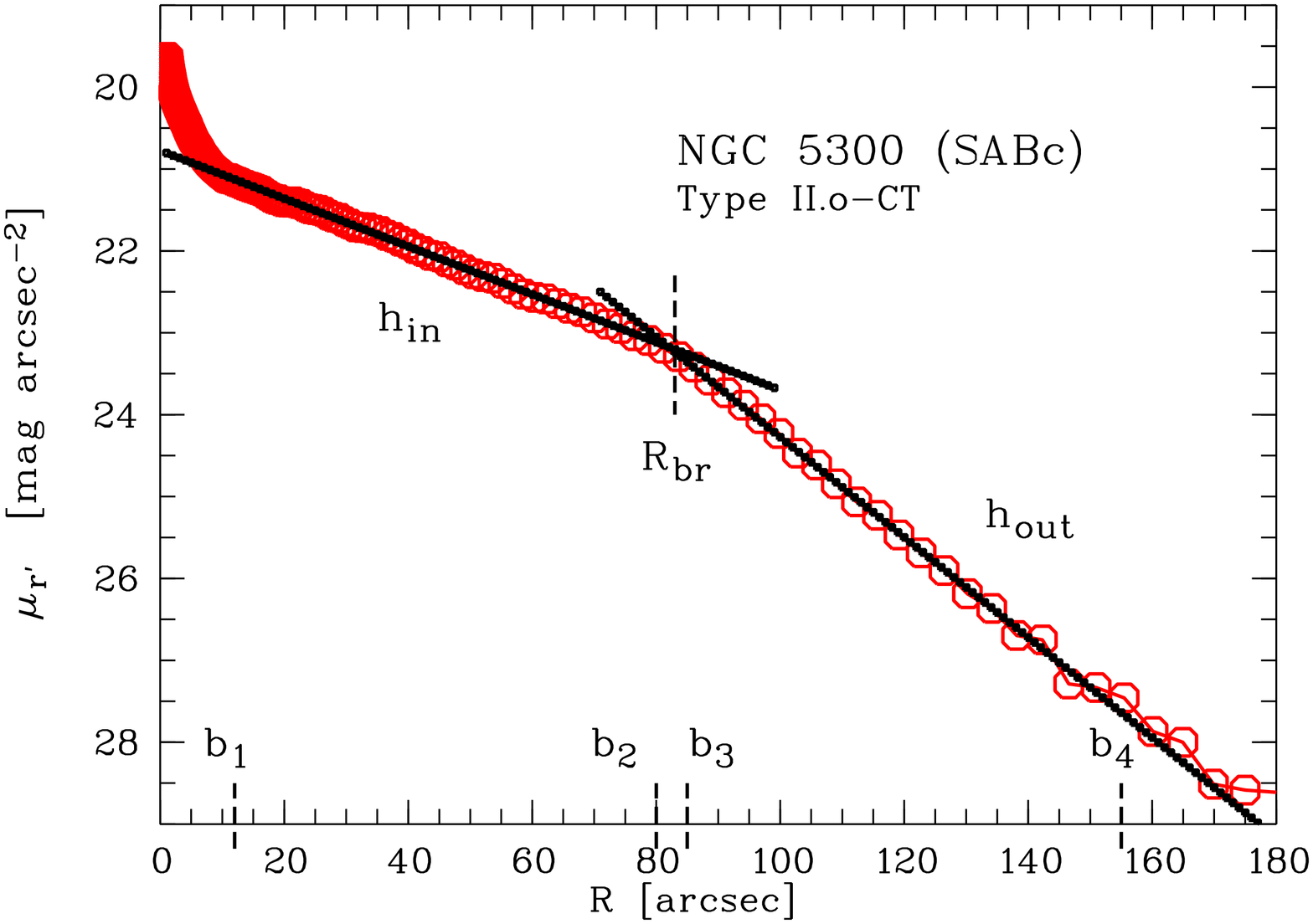}
\includegraphics[width=6.1cm,angle=270]{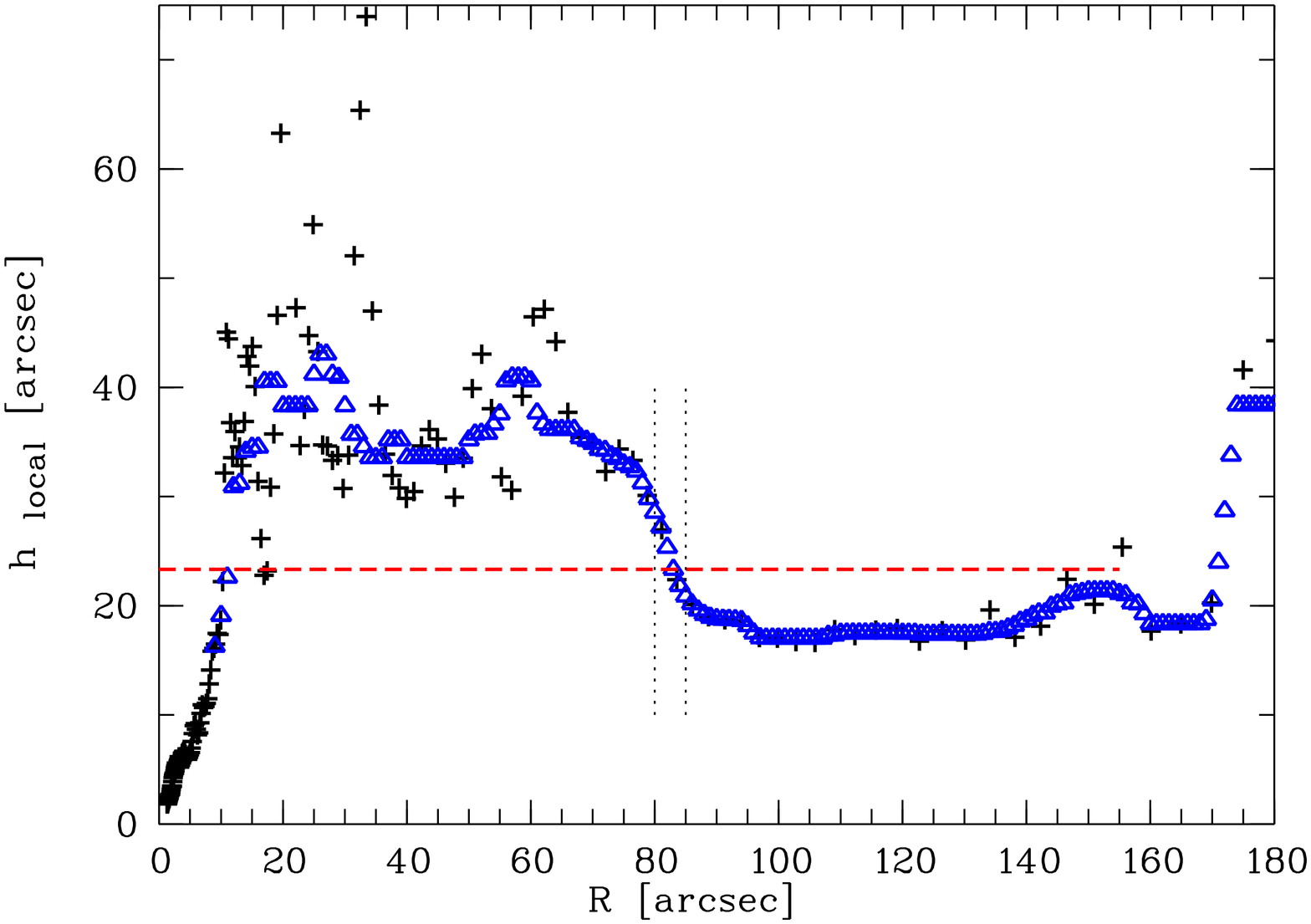}
\caption{Example of how to establish the break radius, the 
break region and the associated scalelengths: {\it (Upper panel):} 
Azimuthally averaged, surface brightness profile, overlayed by the 
best fitting exponentials (with scalelengths \hin and \hout) marking 
the position of the boundaries ($b_{1-4}$) and the break radius \rbr.
{\it (Lower panel):} Approximated derivative {\it (crosses)} of the 
profile providing a local scalelength (\hloc) together with 
the rebinned and smoothed version {\it (triangles)}. The {\it 
horizontal dashed line} marks our characteristic (or mean) value 
which is the median of the $1/\hloc$ distribution. The {\it 
vertical dotted lines} mark the inner and outer boundary of 
the break region. 
\label{derivative}
}
\end{center}
\end{figure}

In addition to our pure identification --break (down- or up-bending) 
{\it yes} or {\it no}-- we have quantified the radial distance \rbr 
where the break occurs, characterised its sharpness and derived 
an inner and outer scalelength (\hin, \hout) (see \fig\ref{derivative}). 
To determine the scalelengths we restricted our fits, to the 1D, 
azimuthally averaged, surface brightness profiles obtained with 
fixed ellipse fits (\cf\sec\ref{ellipse}). Since a detailed bulge 
to disk decomposition is beyond the scope of this paper, we used 
the so called ''marking the disk'' method by applying boundaries 
and fitting simple exponential functions: 
{$\mu (R) = \mun + 1.086 * R/h$}.  
The inner boundary ($b_1$) is chosen manually to exclude the region 
obviously dominated by the bulge plus bar (shoulder) component.
The outer boundary ($b_4$) is taken where the surface brightness 
profile reaches ($3\sigma$) of the noise level obtained from the
sky ellipse fitting described in \sec\ref{skysub}.
If the profile shows a broken exponential we need an additional 
boundary marking the break. This could be placed by eye, but to 
obtain the break radius objectively we used the following approach. 
At first an approximated derivative of the profile is calculated 
at each point, by fitting a straight line including four points 
around each radius. This provides a measure of a local scalelength 
(\hloc) along the radial axis (\cf \fig\ref{derivative}).  
To get a linear spacing of surface brightness measurements along
the radial axis and to decrease the noise in the \hloc distribution 
we first rebinned the profile to linear units using a spline function 
and then median smoothed the \hlocr profile adapting the smoothing 
length to $5\%$ of the full extent of the profile. 
Our general assumption is now that we have two regions with 
nearly constant scalelength together with a change (up- or downbending) 
in the profile. This will be reflected by two plateaus in the plot
of \hlocr (see \fig\ref{derivative}). Depending on the actual shape 
(\eg sharpness) of the transition region \hlocr will move from 
one scalelength region around \hin to another region around \hout. 
To quantify a single radius defining this break we used then 
the radius where the \hlocr profile crosses the horizontal line set 
by a characteristic value. In case the two regions do not cover 
the same radial range, the characteristic value is always taken 
as the median of the $1/\hloc$ distribution, to avoid a bias 
towards the more extended region.
To characterise the shape of the break we defined an additional 
region around the break. Ideally a sharp break would translate into 
a step function in the \hlocr profile whereas a smooth transition 
would cross the line set by the weighted mean value with a finite 
slope. Therefore we defined an inner and outer boundary of the 
break region where the profile, starting from the break, reaches 
a value within two standard deviations away from the mean calculated 
for the inner/outer region separately (\cf \fig\ref{derivative}). 
The derived break radii are almost always consistent with those 
derived by eye, but of course fail if one of the regions is very
small (\eg in the case of IC\,1067, see Appendix \ref{atlas})
or get uncertain if more than two regions with roughly constant 
scalelengths are involved (as for the mixed classifications, see 
below and \sec\ref{results}). 
Using the boundaries of the break region ($b_2$ and $b_3$) 
to mark the separation of an inner and outer exponential disk,
the final scalelengths are derived from two simple exponential fits 
($\mu (R) = \munin + 1.086 * R/\hin$ and $\mu (R) = \munout + 
1.086 * R/\hout$) to the original profile. 
To decrease the noise we give priority to this classical approach 
to determine the scalelength over using a weighted mean from the 
\hloc values inside the now defined inner and outer region. 
Fitting the two exponentials one could alternatively define 
the break radius, as the meeting point of the two exponential
fits. However, since this would add an influence by the actual 
shape of the two disk regions into the positioning of the break 
radius we decided against it.
In cases where the derivative profile (\hlocr) crosses the 
horizontal line set by the characteristic value twice (three 
or more times is considered to be explained by wiggles caused
by asymmetries such as spiral arms) we check if we could apply 
a mixed classification. This implies assigning two breaks associated 
to the individual types as discussed in \sec\ref{classi}
which significantly improve the overall fit. The boundaries for 
both breaks are then used to fit three exponential regions. 
The final results for all the galaxies are given in Table \ref{resultstab}.
We do not provide individual errors for the fitting parameters in 
Table \ref{resultstab}, since their uncertainty is a complex 
combination of random and systematic errors.   
Typical random errors of the exponential fitting routine alone are 
2\% for \hin, 3\% for \hout, $\pm0.04$mag for \munin, 
and $\pm0.17$mag for \munout.
The uncertainty in the sky subtraction, however, translates to 
a clearly systematic error. In the case of over/under-subtraction 
the measured scalelengths are systematically smaller/larger and 
the central surface brightnesses are systematically lower/higher.
To estimate how the error on the sky subtraction alone effects 
our parameters we have over/under-subtracted the measured sky 
level by $\pm 1\sigma$. Typical errors are then only $\pm1\%$ 
for \hin and $\pm0.02$mag for \munin, but $\pm7\%$ for \hout 
and $\pm0.38$mag for \munout.
Another source of uncertainty is generated by the positioning of 
the boundaries ($b_{1-4}$). This uncertainty can not be easily 
quantified since it depends on the shape (up/down-bending) of, 
and features (bulges, bars, etc.) in, the surface brightness 
profiles and cause systematic errors. To minimise their 
contribution in our determination of the structural 
parameters each fit is verified by eye. 
%
\section{Results}
\label{results}
Our final classifications are presented in Table \ref{resultstab}
and discussed in more detail individually in Appendix \ref{atlas}.    
For four galaxies in common with \cite{erwin2006} the classification 
(as well as the final profiles) agree very well.
All features in the surface brightness profiles classified here 
are always consistently found in deep surface photometry profiles 
available (\cf Appendix \ref{atlas}) in the literature 
\cite[e.g.~][]{courteau1996}.  
About 90\% of our sample can be classified into one of
the classes described in \sec\ref{classi}. However, the 
remaining galaxies are better described with two breaks in 
their surface brightness profiles. Galaxies with extra 
breaks are classified as {\it mixed}. 
In almost all cases this is a combination of a downbending 
break, either \typectc, \typeolrc, \typeabc, or \typetic, 
classified as described in \sec\ref{classi}, followed by an 
additional upbending break in the very outer parts. These 
galaxies are classified as \eg \typeolriiic.  
Interestingly, none of the \typeiii profiles show an additional 
outer truncation within our sensitivity limit. 
Only two galaxies (see detailed explanation in Appendix 
\ref{atlas}) exhibit an inner downbending break followed 
by an additional outer downbending break. In one case 
(NGC\,4517A) the outer sharp drop is clearly associated to a 
\typeabc, whereas the inner break is consistent with being a 
\typeolr or \typectc. 
The other case is NGC\,4210, where the inner boundary of an extended, 
exponential, break region is consistent with a \typeolr break leaving 
the outer break to be \typectc. This galaxy could be the ideal test 
object to study the difference between \typeolr and \typect breaks. 
Finally, for only one galaxy, IC\,1067, the classification in 
the $r^{\prime}$ and $g^{\prime}$-band could be different 
(see \sec\ref{classi}). 
%
%
\begin{table*}
\begin{center}
{\normalsize
\begin{tabular}{ l c | rrrrr r r r r r r r r}
\hline
\rule[+0.4cm]{0mm}{0.0cm}
Galaxy
&\multicolumn{1}{c}{Break}
&\multicolumn{1}{c}{$b_1$}
&\multicolumn{1}{c}{$b_2$}
&\multicolumn{1}{c}{$b_3$}
&\multicolumn{1}{c}{$b_4$}
&\multicolumn{2}{c}{\rbr}
&\multicolumn{2}{c}{\hin}
&\multicolumn{1}{c}{\hin}
&\munin
&\munout
&\mubr
&$A_{r^{\prime}/g^{\prime}}$
\\[+0.1cm]
& \multicolumn{1}{c}{type}
& \multicolumn{1}{c}{[\arcsec]}
& \multicolumn{1}{c}{[\arcsec]}
& \multicolumn{1}{c}{[\arcsec]}
& \multicolumn{1}{c}{[\arcsec]}
& \multicolumn{1}{c}{[\arcsec]}
& \multicolumn{1}{c}{[kpc]}
& \multicolumn{1}{c}{[\arcsec]}
& \multicolumn{1}{c}{[kpc]}
& \multicolumn{1}{c}{[\hout]}
& \multicolumn{1}{c}{[\magsqarcsecfrac]}
& \multicolumn{1}{c}{[\magsqarcsecfrac]}
& \multicolumn{1}{c}{[\magsqarcsecfrac]}
& \multicolumn{1}{c}{[mag] }
\\
\rule[-3mm]{0mm}{5mm}
{\scriptsize{\raisebox{-0.7ex}{\it (1)}}}
&\multicolumn{1}{c}{{\scriptsize{\raisebox{-0.7ex}{\it (2)}}}}
&\multicolumn{4}{c}{{\scriptsize{\raisebox{-0.7ex}{\it (3)}}}}
&\multicolumn{2}{c}{{\scriptsize{\raisebox{-0.7ex}{\it (4)}}}}
&\multicolumn{2}{c}{{\scriptsize{\raisebox{-0.7ex}{\it (5)}}}}
&\multicolumn{1}{c}{{\scriptsize{\raisebox{-0.7ex}{\it (6)}}}}
&\multicolumn{2}{c}{{\scriptsize{\raisebox{-0.7ex}{\it (7)}}}}
&\multicolumn{1}{c}{{\scriptsize{\raisebox{-0.7ex}{\it (8)}}}}
&\multicolumn{1}{c}{{\scriptsize{\raisebox{-0.7ex}{\it (9)}}}}
\\
\hline\hline \\[-0.4cm]
\raisebox{-0.7ex}{NGC\,0450}   &   \raisebox{-0.7ex}{II.o-CT}   &   16&   79&   88&  120&   80&       9.5&   31.3&     3.7&  2.7 &21.0 &16.0 &24.0    &0.11 \\[-0.09cm] 
            &        &   16&   78&   89&  128&   79&       9.4&   34.5&     4.1&  2.9 &21.5 &16.3 &24.2    &0.15 \\[+0.02cm] 
\raisebox{-0.7ex}{PGC\,006667} &   \raisebox{-0.7ex}{II-AB}   &   10&   70&   77&  108&   74&       9.7&   22.6&     2.9&  1.9 &21.1 &18.0 &24.7    &0.14 \\[-0.09cm] 
            &        &   10&   72&   78&   96&   73&       9.5&   22.7&     3.0&  1.8 &21.5 &18.7 &25.1    &0.19 \\[+0.02cm] 
\raisebox{-0.7ex}{NGC\,0701}   &  \raisebox{-0.7ex}{II-i}  &   19&  129&    *&    *&    *&         *&   15.4&     1.8&    * &19.0 & 0.0 &19.0    &0.07 \\[-0.09cm] 
            &        &   19&  115&    *&    *&    *&         *&   16.2&     1.9&    * &19.8 & 0.0 &19.8    &0.10 \\[+0.02cm] 
\raisebox{-0.7ex}{NGC\,0853}   &   \raisebox{-0.7ex}{III}  &    0&   43&   48&  100&   45&       4.4&   11.0&     1.1&  0.6 &19.3 &20.9 &23.3    &0.07 \\[-0.09cm] 
            &        &    0&   42&   46&  100&   44&       4.3&   10.3&     1.0&  0.6 &19.6 &21.2 &23.7    &0.10 \\[+0.01cm] 
\raisebox{-0.7ex}{NGC\,0941}   &   \raisebox{-0.7ex}{II-CT}   &    5&   77&   84&  120&   81&       8.6&   19.1&     2.0&  1.4 &20.4 &18.5 &25.0    &0.10 \\[-0.09cm] 
            &        &    5&   76&   80&  107&   77&       8.2&   19.5&     2.1&  1.6 &20.8 &18.0 &25.3    &0.13 \\[+0.01cm] 
\raisebox{-0.7ex}{UGC\,02081}  &   \raisebox{-0.7ex}{II-CT}   &   11&   48&   69&   85&   53&       9.4&   21.0&     3.7&  2.6 &21.6 &16.1 &25.2    &0.07 \\[-0.09cm] 
            &        &   11&   45&   74&   85&   46&       8.1&   22.3&     3.9&  2.8 &22.2 &15.9 &25.6    &0.10 \\[+0.01cm] 
\raisebox{-0.7ex}{NGC\,1042}   &   \raisebox{-0.7ex}{II-AB}   &   39&  122&  122&  219&  122&      10.7&   48.0&     4.2&  1.4 &20.9 &19.9 &23.0    &0.08 \\[-0.09cm] 
            &        &   39&  101&  119&  212&  119&      10.4&   63.9&     5.6&  2.2 &21.7 &19.7 &23.4    &0.11 \\[+0.01cm] 
\raisebox{-0.7ex}{NGC\,1068}   &   \raisebox{-0.7ex}{II.o-OLR}  &  105&  184&  188&  362&  188&      13.9&  118.0&     8.7&  2.0 &21.6 &19.9 &23.4    &0.09 \\[-0.09cm] 
            &        &  105&  183&  187&  384&  189&      14.0&  128.2&     9.5&  2.2 &22.4 &20.4 &24.0    &0.13 \\[+0.01cm] 
\raisebox{-0.7ex}{NGC\,1084}   &   \raisebox{-0.7ex}{III}  &   27&   91&   97&  179&   95&       8.5&   18.5&     1.7&  0.4 &18.3 &21.4 &23.2    &0.08 \\[-0.09cm] 
            &        &   27&   91&  100&  194&   96&       8.6&   17.9&     1.6&  0.4 &18.7 &21.7 &23.7    &0.11 \\[+0.01cm] 
\raisebox{-0.7ex}{NGC\,1087}   &   \raisebox{-0.7ex}{III}  &   49&  139&  142&  238&  140&      14.0&   24.5&     2.4&  0.6 &19.0 &21.4 &24.7    &0.10 \\[-0.09cm] 
            &        &   49&  137&  142&  231&  139&      13.9&   23.6&     2.4&  0.6 &19.4 &22.0 &25.4    &0.13 \\[+0.01cm] 
\raisebox{-0.7ex}{NGC\,1299}   &   \raisebox{-0.7ex}{III}  &   22&   45&   46&   75&   46&       7.0&    8.7&     1.3&  0.7 &18.9 &20.5 &24.5    &0.13 \\[-0.09cm] 
            &        &   22&   47&   52&   77&   49&       7.5&    8.6&     1.3&  0.5 &19.4 &22.5 &25.4    &0.18 \\[+0.01cm] 
\raisebox{-0.7ex}{NGC\,2543}   &  \raisebox{-0.7ex}{II-i}  &   34&  108&    *&    *&   74&      13.3&   15.7&     2.8&    * &19.1 & 0.0 &19.1    &0.19 \\[-0.09cm] 
            &        &   34&  108&    *&    *&   75&      13.5&   14.3&     2.6&    * &19.3 & 0.0 &19.3    &0.27 \\[+0.01cm] 
\raisebox{-0.7ex}{NGC\,2541}   &   \raisebox{-0.7ex}{II-CT}   &   48&  141&  143&  216&  142&       7.2&   51.2&     2.6&  1.6 &21.3 &19.3 &24.5    &0.14 \\[-0.09cm] 
            &        &   48&  143&  147&  223&  145&       7.4&   55.2&     2.8&  2.1 &21.9 &18.6 &25.0    &0.19 \\[+0.01cm] 
\raisebox{-0.7ex}{UGC\,04393}  &   \raisebox{-0.7ex}{I}     &   30&  125&    *&    *&    *&         *&   22.4&     3.6&    * &21.6 & 0.0 &21.6    &0.12 \\[-0.09cm] 
            &        &   30&  114&    *&    *&    *&         *&   21.7&     3.4&    * &21.9 & 0.0 &21.9    &0.16 \\[+0.01cm] 
\raisebox{-0.7ex}{NGC\,2684}   &   \raisebox{-0.7ex}{II-CT}   &    4&   23&   25&   42&   24&       5.1&   10.6&     2.2&  1.9 &19.6 &17.5 &22.1    &0.07 \\[-0.09cm] 
            &        &    4&   23&   25&   42&   24&       5.1&   10.8&     2.3&  2.0 &20.2 &18.0 &22.4    &0.10 \\[+0.01cm] 
                               &                           &   16&   37&   42&   71&   38&       6.7&   24.1&     4.2&  2.9 &20.2 &16.5 &22.1    &0.05 \\[-0.09cm] 
\raisebox{-0.7ex}{NGC\,2701}            &   \raisebox{-0.7ex}{II-AB + III}     &   42&   71&   71&   93&   73&      12.8&    8.3&     1.4&  0.4 &16.5 &22.4 &25.8    &0.05 \\[+0.01cm] 
            &        &   16&   37&   41&   74&   38&       6.7&   27.8&     4.9&  3.3 &20.7 &17.1 &21.9    &0.07 \\[-0.09cm] 
            &        &   41&   74&   76&  102&   75&      13.1&    8.5&     1.5&  0.4 &17.1 &22.1 &26.1    &0.07 \\[+0.01cm] 
\raisebox{-0.7ex}{NGC\,2776}   &   \raisebox{-0.7ex}{I}     &   29&  149&    *&    *&    *&         *&   19.6&     3.8&    * &20.0 & 0.0 &20.0    &0.04 \\[-0.09cm] 
            &        &   29&  134&    *&    *&    *&         *&   20.1&     3.9&    * &20.5 & 0.0 &20.5    &0.05 \\[+0.01cm] 
\raisebox{-0.7ex}{NGC\,2967}   &   \raisebox{-0.7ex}{III-d}  &   10&   59&   90&  156&   73&       9.4&   16.3&     2.1&  0.5 &19.4 &22.0 &24.8    &0.28 \\[-0.09cm] 
            &        &   10&   59&   90&  143&   73&       9.4&   16.9&     2.2&  0.5 &20.1 &22.6 &25.3    &0.38 \\[+0.01cm] 
\raisebox{-0.7ex}{NGC\,3055}   &   \raisebox{-0.7ex}{II.o-CT}   &   10&   53&   57&   88&   54&       6.8&   17.1&     2.1&  1.9 &19.6 &16.3 &23.1    &0.14 \\[-0.09cm] 
            &        &   10&   52&   56&   88&   53&       6.7&   17.5&     2.2&  2.0 &20.1 &16.7 &23.4    &0.20 \\[+0.01cm] 
\raisebox{-0.7ex}{NGC\,3246}   &   \raisebox{-0.7ex}{II-AB}   &    7&   59&   64&   96&   61&       9.0&   20.2&     3.0&  2.0 &20.8 &17.1 &24.3    &0.10 \\[-0.09cm] 
            &        &    7&   60&   62&   93&   60&       8.9&   22.0&     3.3&  2.2 &21.3 &17.3 &24.5    &0.14 \\[+0.01cm] 
\raisebox{-0.7ex}{NGC\,3259}   &   \raisebox{-0.7ex}{III-d}  &    7&   59&   64&  122&   63&       8.4&   11.4&     1.5&  0.4 &19.4 &22.8 &24.7    &0.04 \\[-0.09cm] 
            &        &    7&   59&   63&  138&   63&       8.4&   12.2&     1.6&  0.4 &20.0 &23.1 &25.0    &0.05 \\[+0.01cm] 
\raisebox{-0.7ex}{NGC\,3310}   &   \raisebox{-0.7ex}{III}  &   10&   52&   53&  172&   53&       4.4&   10.6&     0.9&  0.3 &18.2 &21.6 &23.2    &0.06 \\[-0.09cm] 
            &        &   10&   52&   54&  141&   53&       4.4&   10.3&     0.9&  0.3 &18.4 &21.7 &23.3    &0.09 \\[+0.01cm] 
\raisebox{-0.7ex}{NGC\,3359}   &   \raisebox{-0.7ex}{II-AB}   &   29&  183&  204&  266&  184&      16.1&   51.1&     4.5&  1.8 &20.5 &16.9 &24.9    &0.03 \\[-0.09cm] 
            &        &   29&  180&  192&  266&  181&      15.8&   53.0&     4.6&  1.8 &20.9 &17.7 &25.0    &0.04 \\[+0.01cm] 
\raisebox{-0.7ex}{NGC\,3423}   &   \raisebox{-0.7ex}{II-CT}   &   30&   76&   93&  175&   85&       6.1&   52.2&     3.7&  2.8 &20.9 &17.7 &22.6    &0.08 \\[-0.09cm] 
            &        &   30&   88&   97&  160&   95&       6.8&   51.2&     3.7&  3.2 &21.2 &17.1 &23.1    &0.12 \\[+0.01cm] 
\raisebox{-0.7ex}{NGC\,3488}   &   \raisebox{-0.7ex}{II.o-CT}   &   10&   32&   52&   70&   44&       9.8&   16.8&     3.8&  2.5 &20.3 &16.0 &23.2    &0.04 \\[-0.09cm] 
            &        &   10&   29&   53&   72&   43&       9.6&   20.6&     4.6&  3.0 &21.0 &16.3 &23.3    &0.05 \\[+0.01cm] 
\raisebox{-0.7ex}{NGC\,3583}   &   \raisebox{-0.7ex}{III-d}  &   13&   82&   84&  147&   83&      13.5&   16.3&     2.7&  0.7 &19.0 &21.0 &24.7    &0.04 \\[-0.09cm] 
            &        &   13&   81&   82&  138&   82&      13.3&   17.0&     2.8&  0.7 &19.7 &21.4 &25.7    &0.05 \\[+0.01cm] 
\hline
\end{tabular}
}
\caption[]{Results: disk type and exponential disk parameters \newline
{\scriptsize{\it (1)}} Principal name in LEDA, 
{\scriptsize{\it (2)}} profile classification (\cf\sec\ref{classi} ),  
{\scriptsize{\it (3)}} fitting boundaries for the inner and outer exponential
disk region, 
{\scriptsize{\it (4)}} break radius in units of arcsec and kpc, 
{\scriptsize{\it (5)}} inner scalelength in units of arcsec and kpc
{\scriptsize{\it (6)}} inner scalelength in relation to the outer scalelength,    
{\scriptsize{\it (7)}} the central surface brightness of the inner/outer disk,
{\scriptsize{\it (8)}} the surface brightness at the break radius (estimated 
at the crossing point of the two exponential fits),  
{\scriptsize{\it (9)}} galactic extinction according to \cite{schlegel} 
For each galaxy two rows of values are given. The results obtained for 
the $r^{\prime}$ band in the upper row, for the $g^{\prime}$ band in 
the lower row. For galaxies with mixed classification both disk fits 
are shown. 
\label{resultstab} 
}
\end{center}
\end{table*}
\addtocounter{table}{-1}
\begin{table*}
\begin{center}
{\normalsize
\begin{tabular}{ l c | rrrrr r r r r r r r r}
\hline
\rule[+0.4cm]{0mm}{0.0cm}
Galaxy
&\multicolumn{1}{c}{Break}
&\multicolumn{1}{c}{$b_1$}
&\multicolumn{1}{c}{$b_2$}
&\multicolumn{1}{c}{$b_3$}
&\multicolumn{1}{c}{$b_4$}
&\multicolumn{2}{c}{\rbr}
&\multicolumn{2}{c}{\hin}
&\multicolumn{1}{c}{\hin}
&\munin
&\munout
&\mubr
&$A_{r^{\prime}/g^{\prime}}$
\\[+0.1cm]
& \multicolumn{1}{c}{type}
& \multicolumn{1}{c}{[\arcsec]}
& \multicolumn{1}{c}{[\arcsec]}
& \multicolumn{1}{c}{[\arcsec]}
& \multicolumn{1}{c}{[\arcsec]}
& \multicolumn{1}{c}{[\arcsec]}
& \multicolumn{1}{c}{[kpc]}
& \multicolumn{1}{c}{[\arcsec]}
& \multicolumn{1}{c}{[kpc]}
& \multicolumn{1}{c}{[\hout]}
& \multicolumn{1}{c}{[\magsqarcsecfrac]}
& \multicolumn{1}{c}{[\magsqarcsecfrac]}
& \multicolumn{1}{c}{[\magsqarcsecfrac]}
& \multicolumn{1}{c}{[mag] }
\\
\rule[-3mm]{0mm}{5mm}
{\scriptsize{\raisebox{-0.7ex}{\it (1)}}}
&\multicolumn{1}{c}{{\scriptsize{\raisebox{-0.7ex}{\it (2)}}}}
&\multicolumn{4}{c}{{\scriptsize{\raisebox{-0.7ex}{\it (3)}}}}
&\multicolumn{2}{c}{{\scriptsize{\raisebox{-0.7ex}{\it (4)}}}}
&\multicolumn{2}{c}{{\scriptsize{\raisebox{-0.7ex}{\it (5)}}}}
&\multicolumn{1}{c}{{\scriptsize{\raisebox{-0.7ex}{\it (6)}}}}
&\multicolumn{2}{c}{{\scriptsize{\raisebox{-0.7ex}{\it (7)}}}}
&\multicolumn{1}{c}{{\scriptsize{\raisebox{-0.7ex}{\it (8)}}}}
&\multicolumn{1}{c}{{\scriptsize{\raisebox{-0.7ex}{\it (9)}}}}
\\
\hline\hline \\[-0.4cm]
\raisebox{-0.7ex}{NGC\,3589}   &   \raisebox{-0.7ex}{II-CT}   &    0&   37&   38&   72&   38&       5.8&   18.6&     2.9&  1.7 &21.1 &19.7 &23.1    &0.03 \\[-0.09cm] 
            &        &    0&   33&   38&   84&   36&       5.5&   20.5&     3.1&  1.8 &21.5 &20.3 &23.0    &0.04 \\[+0.01cm] 
\raisebox{-0.7ex}{UGC\,06309}  &  \raisebox{-0.7ex}{II-i}  &   23&   74&    *&    *&    *&         *&    9.8&     2.1&    * &18.9 & 0.0 &18.9    &0.05 \\[-0.09cm] 
            &        &   25&   88&    *&    *&    *&         *&   11.1&     2.4&    * &19.9 & 0.0 &19.9    &0.07 \\[+0.01cm] 
\raisebox{-0.7ex}{NGC\,3631}   &   \raisebox{-0.7ex}{I}     &  140&  233&    *&    *&    *&         *&   37.8&     3.6&    * &20.8 & 0.0 &20.8    &0.05 \\[-0.09cm] 
            &        &  140&  226&    *&    *&    *&         *&   36.7&     3.5&    * &21.2 & 0.0 &21.2    &0.06 \\[+0.01cm] 
\raisebox{-0.7ex}{NGC\,3642}   &   \raisebox{-0.7ex}{III-d}  &   30&   75&   79&  229&   79&      10.0&   20.9&     2.6&  0.5 &20.3 &22.3 &24.0    &0.03 \\[-0.09cm] 
            &        &   30&   71&   73&  236&   73&       9.3&   19.7&     2.5&  0.4 &20.6 &22.7 &24.3    &0.04 \\[+0.01cm] 
\raisebox{-0.7ex}{UGC\,06518}  &  \raisebox{-0.7ex}{II-i}  &   10&   43&    *&    *&    *&         *&    6.4&     1.4&    * &19.3 & 0.0 &19.3    &0.04 \\[-0.09cm] 
            &        &   10&   45&    *&    *&    *&         *&    6.1&     1.3&    * &19.6 & 0.0 &19.6    &0.05 \\[+0.01cm] 
\raisebox{-0.7ex}{NGC\,3756}   &   \raisebox{-0.7ex}{II.o-CT}   &   11&   92&  103&  154&   97&      10.2&   33.7&     3.6&  2.0 &20.0 &16.6 &23.2    &0.03 \\[-0.09cm] 
            &        &   11&   78&  104&  149&  100&      10.6&   38.6&     4.1&  2.2 &20.8 &17.7 &23.4    &0.04 \\[+0.01cm] 
\raisebox{-0.7ex}{NGC\,3888}   &   \raisebox{-0.7ex}{I}     &   13&   81&    *&    *&    *&         *&   10.1&     1.9&    * &18.7 & 0.0 &18.7    &0.03 \\[-0.09cm] 
            &        &   13&   84&    *&    *&    *&         *&    9.9&     1.8&    * &19.1 & 0.0 &19.1    &0.04 \\[+0.01cm] 
\raisebox{-0.7ex}{NGC\,3893}   &   \raisebox{-0.7ex}{III-d}  &   15&  134&  134&  184&  134&      11.1&   29.1&     2.4&  0.6 &19.2 &21.7 &24.7    &0.06 \\[-0.09cm] 
            &        &   15&  139&  139&  184&  139&      11.5&   28.3&     2.3&  0.5 &19.6 &22.6 &25.5    &0.08 \\[+0.01cm] 
\raisebox{-0.7ex}{UGC\,06903}  &   \raisebox{-0.7ex}{II.o-CT}   &   21&   59&   72&   98&   60&       8.0&   31.8&     4.2&  3.6 &21.6 &15.8 &23.9    &0.06 \\[-0.09cm] 
            &        &   21&   56&   63&   92&   57&       7.6&   35.0&     4.6&  3.6 &22.2 &17.1 &24.2    &0.08 \\[+0.01cm] 
\raisebox{-0.7ex}{NGC\,3982}   &   \raisebox{-0.7ex}{III}  &   13&   51&   54&   91&   52&       4.9&   10.2&     1.0&  0.7 &18.3 &19.8 &23.8    &0.04 \\[-0.09cm] 
            &        &   13&   49&   51&   99&   50&       4.7&   10.2&     1.0&  0.8 &18.7 &19.8 &23.6    &0.05 \\[+0.01cm] 
\raisebox{-0.7ex}{NGC\,3992}   &   \raisebox{-0.7ex}{II.o-OLR}  &   44&  133&  139&  324&  137&      12.2&   84.0&     7.5&  2.3 &20.5 &18.1 &22.3    &0.08 \\[-0.09cm] 
            &        &   44&  129&  133&  280&  133&      11.8&   96.4&     8.6&  2.7 &21.3 &18.4 &22.9    &0.12 \\[+0.01cm] 
\raisebox{-0.7ex}{NGC\,4030}   &   \raisebox{-0.7ex}{III}  &   37&  147&  164&  266&  152&      15.5&   26.2&     2.7&  0.5 &19.2 &22.1 &25.4    &0.08 \\[-0.09cm] 
            &        &   37&  147&  151&  230&  148&      15.1&   25.5&     2.6&  0.6 &19.6 &22.3 &26.0    &0.11 \\[+0.01cm] 
\raisebox{-0.7ex}{NGC\,4041}   &   \raisebox{-0.7ex}{III-d}  &   26&   76&   80&  192&   80&       8.2&   18.6&     1.9&  0.6 &19.7 &21.8 &24.6    &0.05 \\[-0.09cm] 
            &        &   26&   81&   84&  186&   83&       8.5&   18.2&     1.9&  0.5 &20.1 &22.6 &25.2    &0.07 \\[+0.01cm] 

   & &   40&   57&   57&  116&   57&       4.3&   23.3&     1.7&  1.5 &19.1 &17.8 &21.7    &0.06 \\[-0.09cm] 
\raisebox{-0.7ex}{NGC\,4102}            & \raisebox{-0.7ex}{II.o-OLR + III}       &   57&  116&  119&  165&  118&       8.8&   15.8&     1.2&  0.6 &17.8 &20.9 &26.0    &0.06 \\[+0.01cm] 
            &        &   40&   59&   59&  107&   59&       4.4&   21.6&     1.6&  1.4 &19.5 &18.4 &22.4    &0.08 \\[-0.09cm] 
            &        &   59&  107&  109&  151&  108&       8.1&   15.7&     1.2&  0.7 &18.4 &20.4 &26.1    &0.08 \\[+0.01cm] 

\raisebox{-0.7ex}{NGC\,4108}   &   \raisebox{-0.7ex}{II-AB}   &    7&   44&   46&   66&   44&       8.6&    9.1&     1.8&  1.5 &18.9 &16.2 &24.5    &0.05 \\[-0.09cm] 
            &        &    7&   43&   45&   68&   44&       8.6&    9.8&     1.9&  1.6 &19.5 &16.5 &24.5    &0.07 \\[+0.01cm] 
\raisebox{-0.7ex}{NGC\,4108B}  &   \raisebox{-0.7ex}{I}     &    7&   67&    *&    *&    *&         *&   11.4&     2.2&    * &20.8 & 0.0 &20.8    &0.05 \\[-0.09cm] 
            &        &    7&   71&    *&    *&    *&         *&   11.0&     2.2&    * &21.0 & 0.0 &21.0    &0.07 \\[+0.01cm] 
\raisebox{-0.7ex}{NGC\,4123}   &   \raisebox{-0.7ex}{I}     &   20&  220&    *&    *&    *&         *&   35.5&     3.4&    * &20.6 & 0.0 &20.6    &0.06 \\[-0.09cm] 
            &        &   20&  207&    *&    *&    *&         *&   36.5&     3.5&    * &21.2 & 0.0 &21.2    &0.08 \\[+0.01cm] 
   &  &   15&   35&   37&   50&   36&       7.5&   16.3&     3.4&  1.6 &19.9 &18.6 &22.1    &0.05 \\[-0.09cm] 
\raisebox{-0.7ex}{NGC\,4210}    &  \raisebox{-0.7ex}{II.o-OLR + CT}      &   37&   50&   55&   76&   51&      10.6&   10.2&     2.1&  1.5 &18.6 &15.6 &24.1    &0.05 \\[+0.01cm] 
            &       &   15&   32&   38&   50&   37&       7.7&   19.5&     4.1&  2.0 &20.7 &19.0 &22.4    &0.07 \\[-0.09cm] 
            &        &   38&   50&   52&   73&   51&      10.6&   10.0&     2.1&  1.6 &19.0 &15.6 &24.9    &0.07 \\[+0.01cm] 
  &  &   12&   43&   48&   78&   47&       7.9&   24.5&     4.1&  2.1 &19.8 &17.6 &21.7    &0.05 \\[-0.09cm] 
\raisebox{-0.7ex}{NGC\,4273}             &  \raisebox{-0.7ex}{II-AB + III}      &   48&   78&   79&  119&   78&      13.2&   11.5&     1.9&  0.6 &17.6 &20.3 &25.4    &0.05 \\[+0.01cm] 
            &        &   12&   45&   47&   84&   47&       7.9&   29.7&     5.0&  2.6 &20.5 &17.9 &22.1    &0.07 \\[-0.09cm] 
            &        &   47&   84&   89&  106&   84&      14.2&   11.2&     1.9&  0.7 &17.9 &20.0 &25.6    &0.07 \\[+0.01cm] 
\raisebox{-0.7ex}{NGC\,4480}   &   \raisebox{-0.7ex}{II.o-CT}   &    8&   53&   58&   90&   54&       9.3&   18.8&     3.2&  2.3 &19.8 &15.8 &23.0    &0.06 \\[-0.09cm] 
            &        &    8&   54&   57&   88&   54&       9.3&   21.4&     3.7&  2.6 &20.5 &16.1 &23.3    &0.09 \\[+0.01cm] 
  &   &   25&   76&   77&  139&   77&       8.3&   53.3&     5.8&  1.3 &22.0 &21.4 &23.6    &0.06 \\[-0.09cm] 
\raisebox{-0.7ex}{NGC\,4517A}            &  \raisebox{-0.7ex}{II.o-CT + AB}      &   77&  139&  159&  192&  140&      15.1&   39.7&     4.3&  2.3 &21.4 &15.7 &25.7    &0.06 \\[+0.01cm] 
            &        &   25&   76&   77&  139&   77&       8.3&   59.3&     6.4&  1.4 &22.5 &21.8 &23.9    &0.09 \\[-0.09cm] 
            &        &   77&  139&  159&  192&  140&      15.1&   41.1&     4.4&  2.5 &21.8 &15.5 &26.0    &0.09 \\[+0.01cm] 
\raisebox{-0.7ex}{UGC\,07700}  &   \raisebox{-0.7ex}{II.o-CT}   &   12&   45&   52&   94&   50&      11.2&   26.4&     5.9&  2.2 &22.2 &20.2 &23.8    &0.05 \\[-0.09cm] 
            &        &   12&   49&   58&   74&   50&      11.2&   29.1&     6.5&  3.7 &22.6 &17.5 &24.5    &0.07 \\[+0.01cm] 
\raisebox{-0.7ex}{NGC\,4545}   &   \raisebox{-0.7ex}{II.o-CT}   &    7&   56&   58&   97&   56&      11.6&   17.7&     3.7&  2.0 &20.0 &16.3 &23.8    &0.03 \\[-0.09cm] 
            &        &    7&   49&   57&   89&   55&      11.4&   19.2&     4.0&  2.1 &20.6 &16.7 &24.0    &0.04 \\[+0.01cm] 
\raisebox{-0.7ex}{NGC\,4653}   &   \raisebox{-0.7ex}{II-AB}   &   11&  102&  109&  142&  103&      19.0&   23.0&     4.2&  1.7 &20.7 &17.1 &25.9    &0.06 \\[-0.09cm] 
            &        &   11&  101&  107&  130&  102&      18.8&   24.5&     4.5&  1.8 &21.2 &17.1 &26.0    &0.09 \\[+0.01cm] 
\hline
\end{tabular}
}
\caption[]{(continued): Results
}
\end{center}
\end{table*}
\addtocounter{table}{-1}
\begin{table*}
\begin{center}
{\normalsize
\begin{tabular}{ l c | rrrrr r r r r r r r r}
\hline
\rule[+0.4cm]{0mm}{0.0cm}
Galaxy
&\multicolumn{1}{c}{Break}
&\multicolumn{1}{c}{$b_1$}
&\multicolumn{1}{c}{$b_2$}
&\multicolumn{1}{c}{$b_3$}
&\multicolumn{1}{c}{$b_4$}
&\multicolumn{2}{c}{\rbr}
&\multicolumn{2}{c}{\hin}
&\multicolumn{1}{c}{\hin}
&\munin
&\munout
&\mubr
&$A_{r^{\prime}/g^{\prime}}$
\\[+0.1cm]
& \multicolumn{1}{c}{type}
& \multicolumn{1}{c}{[\arcsec]}
& \multicolumn{1}{c}{[\arcsec]}
& \multicolumn{1}{c}{[\arcsec]}
& \multicolumn{1}{c}{[\arcsec]}
& \multicolumn{1}{c}{[\arcsec]}
& \multicolumn{1}{c}{[kpc]}
& \multicolumn{1}{c}{[\arcsec]}
& \multicolumn{1}{c}{[kpc]}
& \multicolumn{1}{c}{[\hout]}
& \multicolumn{1}{c}{[\magsqarcsecfrac]}
& \multicolumn{1}{c}{[\magsqarcsecfrac]}
& \multicolumn{1}{c}{[\magsqarcsecfrac]}
& \multicolumn{1}{c}{[mag] }
\\
\rule[-3mm]{0mm}{5mm}
{\scriptsize{\raisebox{-0.7ex}{\it (1)}}}
&\multicolumn{1}{c}{{\scriptsize{\raisebox{-0.7ex}{\it (2)}}}}
&\multicolumn{4}{c}{{\scriptsize{\raisebox{-0.7ex}{\it (3)}}}}
&\multicolumn{2}{c}{{\scriptsize{\raisebox{-0.7ex}{\it (4)}}}}
&\multicolumn{2}{c}{{\scriptsize{\raisebox{-0.7ex}{\it (5)}}}}
&\multicolumn{1}{c}{{\scriptsize{\raisebox{-0.7ex}{\it (6)}}}}
&\multicolumn{2}{c}{{\scriptsize{\raisebox{-0.7ex}{\it (7)}}}}
&\multicolumn{1}{c}{{\scriptsize{\raisebox{-0.7ex}{\it (8)}}}}
&\multicolumn{1}{c}{{\scriptsize{\raisebox{-0.7ex}{\it (9)}}}}
\\
\hline\hline \\[-0.4cm]
\raisebox{-0.7ex}{NGC\,4668}   &   \raisebox{-0.7ex}{III}  &    0&   60&   71&  110&   66&       7.6&   11.8&     1.4&  0.5 &19.6 &22.2 &25.5    &0.07 \\[-0.09cm] 
            &        &    0&   65&   71&  104&   68&       7.8&   11.7&     1.3&  0.5 &20.0 &23.2 &25.9    &0.10 \\[+0.01cm] 
\raisebox{-0.7ex}{UGC\,08041}  &   \raisebox{-0.7ex}{II.o-CT}   &   14&   69&   73&  148&   71&       6.8&   36.9&     3.5&  1.4 &21.1 &20.2 &23.1    &0.06 \\[-0.09cm] 
            &        &   14&   69&   75&  157&   70&       6.7&   39.9&     3.8&  1.6 &21.6 &20.5 &23.4    &0.09 \\[+0.01cm] 
\raisebox{-0.7ex}{UGC\,08084}  &   \raisebox{-0.7ex}{II.o-OLR}  &    9&   38&   44&   71&   42&       8.2&   29.0&     5.7&  3.4 &22.4 &18.8 &23.9    &0.08 \\[-0.09cm] 
            &        &    9&   35&   45&   78&   43&       8.4&   31.3&     6.1&  3.9 &22.8 &18.7 &24.2    &0.11 \\[+0.01cm] 
   & &   13&   38&   48&  103&   39&       3.3&   27.6&     2.3&  2.2 &20.2 &18.2 &21.9    &0.07 \\[-0.09cm] 
\raisebox{-0.7ex}{NGC\,4904}            &  \raisebox{-0.7ex}{II.o-OLR + III}      &   48&  103&  116&  157&  104&       8.7&   12.5&     1.0&  0.5 &18.2 &22.4 &26.2    &0.07 \\[+0.01cm] 
            &        &   13&   40&   40&   99&   40&       3.4&   32.7&     2.7&  2.6 &20.9 &18.8 &22.2    &0.10 \\[-0.09cm] 
            &        &   40&   99&   99&  135&   99&       8.3&   12.7&     1.1&  0.3 &18.8 &25.1 &27.4    &0.10 \\[+0.01cm] 
\raisebox{-0.7ex}{UGC\,08237}  &   \raisebox{-0.7ex}{II.o-OLR}  &   14&   25&   29&   50&   26&       5.6&   14.8&     3.2&  2.6 &20.6 &17.8 &22.4    &0.05 \\[-0.09cm] 
            &        &   14&   27&   31&   48&   30&       6.5&   14.0&     3.0&  2.5 &21.1 &18.4 &23.0    &0.06 \\[+0.01cm] 
   & &    5&   29&   32&   72&   30&       2.4&   20.3&     1.6&  1.8 &19.8 &18.3 &21.7    &0.08 \\[-0.09cm] 
  \raisebox{-0.7ex}{NGC\,5147}          &   \raisebox{-0.7ex}{II.o-OLR + III}     &   32&   72&   77&  122&   73&       5.8&   11.0&     0.9&  0.5 &18.3 &22.0 &25.8    &0.08 \\[+0.01cm] 
            &        &    5&   26&   29&   75&   29&       2.3&   22.8&     1.8&  2.2 &20.4 &18.5 &22.0    &0.11 \\[-0.09cm] 
            &        &   29&   75&   79&  108&   76&       6.1&   10.5&     0.8&  0.6 &18.5 &21.6 &26.2    &0.11 \\[+0.01cm] 
\raisebox{-0.7ex}{UGC\,08658}  &   \raisebox{-0.7ex}{II-CT}   &   12&   57&   61&  102&   61&       9.7&   22.4&     3.5&  1.4 &21.0 &20.0 &23.9    &0.05 \\[-0.09cm] 
            &        &   12&   57&   61&  122&   61&       9.7&   24.4&     3.9&  1.5 &21.6 &20.3 &24.4    &0.06 \\[+0.01cm] 
\raisebox{-0.7ex}{NGC\,5300}   &   \raisebox{-0.7ex}{II.o-CT}   &   12&   80&   85&  155&   83&       7.2&   37.1&     3.2&  2.1 &20.8 &18.2 &23.2    &0.06 \\[-0.09cm] 
            &        &   12&   75&   82&  142&   77&       6.6&   41.8&     3.6&  2.6 &21.4 &18.1 &23.5    &0.09 \\[+0.01cm] 
\raisebox{-0.7ex}{NGC\,5334}   &   \raisebox{-0.7ex}{II.o-CT}   &   13&   80&  102&  152&   92&       9.1&   40.2&     4.0&  2.6 &21.0 &16.7 &23.8    &0.13 \\[-0.09cm] 
            &        &   13&  101&  103&  156&  103&      10.2&   40.3&     4.0&  2.8 &21.5 &16.4 &24.4    &0.18 \\[+0.01cm] 
\raisebox{-0.7ex}{NGC\,5376}   &   \raisebox{-0.7ex}{II.o-OLR}  &   10&   34&   35&   97&   35&       5.6&   17.0&     2.7&  1.4 &19.4 &18.6 &21.2    &0.04 \\[-0.09cm] 
            &        &   10&   34&   39&   95&   37&       5.9&   17.9&     2.8&  1.7 &20.1 &18.7 &22.2    &0.05 \\[+0.01cm] 
\raisebox{-0.7ex}{NGC\,5430}   &   \raisebox{-0.7ex}{II.o-OLR}  &   10&   69&   72&  105&   70&      15.7&   15.8&     3.5&  1.4 &19.4 &17.8 &23.9    &0.05 \\[-0.09cm] 
            &        &   10&   69&   71&  102&   69&      15.5&   16.2&     3.6&  1.6 &20.1 &17.6 &24.4    &0.06 \\[+0.01cm] 
\raisebox{-0.7ex}{NGC\,5480}   &   \raisebox{-0.7ex}{III-d}  &    9&   65&   74&  129&   67&       9.8&   13.9&     2.0&  0.5 &19.6 &22.1 &24.9    &0.05 \\[-0.09cm] 
            &        &    9&   67&   76&  122&   68&      10.0&   13.5&     2.0&  0.5 &20.1 &22.6 &25.4    &0.07 \\[+0.01cm] 
\raisebox{-0.7ex}{NGC\,5584}   &   \raisebox{-0.7ex}{II.o-CT}   &   14&   75&  107&  165&   90&      10.6&   39.8&     4.7&  2.0 &20.9 &18.4 &23.3    &0.11 \\[-0.09cm] 
            &        &   14&   68&   92&  155&   84&       9.9&   45.4&     5.4&  2.4 &21.5 &18.7 &23.4    &0.15 \\[+0.01cm] 
\raisebox{-0.7ex}{NGC\,5624}   &   \raisebox{-0.7ex}{III}  &    0&   46&   50&   87&   48&       7.3&    9.9&     1.5&  0.5 &20.1 &22.6 &24.9    &0.05 \\[-0.09cm] 
            &        &    0&   37&   47&   80&   42&       6.4&    8.9&     1.3&  0.4 &20.3 &23.2 &25.4    &0.07 \\[+0.01cm] 
\raisebox{-0.7ex}{NGC\,5660}   &   \raisebox{-0.7ex}{II-CT}   &   12&   63&   70&   90&   67&      12.0&   17.8&     3.2&  1.6 &19.9 &17.3 &24.1    &0.06 \\[-0.09cm] 
            &        &   12&   64&   71&  105&   69&      12.4&   18.5&     3.3&  1.7 &20.4 &17.8 &24.2    &0.08 \\[+0.01cm] 
   & &    *&    *&    *&    *&    *&         *&      *&       *&    * &   * &   * &   *    &0.03 \\[-0.09cm] 
 \raisebox{-0.7ex}{NGC\,5667}           &   \raisebox{-0.7ex}{II-i + III}     &   28&   64&   66&   96&   64&       9.8&   11.0&     1.7&  0.6 &19.1 &21.9 &25.7    &0.03 \\[+0.01cm] 
            &        &    *&    *&    *&    *&    *&         *&      *&       *&    * &   * &   * &   *    &0.04 \\[-0.09cm] 
            &        &   28&   62&   65&  111&   63&       9.7&   10.3&     1.6&  0.4 &19.2 &23.7 &26.4    &0.04 \\[+0.01cm] 
\raisebox{-0.7ex}{NGC\,5668}   &   \raisebox{-0.7ex}{I}     &   20&  178&    *&    *&    *&         *&   26.7&     3.1&    * &20.6 & 0.0 &20.6    &0.10 \\[-0.09cm] 
            &        &   20&  162&    *&    *&    *&         *&   24.3&     2.8&    * &20.7 & 0.0 &20.7    &0.14 \\[+0.01cm] 
\raisebox{-0.7ex}{NGC\,5693}   &   \raisebox{-0.7ex}{II-AB}   &    9&   28&   50&   71&   29&       5.1&   16.7&     2.9&  2.1 &20.9 &17.9 &23.7    &0.10 \\[-0.09cm] 
            &        &    9&   26&   51&   71&   32&       5.6&   20.5&     3.6&  2.2 &21.6 &19.4 &23.5    &0.14 \\[+0.01cm] 
\raisebox{-0.7ex}{NGC\,5713}   &   \raisebox{-0.7ex}{III}  &    8&  102&  114&  164&  103&      14.0&   18.1&     2.5&  0.7 &19.1 &21.0 &25.6    &0.11 \\[-0.09cm] 
            &        &    8&  112&  139&  159&  112&      15.2&   18.1&     2.5&  0.7 &19.6 &21.5 &25.9    &0.15 \\[+0.01cm] 
\raisebox{-0.7ex}{NGC\,5768}   &   \raisebox{-0.7ex}{I}     &    6&   95&    *&    *&    *&         *&   13.1&     1.8&    * &19.9 & 0.0 &19.9    &0.25 \\[-0.09cm] 
            &        &    6&   98&    *&    *&    *&         *&   13.2&     1.8&    * &20.4 & 0.0 &20.4    &0.35 \\[+0.01cm] 
\raisebox{-0.7ex}{ IC\,1067}   &   \raisebox{-0.7ex}{II.o-OLR}  &    *&    *&   40&   99&    *&         *&      *&       *&    * &   * &19.8 &   *    &0.14 \\[-0.09cm] 
            &        &    *&    *&   40&   80&    *&         *&      *&       *&    * &   * &20.0 &   *    &0.19 \\[+0.01cm] 
   & &   43&   81&   86&  116&   86&       9.9&   50.2&     5.8&  2.3 &22.1 &19.9 &23.9    &0.12 \\[-0.09cm] 
 \raisebox{-0.7ex}{NGC\,5774}           &   \raisebox{-0.7ex}{II.o-OLR + III}     &   86&  116&  118&  182&  116&      13.3&   22.2&     2.5&  0.7 &19.9 &21.3 &25.1    &0.12 \\[+0.01cm] 
            &        &   47&   74&   77&  124&   77&       8.8&   60.8&     7.0&  2.7 &22.7 &20.5 &24.1    &0.16 \\[-0.09cm] 
            &        &   77&  124&  125&  182&  124&      14.2&   22.9&     2.6&  0.6 &20.5 &22.3 &25.7    &0.16 \\[+0.01cm] 
\raisebox{-0.7ex}{NGC\,5806}   &   \raisebox{-0.7ex}{III}  &   60&  116&  118&  242&  118&      11.8&   29.9&     3.0&  0.5 &20.3 &22.3 &24.6    &0.14 \\[-0.09cm] 
            &        &   60&  123&  124&  257&  124&      12.4&   31.4&     3.1&  0.5 &21.1 &23.0 &25.3    &0.20 \\[+0.01cm] 
\raisebox{-0.7ex}{NGC\,5850}   &   \raisebox{-0.7ex}{II.o-OLR}  &   96&  136&  140&  233&  137&      25.0&   92.2&    16.8&  2.7 &22.3 &19.7 &23.8    &0.16 \\[-0.09cm] 
            &        &   96&  133&  138&  240&  138&      25.2&  127.1&    23.2&  3.9 &23.3 &20.0 &24.4    &0.21 \\[+0.01cm] 
\hline
\end{tabular}
}
\caption[]{(continued): Results 
}
\end{center}
\end{table*}
\addtocounter{table}{-1}
\begin{table*}
\begin{center}
{\normalsize
\begin{tabular}{ l c | rrrrr r r r r r r r r}
\hline
\rule[+0.4cm]{0mm}{0.0cm}
Galaxy
&\multicolumn{1}{c}{Break}
&\multicolumn{1}{c}{$b_1$}
&\multicolumn{1}{c}{$b_2$}
&\multicolumn{1}{c}{$b_3$}
&\multicolumn{1}{c}{$b_4$}
&\multicolumn{2}{c}{\rbr}
&\multicolumn{2}{c}{\hin}
&\multicolumn{1}{c}{\hin}
&\munin
&\munout
&\mubr
&$A_{r^{\prime}/g^{\prime}}$
\\[+0.1cm]
& \multicolumn{1}{c}{type}
& \multicolumn{1}{c}{[\arcsec]}
& \multicolumn{1}{c}{[\arcsec]}
& \multicolumn{1}{c}{[\arcsec]}
& \multicolumn{1}{c}{[\arcsec]}
& \multicolumn{1}{c}{[\arcsec]}
& \multicolumn{1}{c}{[kpc]}
& \multicolumn{1}{c}{[\arcsec]}
& \multicolumn{1}{c}{[kpc]}
& \multicolumn{1}{c}{[\hout]}
& \multicolumn{1}{c}{[\magsqarcsecfrac]}
& \multicolumn{1}{c}{[\magsqarcsecfrac]}
& \multicolumn{1}{c}{[\magsqarcsecfrac]}
& \multicolumn{1}{c}{[mag] }
\\
\rule[-3mm]{0mm}{5mm}
{\scriptsize{\raisebox{-0.7ex}{\it (1)}}}
&\multicolumn{1}{c}{{\scriptsize{\raisebox{-0.7ex}{\it (2)}}}}
&\multicolumn{4}{c}{{\scriptsize{\raisebox{-0.7ex}{\it (3)}}}}
&\multicolumn{2}{c}{{\scriptsize{\raisebox{-0.7ex}{\it (4)}}}}
&\multicolumn{2}{c}{{\scriptsize{\raisebox{-0.7ex}{\it (5)}}}}
&\multicolumn{1}{c}{{\scriptsize{\raisebox{-0.7ex}{\it (6)}}}}
&\multicolumn{2}{c}{{\scriptsize{\raisebox{-0.7ex}{\it (7)}}}}
&\multicolumn{1}{c}{{\scriptsize{\raisebox{-0.7ex}{\it (8)}}}}
&\multicolumn{1}{c}{{\scriptsize{\raisebox{-0.7ex}{\it (9)}}}}
\\
\hline\hline \\[-0.4cm]
\raisebox{-0.7ex}{UGC\,09741}  &   \raisebox{-0.7ex}{III}  &   17&   24&   29&   62&   25&       4.7&    6.8&     1.3&  0.6 &19.8 &21.2 &23.5    &0.06 \\[-0.09cm] 
            &        &   17&   27&   29&   62&   27&       5.1&    6.9&     1.3&  0.7 &20.4 &21.6 &24.0    &0.09 \\[+0.01cm] 
\raisebox{-0.7ex}{UGC\,09837}  &   \raisebox{-0.7ex}{II-CT}   &    8&   45&   53&   69&   46&       9.4&   17.1&     3.5&  2.4 &21.4 &16.6 &24.6    &0.05 \\[-0.09cm] 
            &        &    8&   44&   46&   69&   46&       9.4&   18.2&     3.7&  2.6 &21.8 &17.0 &24.8    &0.06 \\[+0.01cm]
\raisebox{-0.7ex}{NGC\,5937}   &   \raisebox{-0.7ex}{III}  &   12&   62&   75&  117&   63&      12.5&   12.6&     2.5&  0.6 &19.1 &21.1 &24.6    &0.50 \\[-0.09cm] 
            &        &   12&   68&   70&   98&   68&      13.5&   12.8&     2.5&  0.6 &19.8 &22.2 &25.8    &0.69 \\[+0.01cm] 
\raisebox{-0.7ex}{ IC\,1125}   &   \raisebox{-0.7ex}{I}     &    4&   77&    *&    *&    *&         *&   11.4&     2.3&    * &19.9 & 0.0 &19.9    &0.43 \\[-0.09cm] 
            &        &    4&   68&    *&    *&    *&         *&   11.6&     2.3&    * &20.5 & 0.0 &20.5    &0.59 \\[+0.01cm] 
\raisebox{-0.7ex}{ IC\,1158}   &   \raisebox{-0.7ex}{II.o-CT}   &   14&   53&   67&   98&   65&       9.1&   30.5&     4.3&  2.9 &21.2 &16.9 &23.4    &0.32 \\[-0.09cm] 
            &        &   14&   50&   67&  101&   61&       8.5&   34.8&     4.9&  3.3 &21.9 &17.5 &23.8    &0.44 \\[+0.01cm] 
\raisebox{-0.7ex}{NGC\,6070}   &   \raisebox{-0.7ex}{II.o-CT}   &   20&   87&  108&  162&   97&      14.0&   37.2&     5.4&  2.0 &20.1 &17.3 &22.8    &0.41 \\[-0.09cm] 
            &        &   20&   87&  108&  157&   88&      12.7&   43.4&     6.3&  2.6 &21.0 &17.4 &23.3    &0.56 \\[+0.01cm] 
\raisebox{-0.7ex}{NGC\,6155}   &   \raisebox{-0.7ex}{II-AB}   &    4&   33&   36&   71&   34&       6.3&   12.2&     2.3&  1.5 &19.4 &18.2 &21.9    &0.04 \\[-0.09cm] 
            &        &    4&   33&   37&   63&   34&       6.3&   13.1&     2.4&  1.8 &20.0 &17.9 &22.7    &0.05 \\[+0.01cm] 
\raisebox{-0.7ex}{UGC\,10721}  &   \raisebox{-0.7ex}{III}  &   16&   38&   48&   80&   41&       8.9&    8.3&     1.8&  0.5 &19.3 &21.7 &24.6    &0.10 \\[-0.09cm] 
            &        &   16&   38&   46&   78&   41&       8.9&    7.9&     1.7&  0.6 &19.5 &21.8 &24.8    &0.13 \\[+0.01cm] 
\raisebox{-0.7ex}{NGC\,7437}   &   \raisebox{-0.7ex}{II-CT}   &   12&   39&   47&   91&   43&       6.5&   26.4&     4.0&  1.9 &21.3 &19.7 &23.1    &0.11 \\[-0.09cm] 
            &        &   12&   40&   43&  147&   41&       6.2&   27.1&     4.1&  1.9 &21.8 &20.4 &23.3    &0.15 \\[+0.01cm] 
\raisebox{-0.7ex}{NGC\,7606}   &   \raisebox{-0.7ex}{II-CT}   &   20&   96&  113&  195&   97&      14.7&   43.4&     6.6&  1.9 &19.6 &17.3 &22.3    &0.10 \\[-0.09cm] 
            &        &   20&   92&  114&  187&   93&      14.1&   49.9&     7.6&  2.2 &20.5 &17.7 &22.9    &0.14 \\[+0.01cm] 
\raisebox{-0.7ex}{UGC\,12709}  &   \raisebox{-0.7ex}{II-CT}   &    5&   65&   72&  102&   69&      12.8&   29.8&     5.5&  2.2 &22.4 &19.3 &25.1    &0.11 \\[-0.09cm] 
            &        &    5&   64&   66&  108&   66&      12.2&   33.0&     6.1&  2.4 &23.0 &19.8 &25.2    &0.15 \\[+0.01cm] 
\hline                   
\end{tabular}
}
\caption[]{(continued): Results
}
\end{center}
\end{table*}
%
\subsection{Frequencies}
The vast majority (almost 90\%, see \tab\ref{freqtab}) of the galaxies 
exhibit surface brightness profiles with breaks, only 9 ($11\% \pm 4\%$)
are reasonably well described (allowing sometimes for quite extensive 
deviations) as being purely single-exponential (\typeoc). 
Even adding the barred galaxies showing only a dip inside the bar 
radius (\typetic), their frequency increases up to merely $15\% \pm 4\%$. 
$66\% \pm 5\%$ of the galaxies are classified as \typet with a break 
and downbending profiles ($61\%$ excluding \typetic). The frequencies 
of the three main subgroups are the following: 
a) $33\% \pm 5\%$ (from the full sample) exhibit a classical 
truncations (\typectc), 
b) $15\% \pm 4\%$ (from the full sample) are classified as having 
breaks with downbending profiles that could be (size wise) related to 
the presence of the bar (\typeolrc), and 
c) $13\% \pm 4\%$ (from the full sample) show breaks in their profile 
(\typeabc) that could be originated by a lopsided or asymmetric disk, 
not allowing us to probe for a real, intrinsic break in the stellar 
light distribution using our azimuthally averaged, fixed ellipse fits.  
$33\% \pm 5\%$ of the galaxies show a break with an upbending 
profile (\typeiiic).  
\begin{table}
\begin{center}
{\normalsize
\begin{tabular}{l c r }
\hline
\rule[+0.4cm]{0mm}{0.0cm}
Profile type
& Number [\#]
& Frequency [\%]
\\
\hline\hline \\[-0.4cm]                  
\typeo           &  9     & $11 \pm 3$\\
\typet           &  56    & $66 \pm 5$\\
$-$ \typeti      &  4 (1) & $ 6 \pm 3$ \\
$-$ \typect      & 28 (2) & $32 \pm 5$\\
$-$ \typeolr     & 13 (5) & $15 \pm 4$\\
$-$ \typeab      & 11 (3) & $13 \pm 4$\\
\typeiii         & 28 (7) & $32 \pm 5$\\
\hline
\end{tabular}
}
\caption[]{Frequency of disk types: Listed are the number of galaxies  
per profile type. The values in brackets are those associated to more 
than one break type (mixed classification). Therefore the last column 
gives the frequency for each break type and does not add up to 100\%.}
\label{freqtab}
\end{center}
\end{table}
%
\subsection{Parameter distribution}
In the following we give all measurement of lengths and ratios
as obtained for the $r^{\prime}$ band and excluding the mixed 
classifications if not otherwise stated.
\subsubsection*{No breaks (\typeoc)}
The mean central surface brightness and scalelength of the 9 \typeo 
galaxies are $\mun = 20.2\pm0.8$\,$r^{\prime}$-\magsqarcsec and 
$h = 2.8\pm0.8$ kpc respectively, which are typical values for local galaxies 
\citep[\cf][]{macarthur2003,dejong1996}.
We confidently trace the profiles down to our critical surface 
brightness of $\mu_{\rm crit}\!\sim\!27.0\ r^{\prime}$-\magsqarcsec 
(see \fig\ref{N5300skysub}). So, in terms of scalelength we do
not find a break down to $6-8$ times the measured scalelength. 
\subsubsection*{Classical truncations (\typectc)}
\begin{figure}
\includegraphics[width=6.1cm,angle=270]{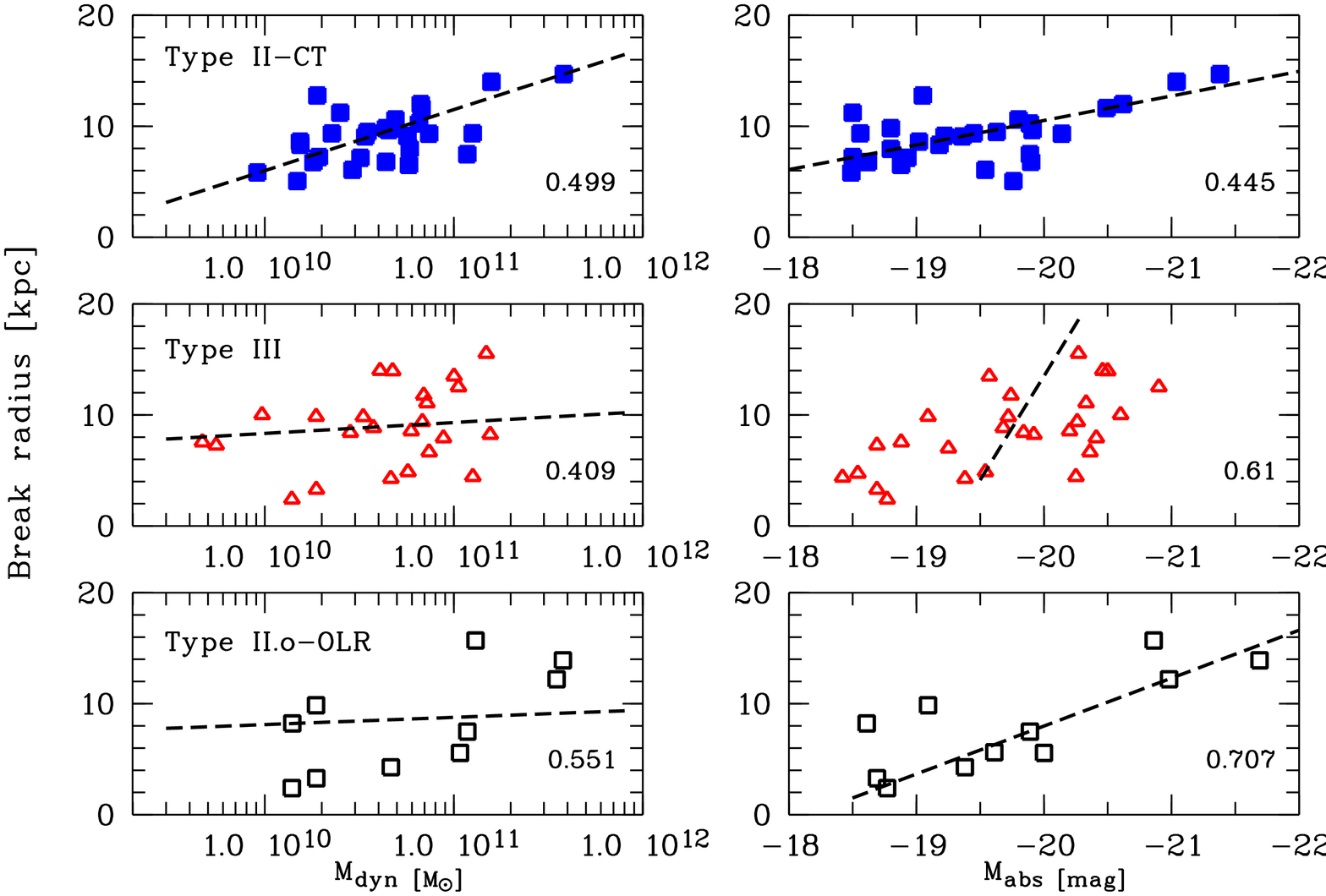}
\caption{Break radius (\rbr) in absolute units (kpc) versus dynamical 
mass ($\mathcal{M}_{\rm dyn}\!=\!2.326\!\cdot\!10^5 \frac{D_{25}}{2} 
\vrot^2$) and absolute magnitude (\mabs) in the B-band according 
to LEDA for the different break types: \typect {\it (upper row)}, 
\typeiii {\it (middle row)}, and \typeolr {\it (bottom row)}. 
Overplotted are robust linear fits {\it (dashed lines)} to guide 
the eye and in the lower right corner of each plot the Spearman 
rank correlation coefficient. The break radius correlates 
in all three cases with absolute magnitude. With the exception 
of the \typectc, the relation of the position of the break with 
the dynamical mass is less tight.} 
\label{f7}
\end{figure}

\begin{figure}
\includegraphics[width=6.1cm,angle=270]{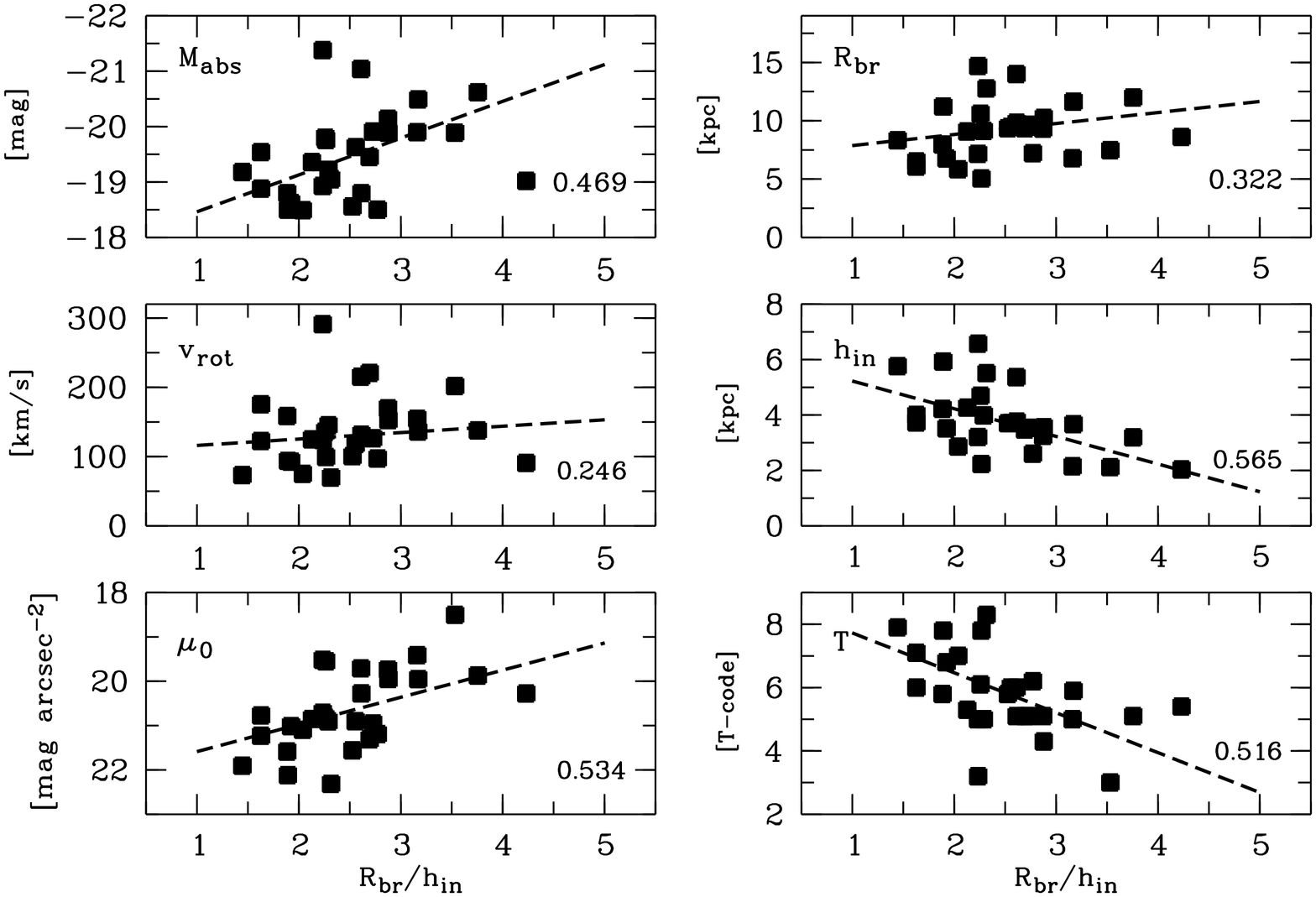}
\caption{Classical truncations (\typectc): Break radius in units of 
inner scalelength versus 
absolute magnitude (\mabs) {\it (left, upper panel)},  
rotational velocity (\vrot) {\it (left, middle panel)}, 
extinction corrected central surface brightness of the 
inner disk (\munin)  {\it (left, lower panel)}, 
break radius in linear unit (\rbr) {\it (right, upper panel)}, 
inner scalelength in linear unit (\hin) {\it (right, middle panel)}, 
and Hubble type ($T$) {\it (right, lower panel)}. Overplotted are 
robust linear fits {\it (dashed lines)} to guide the eye and in the 
lower right corner of each plot the Spearman rank correlation coefficient.
}
\label{f8}
\end{figure}

\begin{figure}
\includegraphics[width=6.1cm,angle=270]{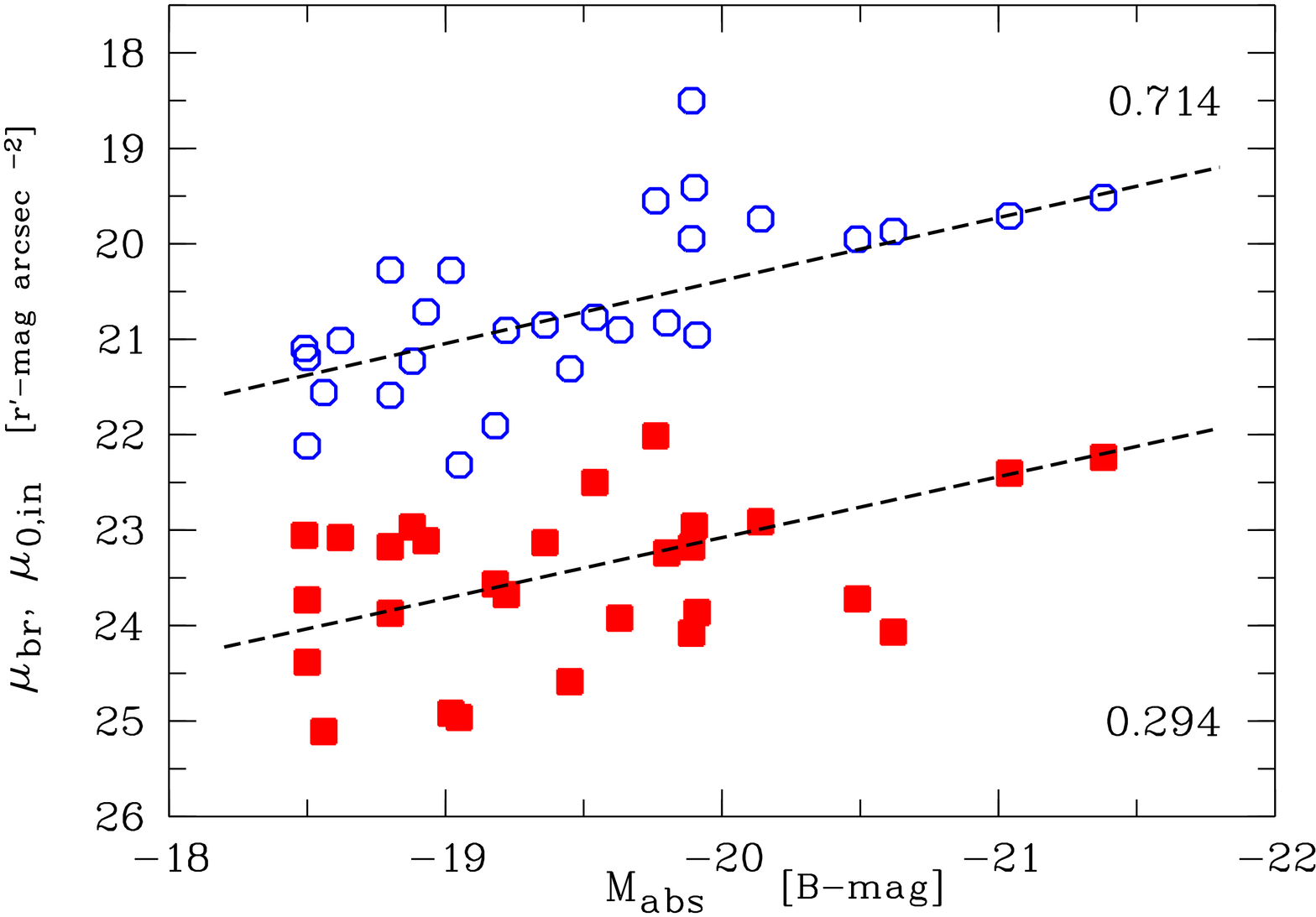}
\caption{Classical truncations (\typectc): Extrapolated central surface 
brightness \mun {\it (open circles)} and surface brightness at the break 
radius \mubr {\it (filled squares)} in the $r^{\prime}$ band versus 
absolute magnitude (\mabs) in B according to LEDA. Overplotted are 
robust linear fits {\it (dashed lines)} to guide the eye and in the 
lower and upper right corner the Spearman rank correlation coefficient
for each data set.}
\label{f9}
\end{figure}

\begin{figure}
\includegraphics[width=6.1cm,angle=270]{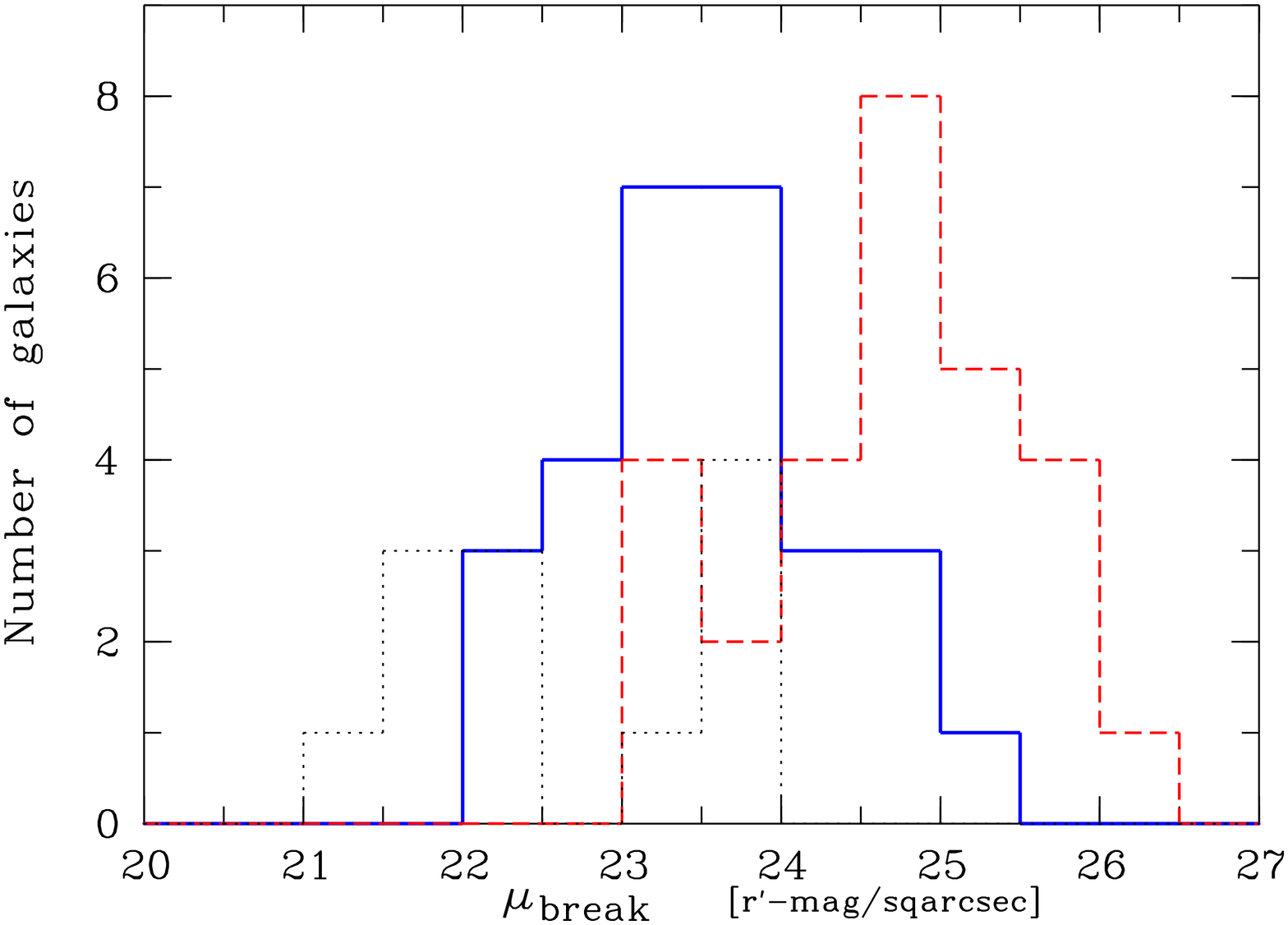}
\caption{Histogram of the surface brightness (in the $r^{\prime}$ band) at the 
break radius for the galaxies of \typect {\it (solid line)}, 
\typeiii {\it (dashed line)}, and \typeolr {\it (dotted line)}. 
\label{mubrHISTO}
}
\end{figure}

The break in the surface brightness profiles appears at $9.2\pm2.4$\,kpc and ranges between $5.1$\,kpc 
and $14.7$\,kpc with a slight trend towards galaxies with higher 
luminosities having larger break radii (\cf\fig\ref{f7}). 
We do not find a systematic difference between the break 
radii as measured in the $r^{\prime}$ or $g^{\prime}$ band. 
In relative units the break is at $2.5\pm0.6$ times the inner 
scalelength with values between $1.4$ and $4.2$, uncorrelated 
with rotational velocity and with a weak trend towards higher 
values for brighter galaxies (\cf\fig\ref{f8}). 
The value found here for the position of the break in terms of
radial scalelength is apparently smaller (although compatible 
within 1$\sigma$) to the one 
quoted for three galaxies by \cite{pohlen2002}: $3.9\pm0.7$. 
According to \cite{pohlen2002}, the one galaxy in common (UGC\,9837) has 
a value of 3.1 for \rbrdhin, which is consistent with our ratio of 2.7 
obtained from the SDSS image. 
The two galaxies with high values (4.3 and 4.2) are too distant to 
be in our sample and both are intrinsically large, having break 
radii of 14.7\,kpc and 21.1\,kpc. 
Fitting the SDSS profile (converted to Johnson R according to 
\sec\ref{phot}, see \fig\ref{SDSSvsCCD}) and using our method 
for one of them (NGC\,5923), we obtain also a consistent value 
\citep[4.1 compared to 4.3 quoted by][]{pohlen2002}.
Due to the completely different masking of a very close extended 
companion the comparison for the third galaxy (NGC\,5434) is 
not straightforward (3.2 compared to 4.2).
So neither their larger intrinsic size nor a systematic error in 
determining the ratio (\rbrdhin) is responsible for the apparent 
mismatch of the mean values for the two samples, suggesting that 
its origin is due to small number statistics.
However, our mean value of $2.5\pm0.6$ is significantly lower compared 
to the $4.5\pm1.0$ value obtained for 16 face-on galaxies by \cite{vdk1988}.
This large offset is almost certainly due to the different 
definition (break versus cut-off) and method used to mark the 
truncation. We determine the truncation at the position of the 
measured break in the profile (so quite far in), whereas 
van der Kruit was looking for a sharp outer boundary estimated 
from isophote maps. 
Our value is supported by \cite{bosma1993} who stated that for 
7 galaxies (with breaks in their surface 
brightness profiles), of the 21 reported by \cite{wevers1984} (so virtually 
the same dataset \cite{vdk1988} used), the mean value of the 
break to scalelength is $2.8\pm 0.4$, 
so very close to our mean value. 
The mean scalelength of the inner disk \hin in our sample 
is $3.8\pm1.2$\,kpc and ranges between $2.0$\,kpc and $6.6$\,kpc, 
which is typical for local galaxies \citep[\cf][]{macarthur2003,dejong1996}.
The central surface brightness (extrapolated from the fitted inner 
exponential, corrected for galactic extinction, but uncorrected 
for inclination) ranges between $\munin = 18.5$\,$r^{\prime}$-\magsqarcsec 
and $\munin = 22.3$\,$r^{\prime}$-\magsqarcsec 
($18.9-22.8$\,$g^{\prime}$-\magsqarcsec) with a clear trend towards galaxies 
with fainter luminosities having a fainter central surface 
brightness (\cf\fig\ref{f9}) as seen by \cite{macarthur2004} 
or \cite{dejong1996}.
However, the surface brightness at the break radius only 
weakly correlates with absolute magnitude (\cf\fig\ref{f9})
and peaks (\cf\fig\ref{mubrHISTO}) at a mean value of 
$\mubr = 23.5\pm 0.8$\,$r^{\prime}$-\magsqarcsec
($\mubr = 23.8\pm 0.8$\,$g^{\prime}$-\magsqarcsec). 
The mean ratio of inner to outer scalelength is $\hindhout=2.1\pm0.5$ 
with a mildly peaked distribution ranging between $\hindhout=1.3-3.6$.
%
\subsubsection*{OLR breaks (\typeolrc)}
The break radii for the \typeolr breaks span a wider range,
compared to \typect breaks, between $2.4$\,kpc and $25.0$\,kpc 
(mean: $9.5\pm6.5$\,kpc) again with a clear trend towards 
galaxies with higher luminosities having larger break 
radii (\cf\fig\ref{f7}). 
In terms of inner scalelength almost all breaks appear in 
the small range between $1.4-2.4$ times \hin (with a mean 
at $1.7\hin$), except for one galaxy (NGC\,5430) where the 
break is at $4.4\hin$ but which could be also classified 
as \typeo with bumps shaped by the prominent bar 
(\cf Appendix \ref{atlas}).  
The surface brightness at the break radius is 
$\mubr = 22.6\pm 1.0$\,$r^{\prime}$-\magsqarcsec. 
Although on the lower end compared to the distribution 
for \rbrdhin and \mubr, neither of these parameters 
works out as a clear discriminatory property for the two different 
downbending break types \typect and \typeolrc.
%
\subsubsection*{Upbending profiles (\typeiiic)}
In relative units the \typeiii break appears further out, compared 
to \typect or \typeolr breaks, at $4.9\pm0.6$ times the inner 
scalelength with values between $3.7$ and $5.8$ (excluding the 
seven galaxies with mixed classifications where an inner scalelength 
is not well defined). 
However, in absolute units they span the same range. The break 
appears at $9.3\pm3.3$\,kpc and ranges between $4.4$\,kpc 
and $15.5$\,kpc having the same trend with 
luminosity (\cf\fig\ref{f7}).
The inner scalelength is on average smaller 
with $1.9\pm0.6$\,kpc ranging only between $0.9$\,kpc and 
$3.0$\,kpc. 
The central surface brightness (extrapolated from the fitted inner 
exponential and corrected for galactic extinction) ranges only 
between $\munin = 18.1$\,$r^{\prime}$-\magsqarcsec and 
$\munin = 20.3$\,$r^{\prime}$-\magsqarcsec (mean: $19.2\pm0.6$\,$r^{\prime}$-\magsqarcsec) 
without showing the clear trend with luminosity as for the 
\typect breaks.
The mean value for the surface brightness at the break radius,  
$\mubr = 24.7\pm 0.8$\,$r^{\prime}$-\magsqarcsec 
($\mubr = 25.6\pm 0.4$\,$r^{\prime}$-\magsqarcsec for the seven galaxies 
with additional downbending breaks inside), is clearly fainter 
compared to the \typect breaks (see \fig\ref{mubrHISTO}). 
%
\subsection{Correlations}
\begin{figure}
\includegraphics[width=6.1cm,angle=270]{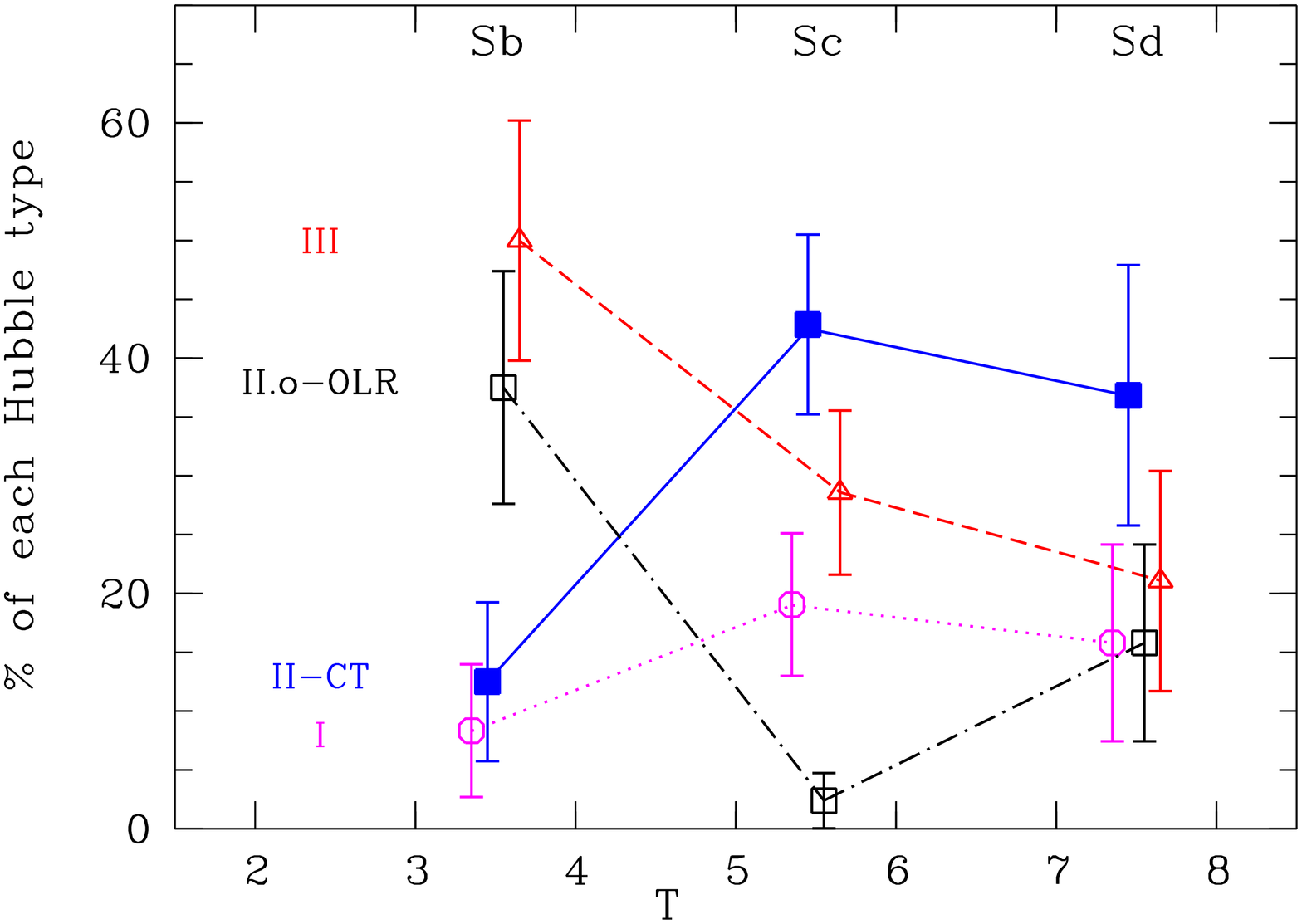}
\caption{Frequency of profile types \typeo {\it (open circles)}, 
\typect {\it (filled squares)}, \typeiii {\it (open triangles)}, and 
\typeolr {\it (open squares)} in relation to the Hubble type. The 
galaxies are merged in three morphological bins ($T$ between $2.5-4.4$, 
$4.5-6.4$, and $6.5-8.4$). The associated points are connected with lines 
and slightly shifted in $T$ to be able to separate them.}
\label{btypeVShtype}
\end{figure}

Plotting the frequency of break types versus the Hubble 
types (grouped in three bins: Sb--Sbc, Sc--Scd, and Sd--Sdm)
reveals a clear correlation (\cf\fig\ref{btypeVShtype}).
Whereas the few \typeo galaxies are evenly distributed 
with Hubble type, the frequency of \typeiii breaks 
decreases, and that of the \typect increases, with later 
Hubble type.
The \typeolr breaks appear to have a minimum for the Sc--Scd
galaxies which might be related to a minimum in the overall 
RC3 strong bar frequency for Sc galaxies as mentioned by \eg
\cite{luettphd}. 
We know that the Hubble type $T$ correlates with absolute magnitude
\mabs (\cf\sec\ref{sample}), but marking the different break 
types in such a plot shows that the correlation of break type 
is more pronounced with the Hubble type compared to \mabs.   
For the classical truncations (\typectc) the position of the break at 
$2.5\pm0.6$ times the inner scalelength shows a trend with Hubble 
type (\cf\fig\ref{f8}) in the sense of later types being 
{\it earlier} truncated. However, it seems to be almost uncorrelated 
with total mass (either measured with \mabs or $\mathcal{M}_{\rm dyn}$, see 
\fig\ref{f7}).
%

\subsection{Environment}
We find for some of our \typeiii galaxies indication for a 
recent interaction (\cf Appendix \ref{atlas}) such as 
close physical companions, streams, or shells. However, using 
the number of neighbours (as determined in \sec\ref{environment}), 
to characterise the environment, we do not find a clear correlation with
the profile type. 
Although it looks like the \typeiii galaxies tend to have more
neighbours within a 1\,Mpc volume (59\% have 3 or more neighbours while for the \typect
the fraction is only 35\%), such a correlation 
is far from being clear-cut. There are still many classical truncations 
among the galaxies having three or more (even up to nine) neighbours as 
well as \typeiii galaxies without any companion. 
However, we have the problem that criteria for environment 
are often too crude and lead even to controversial results. 
For example, our prototypical classical truncation, NGC\,5300, has 
five neighbours in the SDSS database, which is clearly in a denser 
region according to our environment criterion. Using the CfA catalogue 
and applying a different criterion \cite{varela2004} list it as being 
part of their {\it truly isolated} galaxy sample. 
%
\section{Discussion}
\label{discussion}
As reported by \cite{pohlen2004} we also do not find any galaxy with 
a sharp cut-off in the radial light distribution. This means that 
the so-called {\it truncated} galaxies are best described by a broken 
exponential. This seems to be now the intrinsic structure 
associated to the inferred sharp cut-offs proposed by \cite{vdk1979}.    
A similar result is found, using a completely independent approach, by
\cite{fergusonM33} mapping the prototypical truncated galaxy M\,33. They 
traced the profile down to $\mu_{\rm lim}\!\sim\!30$ I-\magsqarcsec 
(transformed to surface brightness) using star-counts.
In general, the shape of the transition regions we found (measured in 
terms of the boundary difference $b_{3}-b_{2}$) is no longer as sharp 
as measured by \cite{pohlen2002}. Only very few cases 
(\eg NGC\,5300 or NGC\,7437) exhibit a very sharp transition between 
the inner and outer disk region. Most often the \typect breaks appear
with a more gradual transition zone up to an almost exponential region 
of size $\sim 4$\,kpc in between. 
Since the discrimination between \typeolr and \typect is solely 
based on bar size measurements one could argue that there 
are some undetected \typeolr breaks among the galaxies classified 
as \typectc. 
However, we carefully use either bar sizes from the literature or in
most cases estimated them from the image to make sure the \typect 
is as good as possible separated from the \typeolr breaks. 
In total, without specifying the origin, most of the galaxies 
($\gtsim 60\%$) exhibit a break followed by a downbending profile
almost consistent with the edge-on results by \cite{kregel2002},
although they probably would have excluded the \typeab galaxies
from their initial sample. 
We do not find truncation breaks beyond $\sim\!25.2\ r^{\prime}$-\magsqarcsec
(\cf\fig\ref{mubrHISTO}). We have explored whether this could be an
observational bias caused by the difficulty of detecting breaks at faint
levels. To do that we simulated 2D galaxies with breaks in their surface
brightness distribution at different surface brightness levels. These 
galaxies are added to the real sky background and we measured the 
breaks as for real galaxies. From this we estimate our limit for detecting 
breaks of $\sim\!26.0\ r^{\prime}$-\magsqarcsec. 
This limit leaves a range of $\sim\!1$mag to reach our critical surface 
brightness of $\mu_{\rm crit}\!\sim\!27.0\ r^{\prime}$-\magsqarcsec (down to 
which the slope 
of the outer profile could be traced  confidently, see \fig\ref{N5300skysub}). 
So we expect to detect all truncations in the region 
$25.2-26.0\ r^{\prime}$-\magsqarcsec if they do exist. Consequently, the 
decline observed in Fig.\ref{mubrHISTO} seems to be real. 
In addition, in terms of scalelength the largest break we find is 
at $\rbrdhin\!=\!4.2$, but we are able to trace the profiles of our 
\typeo galaxies down to $\sim\!6-8$ radial scalelength.
So we think that we do not miss any truncation breaks due to 
the limiting depth of the SDSS data and our \typeo galaxies are 
indeed {\it untruncated}. 
This is consistent with measurements on individual galaxies being 
intrinsically untruncated down to $\sim\!10$ scalelength 
\cite[e.g.][]{weiner2001,ngc300}.  
One of the key questions is now to find why some galaxies show 
truncations and (the minority) does not.
The strong increase for the \typeiii frequency towards earlier Hubble 
types (Sb--Sbc) could be used to argue against the \typeiii feature 
being really related to the disk itself, by explaining the rise 
in the profile as a traditional $R^{1/4}$ bulge component taking 
over from the exponential disk in the outer parts. 
However, \cite{erwin2005a} already showed that this is only true 
for about $\sim\!1/3$ of the earliest types (SB0-SBb, having the 
highest chances of being bulge dominated). We see that $\sim 40\%$ 
of our Type III late-type galaxies exhibit clear signs for spiral arms in the 
outer disk excluding any spheroidal (pressure supported) nature
of the outer structure. For another $\sim 23\%$ of the \typeiii
galaxies we do not find a significant decrease in ellipticity (expected
if produced by a spheroidal component) while fitting free ellipses 
across the break. So we conclude that in most ($> 63\%$) of our 
\typeiii galaxies the outer region is indeed related to the disk.     
This results fits nicely to the extended outer (star-forming) disks 
(\eg NGC\,4625) which have been recently reported by the GALEX 
team \citep{thilker2005,gildepaz2005} as being significantly more 
frequent than thought before. 
Unfortunately, we do not have any galaxy classified as \typeiii on 
our SDSS images in common with those reported by the GALEX team. 
However, the profile shown by \cite{swaters2002} of NGC\,4625 clearly 
suggests a \typeiii classification. 
\section{Summary}
\label{summary}
The main goal of our study is to provide a census of the 
radial disk structure of local ($\vvir < 3250$ km/s) 
late-type galaxies. 
Using the LEDA catalogue we selected a complete sample of 
655 galaxies down to \mabs$=-18.4$ B-mag. 98 (85 with 
useful images) of them are part of the SDSS Second Data 
Release (DR2) which provides the imaging data used here. 
After careful sky subtraction  we 
obtained our radial surface brightness profiles from 
fixed ellipse fits. We classified the resulting profiles 
searching for clear breaks either of the {\it downbending} 
(steeper outer region) or {\it upbending} kind (shallower 
outer region) \citep[following][]{erwin2006} and derived 
scalelength and central surface brightness values for 
exponential fits to the individual regions.  

The main conclusions are as follows: 
\begin{enumerate}
\item
We are able to reliably trace azimuthally averaged, fixed ellipse 
profiles of face-on to intermediate inclined galaxies down to a critical 
surface brightness of $\sim 27.0\ r^{\prime}$-\magsqarcsec using SDSS 
imaging data. 
We have taken extreme care in determining the background {\it (sky)} around 
our objects. Our surface brightness limit value is supported by finding 
similar sky values with different methods. In addition, our profiles match 
those from available deeper photometry and we are able to recover 
simulated profiles added to empty SDSS sky fields down to the above 
surface brightness limit.
\item 
90\% of our galaxies could be classified into one of the following 
classes --\typeo (no break), \typet (downbending break),  and \typeiii 
(upbending break)-- extending Freemans original classification. 
The remaining 10\% of the sample
could be well described being a mix from two individual classes. We introduced
two main sub-classifications for the \typet class connecting the observed break
with possible different physical origins: \typectc, {\it classical truncations},
probably associated with a global, radial star-formation threshold 
(\cf\sec\ref{introduction}), and exclusively for barred galaxies, \typeolr 
{\it OLR breaks}, observed at around twice the bar radius and probably 
associated with the outer Lindblad resonance of the bar 
\citep[see][]{erwin2006}.  
\item 
Surprisingly only $\ltsim 15\%$ of all galaxies have a {\it normal} 
purely exponential disk down 
to our noise limit. A good deal more ($33\%$ of each) have 
profiles with an upbending break (\typeiiic) or a classical 
truncation (\typectc). 
\item 
We find a correlation of break type with morphological type. 
Classical truncations (\typectc) are more frequent in later
types while the fraction of upbending breaks rise towards 
earlier types. 
\item 
Our \typeo galaxies seem to be {\it genuinely} untruncated.  
The exponential profiles extend reliably down to a surface 
brightness of $\sim\!27.0\ r^{\prime}$-\magsqarcsec (equivalent to 
$\sim6-8$ times the scalelength). 
\item 
We do not find any galaxy with a sharp (or complete) 
cut-off in the radial light distribution. This means {\it truncated} 
galaxies are in fact best described by a broken exponential with a 
shallow inner and a steeper outer exponential region separated at a 
more or less well defined break radius. 
\item
For \typect galaxies the break happens already at $2.5\pm0.6$ 
times the inner scalelength at a typical surface brightness 
of $\mubr = 23.5\pm 0.8$\,$r^{\prime}$-\magsqarcsec. The position of 
the break (in kpc) seems to be correlated with absolute 
magnitude: the more {\it luminous} the {\it larger} the inner disk of the
galaxy. 
\item 
For \typeiii galaxies the break happens typically further
out at $4.9\pm0.6$ times the inner scalelength, and at a
lower surface brightness of 
$\mubr = 24.7\pm 0.8$\,$r^{\prime}$-\magsqarcsec.
For more than $60\%$ of these galaxies we find good indication that the 
outer upbending part is a disk-like structure (e.g. by finding 
spiral arms). Close physical neighbours and slightly disturbed morphology
suggest in several cases interaction as a possible origin. 
\end{enumerate}
%
\begin{acknowledgements}
We would like to thank Peter Erwin and John Beckman for their 
stimulating discussions and useful suggestions during this work 
and especially Peter Erwin for helping with the ellipse task and 
for carefully reading parts of this paper. 
Thanks also to Reynier Peletier for reading the paper and for suggesting 
to use the derivative method for objectively quantifying the break 
radius. We also thank Marco Barden for very stimulating discussions
and we would like to thank the anonymous referee for detailed
comments which helped us to improve the quality of the manuscript.
Part of this work was supported by a Marie Curie Intra-European 
Fellowship within the 6th European Community Framework Programme.
This research has made use the Lyon/Meudon Extragalactic Database 
(LEDA, {\ttfamily http://leda.univ-lyon1.fr}) and the 
NASA/IPAC Extragalactic Database (NED) which is operated by 
the Jet Propulsion Laboratory, 
California Institute of Technology, under contract with the National 
Aeronautics and Space Administration. 
This research has made use of NASA's Astrophysics Data 
System Bibliographic Services.
Funding for the creation and distribution of the SDSS Archive has been 
provided by the Alfred P. Sloan Foundation, the Participating Institutions, 
the National Aeronautics and Space Administration, the National Science 
Foundation, the U.S. Department of Energy, the Japanese Monbukagakusho, 
and the Max Planck Society. The SDSS Web site is 
{\ttfamily http://www.sdss.org/}.
The SDSS is managed by the Astrophysical Research Consortium (ARC) for 
the Participating Institutions. The Participating Institutions are The 
University of Chicago, Fermilab, the Institute for Advanced Study, the 
Japan Participation Group, The Johns Hopkins University, the Korean 
Scientist Group, Los Alamos National Laboratory, the Max-Planck-Institute 
for Astronomy (MPIA), the Max-Planck-Institute for Astrophysics (MPA), 
New Mexico State University, University of Pittsburgh, University of 
Portsmouth, Princeton University, the United States Naval Observatory, 
and the University of Washington.
\end{acknowledgements}

%
\newpage
\appendix
\section{Galaxy Atlas}
\label{atlas}
The atlas presented here contain the 85 (from 98) galaxies 
with useful images in the SDSS Second Data Release (DR2)
as described in \sec\ref{sample}.
The data for every two galaxies is presented on a single page. 
On the left side we give detailed comments on the individual 
galaxies concerning the applied classification and global 
characteristics (such as distinct morphological features 
or environment). 
The headline gives the name and classification type. Below 
we reproduce the main parameters: Position \texttt{(J2000)}, 
RC3 Hubble-type, coded LEDA Hubble parameter \texttt{T}, 
absolute B band magnitude \texttt{M$_{\texttt{abs}}$ [B-mag]}, 
apparent diameter \texttt{[']}, virgocentric radial velocity 
\texttt{v$_{\texttt{vir}}$ [km$\,$s$^{-1}$]}.
On the right side we reproduce the JPEG 800x600 colour mosaics 
obtained with the SDSS 'Finding Chart' 
tool\footnote{\texttt{http://cas.sdss.org/dr4/en/tools/chart/chart.asp}}. 
Overlayed are a N-S, E-W grid through the center, a $1\arcmin$ scale 
bar, and the outline of the individual SDSS fields. 
In addition we show azimuthally averaged, radial surface 
brightness profiles in the $g^{\prime}$ {\it (triangles)} and 
$r^{\prime}$ {\it (circles)} band obtained from fixed ellipse
fits. 
The $r^{\prime}$-band is overlayed by the best exponential fits 
(\cf\sec\ref{expfit}) to the individual regions: single disk, 
inner and outer disk, or three fits in the case of a mixed 
classification.  
The boundaries ($b_{1-4}$) used for the fit are marked with 
the short {\it vertical, dashed lines} on the abscissa.  
Following \cite{courteau1996} we also chose not to show error 
bars on all the profiles to avoid overcrowding of the plots. 
Instead, we show for the $g^{\prime}$-band {\it (horizontal, 
dashed line)} and for the $r^{\prime}$-band {\it (horizontal, 
dotted line)} the critical surface brightness ($\mu_{\rm crit}$) 
down to which the profile is reliable. This limit is placed where 
profiles obtained by using $\pm 1\sigma sky$ deviate by more than 
0.2\,mag from the one with our mean sky value (\cf\sec\ref{skysub}).
In the upper right corner of the surface brightness profiles 
we note the galaxy name, the Hubble type according to RC3 
(for galaxies without or uncertain (S?) classification from LEDA), 
and the profile type.   
\newpage
\onecolumn
\begin{minipage}[t][\textheight][t]{\textwidth}
\parbox[t][0.5\height][t]{0.47\textwidth}
{
\noindent {\bf  IC\,1067      :}        \typeolr  (possible \typeoc)         \\ 
\texttt{J145305.2+031954 .SBS3.  3.0 -18.97   2.0 1665 } \\[0.25cm]
Galaxy close to the border of the SDSS field but almost complete 
with a nearby, physical companion (IC\,1066, $v=1577$\kms) only 
$2.2\arcmin$ away, which slightly influences the background 
estimation (together with two nearby bright stars). 
The strong peak at $\sim 20\arcsec$ in the final profile is related 
to the ring around the prominent bar and the dip at $\sim 35$\arcsec
to the inter-arm region of the two symmetric spiral arms peaking 
at $\sim 40$\arcsec.
Although possible to fit (at least for the $r^{\prime}$ band profile) with 
a single exponential (\typeoc), we classify the galaxy as \typeolrc, 
with a break around twice the bar radius at $\sim 40\arcsec$, having 
an inner disk region too small to be detected with our automatic 
break searching routine. 
This is consistent with the classification and profile by 
\cite{erwin2006}. 
The region inside $\sim 40\arcsec$ is excluded for the fit 
of the outer scalelength. 
}
\hfill 
\parbox[t][0.5\height][t]{0.47\textwidth}
{
\includegraphics[width=5.7cm,angle=270,]{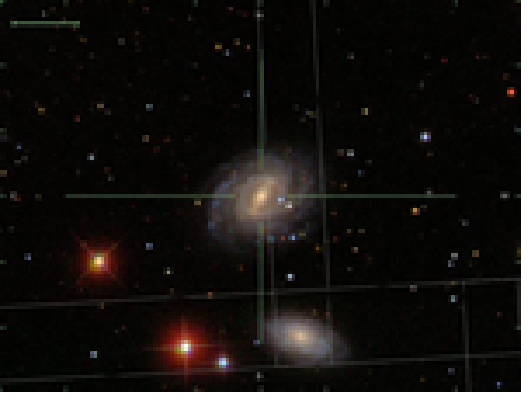}
\hspace*{-0.8cm}
\includegraphics[width=6.1cm,angle=270]{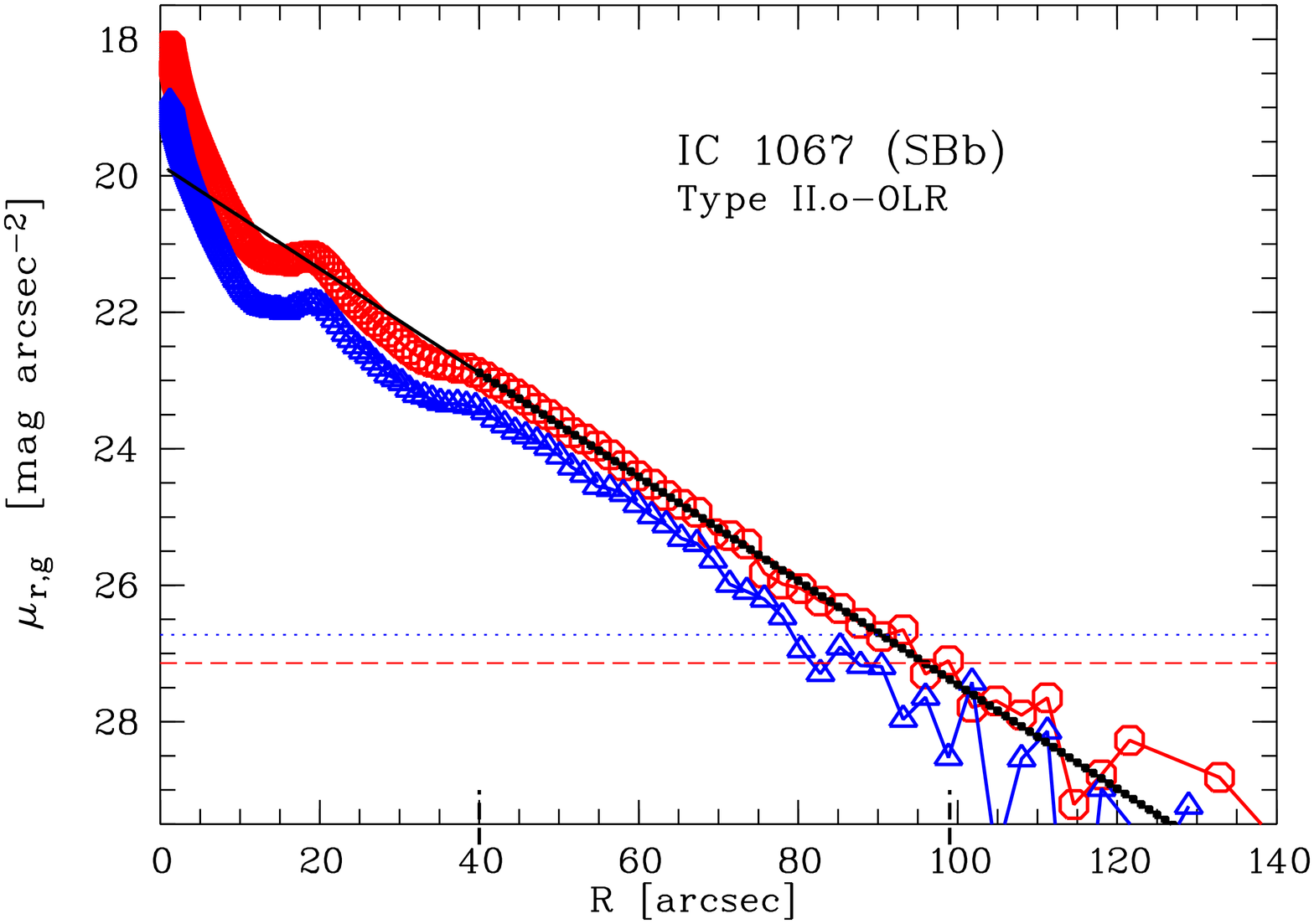}
}
\vfill
\parbox[b][0.5\height][t]{0.47\textwidth}
{
\noindent {\bf  IC\,1125      :}        \typeo                \\   
\texttt{J153305.6-013742 .S..8*  7.7 -20.05   1.7 2868}\\[0.25cm]
According to the coordinates in NED the companion (KPG\,467A) in 
Karachentsev's Isolated Pairs of Galaxies Catalogue \cite{kara} 
corresponds only to a faint starlike object about $11\arcsec$ away. 
The small bump in the radial profile at $\sim\!40\arcsec$ corresponds 
to an outer spiral arms. Otherwise we see no obvious deviation from a 
single exponential and therefore the galaxy is classified as \typeoc.  
}
\hfill 
\parbox[b][0.5\height][b]{0.47\textwidth}
{
\includegraphics[width=5.7cm,angle=270]{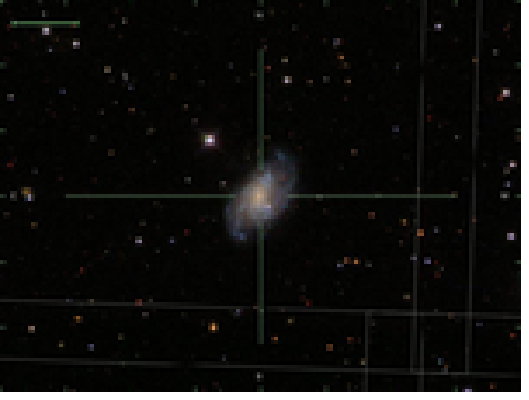}
\hspace*{-0.8cm}
\includegraphics[width=6.1cm,angle=270]{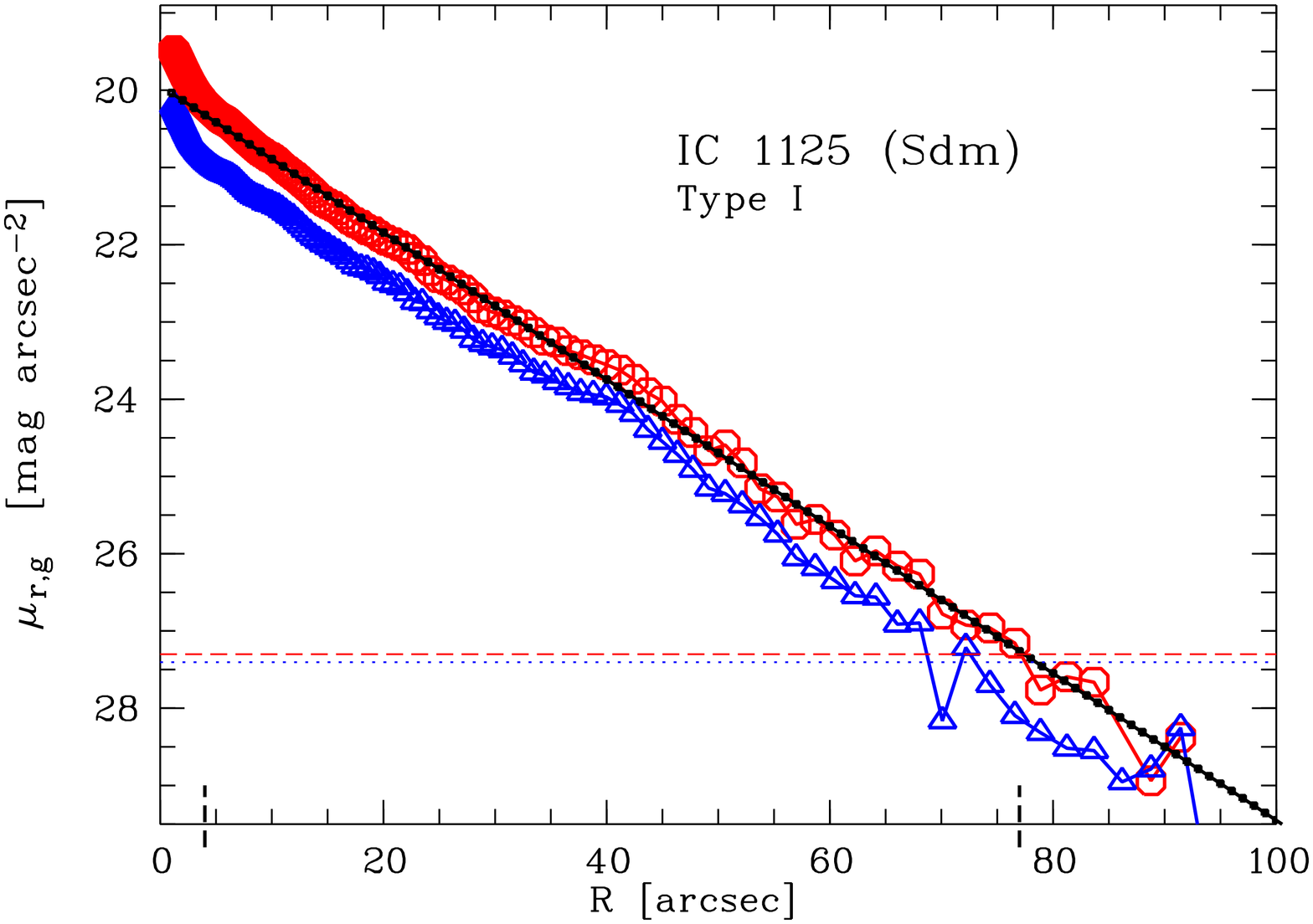}
}
\end{minipage}
\newpage
\onecolumn
\begin{minipage}[t][\textheight][t]{\textwidth}
\parbox[t][0.5\height][t]{0.47\textwidth}
{
\noindent {\bf  IC\,1158      :}        \typetoct               \\          
\texttt{J160134.1+014228 .SXR5*  5.3 -19.36   2.2 2018}\\[0.25cm]
Galaxy shows clearly a truncated profile with an extended break region, 
starting approximately at the end of the spiral arms. The center looks 
like a secondary bar with a small ring embedded in larger bar of size 
roughly $\ltsim 22\arcsec$ (without any noticeable feature in the 
final profile). So the break with downbending profile at $\sim 60\arcsec$ 
is well beyond a typical \typeolr break.   
}
\hfill 
\parbox[t][0.5\height][t]{0.47\textwidth}
{
\includegraphics[width=5.7cm,angle=270,]{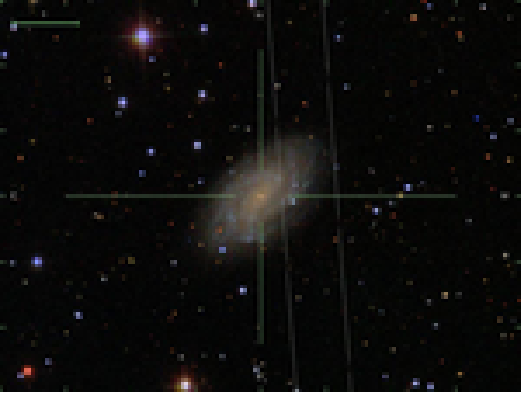}
\hspace*{-0.8cm}
\includegraphics[width=6.1cm,angle=270]{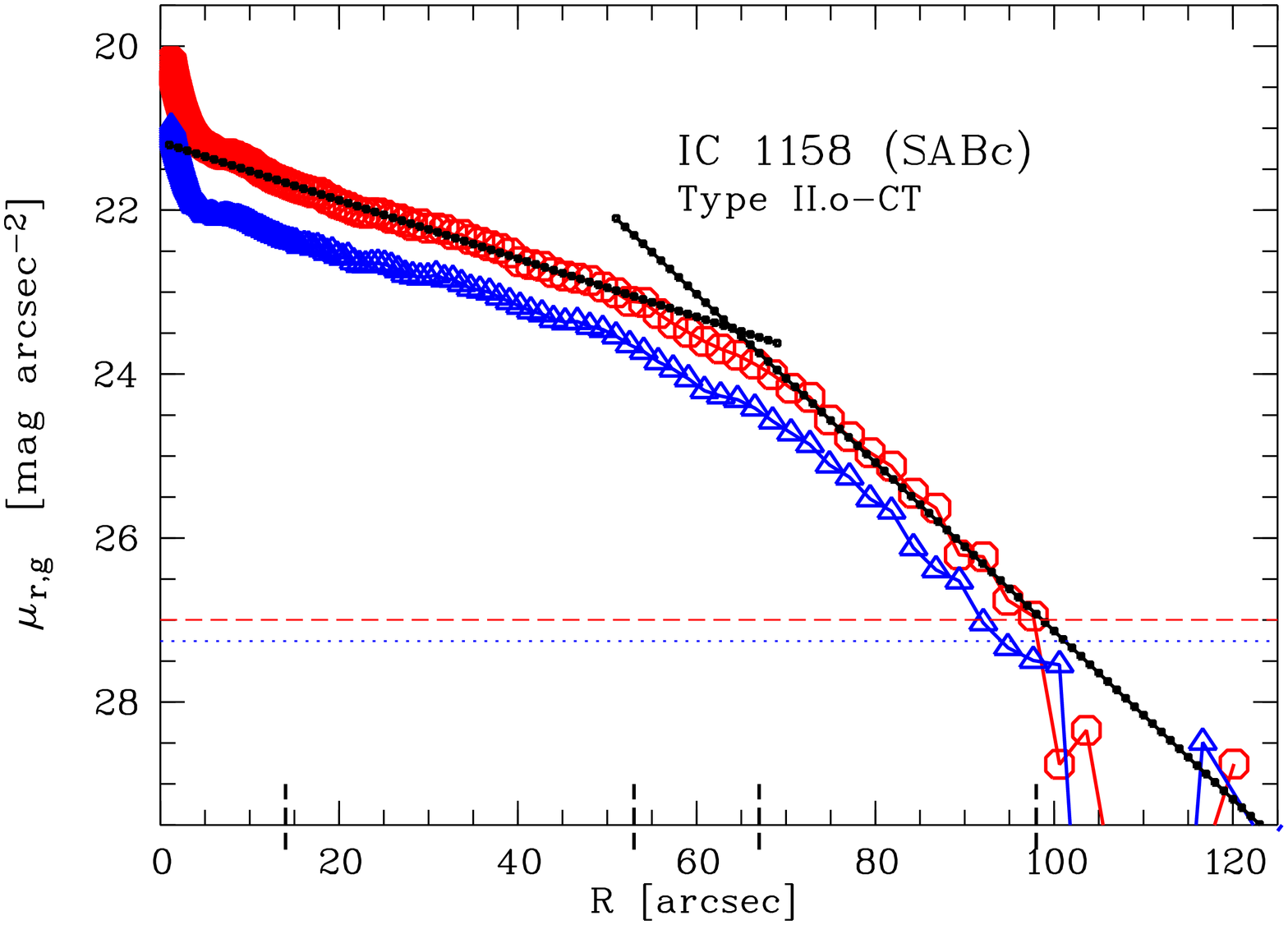}
}
\vfill
\parbox[b][0.5\height][t]{0.47\textwidth}
{
\noindent {\bf NGC\,0450     :}        \typetoct                \\          
\texttt{J011530.8-005138 .SXS6*  6.0 -19.63   3.0 1712}\\[0.25cm]
A background galaxy (UGC\,00807, 11587\kms) is superimposed. The mean 
ellipticity of the outer disk is difficult to determine. The inner 
disk looks fairly round whereas the outer disk looks rather elliptical. 
The size of the slightly more elliptical inner part, associated 
to be the bar (although without any noticeable feature in the 
final profile), is roughly $\sim\!8$\arcsec, so the break with 
the downbending profile at $\sim\!84\arcsec$ is well beyond a 
typical \typeolr break and roughly outside the spiral arm region.  

}
\hfill 
\parbox[b][0.5\height][b]{0.47\textwidth}
{
\includegraphics[width=5.7cm,angle=270]{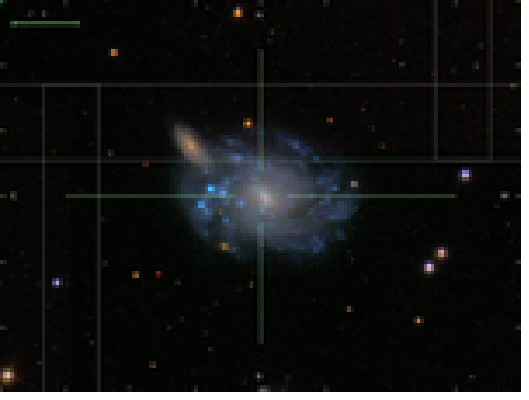}
\hspace*{-0.8cm}
\includegraphics[width=6.1cm,angle=270]{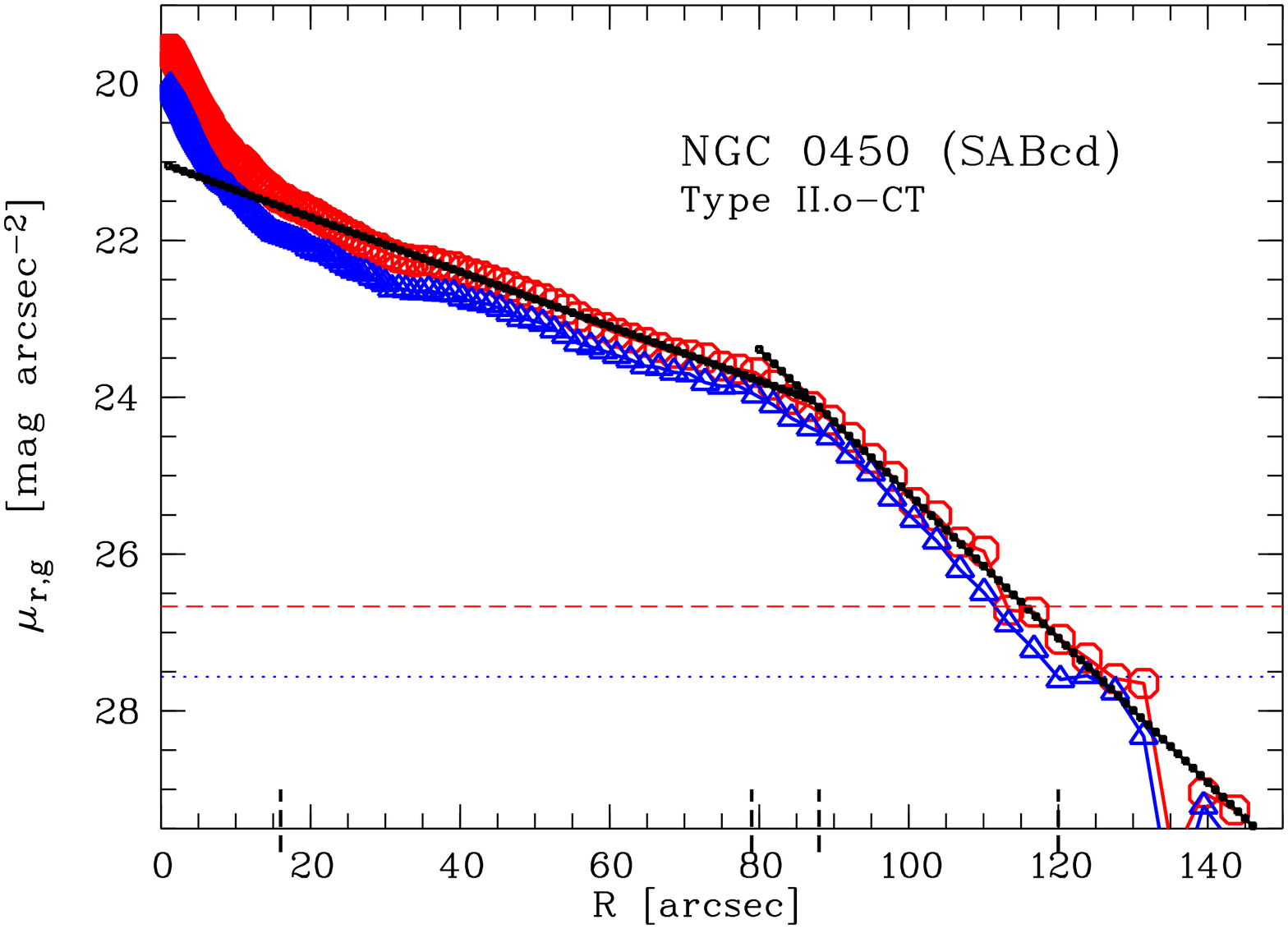}
}
\end{minipage}

\newpage
\onecolumn
\begin{minipage}[t][\textheight][t]{\textwidth}
\parbox[t][0.5\height][t]{0.47\textwidth}
{
\noindent {\bf NGC\,0701     :}        \typeti (possible \typeoc)      \\          
\texttt{J015103.8-094209 .SBT5.  5.1 -19.74   2.5 1729}\\[0.25cm]
Galaxy is in a multiple system together with NGC\,0681 and 
NGC\,4594 (where PGC\,006667 (see below) is also part of) with an 
additional small physical companion (IC\,1738, $v=1750$\kms) 
about $6\arcmin$ away.
Galaxy close to our high axis ratio limit so the dust may have some 
influence. The inner region shows no nucleus, so the centering 
is done from the outer isophotes. 
The extend of the bar is unknown, but the visible bump at $45\arcsec$ 
is probably not related to a bar. It is possible to construct a break 
at $45\arcsec$ (which would be more pronounced in the $g^{\prime}$ band), 
but the inner drop could be also due to the higher inclination with dust 
and/or patchy star formation affecting the profile. 
}
\hfill 
\parbox[t][0.5\height][t]{0.47\textwidth}
{
\includegraphics[width=5.7cm,angle=270,]{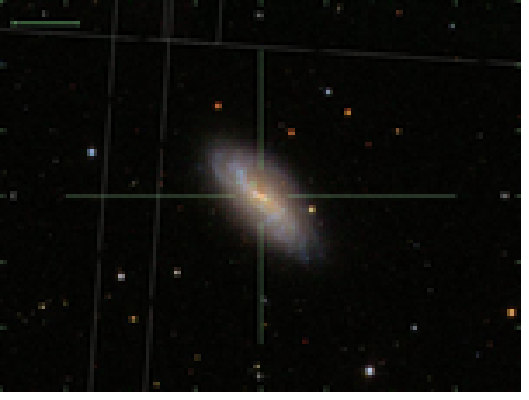}
\hspace*{-0.8cm}
\includegraphics[width=6.1cm,angle=270]{N0701_radn.ps}
}
\vfill
\parbox[b][0.5\height][t]{0.47\textwidth}
{
\noindent {\bf NGC\,0853     :}        \typeiii                 \\        
\texttt{J021141.5-091817 .S..9P  8.3 -18.42   1.5 1405}\\[0.25cm]
The inner region shows no clear nucleus but many \hii regions, so the 
centering is done from the outer isophotes. The inner disk is slightly 
asymmetric with an additional light patch in the western part, therefore 
mean ellipticity and PA difficult to fix. Almost continuously upbending 
profile beyond the peak at $15$\arcsec, which includes all \hii regions, 
getting flat at $\sim\!42$\arcsec. The inner profile is rather curved and 
not well approximated with a single exponential. The apparent outer 
break at $\sim\!100\arcsec$ is only due to the improper fixed ellipse fit 
in the outer parts. 

}
\hfill 
\parbox[b][0.5\height][b]{0.47\textwidth}
{
\includegraphics[width=5.7cm,angle=270]{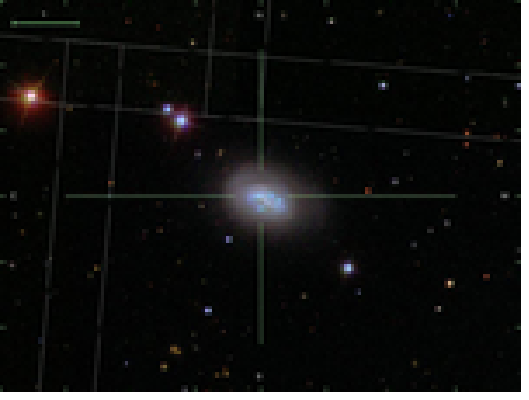}
\hspace*{-0.8cm}
\includegraphics[width=6.1cm,angle=270]{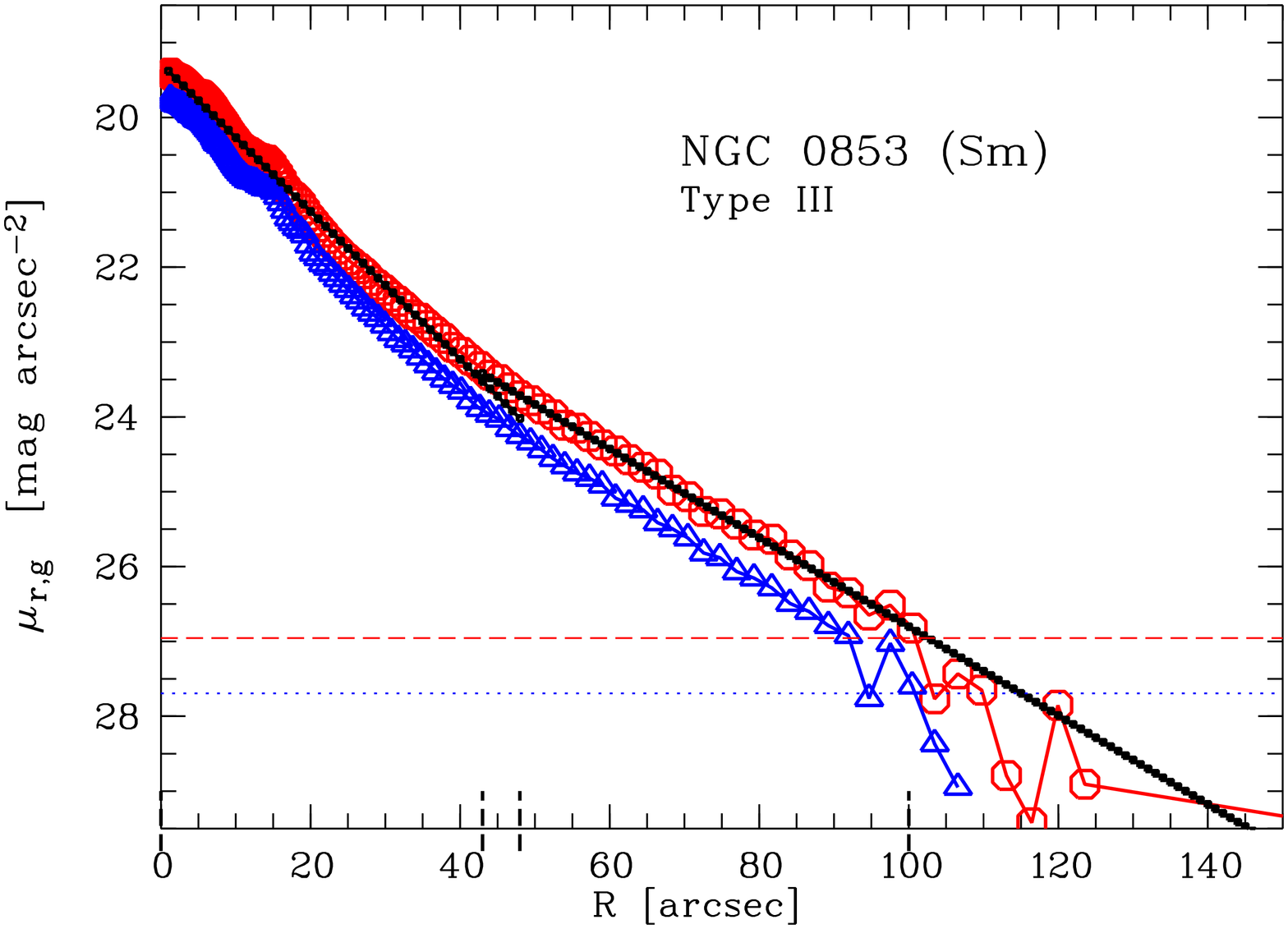}
}
\end{minipage}

\newpage
\onecolumn
\begin{minipage}[t][\textheight][t]{\textwidth}
\parbox[t][0.5\height][t]{0.47\textwidth}
{
\noindent {\bf NGC\,0941     :}        \typect                \\          
\texttt{J022827.9-010906 .SXT5.  5.4 -19.02   2.5 1535}\\[0.25cm]
Galaxy shows only a very small, point-like nucleus and is in a small 
group of galaxies (including NGC\,936 and NGC\,955). The small dip 
at $\sim\!40\arcsec$ is associated with the inner spiral arm region. 
Although classified as SAB no real bar visible on image or in the 
profile thus the downbending break is classified as \typectc.

}
\hfill 
\parbox[t][0.5\height][t]{0.47\textwidth}
{
\includegraphics[width=5.7cm,angle=270,]{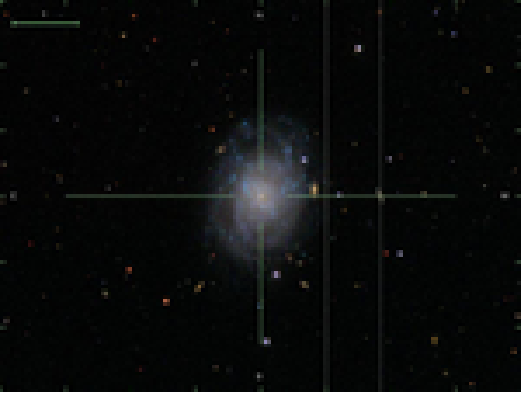}
\hspace*{-0.8cm}
\includegraphics[width=6.1cm,angle=270]{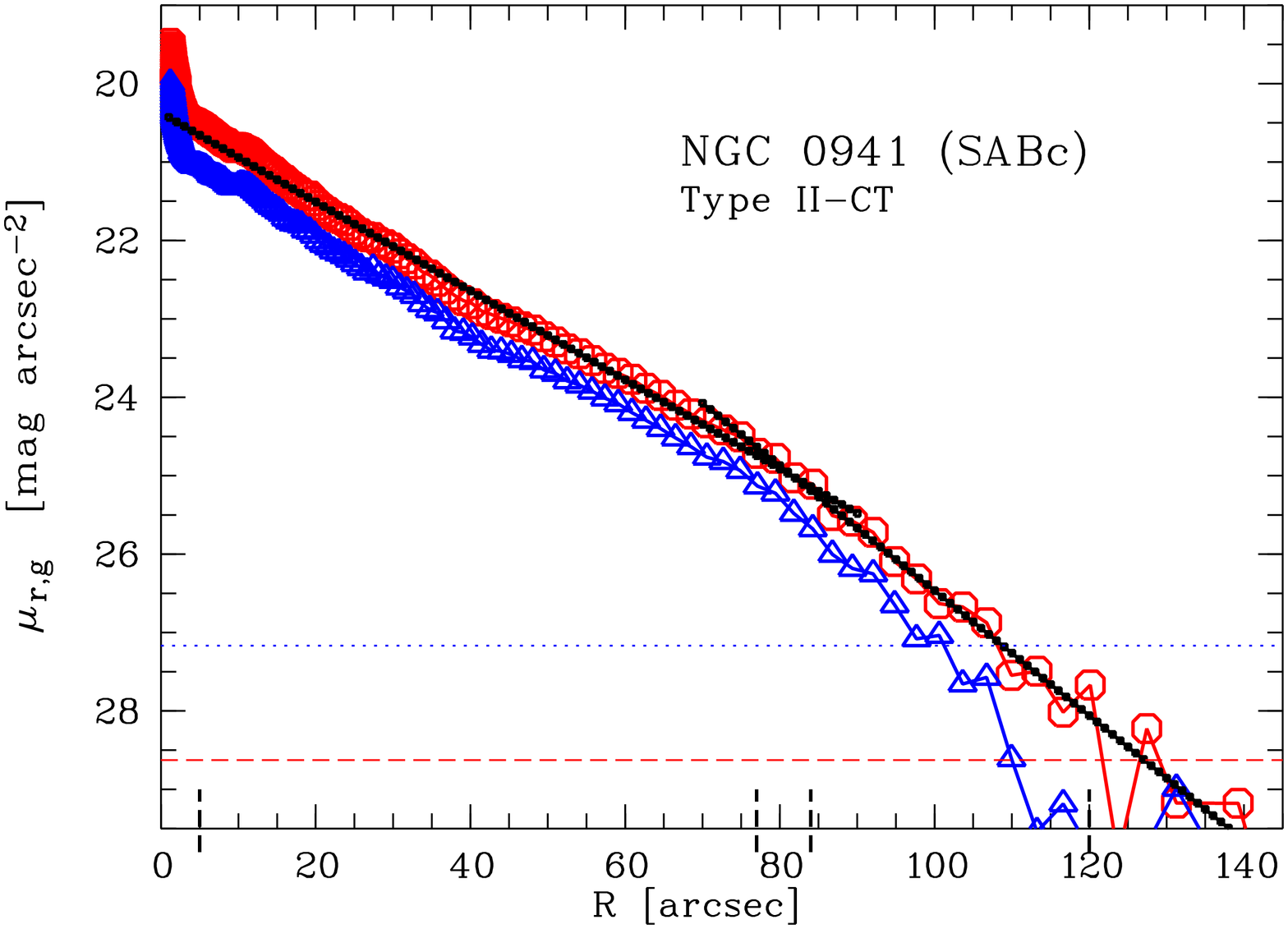}
}
\vfill
\parbox[b][0.5\height][t]{0.47\textwidth}
{
\noindent {\bf NGC\,1042     :}        \typeab                \\          
\texttt{J024023.9-082558 .SXT6.  6.1 -19.83   4.4 1264}\\[0.25cm]
A large scale gradient (from top to bottom) in the background of 
the $r^{\prime}$ and $g^{\prime}$ band image is removed with a linear fit.
Only partly fitted since $\ltsim 1/3$ of galaxy is beyond 
SDSS field.   
Galaxy shows an asymmetric (frayed) extension towards the 
south-east and a sharper edge towards the north-west, which makes 
it not really lopsided but looking rather like being affected by 
moving in a dense IGM. The extended (frayed) spiral arms build the 
outer disk, so ellipticity and PA very difficult to fix. The inner 
bump in the profile at $\sim\!90\arcsec$ is due to spiral arms. The 
inner profile is below the inward extrapolation of the outer disk, 
but due to the asymmetric shape of the disk we call this galaxy 
only \typeabc. 

}
\hfill 
\parbox[b][0.5\height][b]{0.47\textwidth}
{
\includegraphics[width=5.7cm,angle=270]{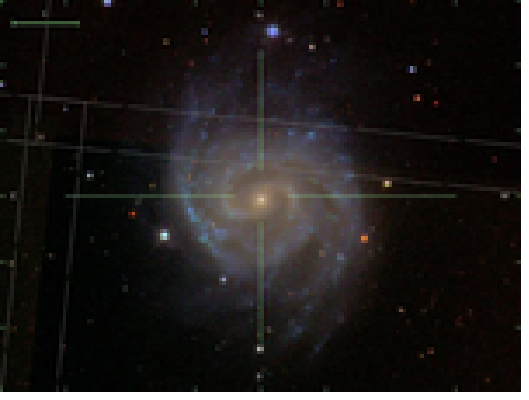}
\hspace*{-0.8cm}
\includegraphics[width=6.1cm,angle=270]{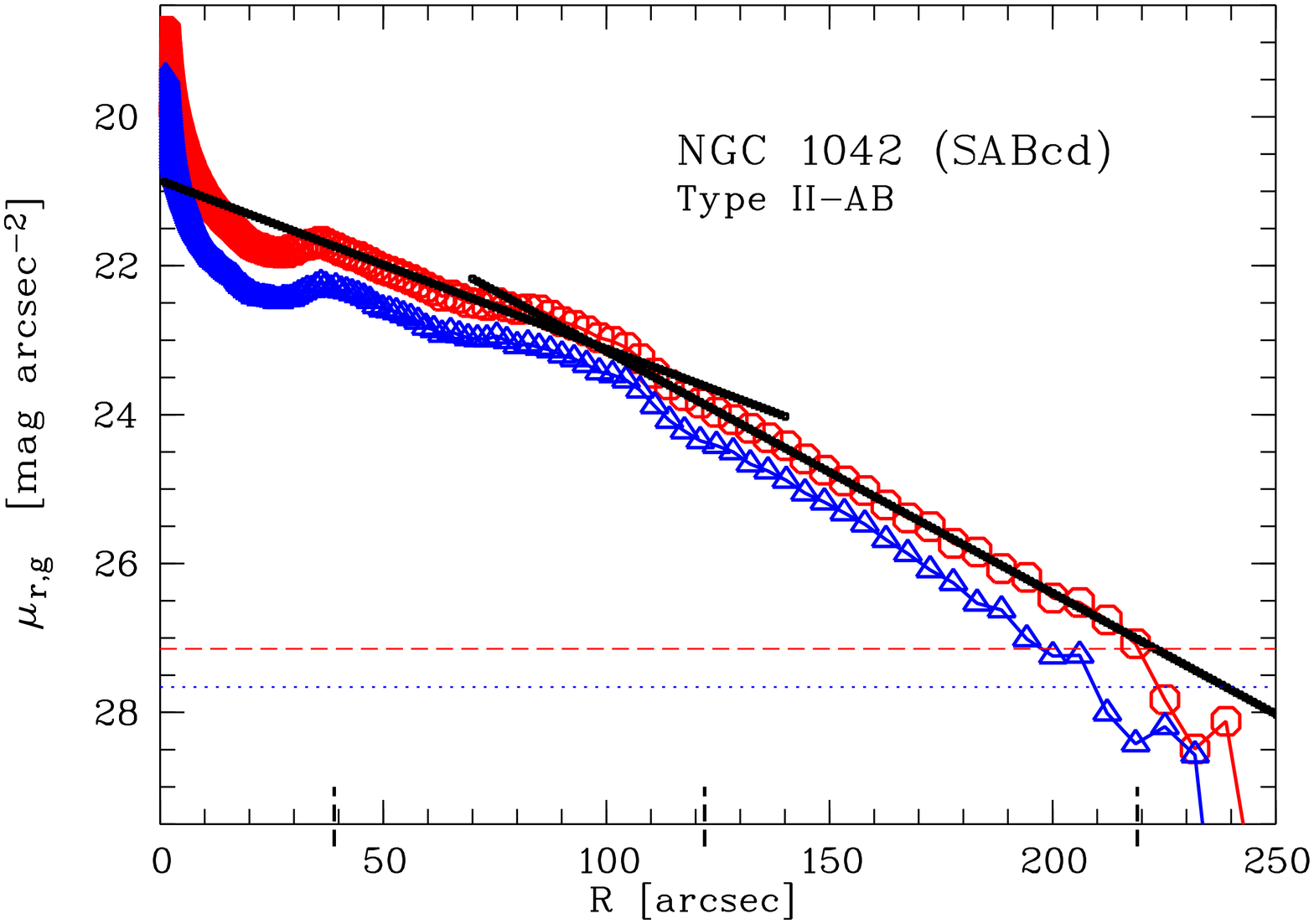}
}
\end{minipage}

\newpage
\onecolumn
\begin{minipage}[t][\textheight][t]{\textwidth}
\parbox[t][0.5\height][t]{0.47\textwidth}
{
\noindent {\bf NGC\,1068 $\equiv$ M\,77    :}        \typeolr          \\  
\texttt{J024240.8-000048 RSAT3.  3.0 -21.69   7.6 1068}\\[0.25cm]
Only partly fitted since \ltsim $1/3$ of galaxy is beyond the SDSS field.
Galaxy appears to have an inner disk, elongated into a bar-like structure
sitting in an outer disk with clearly different PA and with indication 
for an additional outer ring at $\sim\!200$\arcsec. According to 
\cite{erwin2004} it is a double-barred galaxy with an inner bar 
of size $\sim\!15--20\arcsec$ and an outer bar which is very oval 
(not very strong) and very large (extending out to \ltsim$100$\arcsec; 
as seen by \eg \cite{schinnerer2000}). The galaxy looks in this 
sense similar to NGC 5248 \cite[]{jogee_n5248}. 
This peculiar shape makes its very difficult to decide on a mean 
ellipticity and PA of the outer disk from the photometry. The 
position of the outer ring corresponds to the downbending break at 
$\sim\!190\arcsec$ and is therefore roughly twice the bar radius. 
This suggests the \typeolr classification, although the profile
looks at first glance more like a \typeiii with a break at 
$\sim\!100$\arcsec. 

}
\hfill 
\parbox[t][0.5\height][t]{0.47\textwidth}
{
\includegraphics[width=5.7cm,angle=270,]{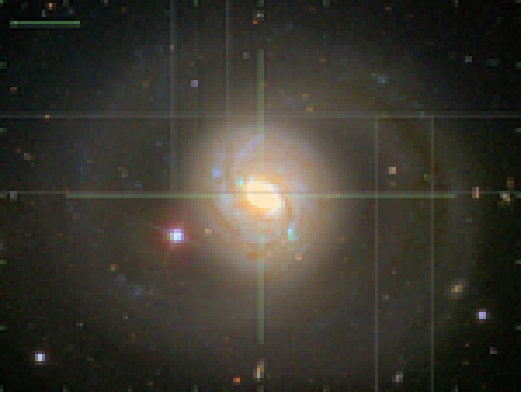}
\hspace*{-0.8cm}
\includegraphics[width=6.1cm,angle=270]{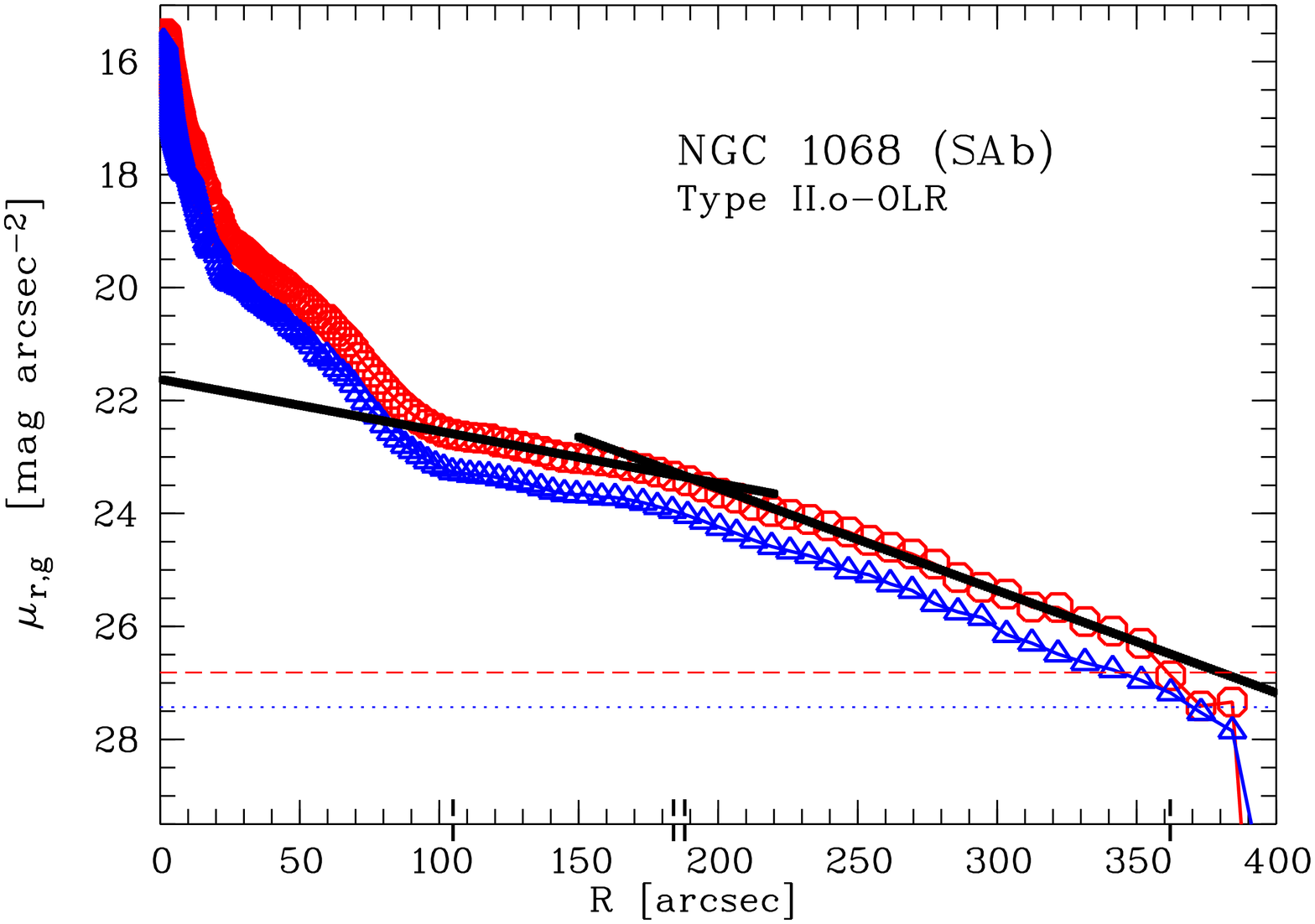}
}
\vfill
\parbox[b][0.5\height][t]{0.47\textwidth}
{
\noindent {\bf NGC\,1084     :}        \typeiii                 \\        
\texttt{J024559.7-073437 .SAS5.  5.1 -20.20   3.2 1299}\\[0.25cm]
A large scale gradient (from top to bottom) in the background of 
the $r^{\prime}$ band image is removed with a linear fit.
Slightly inclined galaxy sitting clearly in a more roundish outer 
envelope, beyond which some stream-like remnants of a possible recent 
interaction are visible. Small edge-on galaxy along the stream path 
is confirmed to be only a background galaxy. According to SDSS 
spectroscopy there is another object inside the disk 
(RA: 02\,46\,00.3, DEC $-07$\,34\,17) with similar velocity and 
associated to a visible light concentration.
The peculiar shape of the outer disk region makes it very difficult 
to decide on a mean ellipticity and PA from the photometry. The 
resulting profile inside the break at $\sim\!90\arcsec$ is rather 
curved towards the center with additional kinks ($\sim\!25\arcsec$ 
and $\sim\!50$\arcsec), so it is not clear what the galaxy might be 
inside the \typeiii envelope. 

}
\hfill 
\parbox[b][0.5\height][b]{0.47\textwidth}
{
\includegraphics[width=5.7cm,angle=270]{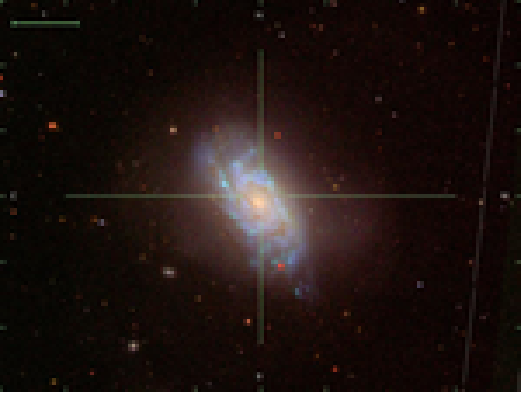}
\hspace*{-0.8cm}
\includegraphics[width=6.1cm,angle=270]{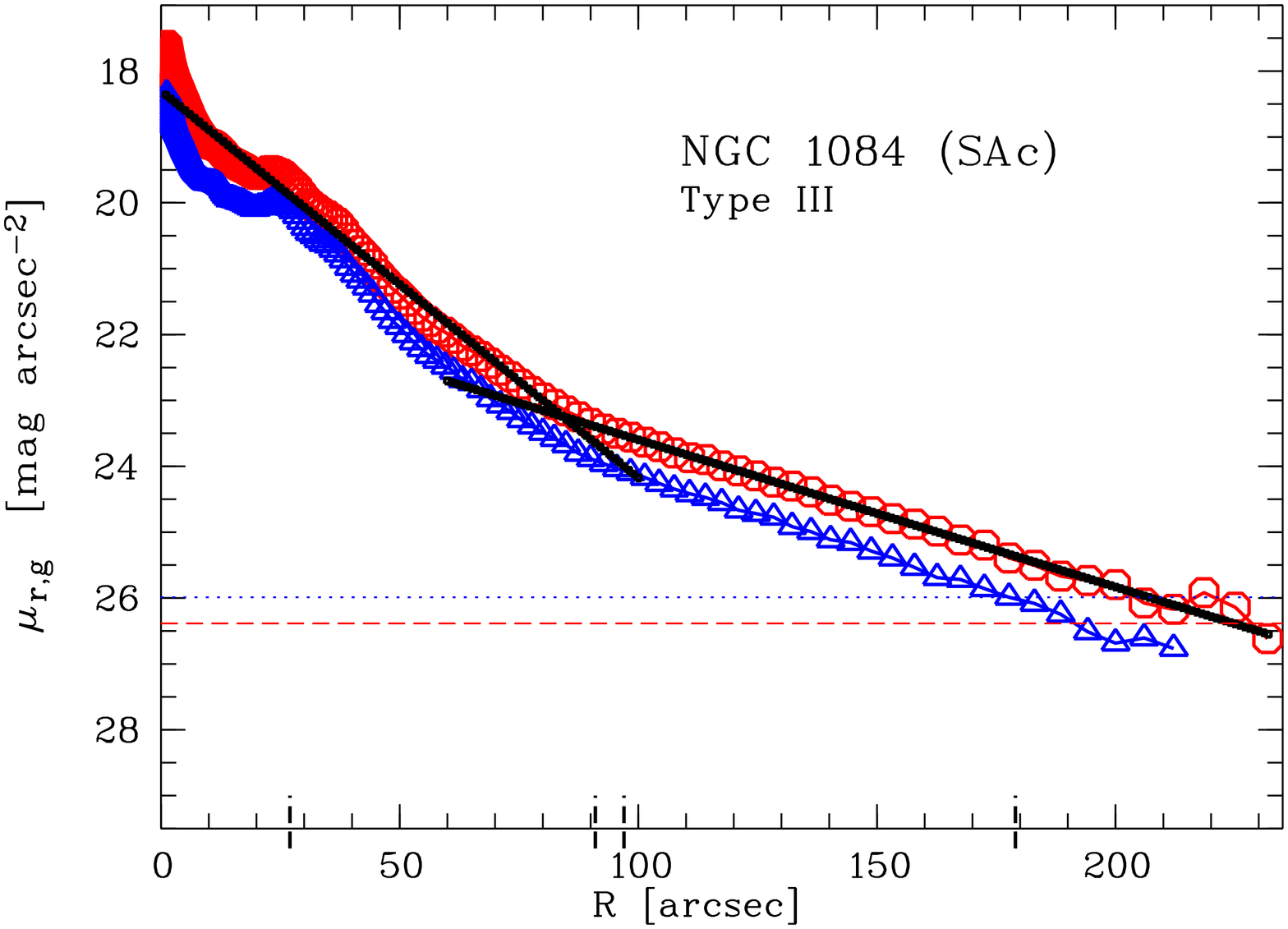}
}
\end{minipage}

\newpage
\onecolumn
\begin{minipage}[t][\textheight][t]{\textwidth}
\parbox[t][0.5\height][t]{0.47\textwidth}
{
\noindent {\bf NGC\,1087     :}        \typeiii                 \\        
\texttt{J024625.2-002955 .SXT5.  5.3 -20.46   3.7 1443}\\[0.25cm]
An apparent faint low surface brightness (center-less) structure 
towards the North is only scattered light from a nearby star. 
The galaxy center is not well defined, it shows only a small bar-like 
structure with a double nucleus, so we used the slightly lopsided 
(but symmetric) outer disk for centering, which accounts for the 
central drop inside $\sim\!50\arcsec$ in the final profile. There is 
no spiral structure visible in the outer disk. The transition seems 
to be rather sharp, but the inner profile is almost curved, so it is 
not clear what its break type is inside the \typeiii envelope. 

}
\hfill 
\parbox[t][0.5\height][t]{0.47\textwidth}
{
\includegraphics[width=5.7cm,angle=270,]{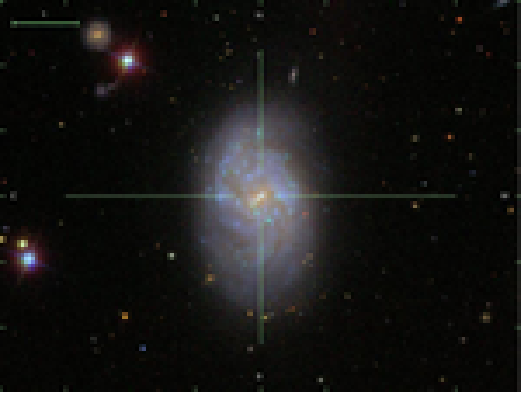}
\hspace*{-0.8cm}
\includegraphics[width=6.1cm,angle=270]{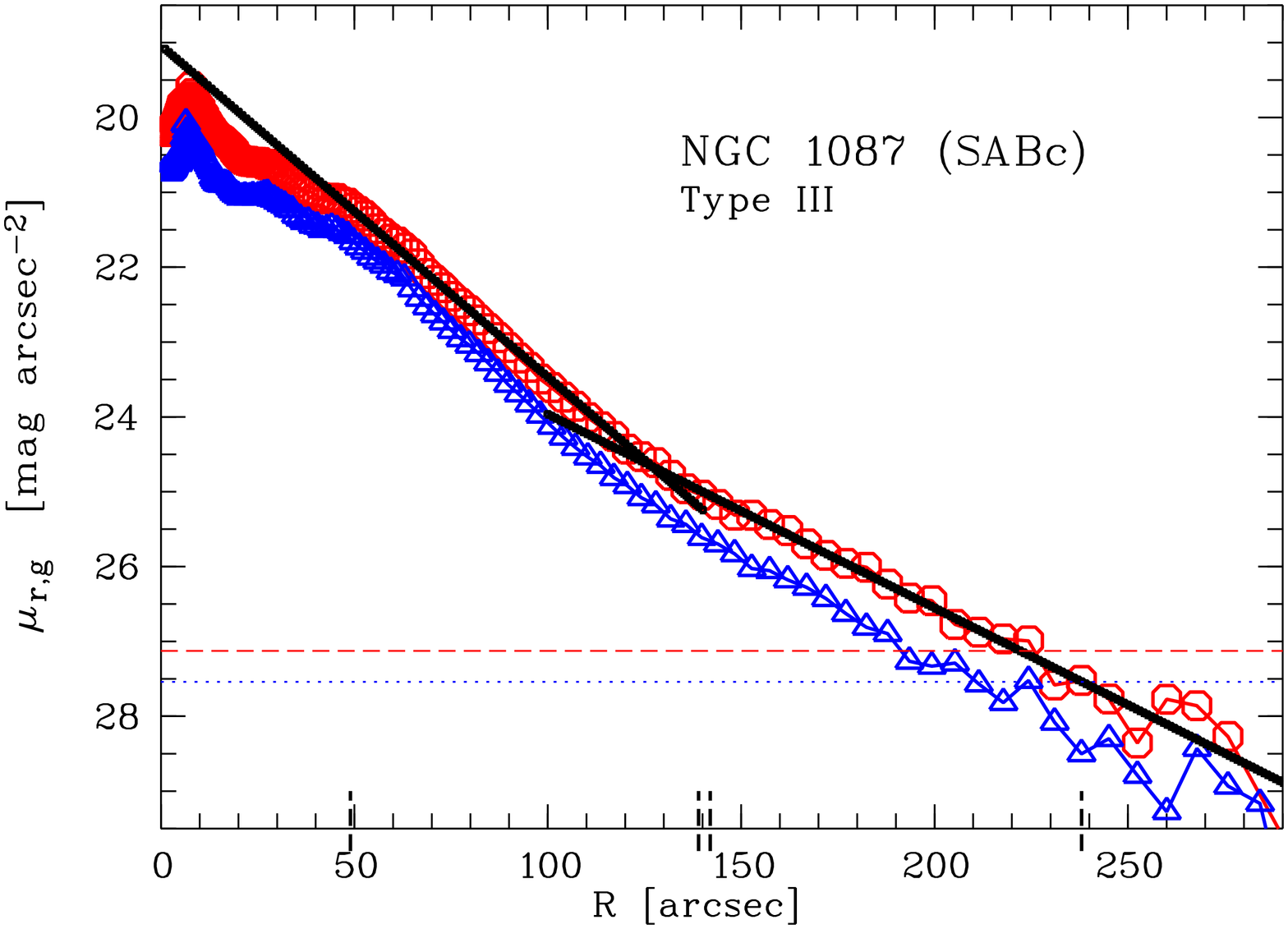}
}
\vfill
\parbox[b][0.5\height][t]{0.47\textwidth}
{
\noindent {\bf NGC\,1299     :}        \typeiii                 \\        
\texttt{J032009.4-061550 .SBT3?  3.2 -19.25   1.2 2197}\\[0.25cm]
Galaxy is small and close to our high axis ratio limit. The background 
exhibits a gradient but the galaxy is small enough to avoid fitting it. 
SDSS spectroscopy detects an object 0.2\arcmin away from center with 
similar velocity. The rather symmetric outer disk shows no spiral 
structure but has slightly different ellipticity and PA compared to 
the inner part. The shoulder in the profile at $\sim\!15$\arcsec
is related to the inner (bar-like) region with changing PA.    

}
\hfill 
\parbox[b][0.5\height][b]{0.47\textwidth}
{
\includegraphics[width=5.7cm,angle=270]{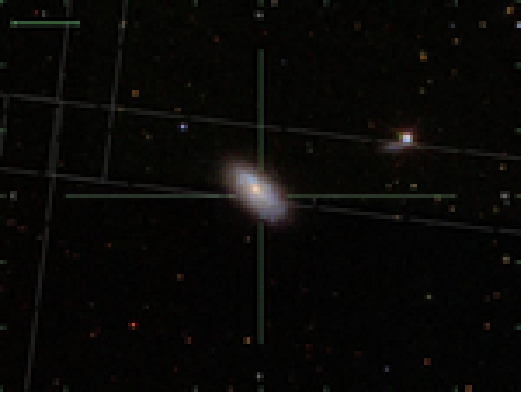}
\hspace*{-0.8cm}
\includegraphics[width=6.1cm,angle=270]{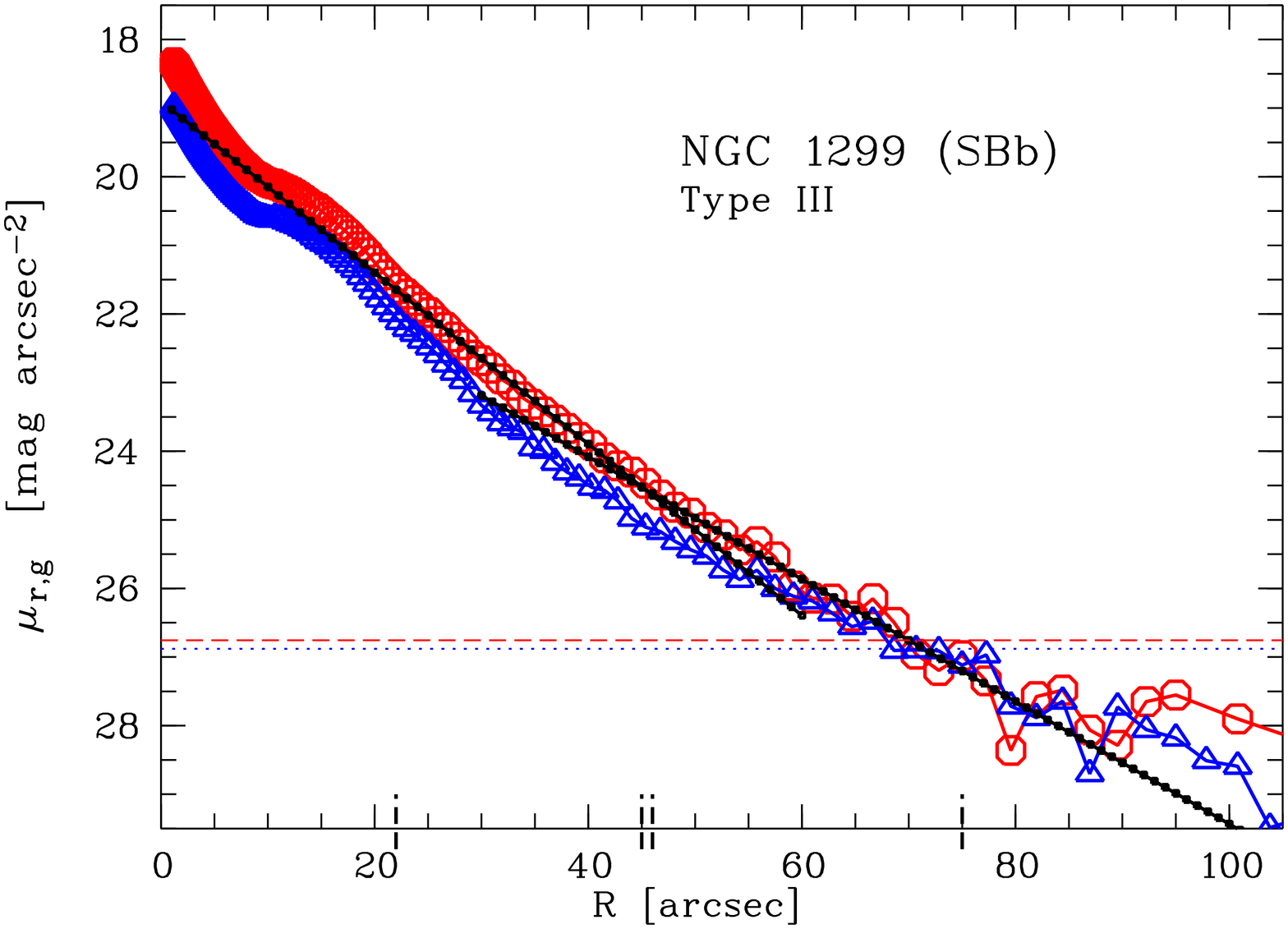}
}
\end{minipage}

\newpage
\onecolumn
\begin{minipage}[t][\textheight][t]{\textwidth}
\parbox[t][0.5\height][t]{0.47\textwidth}
{
\noindent {\bf NGC\,2541     :}        \typect                \\          
\texttt{J081440.1+490341 .SAS6.  6.2 -18.50   5.8  734}\\[0.25cm]
Galaxy with low surface brightness disk, having many patchy SF 
knots and an asymmetric (not well described with an ellipse), 
lopsided outer disk, which makes the chosen ellipticity 
and PA of the disk uncertain. 
The central peak is off by $\sim8\arcsec$ compared to the outer 
disk where the centering is done, which causes the 
dip in the final profile close to the center. 
The downbending break at $\sim\!140\arcsec$ is still inside the
nearly symmetric part of the disk, thus classified as \typect but 
this should be taken with some caution.
This galaxy was classified as SB in our original LEDA catalogue but 
has been recently reclassified being now SAB. However, NED lists this 
galaxy as SA, using the RC3 classification, which is consistent with 
no bar being visible on the image or in the profile, so the downbending 
break is not a \typeolr break. 

}
\hfill 
\parbox[t][0.5\height][t]{0.47\textwidth}
{
\includegraphics[width=5.7cm,angle=270,]{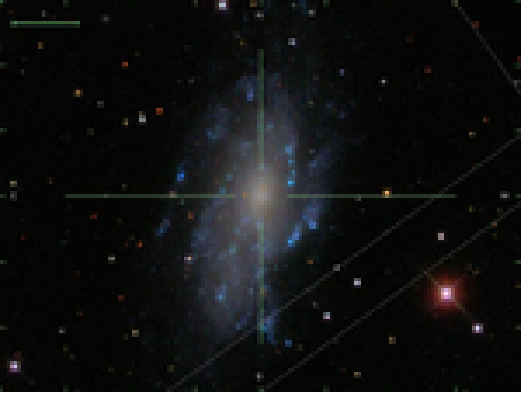}
\hspace*{-0.8cm}
\includegraphics[width=6.1cm,angle=270]{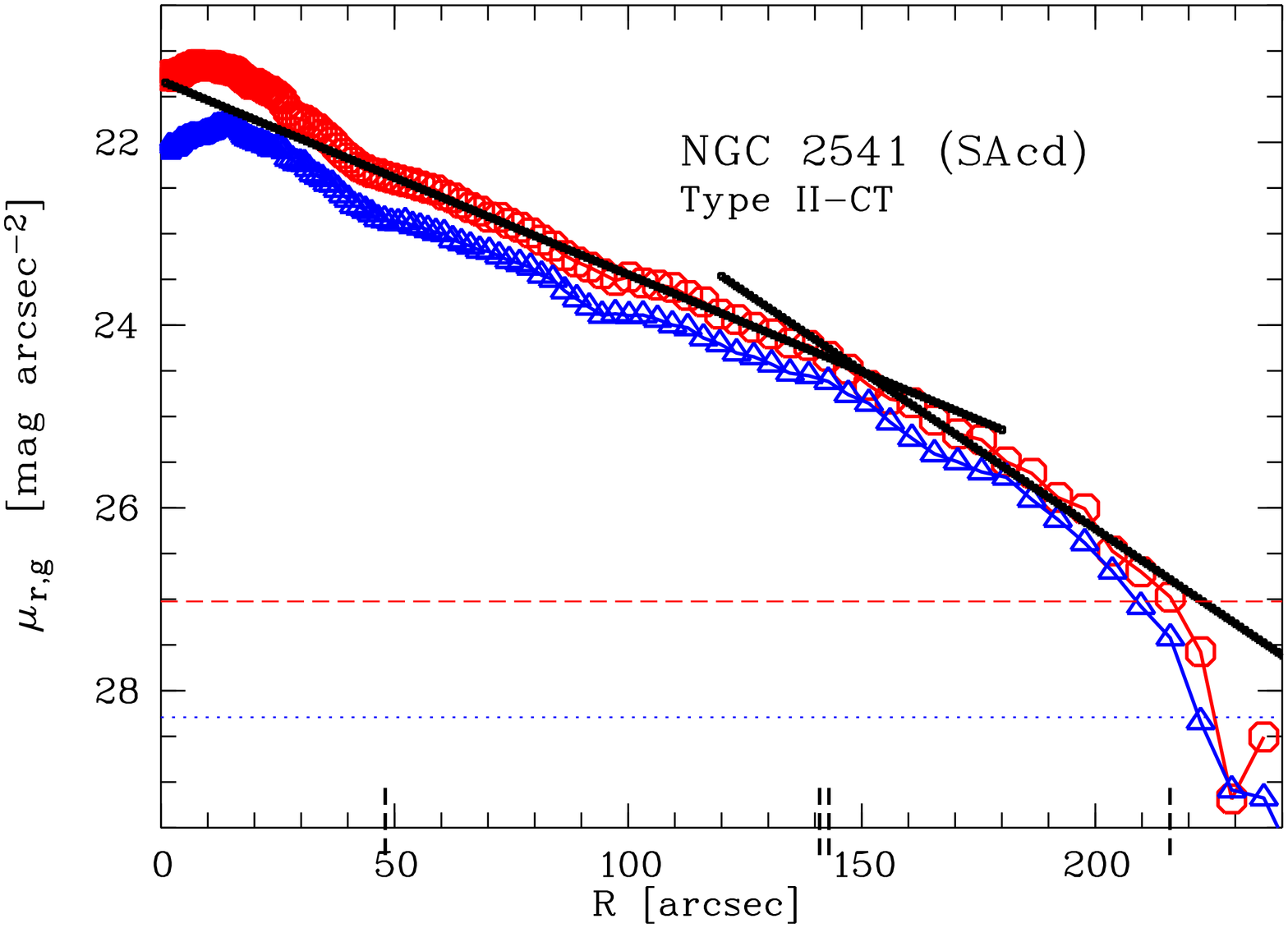}
}
\vfill
\parbox[b][0.5\height][t]{0.47\textwidth}
{
\noindent {\bf NGC\,2543     :}        \typeti                \\         
\texttt{J081258.0+361516 .SBS3.  3.1 -20.66   2.5 2590}\\[0.25cm]
According to NED there is a dwarf elliptical galaxy (KUG\,0809+363) at 
a similar distance. On the SDSS image (confirmed with DSS) there is an 
additional, low surface brightness structure visible at 
RA\,08\,13\,08.5 and DEC\,$+$36\,16\,35, which is possibly another 
dwarf companion with an unknown distance.
The galaxy is clearly double barred with a $R\sim9\arcsec$ long 
secondary bar. Two bright, s-shaped spiral arms dominate the inner 
disk. The extended spiral arm towards the north-east is responsible 
for a slightly asymmetric outer disk. This structure makes the 
photometric measurement of the inclination (ellipticity) and PA 
very difficult. The bar and pseudo-ring build by the inner spiral 
arms are responsible for the prominent bump at $\sim\!35\arcsec$ in 
the profile, followed by a dip inside, which leads to the 
\typeti classification.  

}
\hfill 
\parbox[b][0.5\height][b]{0.47\textwidth}
{
\includegraphics[width=5.7cm,angle=270]{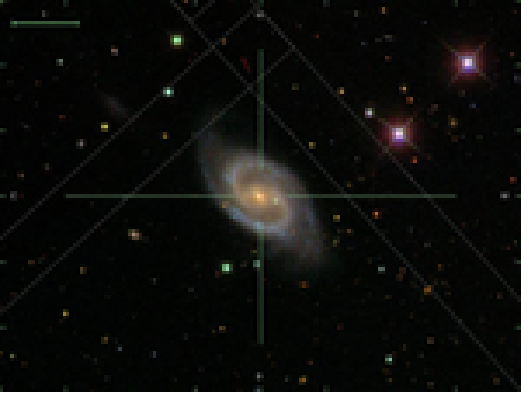}
\hspace*{-0.8cm}
\includegraphics[width=6.1cm,angle=270]{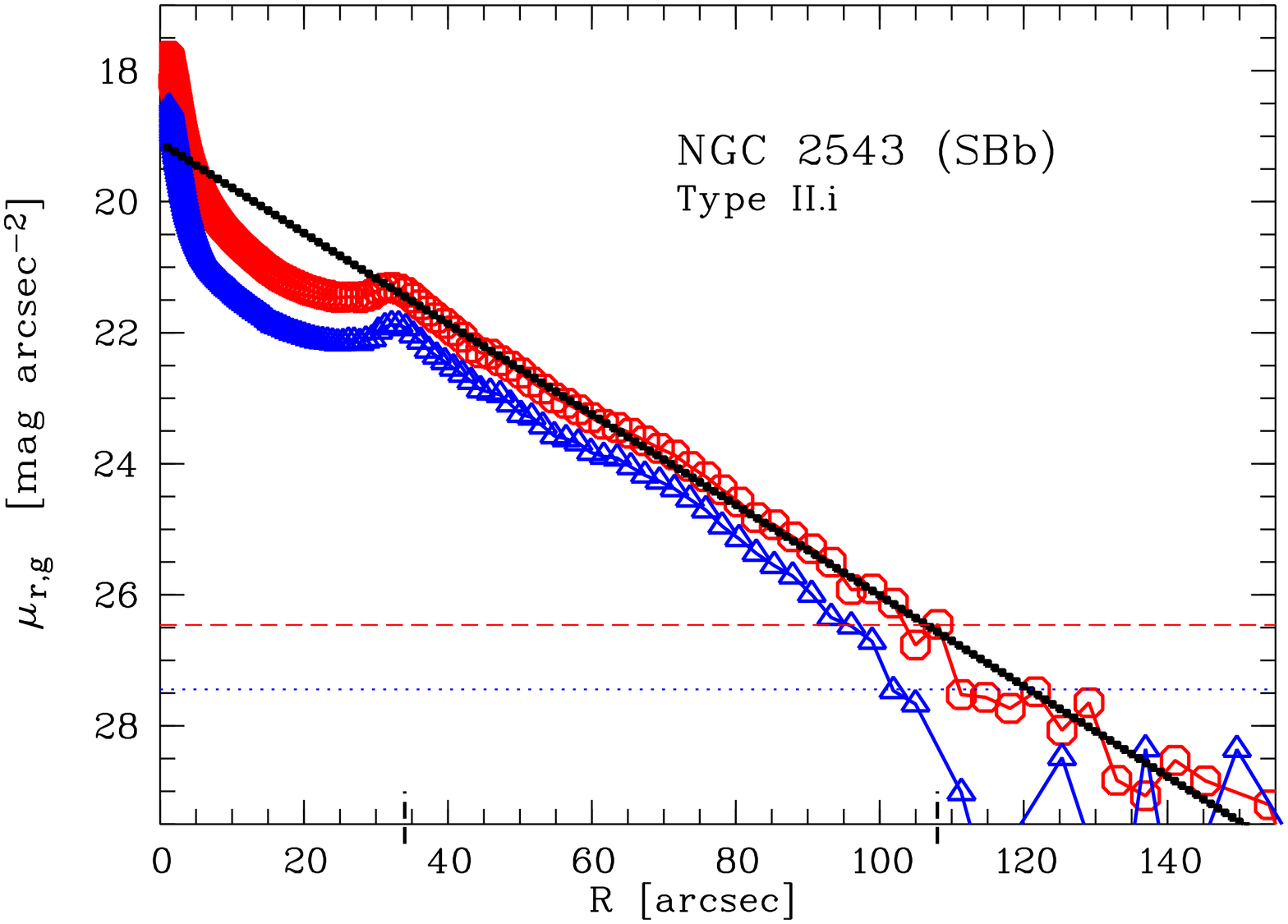}
}
\end{minipage}

\newpage
\onecolumn
\begin{minipage}[t][\textheight][t]{\textwidth}
\parbox[t][0.5\height][t]{0.47\textwidth}
{
\noindent {\bf NGC\,2684     :}        \typect                 \\          
\texttt{J085454.1+490937 .S?...  7.8 -19.76   1.0 3043}\\[0.25cm]
The rich cluster in background, with significantly redder colour, forces 
a large background mask covering a small part of the galaxy. Slightly 
asymmetric disk with bright center and no coherent spiral arms makes 
ellipticity and PA uncertain. Bump at $\sim\!12\arcsec$ in profile is 
due to an inner ring-like structure, without an associated obvious bar 
structure. The break at $\sim\!25\arcsec$ is not related to a 
morphological feature in the disk and, although at about twice the 
ring radius, classified as \typectc.
The upbending profile starting at $\sim\!42\arcsec$ is most probably 
due to an improper sky subtraction caused by extended light from 
the background cluster.  

}
\hfill 
\parbox[t][0.5\height][t]{0.47\textwidth}
{
\includegraphics[width=5.7cm,angle=270,]{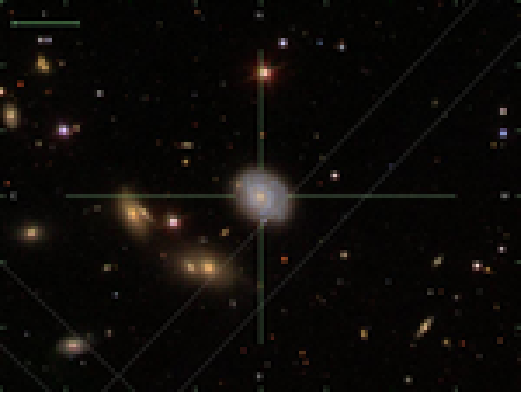}
\hspace*{-0.8cm}
\includegraphics[width=6.1cm,angle=270]{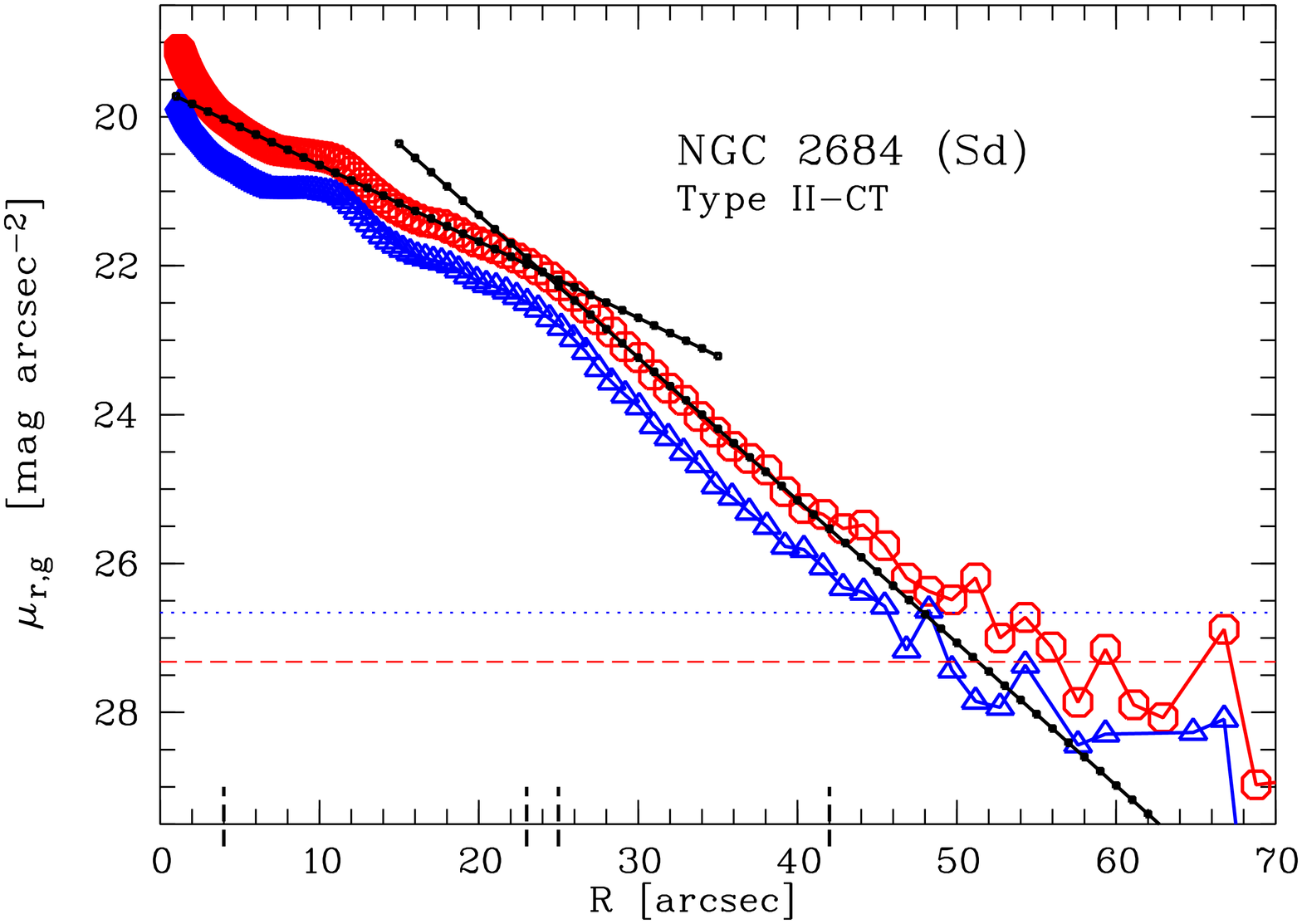}
}
\vfill
\parbox[b][0.5\height][t]{0.47\textwidth}
{
\noindent {\bf NGC\,2701     :}        \typeabiii     \\
\texttt{J085905.9+534616 .SXT5*  5.1 -20.36   2.1 2528}\\[0.25cm]
According to NED there is a small companion (with similar velocity) 
superimposed on the galaxy disk towards the South at the end of 
a spiral arm which is masked here in addition to the bright star on 
the disk towards the West. The galaxy has a lopsided disk with the 
bar being offcenterd compared to the outer isophotes, where the 
centering is done. This is responsible for the dip in the center 
an possibly for the break at $\sim\!42$\arcsec, which is therefore 
called \typeabc. 
The region beyond the second break at $\sim\!70\arcsec$ corresponds to some 
faint, almost symmetric light around the galaxy with some large extension 
towards north-east, adding the classification \typeiiic. This is possibly 
triggered by interaction with the dwarf companion at the other side. 

}
\hfill 
\parbox[b][0.5\height][b]{0.47\textwidth}
{
\includegraphics[width=5.7cm,angle=270]{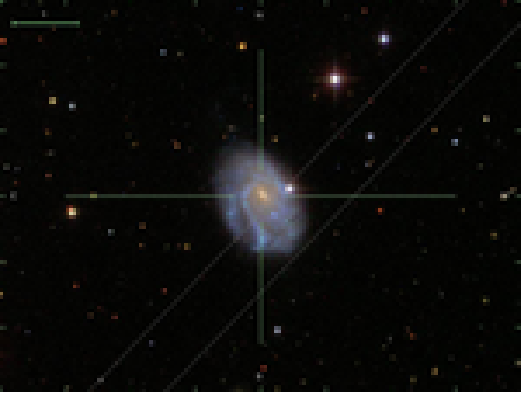}
\hspace*{-0.8cm}
\includegraphics[width=6.1cm,angle=270]{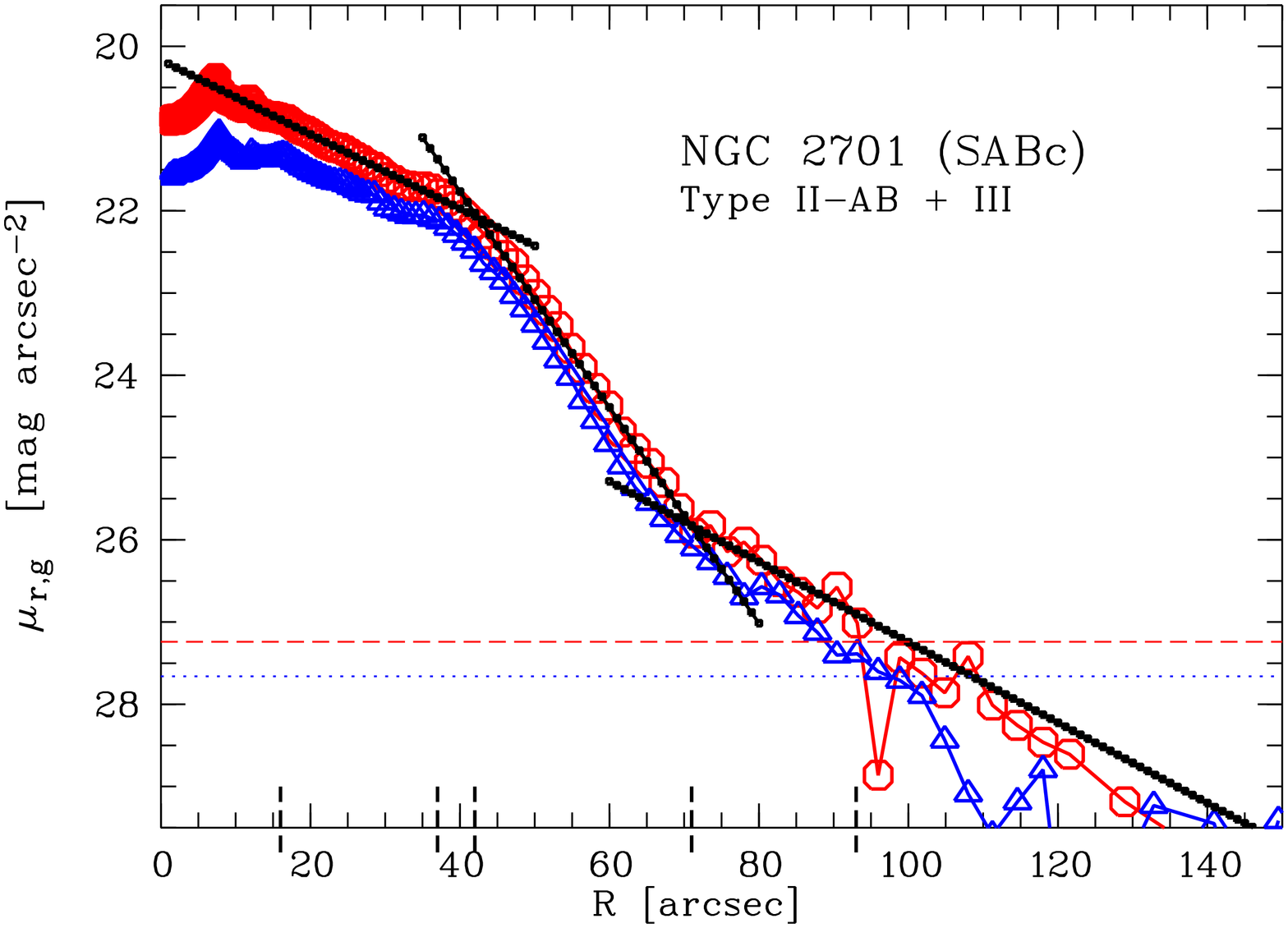}
}
\end{minipage}

\newpage
\onecolumn
\begin{minipage}[t][\textheight][t]{\textwidth}
\parbox[t][0.5\height][t]{0.47\textwidth}
{
\noindent {\bf NGC\,2776     :}        \typeo                 \\          
\texttt{J091214.3+445717 .SXT5.  5.1 -21.04   2.9 2796}\\[0.25cm]
The galaxy is classified as SAB but a bar is not obvious on the image. 
The background estimate is uncertain due to a nearby (off-field), 
extremely bright star. For the final profile we used a round ellipticity 
instead of the usual value at $1\sigma$ (\cf\sec\ref{ellipse}) which 
gives an almost prototypical \typeo profile with some minor wiggles, 
a bump at $\sim25\arcsec$ due to the inner spiral arms, and some outer 
bumps due to some very weak spiral arm structures in the outer disk.  

}
\hfill 
\parbox[t][0.5\height][t]{0.47\textwidth}
{
\includegraphics[width=5.7cm,angle=270,]{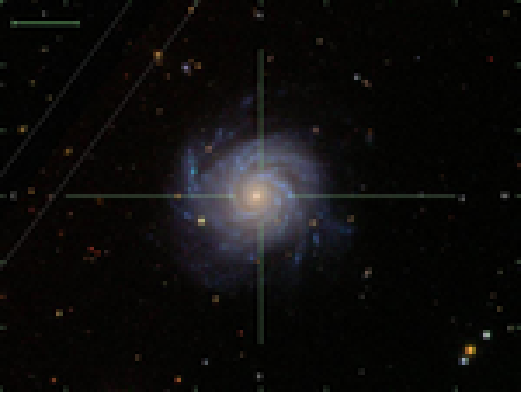}
\hspace*{-0.8cm}
\includegraphics[width=6.1cm,angle=270]{N2776_radn.ps}
}
\vfill
\parbox[b][0.5\height][t]{0.47\textwidth}
{
\noindent {\bf NGC\,2967     :}        \typeiiid                 \\        
\texttt{J094203.3+002011 .SAS5.  4.8 -20.26   2.6 1858}\\[0.25cm]
The galaxy is close to the edge of the SDSS field in an area with an 
increased background which is masked. Very faint, extended, but symmetric 
spiral arms inside the outer disk at $\sim100\arcsec$ are visible, which 
do not continue inwards but start beyond the bump in the transition region 
at $\sim\!75$\arcsec. The exact break radius is therefore uncertain. 

}
\hfill 
\parbox[b][0.5\height][b]{0.47\textwidth}
{
\includegraphics[width=5.7cm,angle=270]{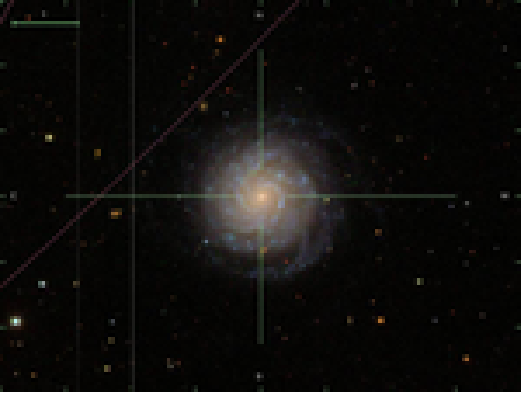}
\hspace*{-0.8cm}
\includegraphics[width=6.1cm,angle=270]{N2967_radn.ps}
}
\end{minipage}

\newpage
\onecolumn
\begin{minipage}[t][\textheight][t]{\textwidth}
\parbox[t][0.5\height][t]{0.47\textwidth}
{
\noindent {\bf NGC\,3055     :}        \typetoct                \\          
\texttt{J095517.9+041611 .SXS5.  5.0 -19.90   2.1 1816}\\[0.25cm]
Galaxy is classified as SAB and shows a narrow, thin bar, 
with the spiral arm structure wrapped around, which is responsible 
for the bump in the final profile at $\sim\!18$\arcsec. The break 
with a downbending profile at $\sim\!55\arcsec$ is well beyond a 
typical \typeolr break.   

}
\hfill 
\parbox[t][0.5\height][t]{0.47\textwidth}
{
\includegraphics[width=5.7cm,angle=270,]{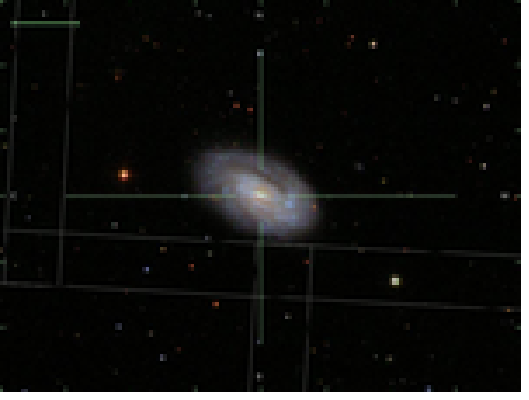}
\hspace*{-0.8cm}
\includegraphics[width=6.1cm,angle=270]{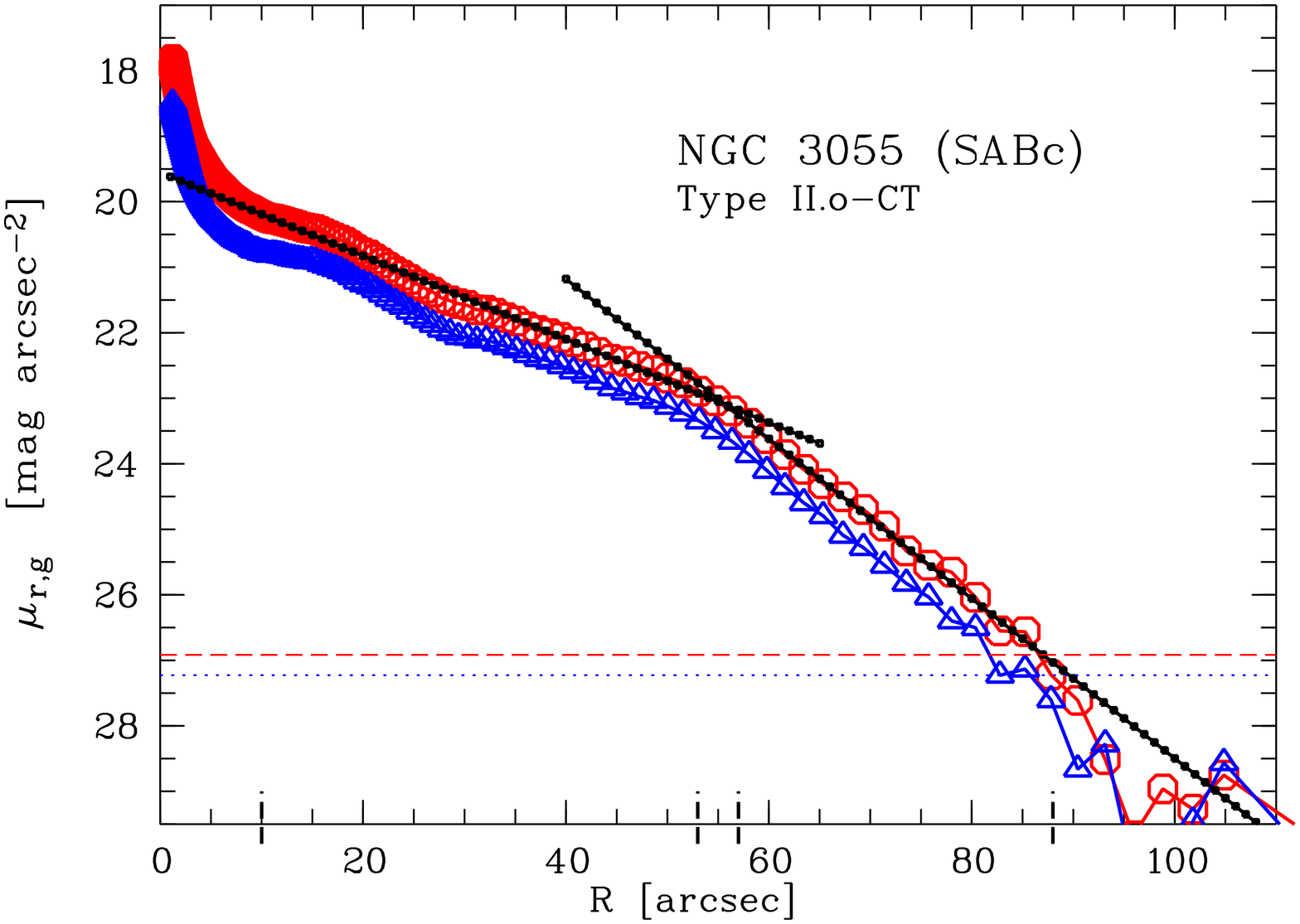}
}
\vfill
\parbox[b][0.5\height][t]{0.47\textwidth}
{
\noindent {\bf NGC\,3246     :}        \typeab                \\          
\texttt{J102641.8+035143 .SX.8.  8.0 -18.91   2.2 2138}\\[0.25cm]
Galaxy is selected from the NGC catalogue \cite{tully1988} 
by \cite[]{pisano1999} as being isolated. Their \hi measurements 
show a lopsided disk which resembles the one on the present SDSS
image. The central bar region is clearly offcenterd compared to 
the outer isophote, which is used for the centering. The downbending 
break in the final profile and the downbending at the very center
is therefore most probably only an apparent break due to the fixed 
ellipse fitting. Thus the galaxy is classified as \typeabc. 
The background is rather disturbed by a bright star, 
but the galaxy is small enough not to be significantly influenced.    

}
\hfill 
\parbox[b][0.5\height][b]{0.47\textwidth}
{
\includegraphics[width=5.7cm,angle=270]{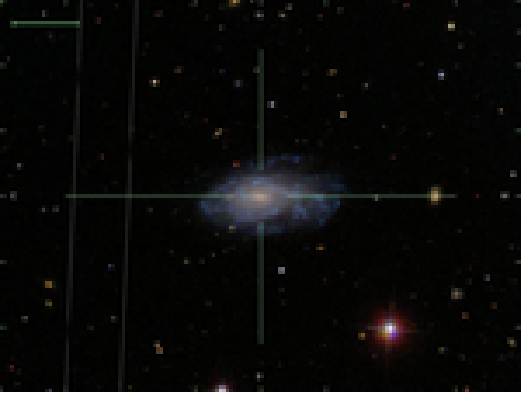}
\hspace*{-0.8cm}
\includegraphics[width=6.1cm,angle=270]{N3246_radn.ps}
}
\end{minipage}

\newpage
\onecolumn
\begin{minipage}[t][\textheight][t]{\textwidth}
\parbox[t][0.5\height][t]{0.47\textwidth}
{
\noindent {\bf NGC\,3259     :}        \typeiiid                 \\        
\texttt{J103234.8+650228 .SXT4*  4.0 -19.84   2.1 1929}\\[0.25cm]
The background is rather disturbed by a bright star, but the galaxy 
is small enough not to be significantly influenced. Galaxy is isolated 
according to \cite{prada2003}, but has one SDSS-detected dwarf companion 
($\sim18.0$ $g^{\prime}$-mag) about 10\arcmin away with similar 
velocity ($v=1744$\kms) and another possible, dwarf companion (or large 
\hii region) projected on the south part of the disk (no counterpart in NED), 
which is partly covered by a foreground star, but also clearly present on the 
DSS image. 
The break region of the upbending profile at $\sim55\arcsec$ is 
fairly extended and curved. The inner profile shows wiggles at 
$\sim\!15\arcsec$ and $\sim\!30$\arcsec, but the resolution is too 
low to identify a bar or ring. The spiral structure apparently 
extends into the outer disk well beyond the break, with two, 
thin symmetric spiral arms (plus more fluffy ones) starting at 
the defined break radius. However, it is not clear if they are 
really the continuation of the inner arms.
The upbending characteristic is confirmed by a profile shown in 
\cite{courteau1996} (see UGC\,05717).

}
\hfill 
\parbox[t][0.5\height][t]{0.47\textwidth}
{
\includegraphics[width=5.7cm,angle=270,]{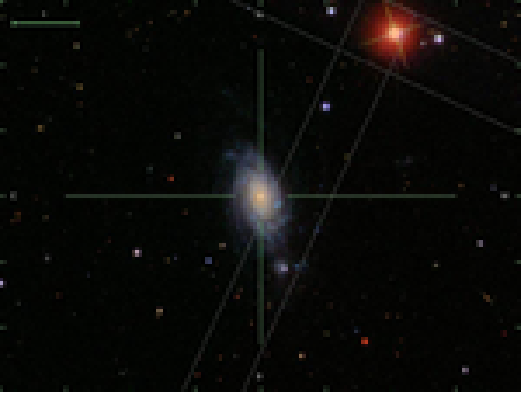}
\hspace*{-0.8cm}
\includegraphics[width=6.1cm,angle=270]{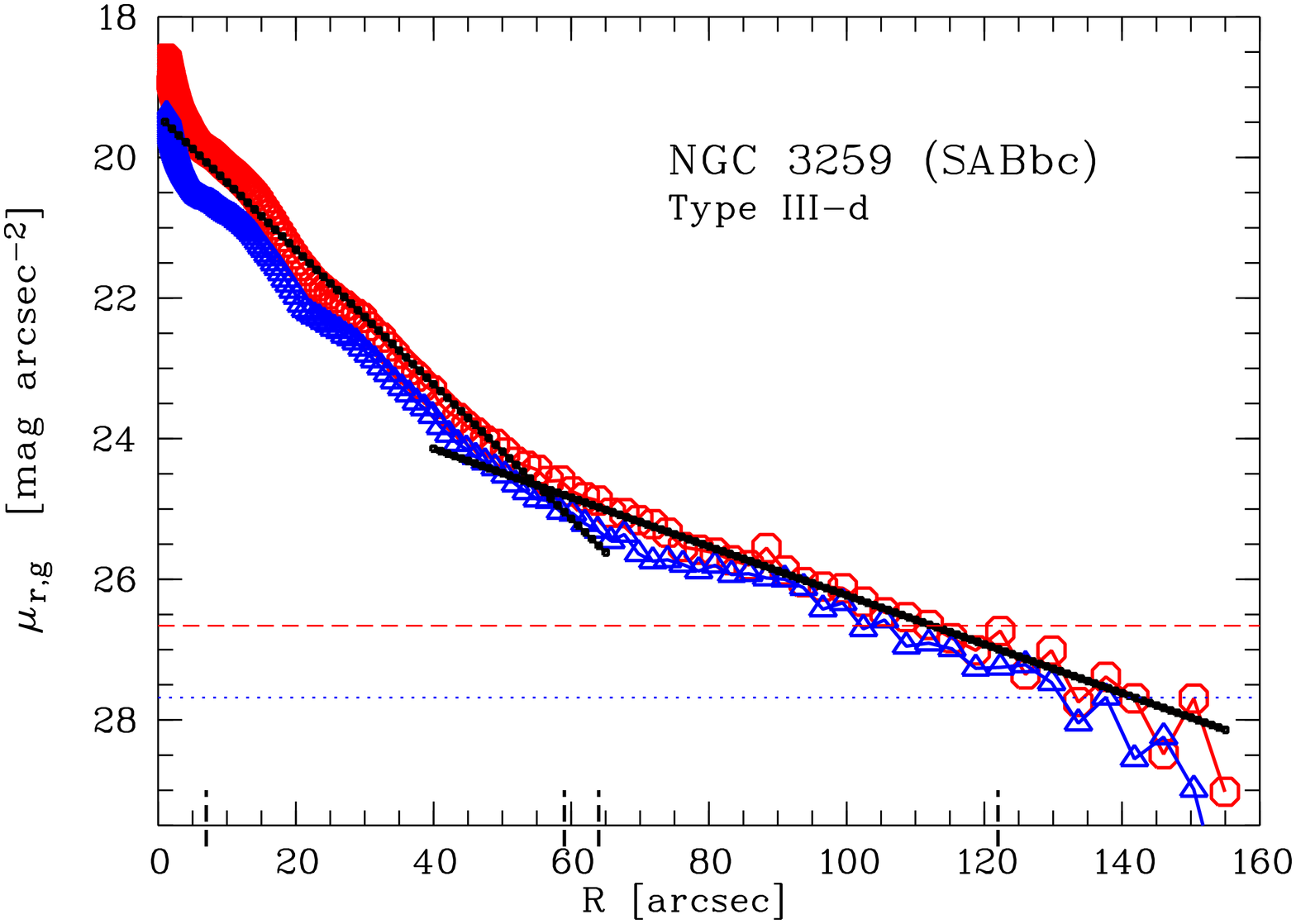}
}
\vfill
\parbox[b][0.5\height][t]{0.47\textwidth}
{
\noindent {\bf NGC\,3310     :}        \typeiii                 \\        
\texttt{J103845.8+533012 .SXR4P  4.0 -20.25   2.8 1208}\\[0.25cm]
The galaxy is clearly disturbed with a shell-like structure in the 
outer disk, which is probably the result of a recent merger with a 
smaller galaxy \cite[\cf]{conselice2000}. The inner region is also 
slightly asymmetric, but still used for centering. The inner region, 
corresponding to a high surface brightness, tightly wound spiral, inside 
the break at $\sim\!50$\arcsec, is curved starting at $\sim\!20$\arcsec. 
The wiggles in the outer disk are due to the shells. An additional 
upbending beyond $\sim\!150\arcsec$ is due to a prominent outer shell.
The very outer parts are in addition influenced by an inhomogeneous 
sky and a large mask.

}
\hfill 
\parbox[b][0.5\height][b]{0.47\textwidth}
{
\includegraphics[width=5.7cm,angle=270]{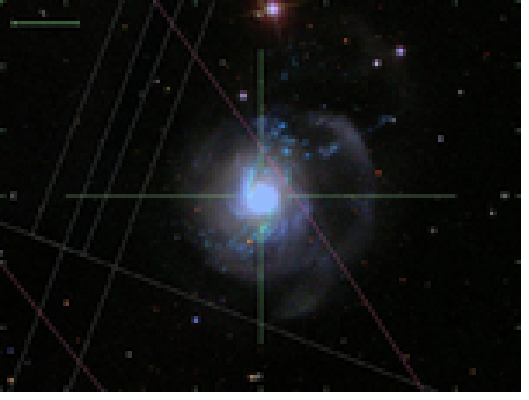}
\hspace*{-0.8cm}
\includegraphics[width=6.1cm,angle=270]{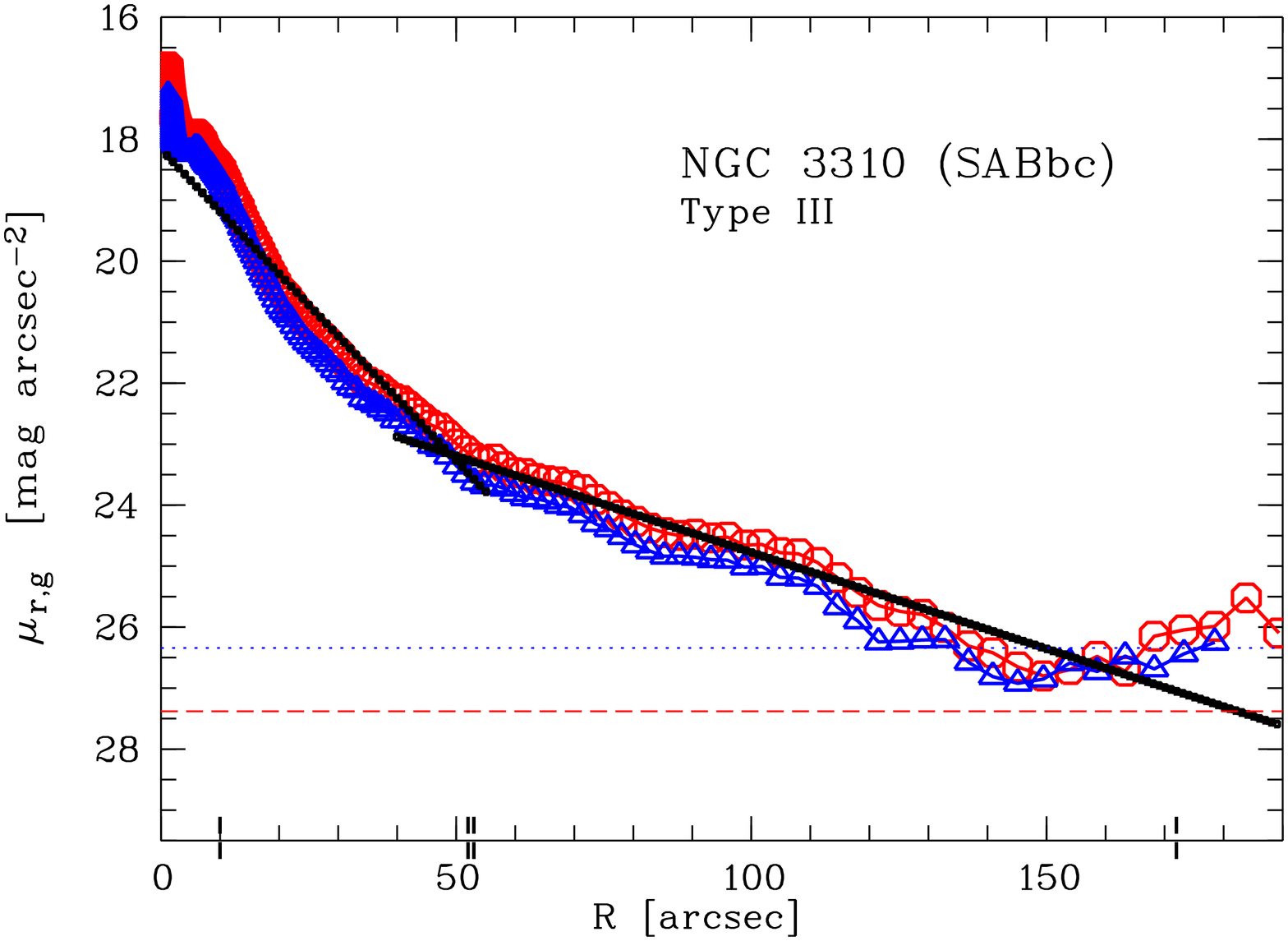}
}
\end{minipage}

\newpage
\onecolumn
\begin{minipage}[t][\textheight][t]{\textwidth}
\parbox[t][0.5\height][t]{0.47\textwidth}
{
\noindent {\bf NGC\,3359     :}        \typeab                \\          
\texttt{J104636.3+631328 .SBT5.  5.0 -20.42   7.2 1262}\\[0.25cm]
The inner disk shows an extended bar (roughly $R\sim\!45$\arcsec), which 
is not visible in the final profile, and two bright spiral arms. 
The broad extension of this arms build the outer disk, so there 
is no clear elliptical outer structure visible which makes the 
photometric inclination (ellipticity) and PA measurements highly 
uncertain. 
The outer slope is influenced by a rather uncertain sky estimate 
in both bands, but the apparent downbending break at $\sim\!200$\arcsec
is not due to a sky error, but coincides with the region where the 
two outer arms dominate. Thus the galaxy is classified as \typeabc.  

}
\hfill 
\parbox[t][0.5\height][t]{0.47\textwidth}
{
\includegraphics[width=5.7cm,angle=270,]{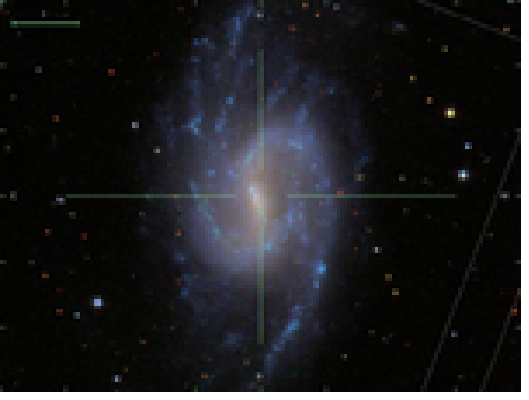}
\hspace*{-0.8cm}
\includegraphics[width=6.1cm,angle=270]{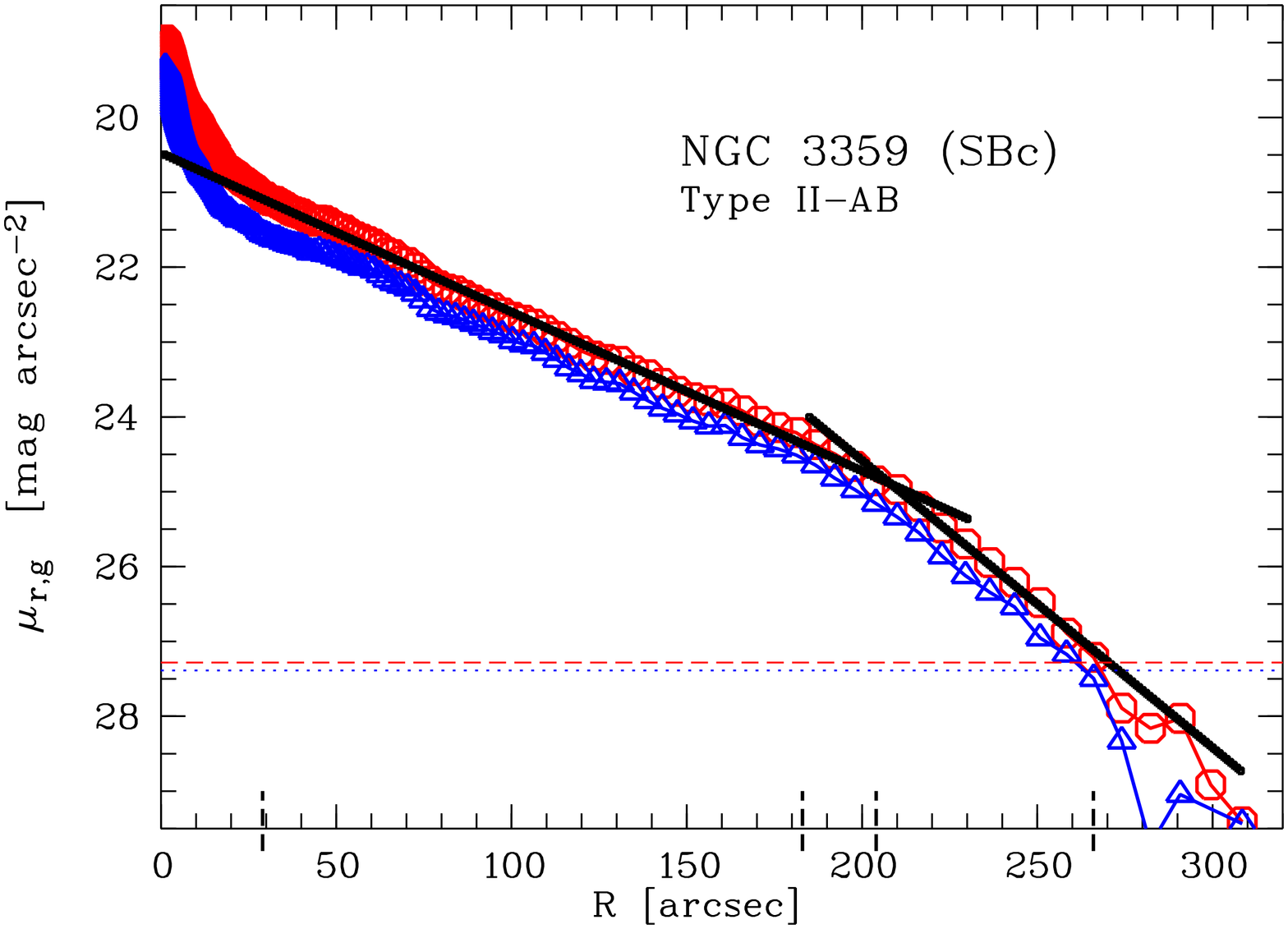}
}
\vfill
\parbox[b][0.5\height][t]{0.47\textwidth}
{
\noindent {\bf NGC\,3423     :}        \typect                \\          
\texttt{J105114.3+055024 .SAS6.  6.0 -19.54   3.9 1032}\\[0.25cm]
Bulge-like inner region ($\ltsim30$\arcsec) coincides with inner high
surface brightness disk with spiral structure. Final profile is clearly
downbending but the exact break radius is difficult to place due to an 
extended feature around $\sim\!70\arcsec$ corresponding to an aligned spiral 
arm (with a possible straight structural element towards the north-east,
called \cite[]{chernin1999} spiral-arm row). 

}
\hfill 
\parbox[b][0.5\height][b]{0.47\textwidth}
{
\includegraphics[width=5.7cm,angle=270]{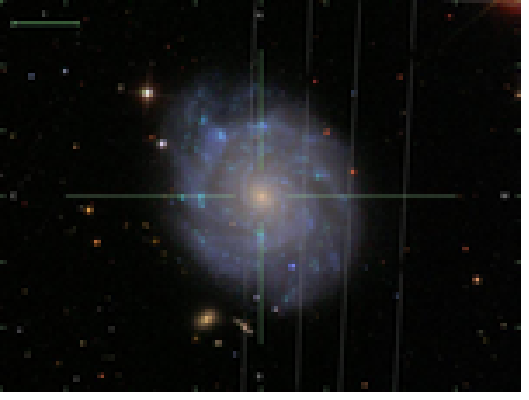}
\hspace*{-0.8cm}
\includegraphics[width=6.1cm,angle=270]{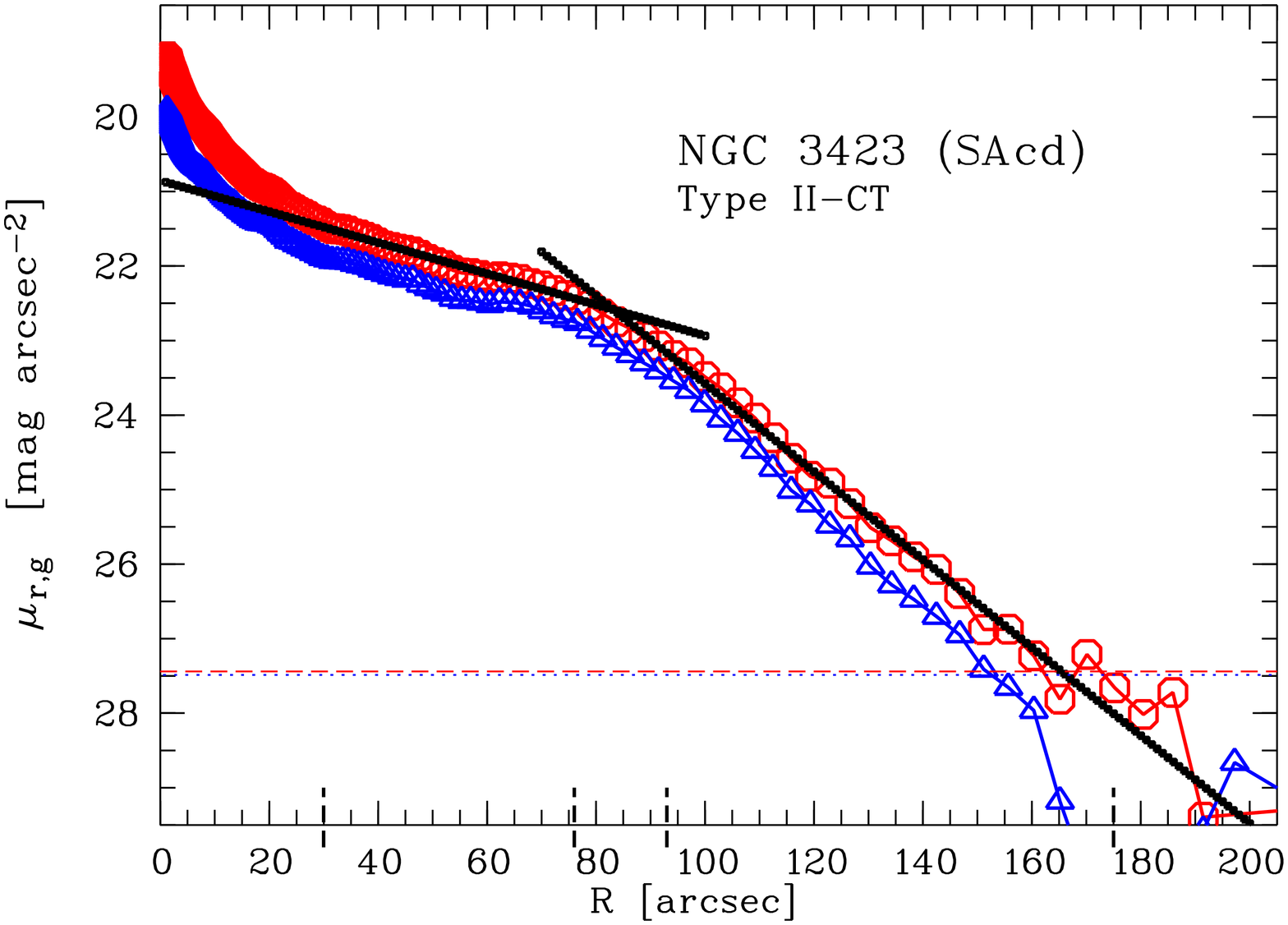}
}
\end{minipage}

\newpage
\onecolumn
\begin{minipage}[t][\textheight][t]{\textwidth}
\parbox[t][0.5\height][t]{0.47\textwidth}
{
\noindent {\bf NGC\,3488     :}        \typetoct                \\          
\texttt{J110123.6+574039 .SBS5*  5.1 -18.80   1.6 3226}\\[0.25cm]
Galaxy is isolated according to \cite{prada2003}. A bright star 
covering part of the outer disk needs an extended mask. A small bar 
of roughly $R\sim\!6\arcsec$ size is visible as an elongated isophote 
on the image. The final profile is clearly downbending, but the exact 
break radius is difficult to place. The extended break region resembles
in this case again a straight line, thus one could also define two 
break radii at $\sim\!32\arcsec$ and $\sim\!58$\arcsec. This could 
be caused by the extended mask, whereas the spiral structure extending 
to $\sim\!40\arcsec$ seems not to be responsible for this behaviour. 
The bar is to small for its OLR to be associated with the first break,
thus the galaxy is classified \typetoctc. The light beyond 
$\sim\!58\arcsec$ is still symmetric and not affected by sky errors. 
The peak at $\sim\!20\arcsec$ corresponds to the inner two arms.  

}
\hfill 
\parbox[t][0.5\height][t]{0.47\textwidth}
{
\includegraphics[width=5.7cm,angle=270,]{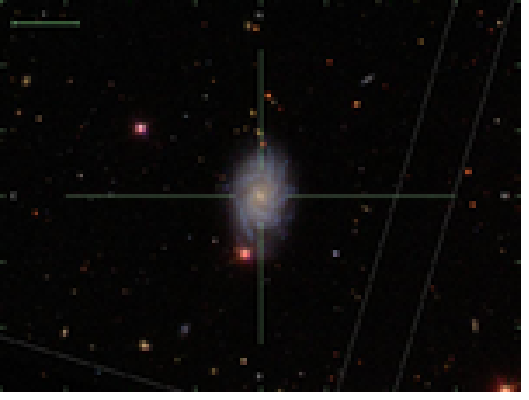}
\hspace*{-0.8cm}
\includegraphics[width=6.1cm,angle=270]{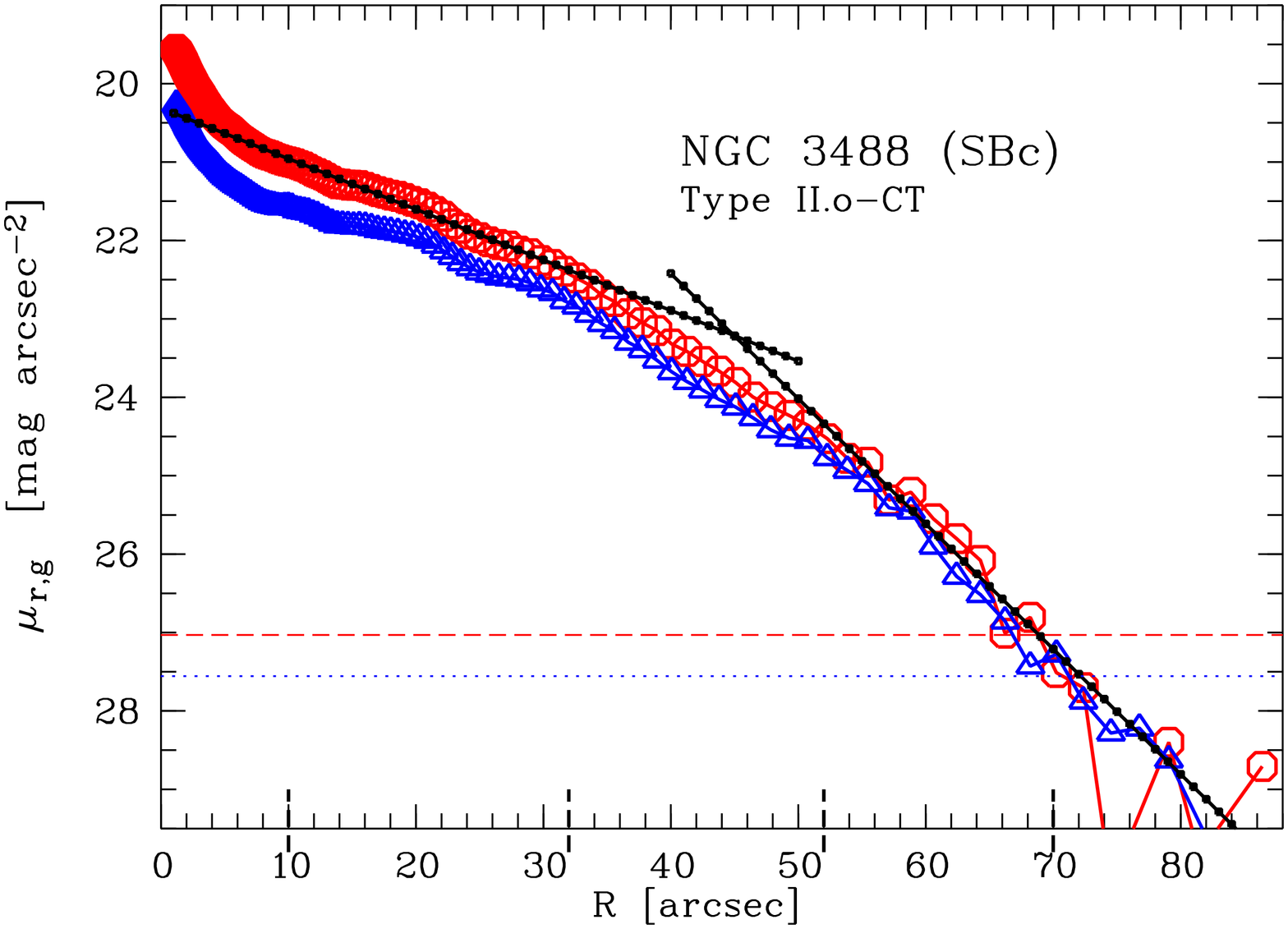}
}
\vfill
\parbox[b][0.5\height][t]{0.47\textwidth}
{
\noindent {\bf NGC\,3583     :}        \typeiiid                 \\        
\texttt{J111411.0+481906 .SBS3.  3.2 -19.57   2.3 2347}\\[0.25cm]
The apparent companion, a spiral galaxy (NGC\,3577) towards the 
south-west is non-physical  with $v=5336$\kms. The galaxy shows a
stream-like arm extending towards the north-west outside of the 
galaxy connected to a small superimposed E0 satellite at $0.9\arcmin$ 
north (not in NED). Due to this structure the outer disk seems to 
have a slightly different ellipticity and PA. The bump at $\sim\!30$\arcsec
in the final profile is due to the end of the bar and beginning of 
spiral arms. The inner profile with an extended wiggle at $\sim\!55\arcsec$ 
(corresponding to the spiral arms) makes it difficult to characterise the 
break region (starting at $\sim\!85\arcsec$ with an upbending profile) as 
being curved or sharp.

}
\hfill 
\parbox[b][0.5\height][b]{0.47\textwidth}
{
\includegraphics[width=5.7cm,angle=270]{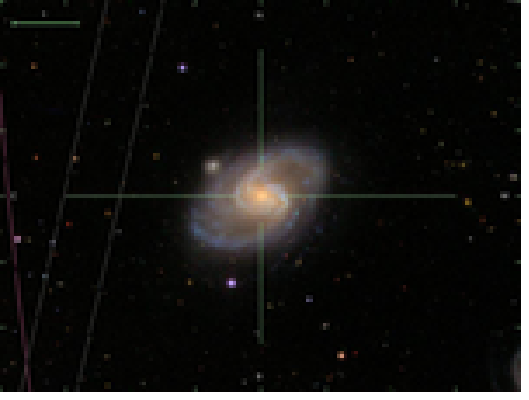}
\hspace*{-0.8cm}
\includegraphics[width=6.1cm,angle=270]{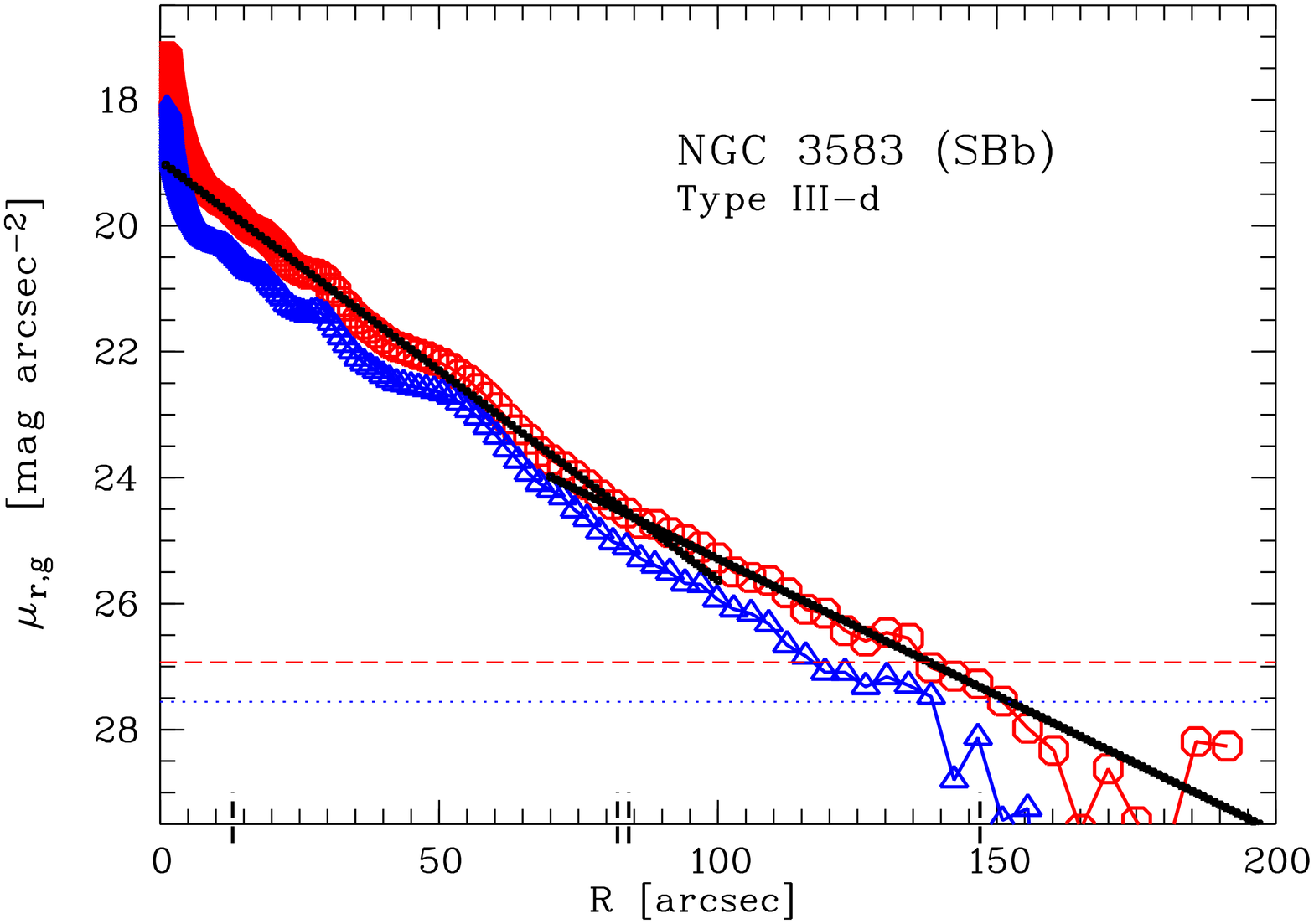}
}
\end{minipage}

\newpage
\onecolumn
\begin{minipage}[t][\textheight][t]{\textwidth}
\parbox[t][0.5\height][t]{0.47\textwidth}
{
\noindent {\bf NGC\,3589     :}        \typect                \\          
\texttt{J111513.2+604201 .S..7*  7.0 -18.49   1.5 2217}\\[0.25cm]
Due to the higher inclination and extended elongated structure the size 
of the bar is not well defined. The center is also not well defined, but 
centering on the outer isophote coincides roughly with the brightest pixel.
The sky estimate is rather uncertain due to the bright star in the FOV.
The final profile shows a downbending break at $\sim\!40$\arcsec, classified 
as \typect but also consistent with being \typeolr allowing the bar 
to be of size $\sim\!20\arcsec$ which is consistent with the image. 
The small bump $\sim\!22\arcsec$ in the profile just corresponds to 
some aligned \hii regions.  

}
\hfill 
\parbox[t][0.5\height][t]{0.47\textwidth}
{
\includegraphics[width=5.7cm,angle=270,]{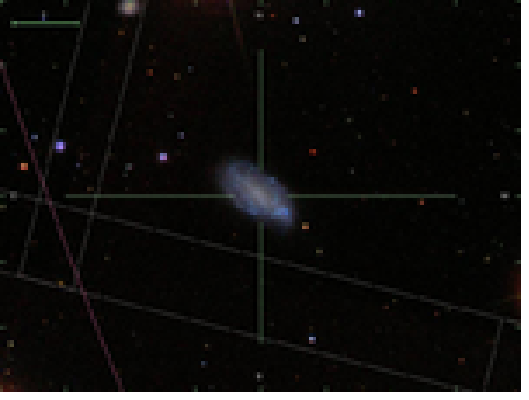}
\hspace*{-0.8cm}
\includegraphics[width=6.1cm,angle=270]{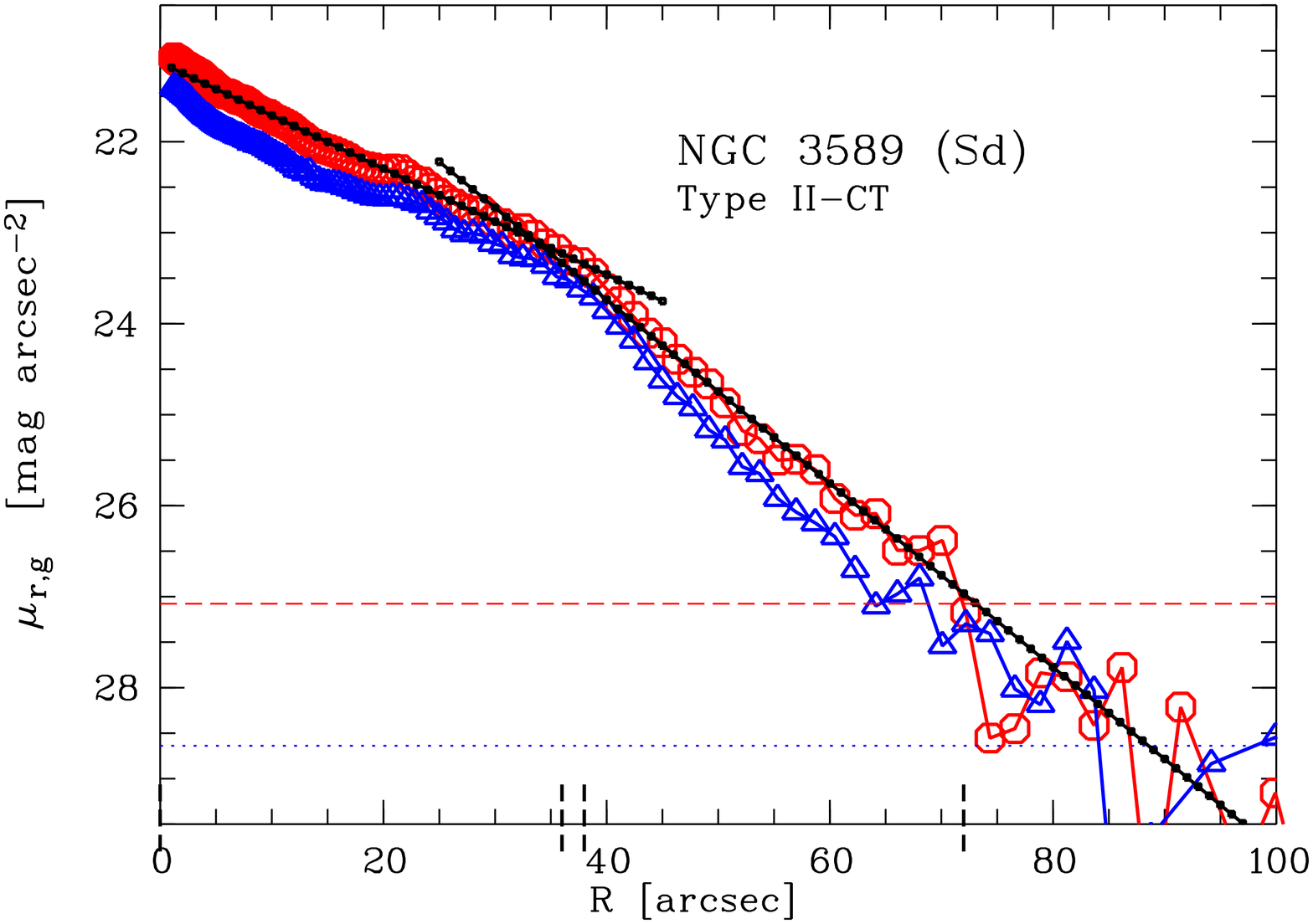}
}
\vfill
\parbox[b][0.5\height][t]{0.47\textwidth}
{
\noindent {\bf NGC\,3631     :}        \typeo                 \\          
\texttt{J112102.4+531008 .SAS5.  5.1 -20.62   4.9 1388}\\[0.25cm]
The galaxy forms a nearby group together with NGC\,3178, NGC\,3898, 
NGC\,3953 and NGC\,3992 (also part of our sample) in the Ursa Major 
cluster and shows plenty of substructure.  
It has two prominent spiral arms with different pitch angles, one 
of them turns nearly straight, classified by \cite{arp1966} as a 
spiral galaxy with one heavy arm. 
In addition, it has a faint extended spiral arm 
(or maybe stream-like structure) in the far outer disk towards North,
therefore the outer disk is slightly asymmetric and the mean ellipticity 
and PA are difficult to fix. There is an extended light patch on the 
image visible towards south-west. Although there is no bright star 
nearby which could create such an artefact it is unlikely a dwarf 
companion, since it is not visible on the DSS. 
The final radial light profile is quite unusual and cumbersome to 
classify, since it exhibits an extended bump between 
$\sim\!50-140$\arcsec, which corresponds to the begin of the two 
spirals to the end of the inner (undisturbed) disk. 
Beyond $\sim\!140\arcsec$ the extended, slightly offcenterd outer 
disk with the spiral arm structure (or stream) starts. 
If we do not exclude the bump from the fit, one can argue for a 
downbending break at $\sim\!75$\arcsec, or a \typeo profile with
an extreme wiggle (this would be a \typeti if the galaxy would
be barred). We decided to exclude the bump and the inner region 
from the fit ($R \gtsim 140$\arcsec) which gives also a \typeo 
profile where the region around $\sim\!35\arcsec$ is also roughly 
fitted. 
Due to this problems the classification should be taken with caution. 

}
\hfill 
\parbox[b][0.5\height][b]{0.47\textwidth}
{
\includegraphics[width=5.7cm,angle=270]{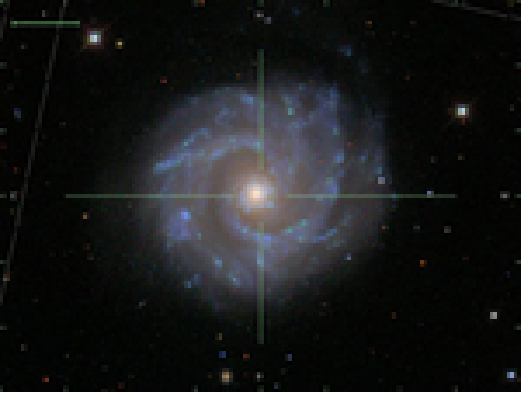}
\hspace*{-0.8cm}
\includegraphics[width=6.1cm,angle=270]{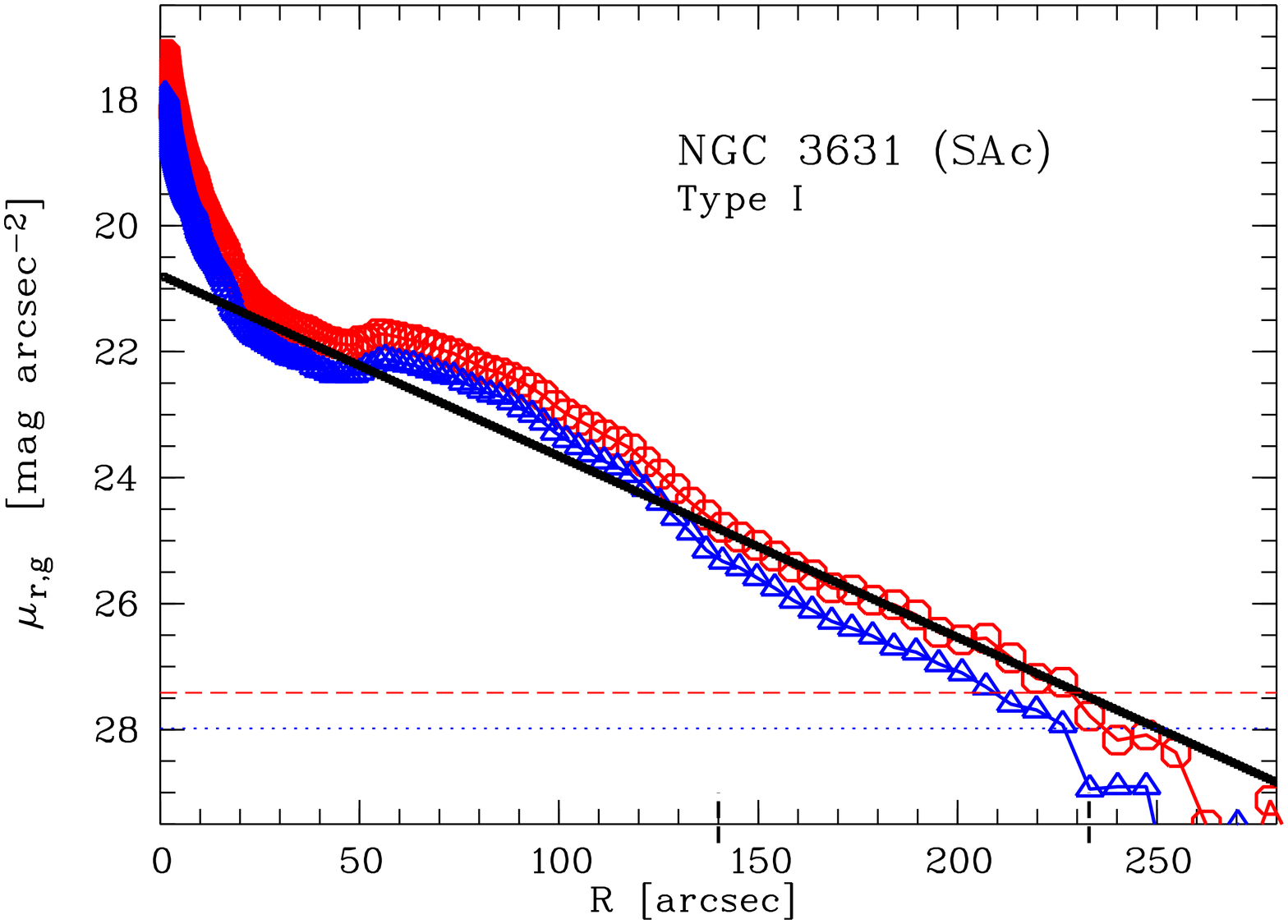}
}
\end{minipage}

\newpage
\onecolumn
\begin{minipage}[t][\textheight][t]{\textwidth}
\parbox[t][0.5\height][t]{0.47\textwidth}
{
\noindent {\bf NGC\,3642     :}        \typeiiid                 \\        
\texttt{J112218.0+590427 .SAR4*  4.0 -20.60   5.5 1831}\\[0.25cm]
The background is slightly inhomogeneous and the galaxy is rather 
extended. From the nearly round inner disk a single, star forming,
spiral arm starts towards the outer disk with slightly more than one 
revolution, forming the strongly asymmetric, lopsided outer disk. 
Therefore the ellipticity and PA is determined further inside. 
The final profile looks fairly curved and with extended wiggles 
in the outer disk at $\sim\!100\arcsec$ and  $\sim\!155$\arcsec
due to the outer spiral arm passing the ellipse.  
There is a steeper inner exponential part visible between 
$\sim\!25-75\arcsec$ followed outwards by an upbending profile, 
which is roughly at the position where the extended spiral arm 
starts to unwrap, which leads to the \typeiii classification.
Inwards of $\sim\!25\arcsec$ there is again another steeper 
exponential part visible together with a possible bulge
component.    

}
\hfill 
\parbox[t][0.5\height][t]{0.47\textwidth}
{
\includegraphics[width=5.7cm,angle=270,]{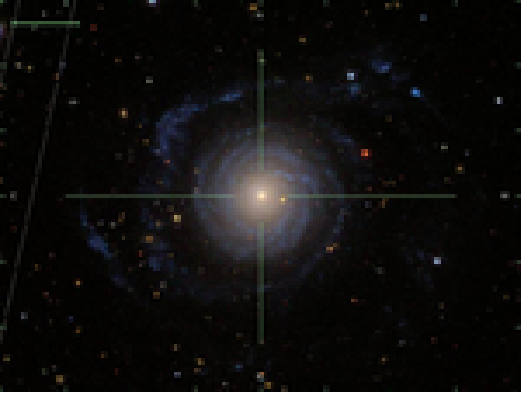}
\hspace*{-0.8cm}
\includegraphics[width=6.1cm,angle=270]{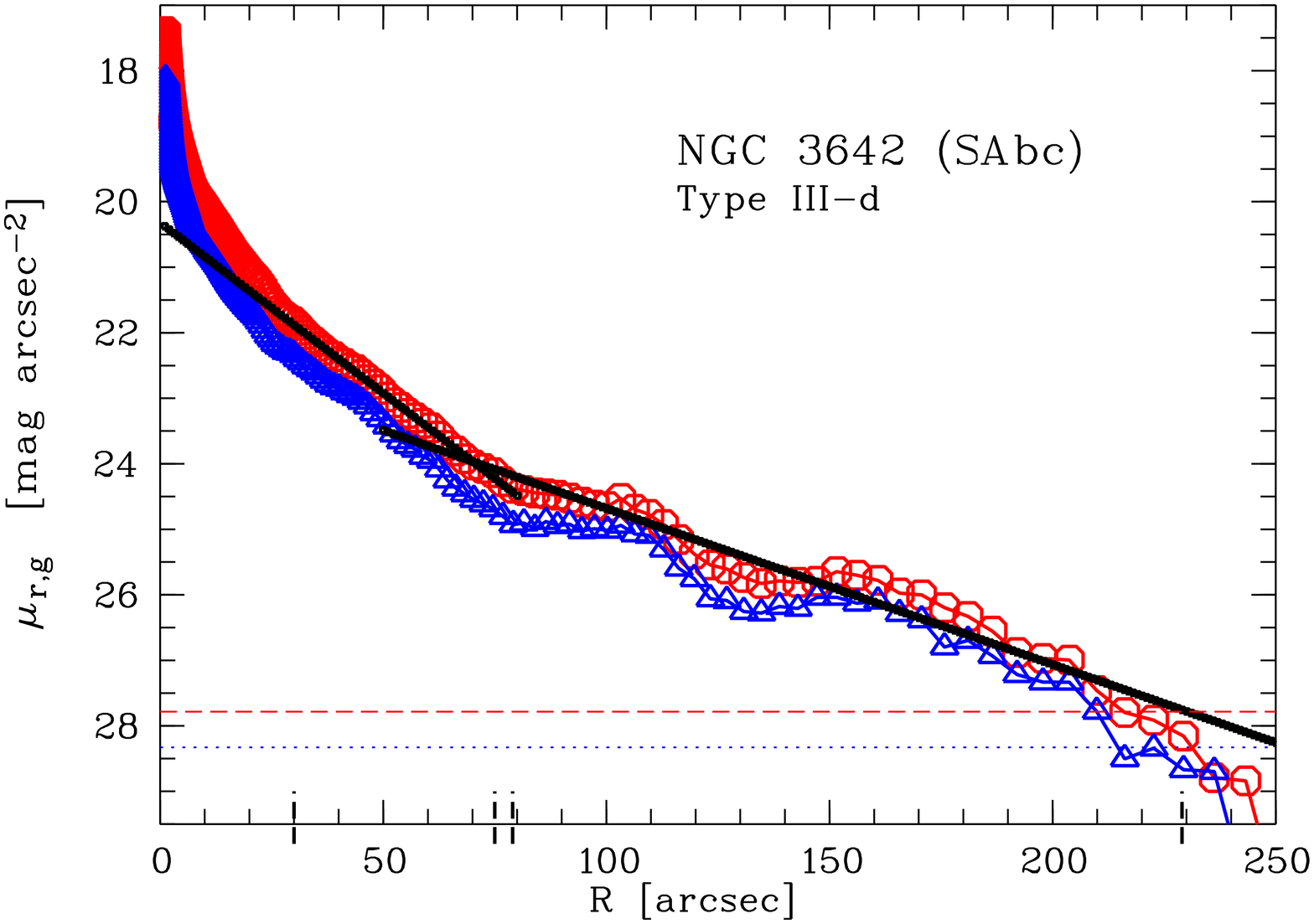}
}
\vfill
\parbox[b][0.5\height][t]{0.47\textwidth}
{
\noindent {\bf NGC\,3756     :}        \typetoct                \\          
\texttt{J113648.3+541737 .SXT4.  4.3 -19.89   3.8 1525}\\[0.25cm]
Only partly fitted since \ltsim $1/4$ of galaxy is beyond field.
Known three armed spiral galaxy which shows one wide spiral arm 
starting at $\sim\!60\arcsec$ extending roughly towards the downbending 
break at $\sim\!100$\arcsec. The extended bump $\sim\!45\arcsec$ is due 
to the three inner tightly wound spiral arms and makes it, similar to 
NGC\,3631, difficult to fit the inner region. However, since the bump is 
less strong we decided to include it. The galaxy is classified as SAB 
(RC3) but the bar is not obvious on the image and its maximum 
size would be $R \ltsim 25\arcsec$ thus the break is well outside the 
range for typical \typeolr breaks and is classified as \typetoctc.

}
\hfill 
\parbox[b][0.5\height][b]{0.47\textwidth}
{
\includegraphics[width=5.7cm,angle=270]{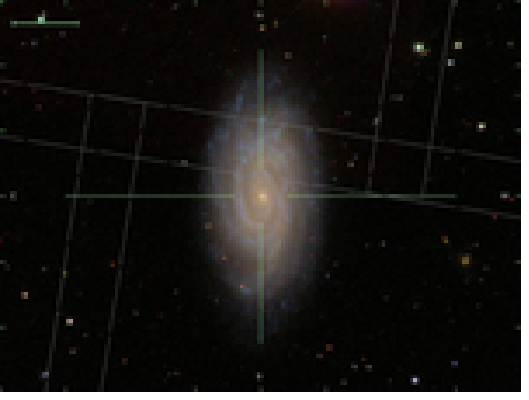}
\hspace*{-0.8cm}
\includegraphics[width=6.1cm,angle=270]{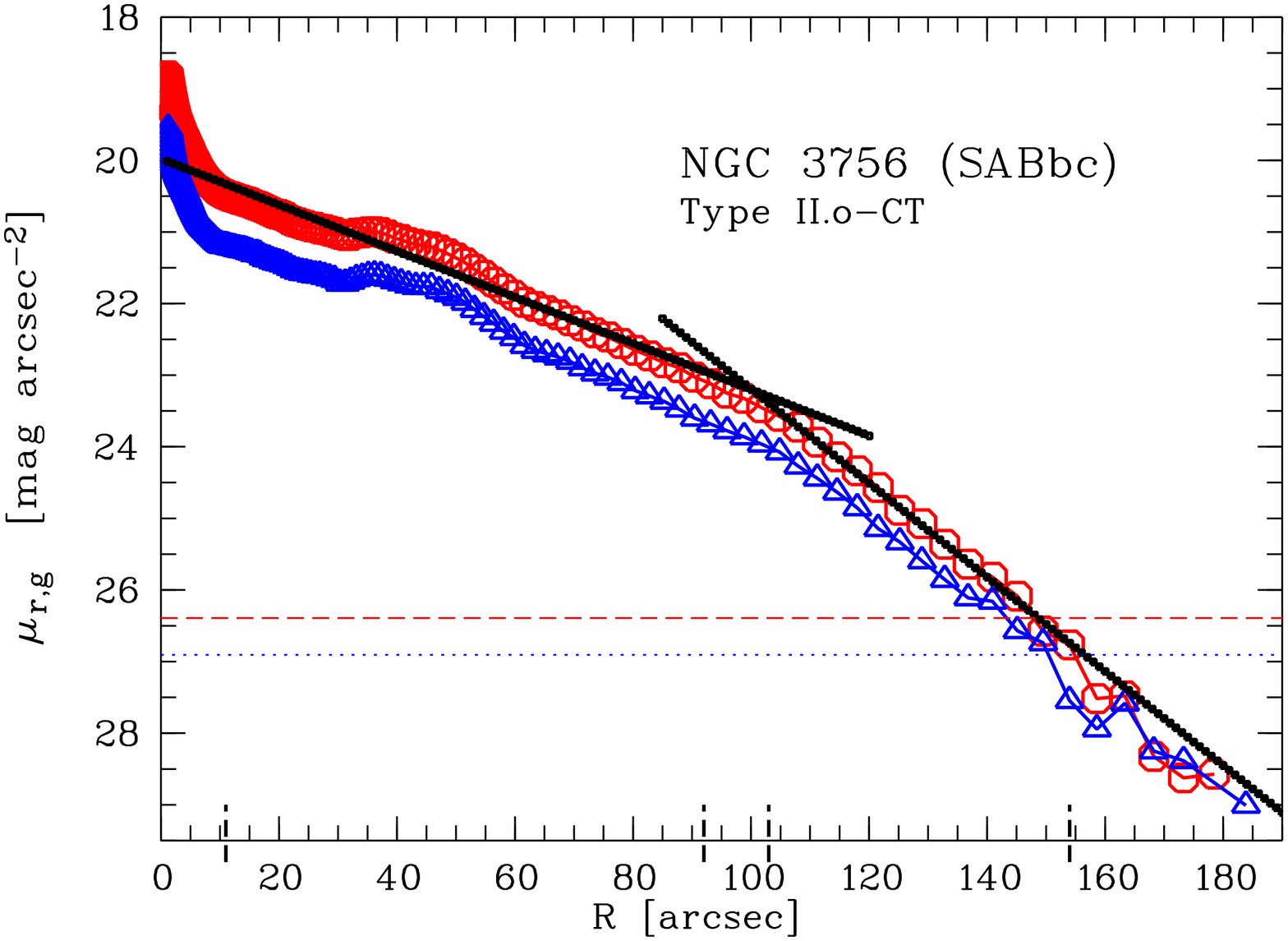}
}
\end{minipage}

\newpage
\onecolumn
\begin{minipage}[t][\textheight][t]{\textwidth}
\parbox[t][0.5\height][t]{0.47\textwidth}
{
\noindent {\bf NGC\,3888     :}        \typeo                 \\          
\texttt{J114734.7+555800 .SXT5.  5.3 -20.43   1.8 2648}\\[0.25cm]
The background on the image shows a large gradient due to a bright star
off the field, but the galaxy is small enough to be not affected.  
The galaxy has one 'malformed' (kinked) spiral arm across the disk.
There is a possible dwarf, irregular companion visible towards the 
south-west, but without any distance information.  
The final profile exhibits an extended wiggle (with an unusual integral 
sign shape) at $\sim\!33$\arcsec, looking like two shifted exponential
regions, but there is no point in placing a break there.  
The peak at $\sim\!22\arcsec$ is related to an aligned spiral arm, 
whereas the peak at $\sim\!12\arcsec$ could be due to the end of a 
possible bar or inner ring (no obvious on the image).
Therefore the galaxy is classified as \typeo and not as 
\typetic, although the dip in the profile is clearly visible. It
could be explained by dust lanes close to the center, which also 
explains the dip being more prominent in the $g^{\prime}$ band.   

}
\hfill 
\parbox[t][0.5\height][t]{0.47\textwidth}
{
\includegraphics[width=5.7cm,angle=270,]{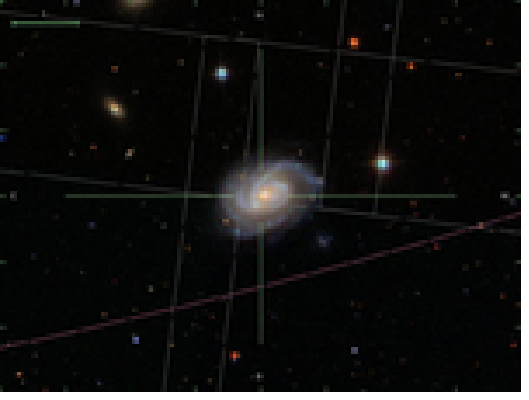}
\hspace*{-0.8cm}
\includegraphics[width=6.1cm,angle=270]{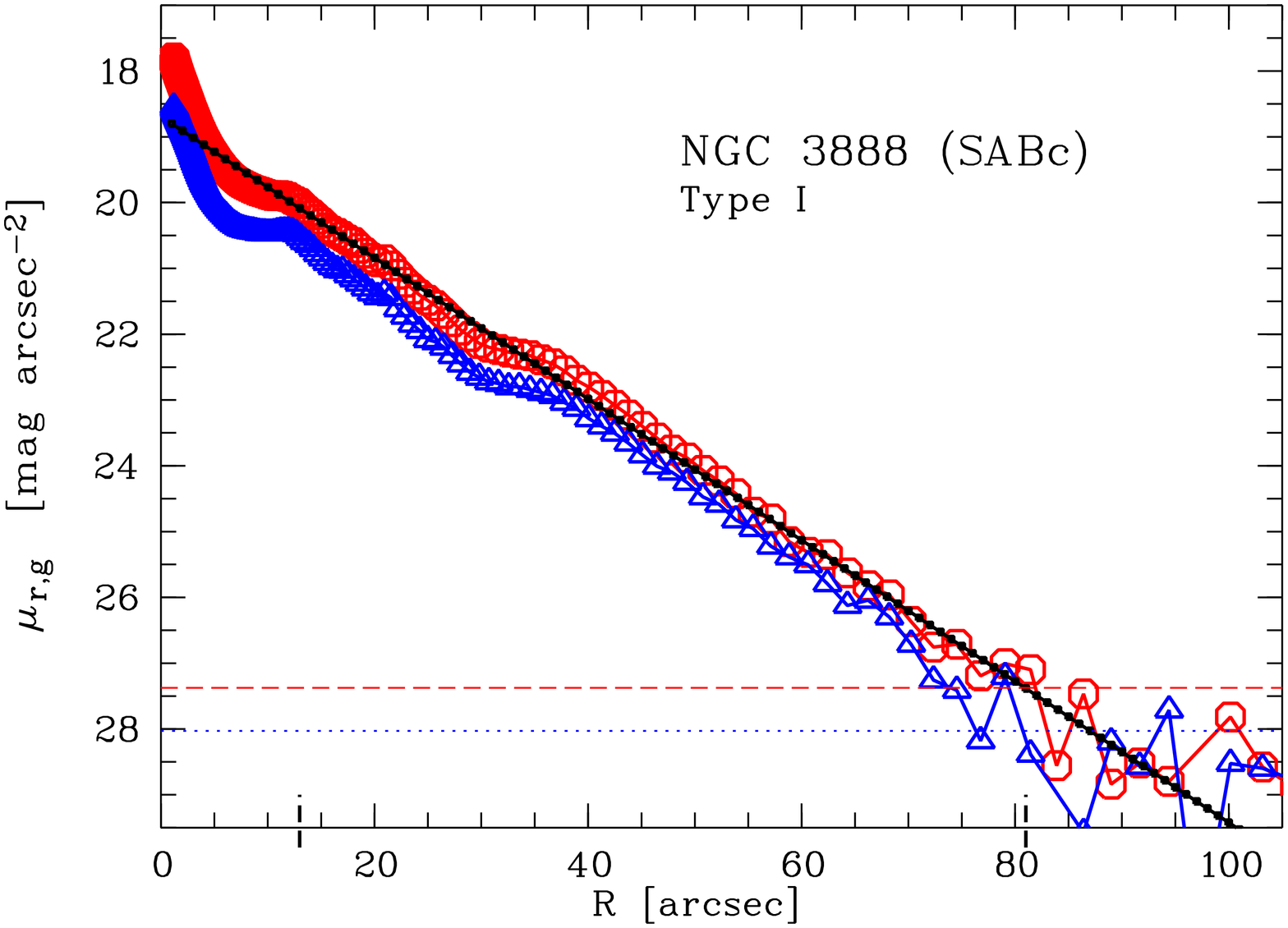}
}
\vfill
\parbox[b][0.5\height][t]{0.47\textwidth}
{
\noindent {\bf NGC\,3893     :}        \typeiiid                 \\        
\texttt{J114838.4+484234 .SXT5*  4.8 -20.33   4.1 1193}\\[0.25cm]
The background around the galaxy is affected by a nearby, bright star 
and a close (3.7\arcmin away, with similar velocity $v=905$\kms, 
SB0/a:pec), smaller, companion galaxy (NGC\,3896), both members of 
the Ursa Major Cluster. 
Galaxy exhibits a wide, extended spiral arm like structure towards 
the east, which could also be a stream or tidal feature. 
Due to this structure it is not possible to derive the ellipticity 
and PA from the outer disk, so their values are more uncertain.  
The final profile clearly shows an \typeiii upbending profile 
starting at $\sim\!150\arcsec$ corresponding to the beginning 
of the extended structure to one side and some more symmetric 
faint light on the other side. The inner disk shows a prominent 
wiggle $\sim\!90\arcsec$ which is associated to a pseudoring of 
wrapped spiral arms.  

}
\hfill 
\parbox[b][0.5\height][b]{0.47\textwidth}
{
\includegraphics[width=5.7cm,angle=270]{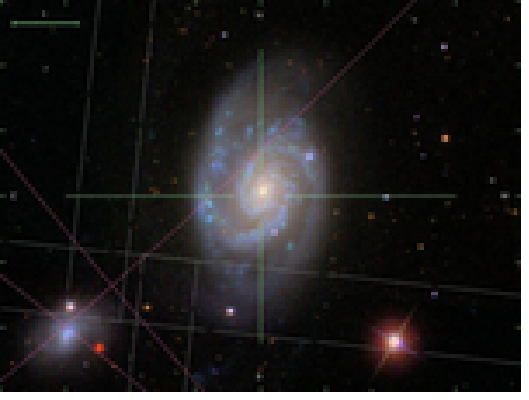}
\hspace*{-0.8cm}
\includegraphics[width=6.1cm,angle=270]{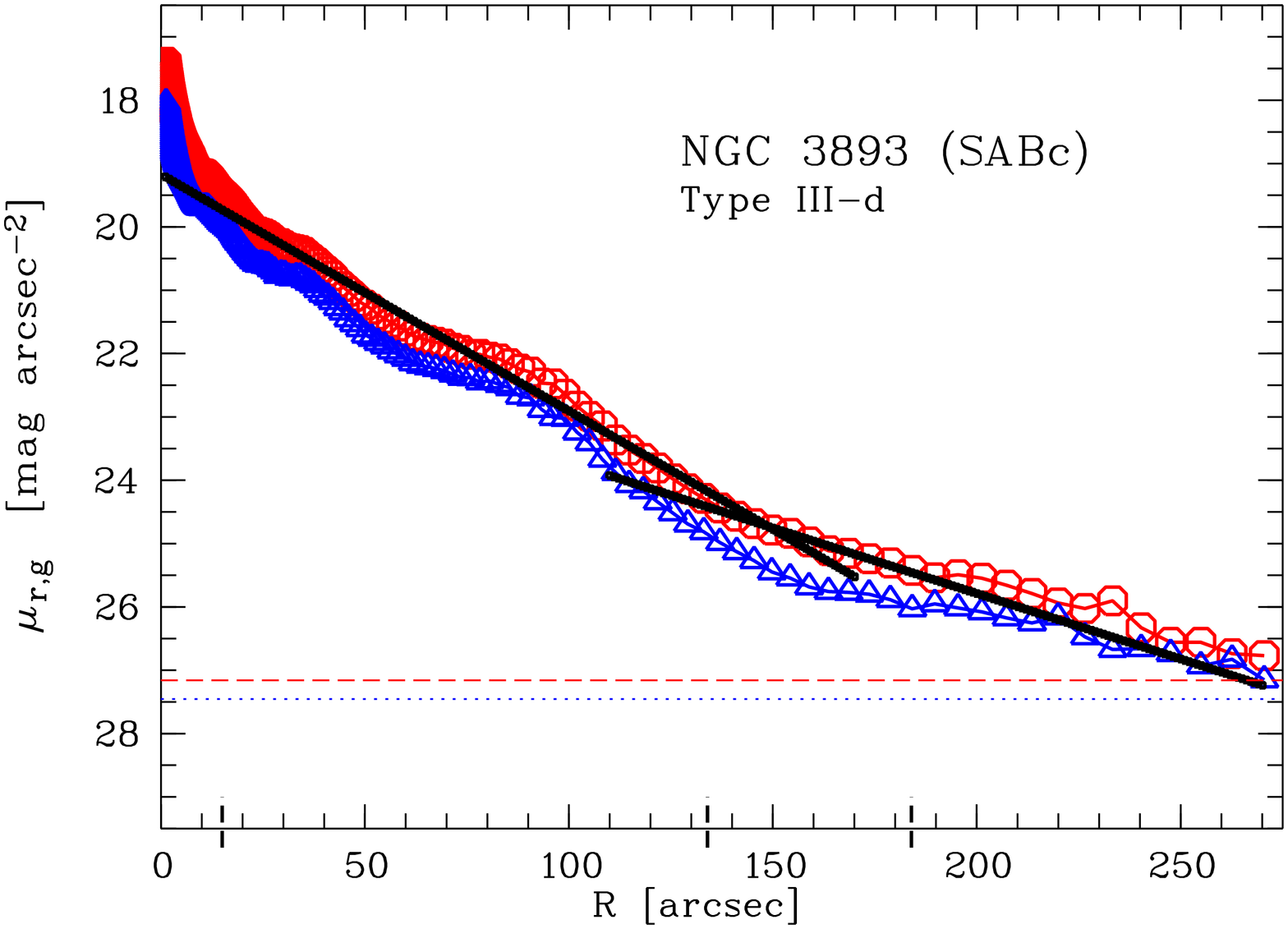}
}
\end{minipage}

\newpage
\onecolumn
\begin{minipage}[t][\textheight][t]{\textwidth}
\parbox[t][0.5\height][t]{0.47\textwidth}
{
\noindent {\bf NGC\,3982     :}        \typeiii                 \\        
\texttt{J115628.3+550729 .SXR3*  3.2 -19.54   2.3 1351}\\[0.25cm]
Close to Ursa Major Cluster but membership questionable 
\cite[]{tully1996}. Galaxy exhibits a detached, narrow, starforming 
spiral arm embedded into the very outer, symmetric disk towards 
south-west.  
The final profile beyond the central part is smooth and straight down 
to $\sim\!55\arcsec$ where it clearly starts upbending, classified 
as \typeiiic, which is consistent with the profile by \cite[]{tully1996}.
The break corresponds to the end of the inner disk, excluding the 
faint outer arm, which produces the bump at $\sim\!75\arcsec$ (more
pronounced in the $g^{\prime}$ band). The transition zone is rather sharp. 
Dip-peak structure between $\sim\!5-12\arcsec$ corresponds to the inner 
bar/ring region. The exact size of the bar is not well determined
on the SDSS image. \cite{erwin2005a}, using HST images, gives 5\arcsec 
and classifies the galaxy therefore as \typeolriii (with an 'extreme' 
OLR break). We exclude the inner $R\ltsim 13\arcsec$ from our fit 
and leave it with a \typeiii classification.  

}
\hfill 
\parbox[t][0.5\height][t]{0.47\textwidth}
{
\includegraphics[width=5.7cm,angle=270,]{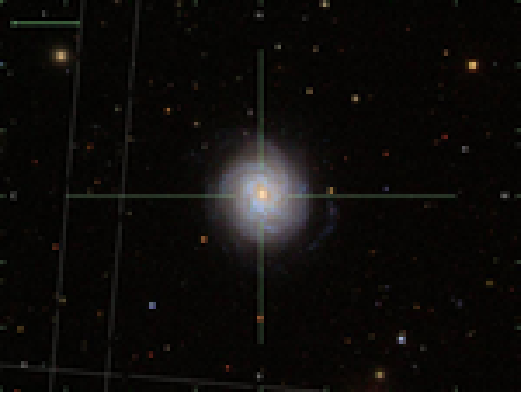}
\hspace*{-0.8cm}
\includegraphics[width=6.1cm,angle=270]{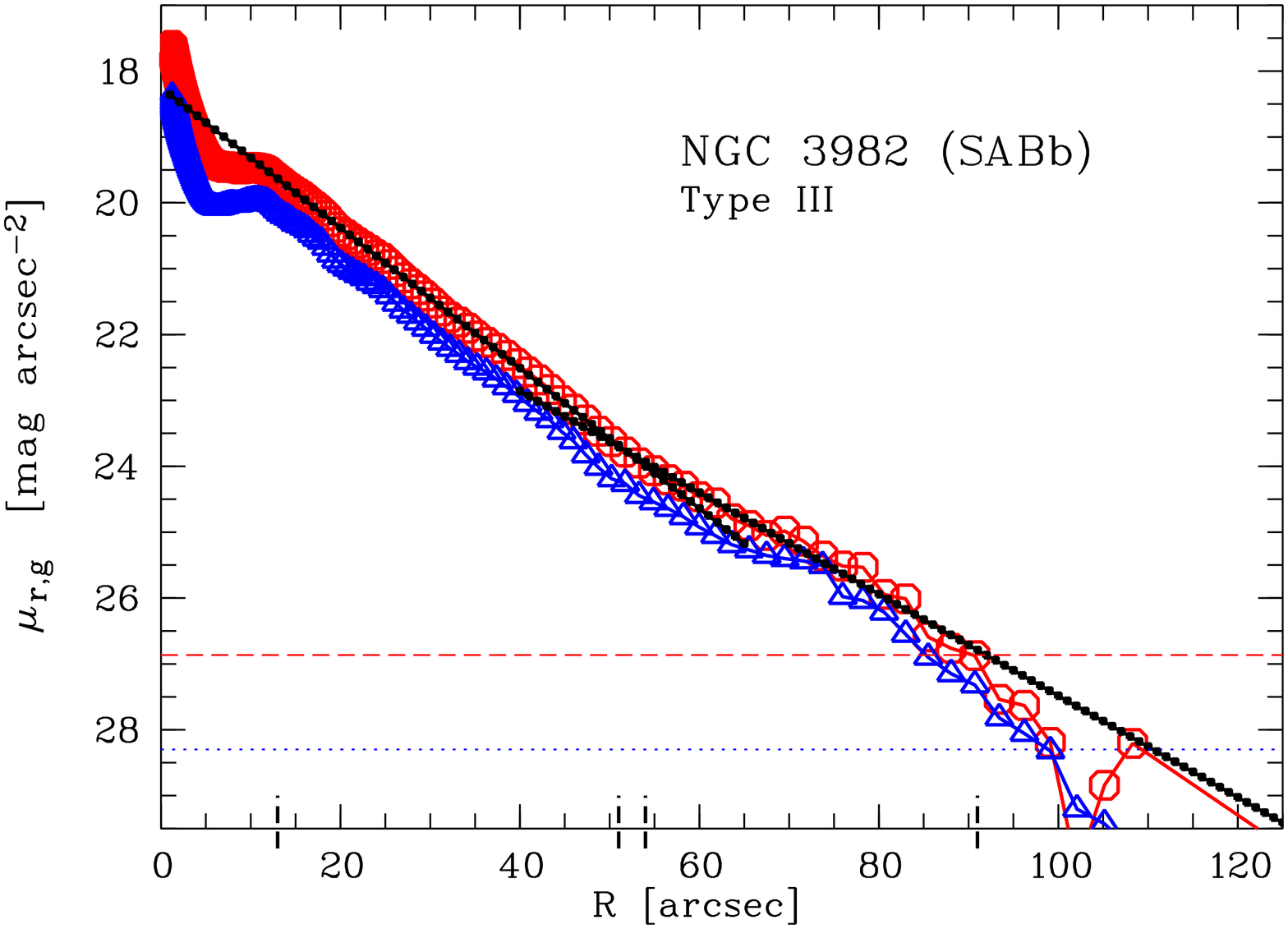}
}
\vfill
\parbox[b][0.5\height][t]{0.47\textwidth}
{
\noindent {\bf NGC\,3992  $\equiv$ M\,109:}     \typeolr   \\           
\texttt{J115735.9+532235 .SBT4.  3.8 -20.98   6.9 1286}\\[0.25cm]
Large galaxy with a small part off the SDSS field. Member of Ursa 
Major Cluster \cite[]{tully1996}. Embedded in the large bar of size 
roughly $R\sim70\arcsec$ (as measured on the image) is possibly a 
secondary bar half the size. 
Unusual very faint extension visible in both bands (more pronounced 
in $g^{\prime}$ band) towards north-east, which is possibly but unlikely 
stray-light from a nearby bright, red star. The one-sigma ellipse 
is therefore not representative so the mean ellipticity and PA 
are determined further inside. 
Final profile shows a clear downbending break at $\sim\!140$\arcsec
at about twice the bar radius corresponding roughly to a pseudoring 
of the spiral arms, therefore classified as \typeolrc.
Bump at $\sim\!200\arcsec$ does not coincide with a spiral arm.
 
}
\hfill 
\parbox[b][0.5\height][b]{0.47\textwidth}
{
\includegraphics[width=5.7cm,angle=270]{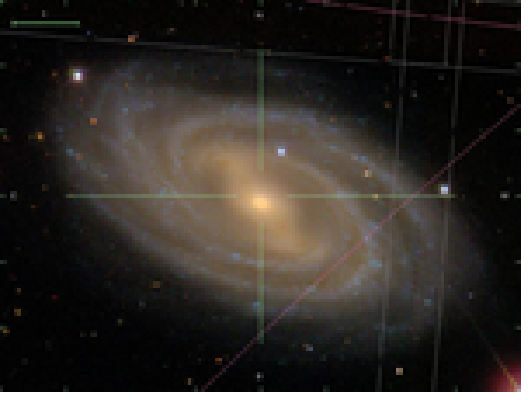}
\hspace*{-0.8cm}
\includegraphics[width=6.1cm,angle=270]{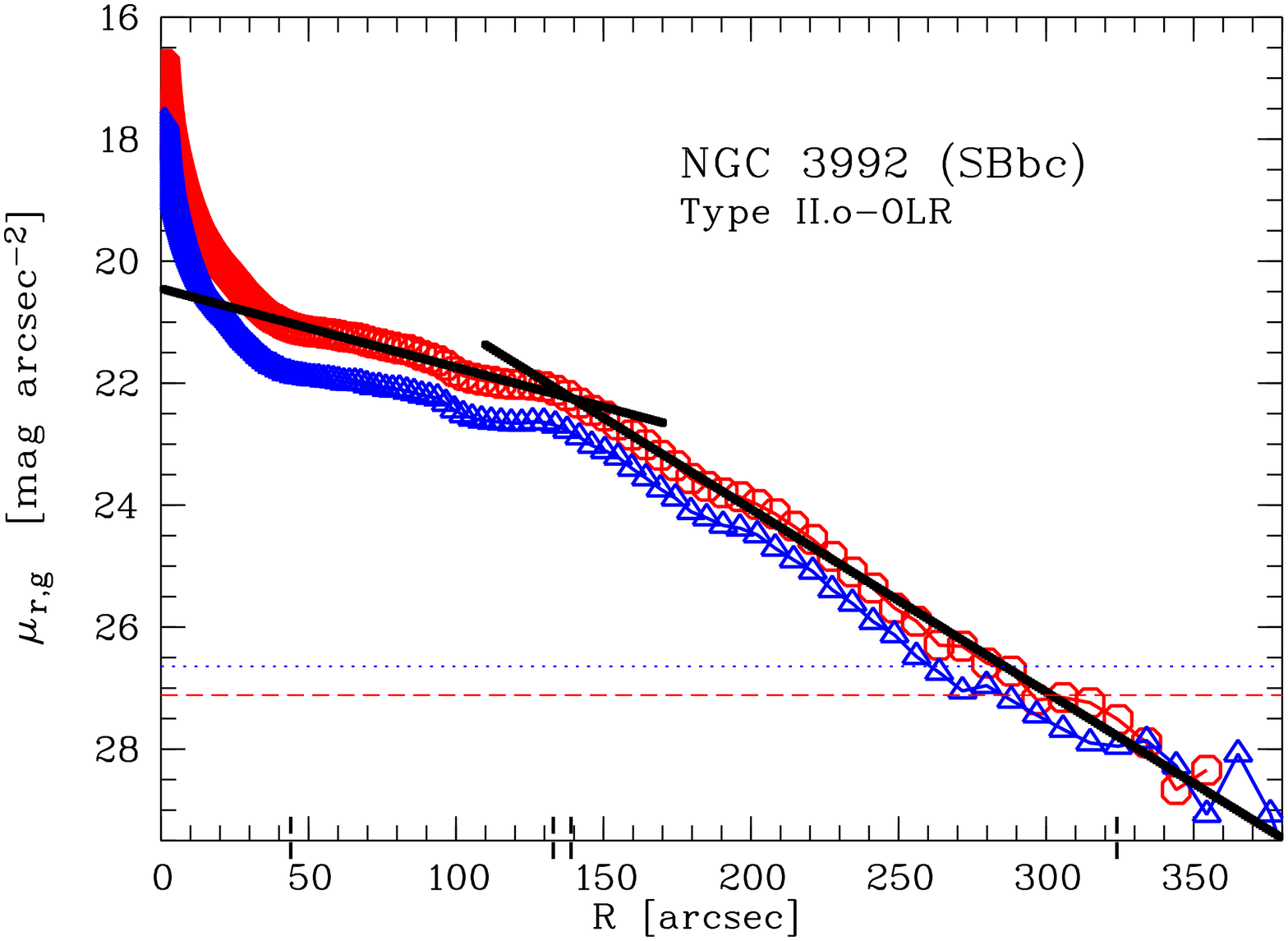}
}
\end{minipage}

\newpage
\onecolumn
\begin{minipage}[t][\textheight][t]{\textwidth}
\parbox[t][0.5\height][t]{0.47\textwidth}
{
\noindent {\bf NGC\,4030     :}        \typeiii                 \\        
\texttt{J120023.4-010603 .SAS4.  4.1 -20.27   3.9 1476}\\[0.25cm]
Two superimposed bright stars are masked. According to \cite{zaritsky1993}
it is an isolated galaxy with a satellite system (largest UGC\,06970, 
SB(s)m, with $\mabs=-17.7$). The galaxy shows two prominent spiral 
arms beyond a more flocculent center starting at $\sim\!30$\arcsec, 
followed by a single arm extending slightly further out towards 
south-west. 
Final profile shows a \typeiii break at $\sim\!150\arcsec$ where the transition 
zone is rather sharp, followed by a fairly symmetric outer region without
spiral structure. The bump at $\sim\!80\arcsec$ corresponds to the end of 
inner spiral arms. 

}
\hfill 
\parbox[t][0.5\height][t]{0.47\textwidth}
{
\includegraphics[width=5.7cm,angle=270,]{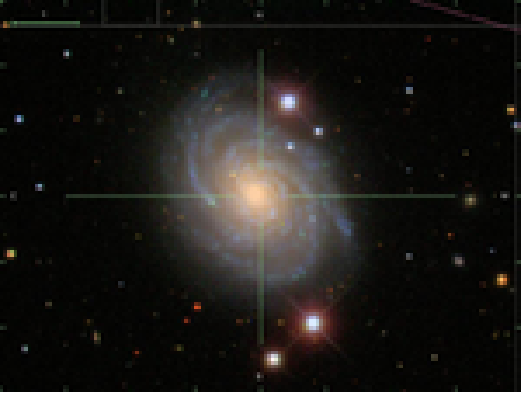}
\hspace*{-0.8cm}
\includegraphics[width=6.1cm,angle=270]{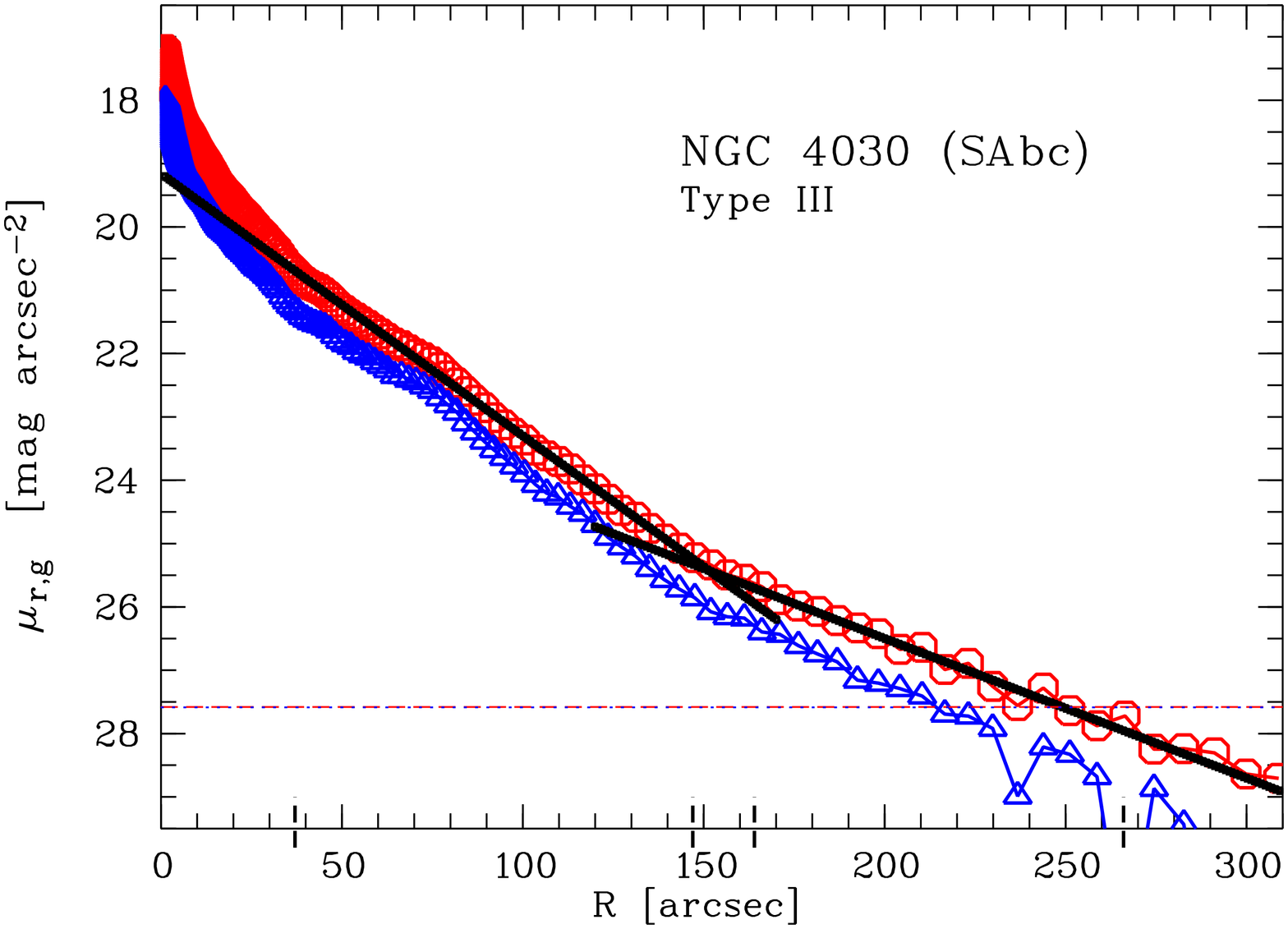}
}
\vfill
\parbox[b][0.5\height][t]{0.47\textwidth}
{
\noindent {\bf NGC\,4041     :}        \typeiiid                \\  
\texttt{J120212.2+620814 .SAT4*  4.3 -19.92   2.7 1486}\\[0.25cm]
According to \cite{sandage1994} this galaxy forms a kinematic triplet 
with NGC\,4036 and UGC\,7009 and is probably member of the Ursa Major 
Cluster. However, \cite{tully1996} does not list any of these as members.  
Galaxy shows a detached, extended, star-forming spiral arm structure in 
the outer, asymmetric disk, similar to NGC\,2967, but smaller and only 
one-sided, which makes the photometric inclination (ellipticity) and 
PA uncertain. 
In the final profile the disk inside the \typeiii break at $\sim\!75\arcsec$ 
corresponds roughly to the region with the spiral arms, which are also 
responsible for the extended bump between $\sim\!30-75\arcsec$ (more 
pronounced in $g^{\prime}$ band).  
The very inner region looks bar-like, but is probably only the spiral 
structure continuing into the center.

}
\hfill 
\parbox[b][0.5\height][b]{0.47\textwidth}
{
\includegraphics[width=5.7cm,angle=270]{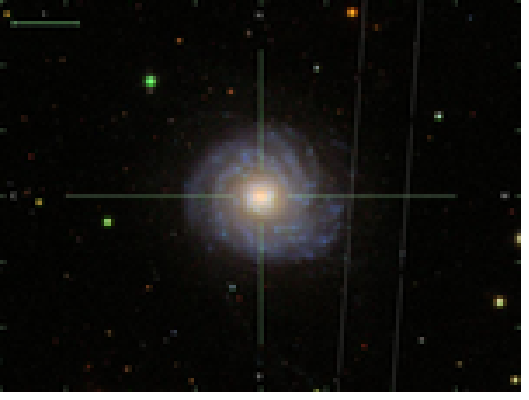}
\hspace*{-0.8cm}
\includegraphics[width=6.1cm,angle=270]{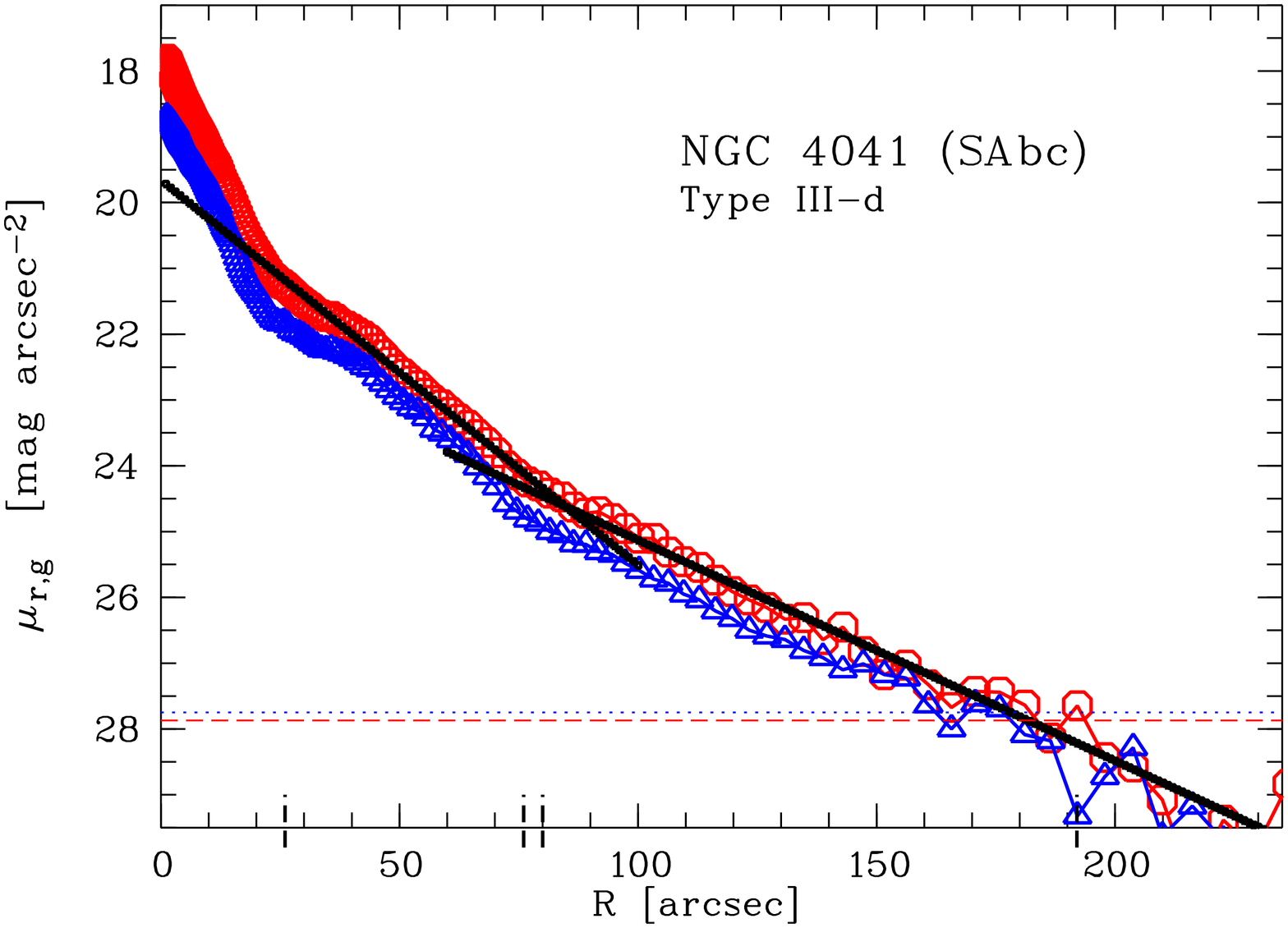}
}
\end{minipage}

\newpage
\onecolumn
\begin{minipage}[t][\textheight][t]{\textwidth}
\parbox[t][0.5\height][t]{0.47\textwidth}
{
\noindent {\bf NGC\,4102     :}        \typeolriii    \\
\texttt{J120623.1+524239 .SXS3\$  3.0 -19.38   3.1 1082}\\[0.25cm]
Galaxy belongs to the Ursa Major Cluster \cite{tully1996}.
On the image an elongated inner bar region inside a starforming 
ring is visible which is responsible for the dip and bump structure
between $\sim\!25-35\arcsec$ in the final profile. \cite{erwin2005a}
measures the size of the bar to $\sim\!15\arcsec$ and the outer ring
to $\sim\!35$\arcsec, so the apparent break at $\gtsim 40\arcsec$ 
(although measured from outside, it starts only at $\sim\!58$\arcsec)
is classified as a \typeolrc. The apparent upbending at $\sim\!120\arcsec$  
is unlikely a sky error, visible around the whole galaxy and 
confirmed by the profile in \cite{courteau1996} (see UGC\,07096)
and also visible in \cite{erwin2006}. This adds the \typeiii 
classification.  
}
\hfill 
\parbox[t][0.5\height][t]{0.47\textwidth}
{
\includegraphics[width=5.7cm,angle=270,]{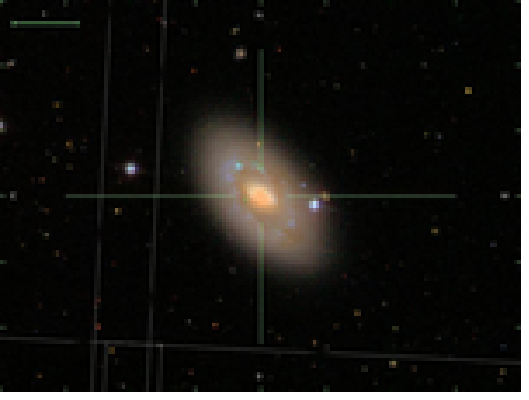}
\hspace*{-0.8cm}
\includegraphics[width=6.1cm,angle=270]{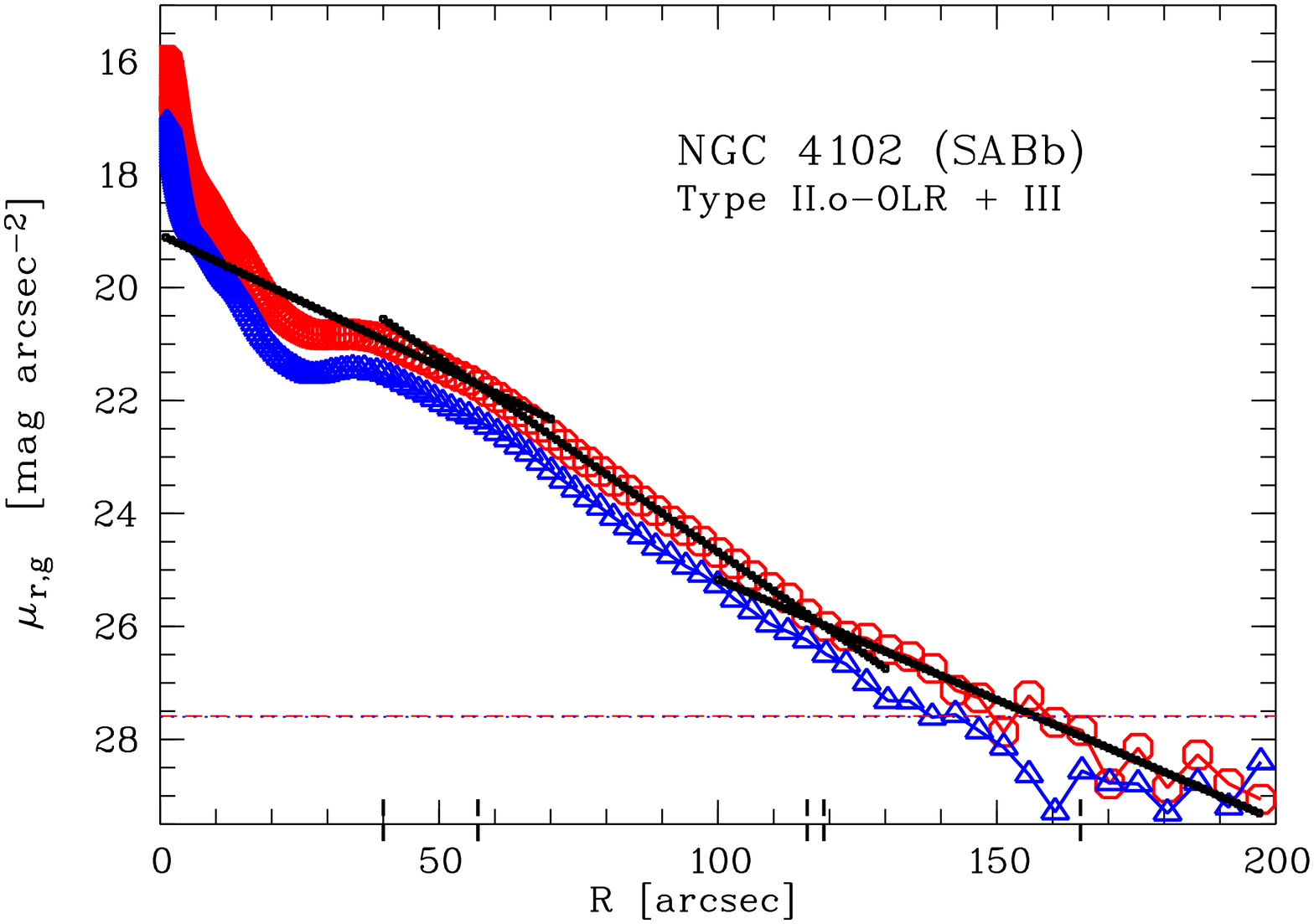}
}
\vfill
\parbox[b][0.5\height][t]{0.47\textwidth}
{
\noindent {\bf NGC\,4108     :}        \typeab                \\          
\texttt{J120644.0+670944 PSA.5*  5.0 -20.13   1.7 2828}\\[0.25cm]
Galaxy in a triple system with  NGC\,4108A ($v=2256$\kms) and 
NGC\,4108B ($v=2673$\kms), the latter is also part of the present 
sample. According to \cite{nordgren1997} NGC\,4108 and NGC\,4108B 
appear to be possibly joined by an \hi bridge. 
Very inner center $\ltsim 5\arcsec$ looks like small bar, or the 
spiral structure starts from the very center.
The outer disk is asymmetric with one side rather sharp and the 
other more shallow with indication for a single spiral arm. The 
disk is lopsided, so the outer contour is used for centering 
(center $\sim\!4\arcsec$ off the brightest pixel, causing the 
central dip in the profile), which implies that the photometric 
inclination (ellipticity) and PA are not well defined.   
The apparent break at $\sim\!44\arcsec$ is related to the transition 
between the symmetric outer disk going inwards to the offcenterd 
inner disk and therefore most probably not intrinsic but a \typeab
break. 

}
\hfill 
\parbox[b][0.5\height][b]{0.47\textwidth}
{
\includegraphics[width=5.7cm,angle=270]{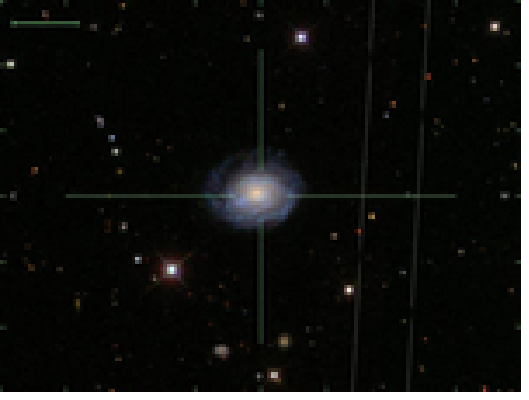}
\hspace*{-0.8cm}
\includegraphics[width=6.1cm,angle=270]{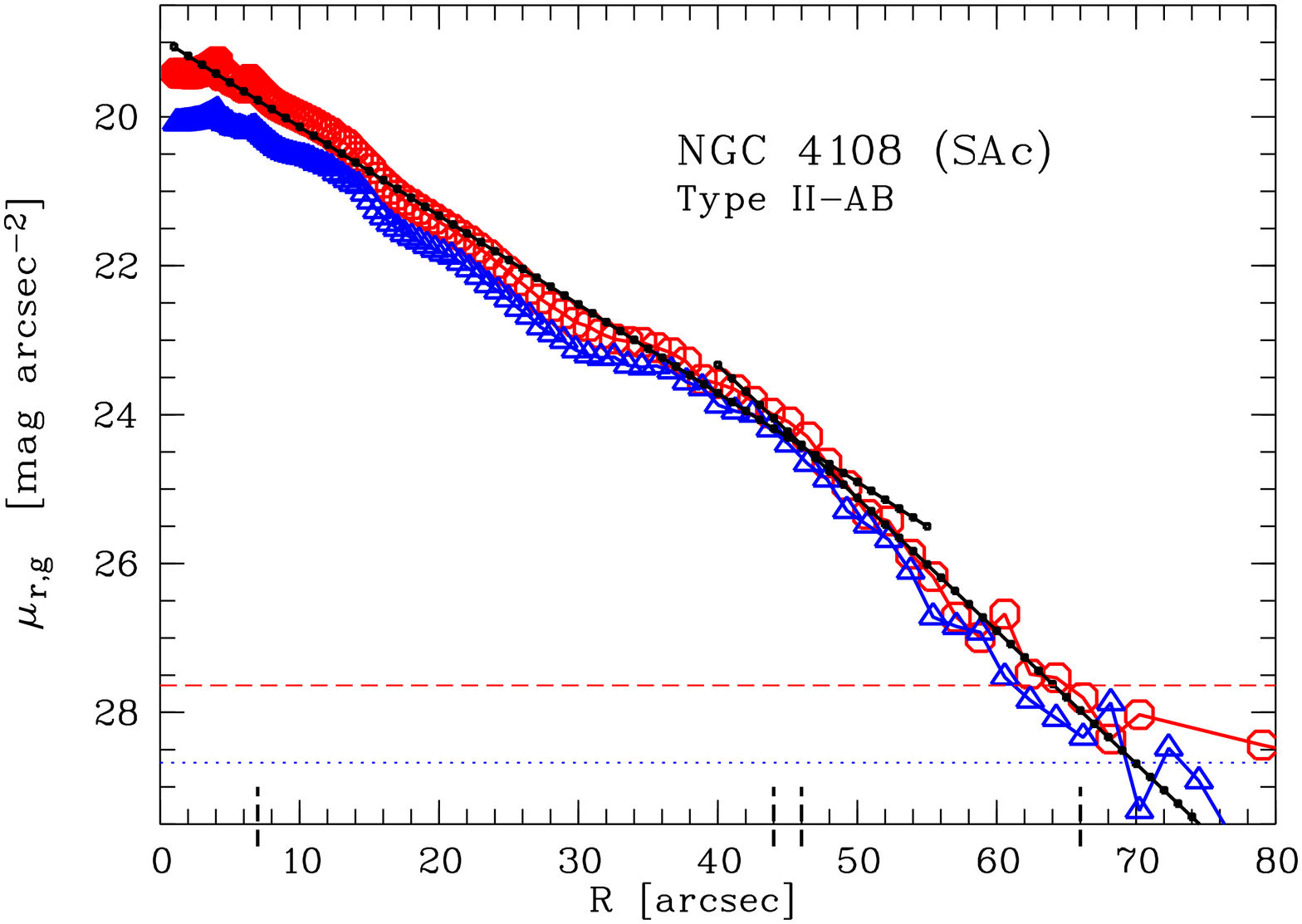}
}
\end{minipage}

\newpage
\onecolumn
\begin{minipage}[t][\textheight][t]{\textwidth}
\parbox[t][0.5\height][t]{0.47\textwidth}
{
\noindent {\bf NGC\,4108B    :}        \typeo                 \\          
\texttt{J120711.3+671410 .SXS7P  7.0 -18.46   1.3 2840}\\[0.25cm]
Companion to NGC\,4108 (see above). Galaxy exhibits an offcentered 
bar ($\sim\!5\arcsec$ off compared to the center derived from the 
outer isophote). Very outer disk slightly asymmetric towards 
north-east (outer spiral arm?). In the final profile the central 
dip and inner wiggle are due to the the offcentered bar, the peak 
at $\sim\!22\arcsec$ corresponds to an aligned spiral arm-like 
region. Although a downbending break at $\sim\!32\arcsec$ with low 
scalelength contrast is possible, \typeabc, the galaxy is consistent 
and classified as \typeo due to the wiggles in the center.  

}
\hfill 
\parbox[t][0.5\height][t]{0.47\textwidth}
{
\includegraphics[width=5.7cm,angle=270,]{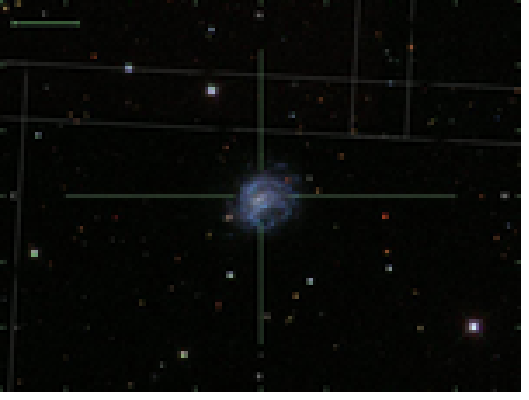}
\hspace*{-0.8cm}
\includegraphics[width=6.1cm,angle=270]{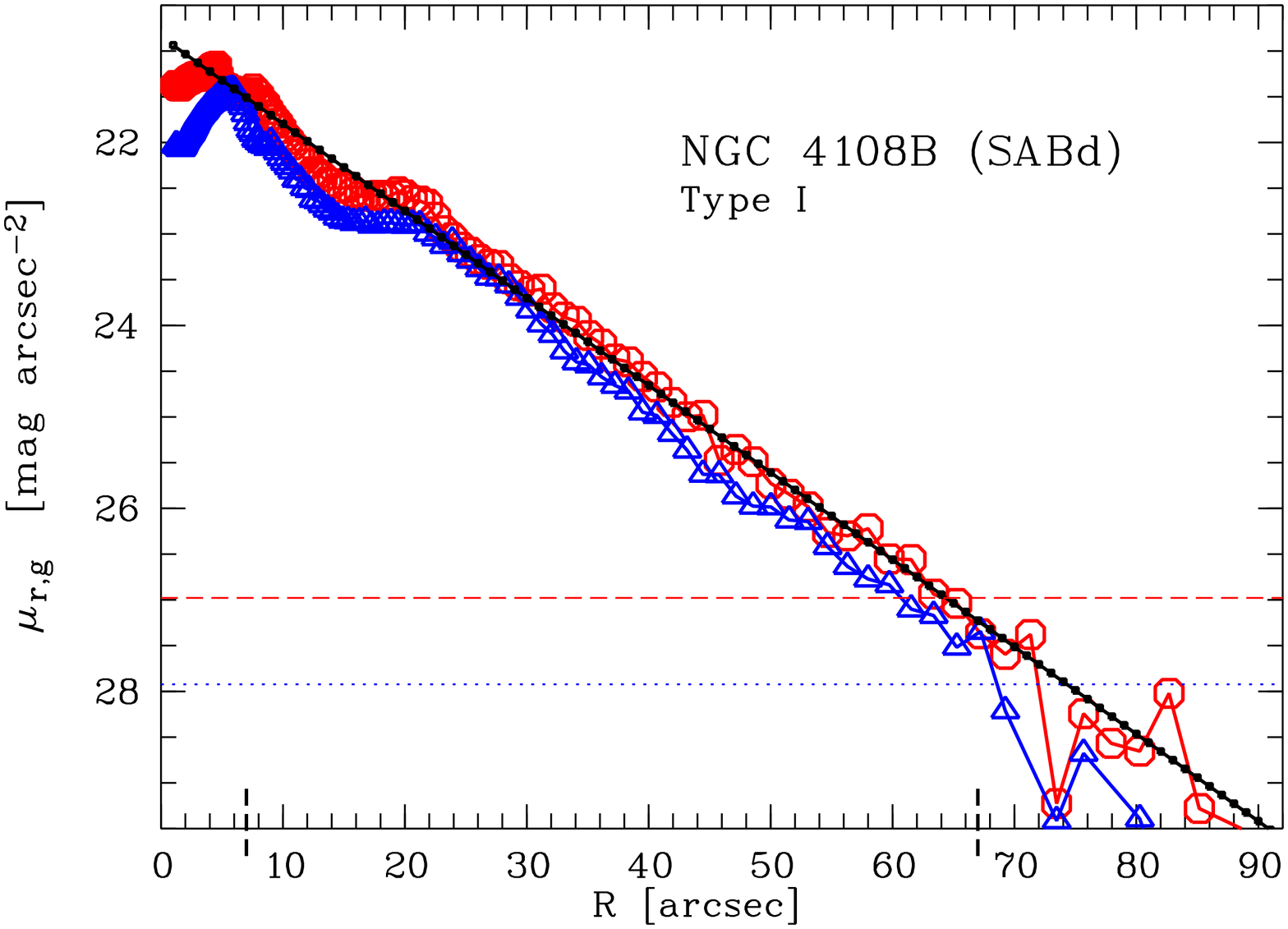}
}
\vfill
\parbox[b][0.5\height][t]{0.47\textwidth}
{
\noindent {\bf NGC\,4123     :}        \typeo                 \\          
\texttt{J120811.1+025242 .SBR5.  4.8 -19.55   4.1 1364}\\[0.25cm]
Galaxy with extended bar ($R\sim\!55$\arcsec) studied in detail 
by \cite{weiner2001} using a surface brightness profile which is 
consistent with the SDSS one.   
The outer disk is slightly asymmetric, more rectangular than 
elliptical, with an extended starforming spiral arm structure 
towards south-west, which makes the photometric inclination 
(ellipticity) and PA uncertain.  
The final profile exhibits extended wiggles. The bar and associated 
spiral arms are visible between $\sim\!30-110\arcsec$ and the bump 
at $\sim\!170\arcsec$ corresponds to a starforming outer spiral arm. 
Although being far from a prototypical \typeo we used this 
classification, since the profile shows too many changes to 
reasonably argue for any other type. 

}
\hfill 
\parbox[b][0.5\height][b]{0.47\textwidth}
{
\includegraphics[width=5.7cm,angle=270]{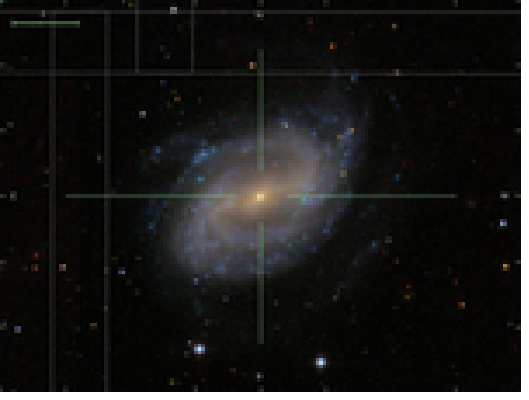}
\hspace*{-0.8cm}
\includegraphics[width=6.1cm,angle=270]{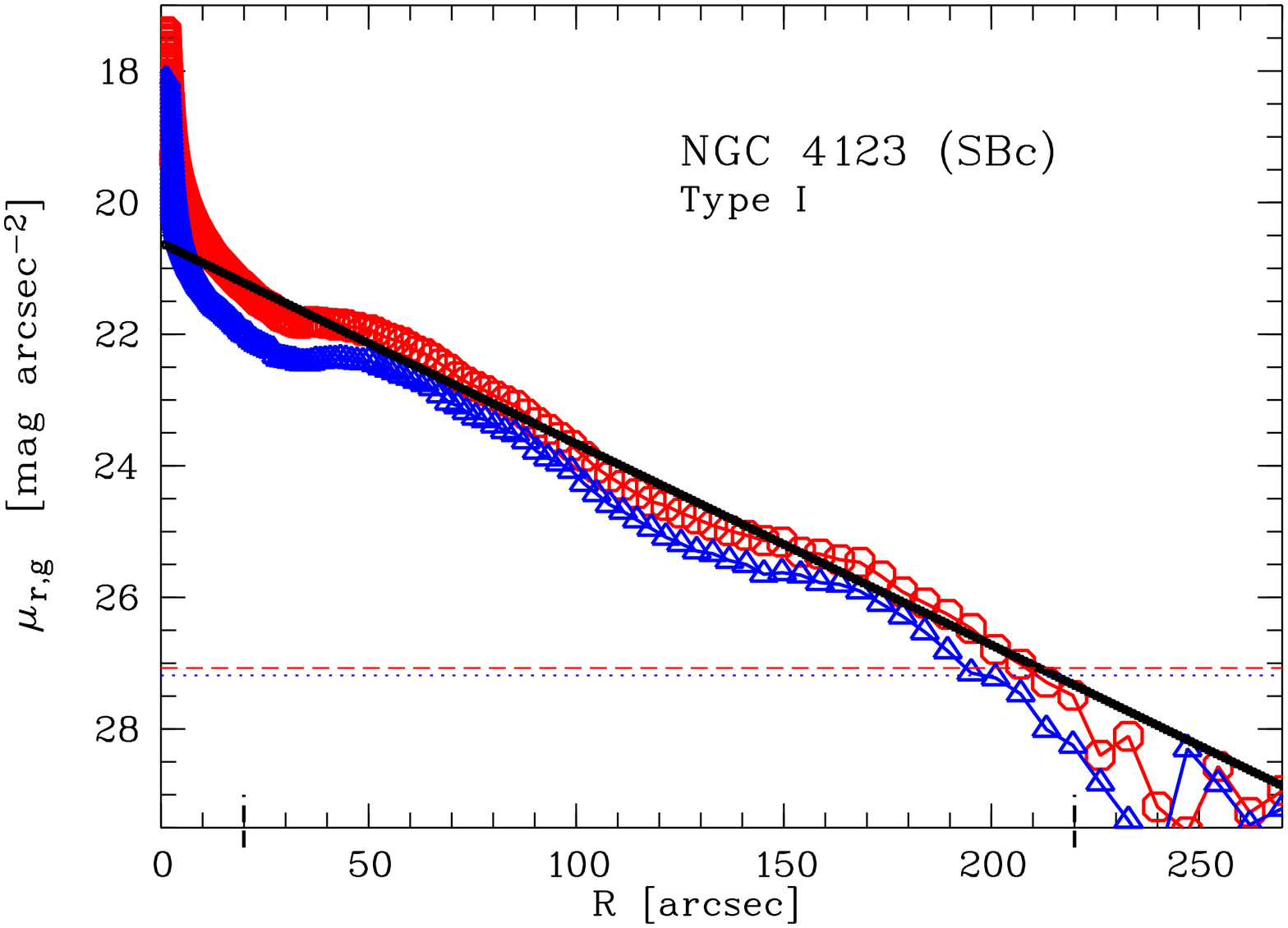}
}
\end{minipage}

\newpage
\onecolumn
\begin{minipage}[t][\textheight][t]{\textwidth}
\parbox[t][0.5\height][t]{0.47\textwidth}
{
\noindent {\bf NGC\,4210     :}        \typeolrct                \\          
\texttt{J121515.9+655908 .SBR3.  3.0 -19.89   2.0 3000}\\[0.25cm]
Companion galaxy (CGCG\,315-030, $m=15.5$) 13\arcmin away with similar 
velocity ($v=2679$\kms). Gradient in the background due to a very 
bright star (off field), but galaxy small enough to be not significantly 
influenced.   
The final profile shows an inner bump at $\sim\!15\arcsec$ corresponding 
to a pseudoring of wrapped spiral arms (slightly further out than the bar).
The apparent dip $\sim\!8\arcsec$ is due to the bar orientation being 
perpendicular to the major axis of the galaxy.
Similar to NGC\,3488 the profile shows a clear downbending. However, 
the break radius is uncertain due to the extended break region, which 
resembles in this case again a straight line. Thus one could also 
define two break radii at $\sim\!33$\arcsec, corresponding roughly to
end of inner spiral arm structure, and $\sim\!55$\arcsec, not related 
to sky errors and symmetric around the disk.  
In contrast to NGC\,3488 the bar is here large enough to argue for 
a combination of a \typeolr and a \typetoct break, this should be 
confirmed with a more detailed study on the bar size of NGC\,4210. 

}
\hfill 
\parbox[t][0.5\height][t]{0.47\textwidth}
{
\includegraphics[width=5.7cm,angle=270,]{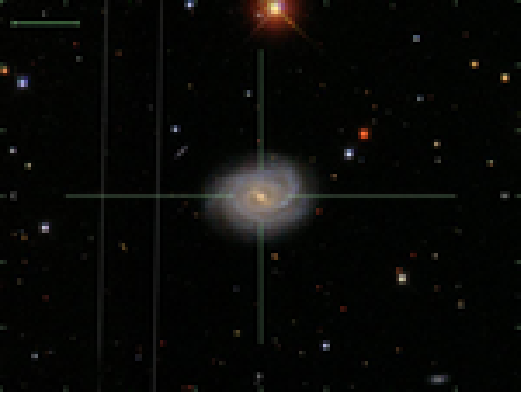}
\hspace*{-0.8cm}
\includegraphics[width=6.1cm,angle=270]{N4210_radn.ps}
}
\vfill
\parbox[b][0.5\height][t]{0.47\textwidth}
{
\noindent {\bf NGC\,4273     :}        \typeabiii     \\
\texttt{J121956.2+052036 .SBS5.  5.1 -20.41   2.1 2435}\\[0.25cm]
Galaxy in south-western corner of the Virgo Cluster with three 
similar sized neighbours, NGC\,4268, NGC\,4281, and closest
and slightly overlapping at the very outer disk NGC\,4277. 
The disk is lopsided with one side sharp edged and the other side 
with a wide spiral arm feature towards south-east (embedded in 
the outer disk), similar to NGC\,4108. Centering on the outer 
disk causes the dip at $\sim\!10\arcsec$ and hides the very bright 
nucleus sitting in a bar-like structure which is offcentered 
by $\sim\!16$\arcsec.
The break at $\sim\!45\arcsec$ in the final profile is clearly 
related to the lopsided disk and therefore classified as \typeabc. 
In addition, there is an \typeiii break with an upbending profile 
visible beyond $\sim\!80\arcsec$ due to the influence of the 
companion (possibly intrinsic and not just projected). 
 
}
\hfill 
\parbox[b][0.5\height][b]{0.47\textwidth}
{
\includegraphics[width=5.7cm,angle=270]{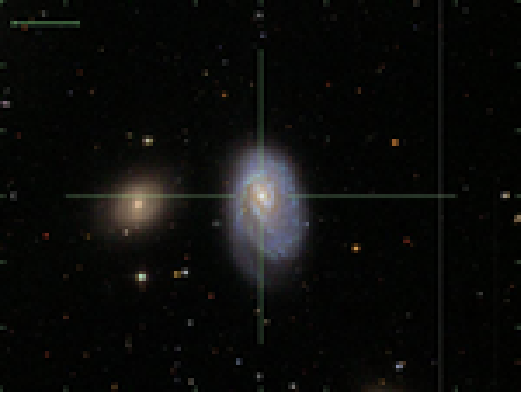}
\hspace*{-0.8cm}
\includegraphics[width=6.1cm,angle=270]{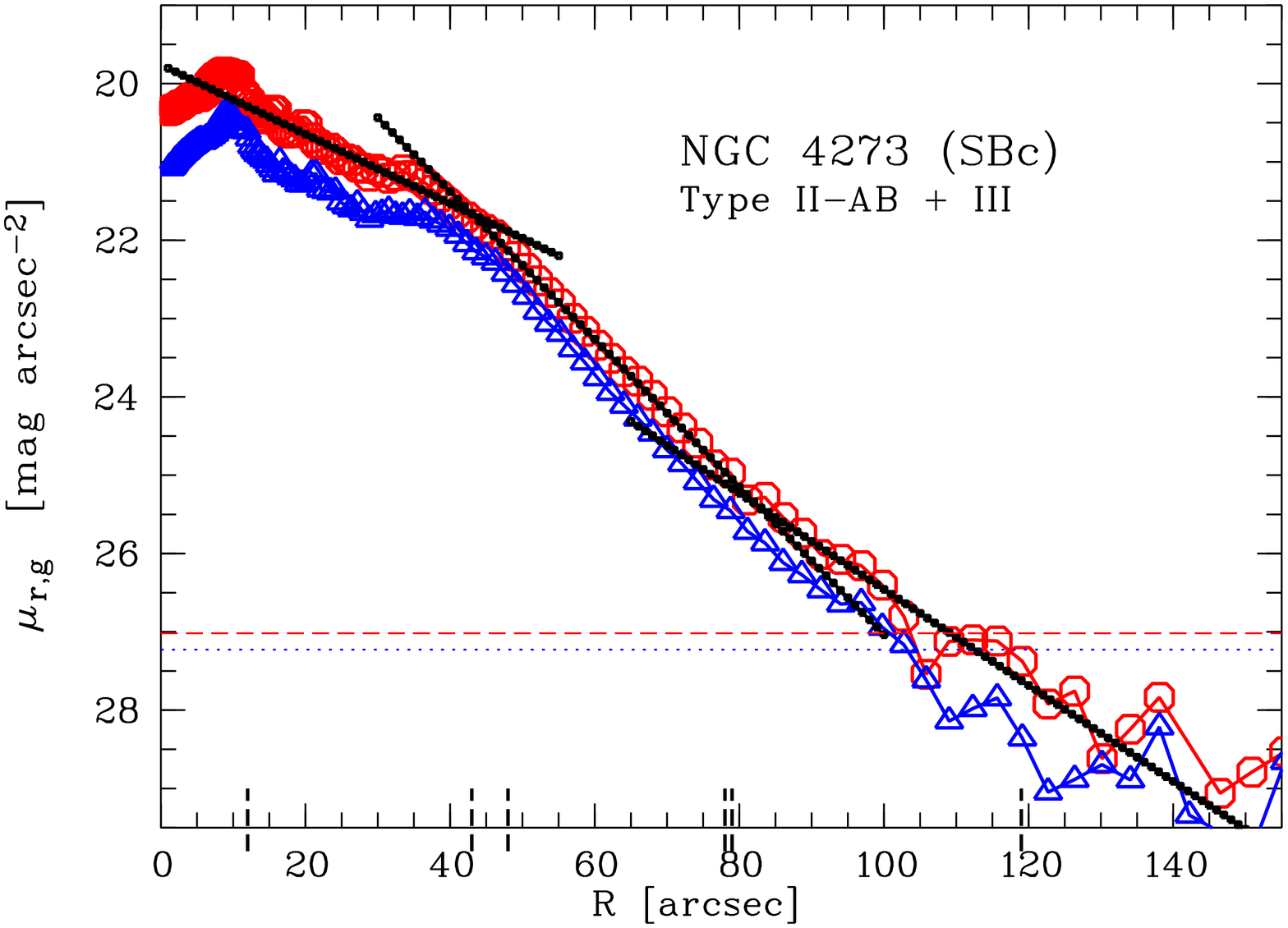}
}
\end{minipage}

\newpage
\onecolumn
\begin{minipage}[t][\textheight][t]{\textwidth}
\parbox[t][0.5\height][t]{0.47\textwidth}
{
\noindent {\bf NGC\,4480     :}        \typetoct                \\          
\texttt{J123026.8+041447 .SXS5.  5.1 -20.14   2.1 2494}\\[0.25cm]
The galaxy is small, rather inclined, and classified as SAB with 
bar size of roughly $R\sim\!10\arcsec$ (measured from the image). 
The final profile shows a clear downbending break at 
$\sim\!55$\arcsec, significantly further out compared to a 
typical \typeolr break thus classified as \typetoctc, corresponding 
roughly to the end of the spiral arms. The inner profile exhibits 
an extended bump at $\sim\!35\arcsec$ (more pronounced in the $g^{\prime}$ 
and less in the $i^{\prime}$ band) most probably related to the star formation
in the spiral arms and not to a possible \typeolrct as for NGC\,4210,
which shows the same structure in both bands. 

}
\hfill 
\parbox[t][0.5\height][t]{0.47\textwidth}
{
\includegraphics[width=5.7cm,angle=270,]{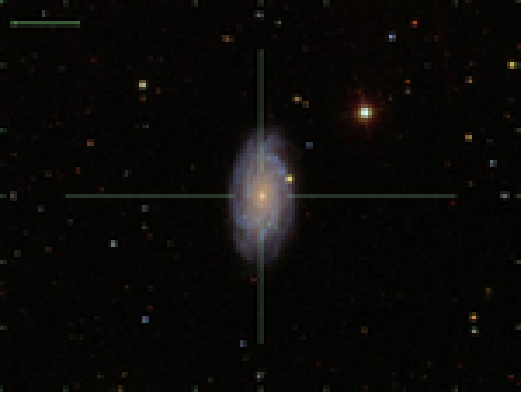}
\hspace*{-0.8cm}
\includegraphics[width=6.1cm,angle=270]{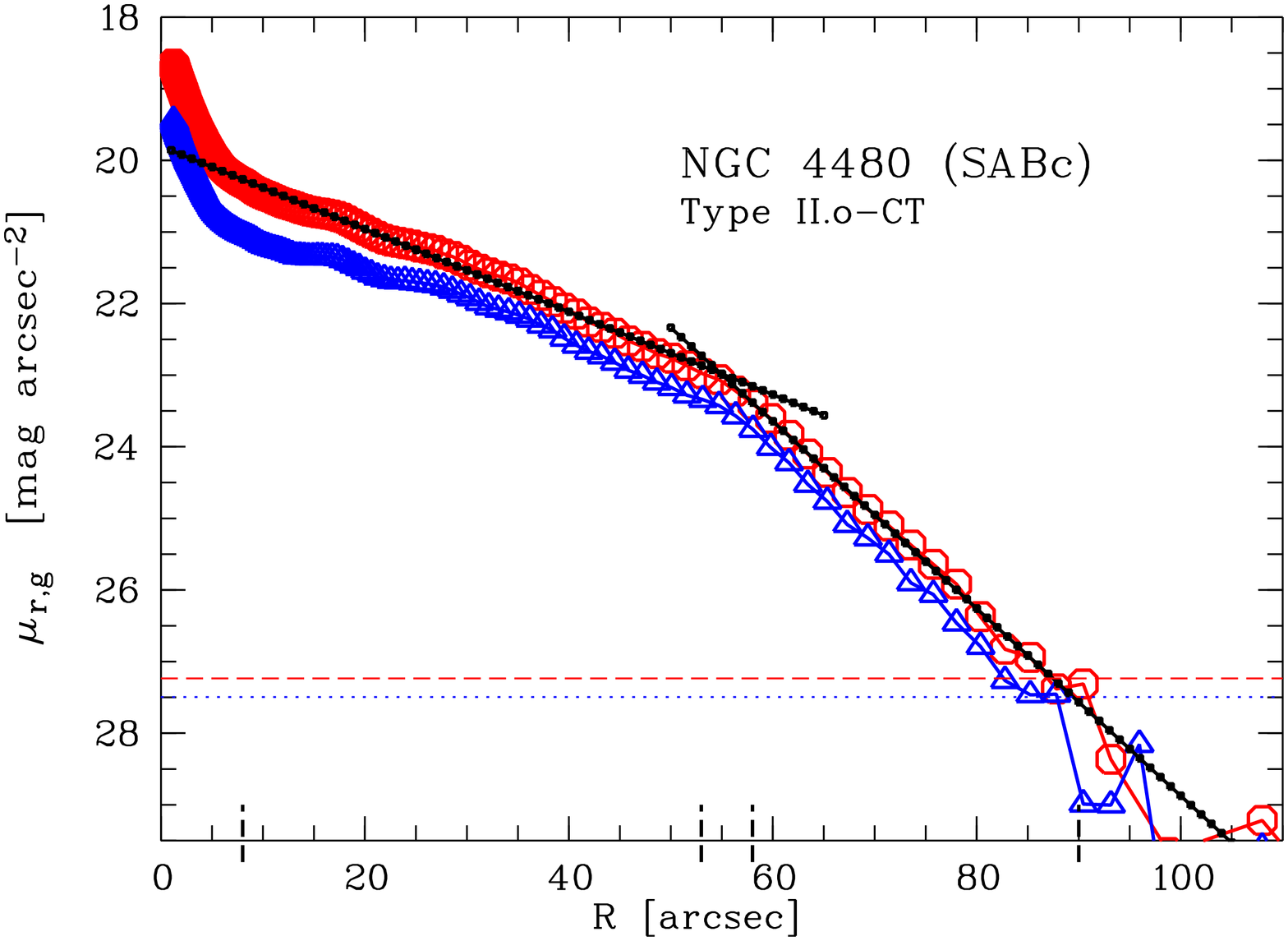}
}
\vfill
\parbox[b][0.5\height][t]{0.47\textwidth}
{
\noindent {\bf NGC\,4517A    :}        \typetoctab    \\
\texttt{J123228.1+002325 .SBT8*  7.9 -19.18   3.8 1562}\\[0.25cm]
This low surface brightness galaxy has an elongated, barred center,
of size $R\sim\!30$\arcsec, with a foreground star very close and 
a slightly asymmetric, lopsided outer disk. 
The final profile exhibits a sharp break at $\sim\!150$\arcsec
due to the asymmetric outer disk, which is not well described 
with the mean ellipticity and PA used, thus classified as \typeabc. 
There is an additional feature (bump or break) visible around 
$\sim\!80$\arcsec, which is not obviously related to an aligned 
spiral arm and could be explained by a weak downbending break. 
Since the bar appears to be only weak and the break at more 
than twice the bar radius it is classified as \typetoctc, which 
should be taken with caution.
 
}
\hfill 
\parbox[b][0.5\height][b]{0.47\textwidth}
{
\includegraphics[width=5.7cm,angle=270]{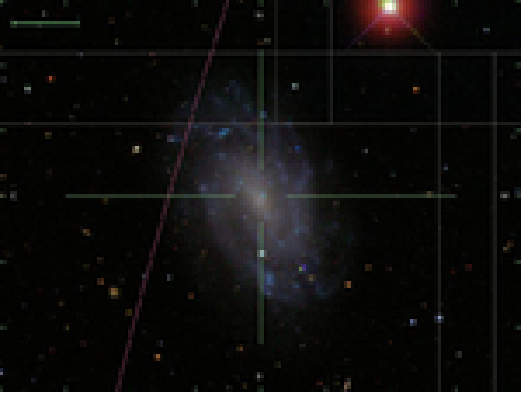}
\hspace*{-0.8cm}
\includegraphics[width=6.1cm,angle=270]{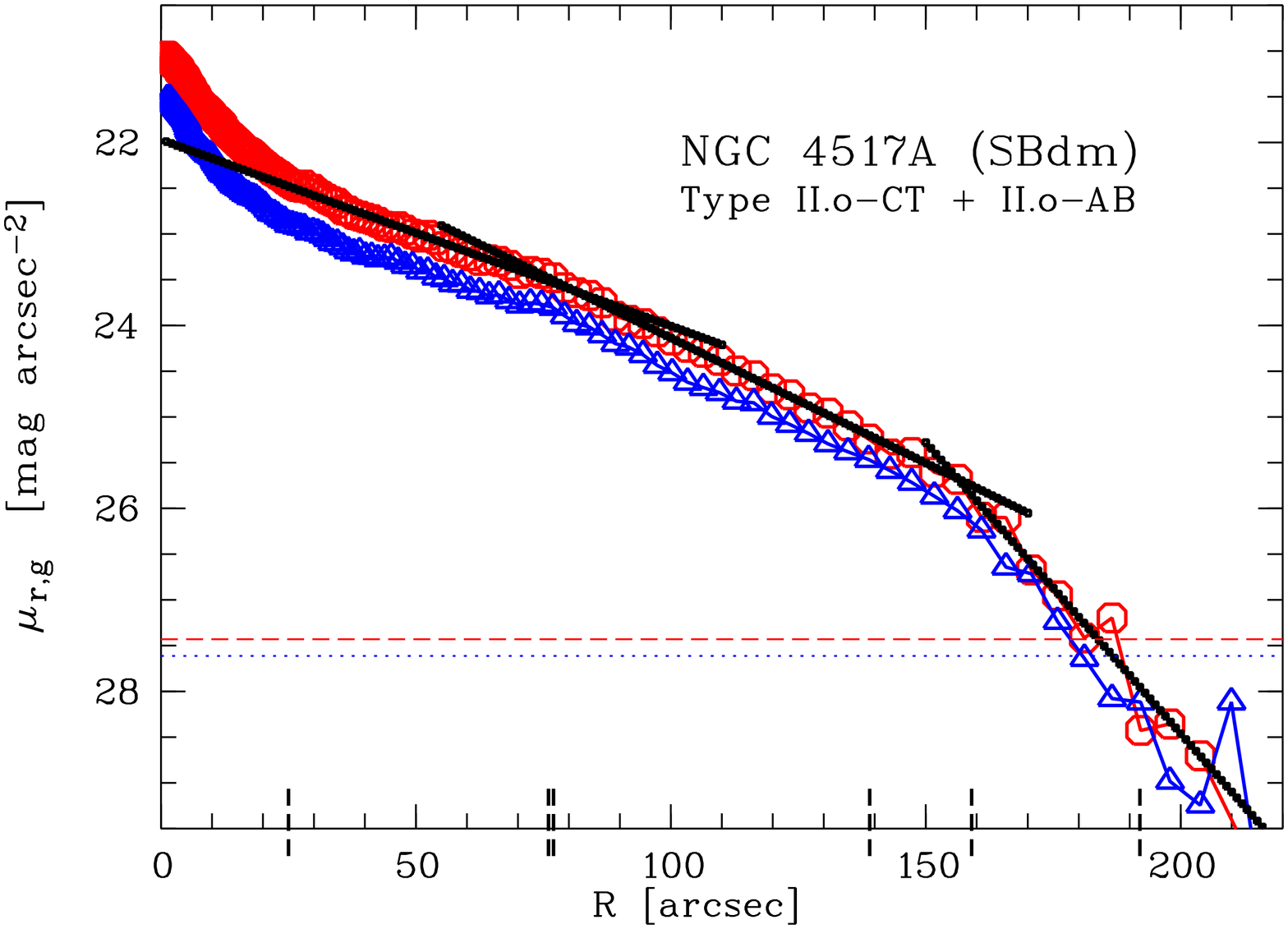}
}
\end{minipage}

\newpage
\onecolumn
\begin{minipage}[t][\textheight][t]{\textwidth}
\parbox[t][0.5\height][t]{0.47\textwidth}
{
\noindent {\bf NGC\,4545     :}        \typetoct                \\    
\texttt{J123434.2+633130 .SBS6*  5.9 -20.49   2.5 3000}\\[0.25cm]
Small galaxy with an s-shaped inner region of size $\sim\!12\arcsec$ 
(associated to be the bar) leading to two (or three) tightly wrapped, 
spiral arms.  
Final profile exhibits a break at $\sim\!58$\arcsec, being too far 
out for a typical \typeolr thus classified as \typetoctc.  
The dip at $\sim\!8\arcsec$ corresponds to the region where the spiral 
arms start from the bar-structure which is nearly perpendicular to the 
ellipse major axis.

}
\hfill 
\parbox[t][0.5\height][t]{0.47\textwidth}
{
\includegraphics[width=5.7cm,angle=270,]{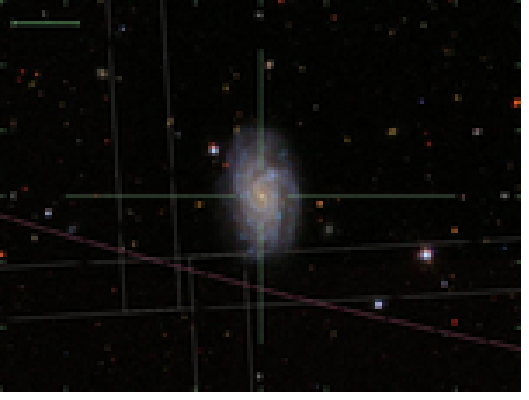}
\hspace*{-0.8cm}
\includegraphics[width=6.1cm,angle=270]{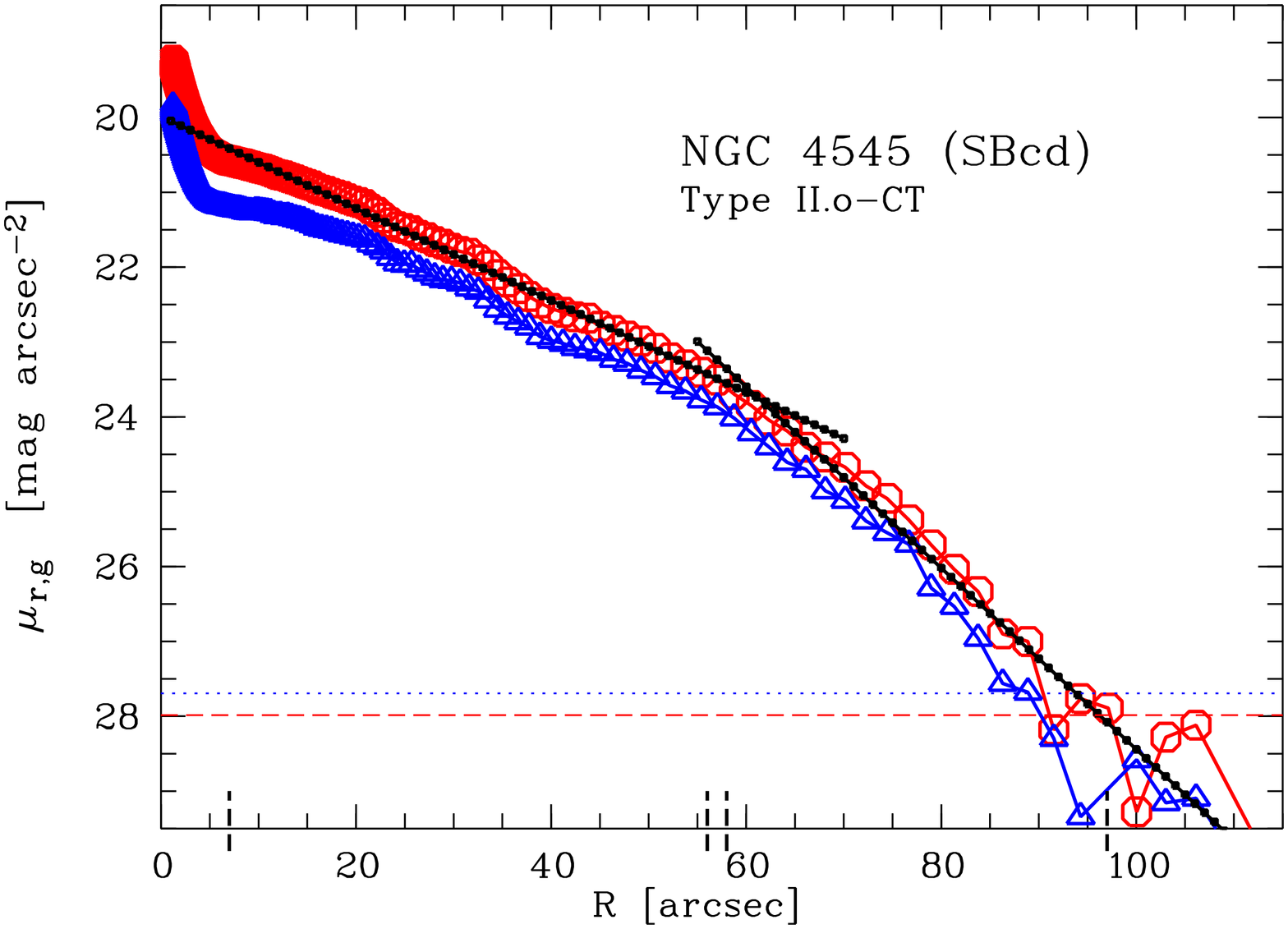}
}
\vfill
\parbox[b][0.5\height][t]{0.47\textwidth}
{
\noindent {\bf NGC\,4653     :}        \typeab                \\          
\texttt{J124350.9-003341 .SXT6.  6.1 -20.37   3.0 2658}\\[0.25cm]
Paired galaxy with NGC\,4642 ($v=2471$\kms, $\sim\!10\arcmin$ away)
while apparent dwarf companion on the image towards south-east 
is a confirmed background galaxy ($v=7843$\kms).
Extended mask used to avoid straylight by bright stars.  
Mean ellipticity and PA difficult to determine, since it is 
different for the inner disk (used for centering) with two, 
very regular spiral arms inside $\ltsim 30$\arcsec, compared to 
the outer disk with more frayed spiral fragments, starting to be 
asymmetric on larger scales at $\ltsim 80\arcsec$ (mainly towards 
south-west). This asymmetry causes also the apparent break at 
$\sim\!105\arcsec$ thus the galaxy is classified as \typeabc. 

}
\hfill 
\parbox[b][0.5\height][b]{0.47\textwidth}
{
\includegraphics[width=5.7cm,angle=270]{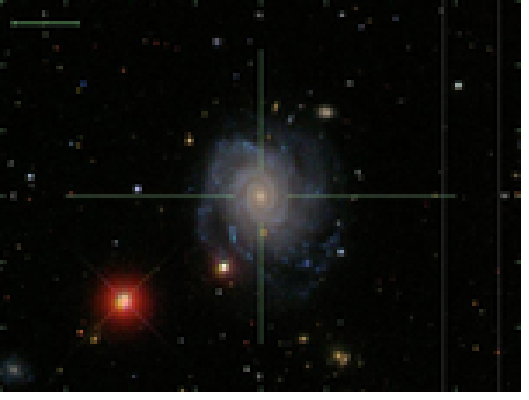}
\hspace*{-0.8cm}
\includegraphics[width=6.1cm,angle=270]{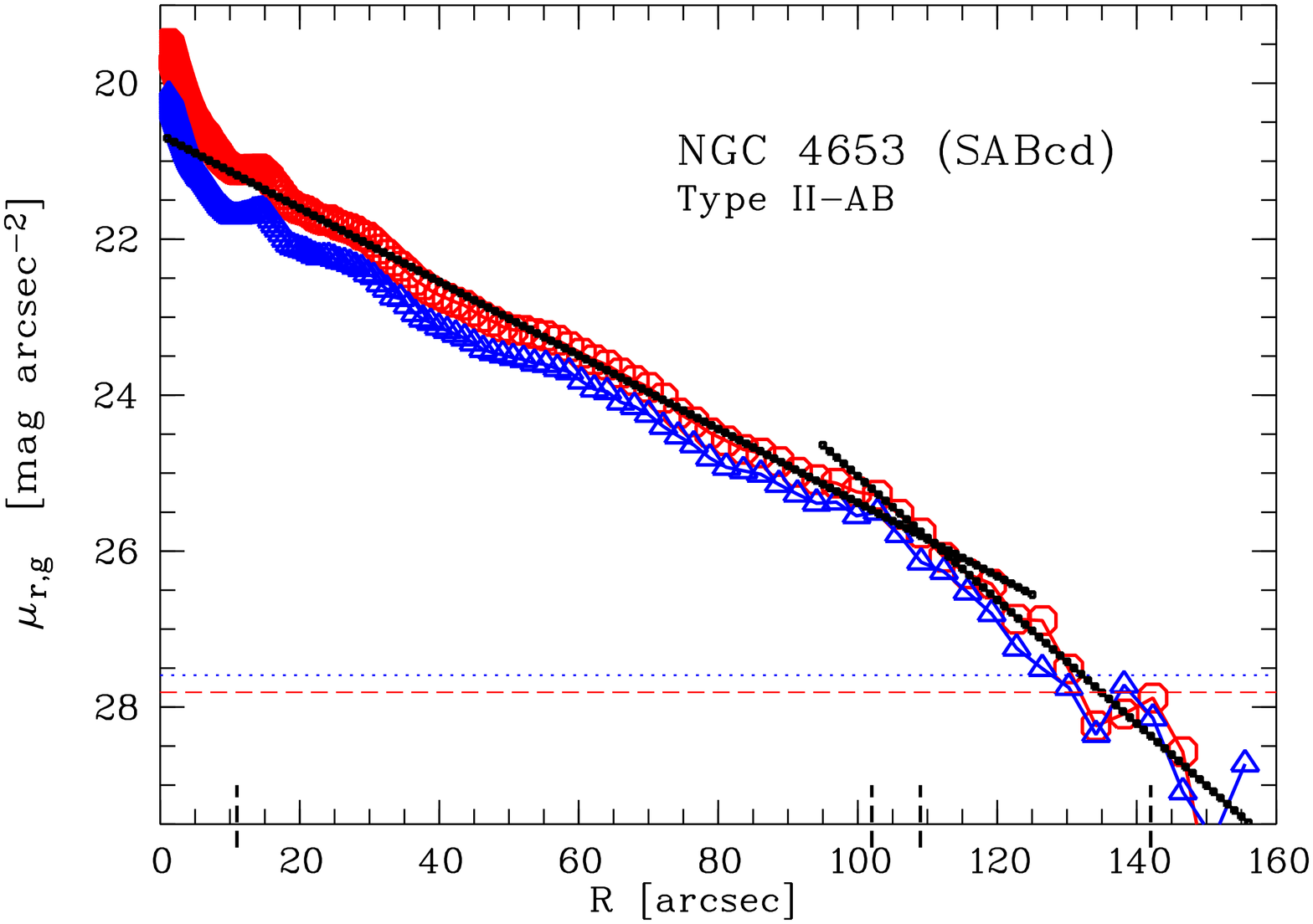}
}
\end{minipage}

\newpage
\onecolumn
\begin{minipage}[t][\textheight][t]{\textwidth}
\parbox[t][0.5\height][t]{0.47\textwidth}
{
\noindent {\bf NGC\,4668     :}        \typeiii                 \\        
\texttt{J124532.0-003209 .SBS7*  7.1 -18.88   1.5 1654}\\[0.25cm]
Small galaxy close to our high axis ratio limit, which forms a close pair 
with NGC\,4666 ($v=1474$\kms) about $7\arcmin$ away and also has a wider 
association with NGC\,4632 ($v=1557$\kms) according to \cite{sandage1994}.
The final profile exhibits a break at $\sim\!70\arcsec$ followed 
by an upbending profile corresponding to a slightly asymmetric, 
faint outer disk and although close to, most probably not related 
to a sky error, thus classified as \typeiiic.  
The full extend of the bar is uncertain, although the bump around 
$\sim\!20\arcsec$ is certainly related to the bar and a possible 
spiral arm. The whole inner region is included in the inner 
disk fit.

}
\hfill 
\parbox[t][0.5\height][t]{0.47\textwidth}
{
\includegraphics[width=5.7cm,angle=270,]{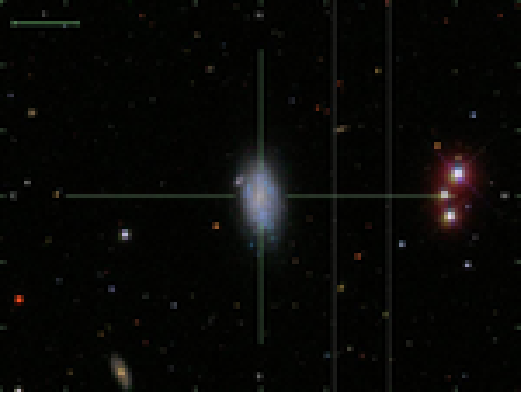}
\hspace*{-0.8cm}
\includegraphics[width=6.1cm,angle=270]{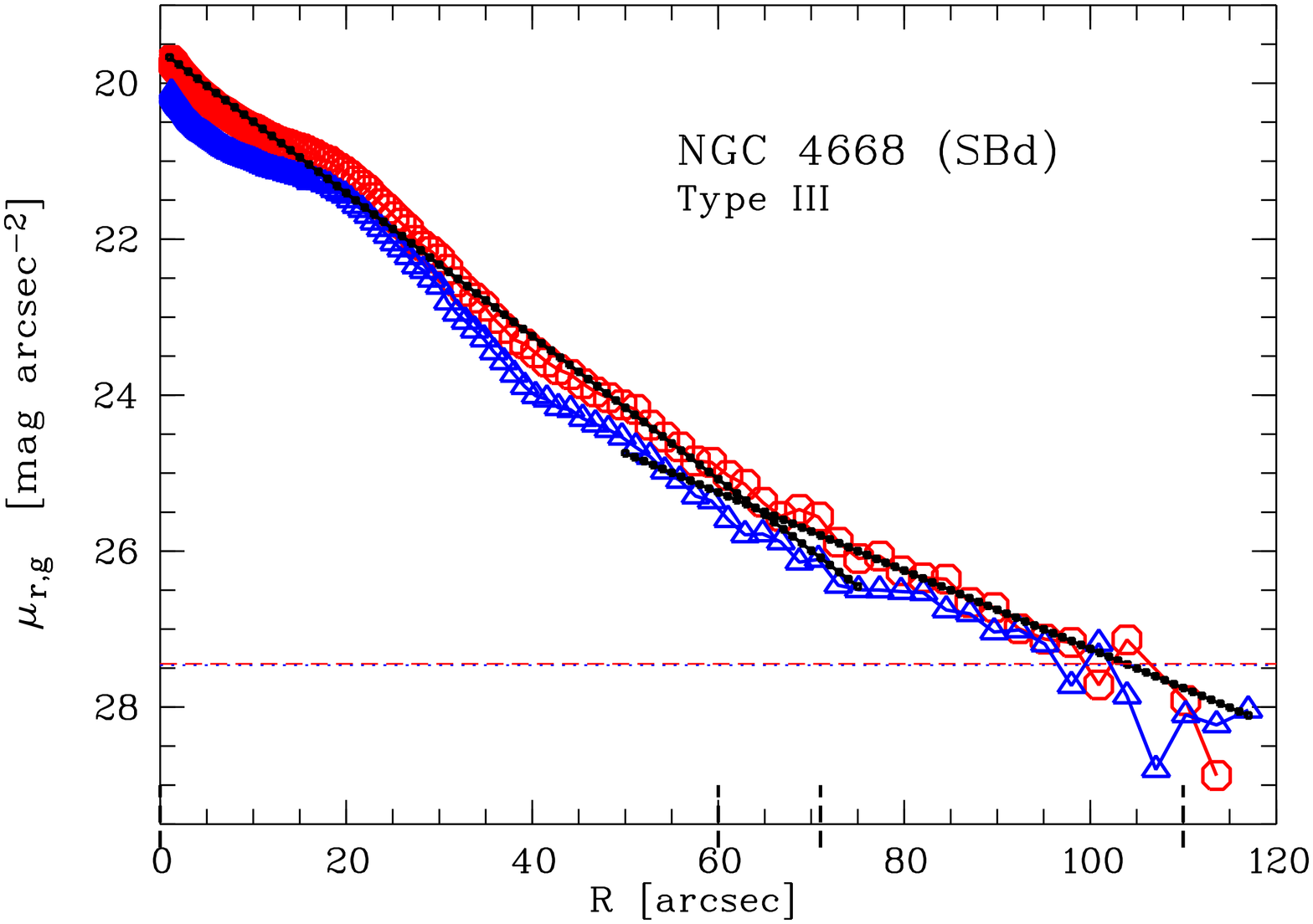}
}
\vfill
\parbox[b][0.5\height][t]{0.47\textwidth}
{
\noindent {\bf NGC\,4904     :}        \typeolriii    \\
\texttt{J130058.6-000138 .SBS6.  5.7 -18.69   2.1 1213}\\[0.25cm]
Only partly fitted since \ltsim $1/5$ of galaxy is beyond field.
The galaxy has a dominating, thick bar of size $R\gtsim 20$\arcsec
whereas the center along the bar is not well defined (centering 
on brightest pixel). It has an unusual spiral arm structure 
with one extending nearly straight towards south-west forming 
more droplet shaped isophotes.
The image shows indication for an additional, extremely faint, 
extended patch of light outside the disk towards south-west.
Final profile shows a clear downbending break at $\sim\!40$\arcsec,
associated on one side with a ring-like structure which is about 
twice the bar radius, thus classified as \typeolrc. 
The apparent upbending at $\sim\!90\arcsec$ is almost certainly
produced by the additional patch of light unrelated to any 
artificial source, thus adding the \typeiii classification.  

}
\hfill 
\parbox[b][0.5\height][b]{0.47\textwidth}
{
\includegraphics[width=5.7cm,angle=270]{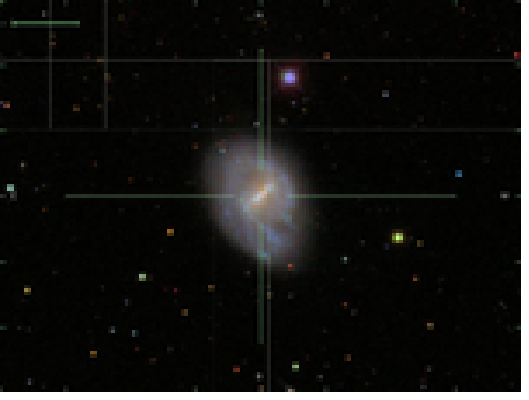}
\hspace*{-0.8cm}
\includegraphics[width=6.1cm,angle=270]{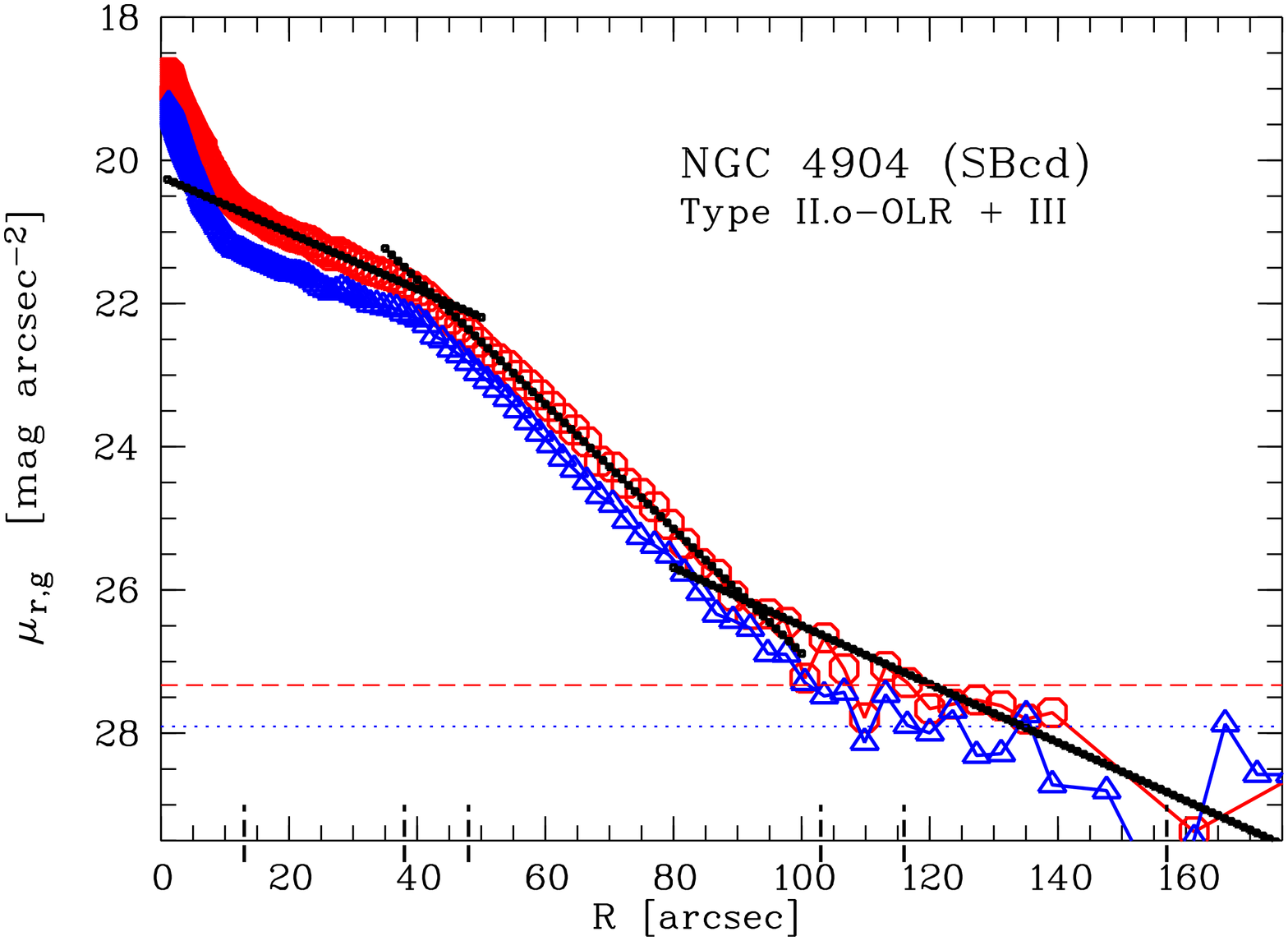}
}
\end{minipage}

\newpage
\onecolumn
\begin{minipage}[t][\textheight][t]{\textwidth}
\parbox[t][0.5\height][t]{0.47\textwidth}
{
\noindent {\bf NGC\,5147     :}        \typeolriii     \\
\texttt{J132619.2+020604 .SBS8.  7.9 -18.77   1.8 1154}\\[0.25cm]
The medium-sized star superimposed on the galaxy close to its 
center is masked. The bar size is uncertain but could be associated 
to an elliptical region of size $R\ltsim 15\arcsec$ on the image not 
reflected in the final profile.  
The extended bump between $\sim\!25-45\arcsec$ is associated 
to the inner disk containing the spiral arms with some 
ring-like, distributed SF regions. Excluding the full 
region allows for a crude \typeo classification. 
However, using a break at $\sim\!30\arcsec$ (about twice 
the bar radius thus classified as \typeolrc) followed by a downbending 
profile with another break at $\sim\!75\arcsec$ leading to an upbending 
profile (therefore classified as \typeiiic) seems more reasonable,
even so the upbending is very close to our detection limit.  
The disk beyond the ring-like structure is slightly asymmetric with 
a single, wide, and faint spiral arm-like structure towards north-east
just inside the outer break. The outermost isophote is again 
nearly round and symmetric. 

}
\hfill 
\parbox[t][0.5\height][t]{0.47\textwidth}
{
\includegraphics[width=5.7cm,angle=270,]{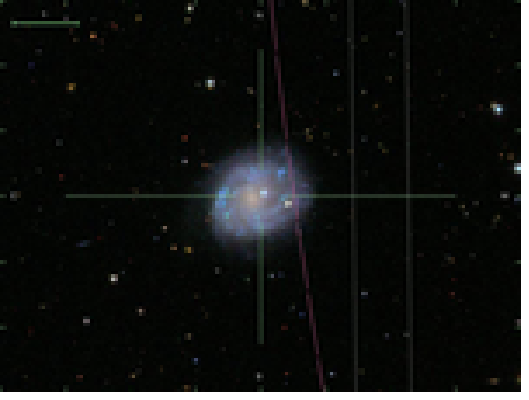}
\hspace*{-0.8cm}
\includegraphics[width=6.1cm,angle=270]{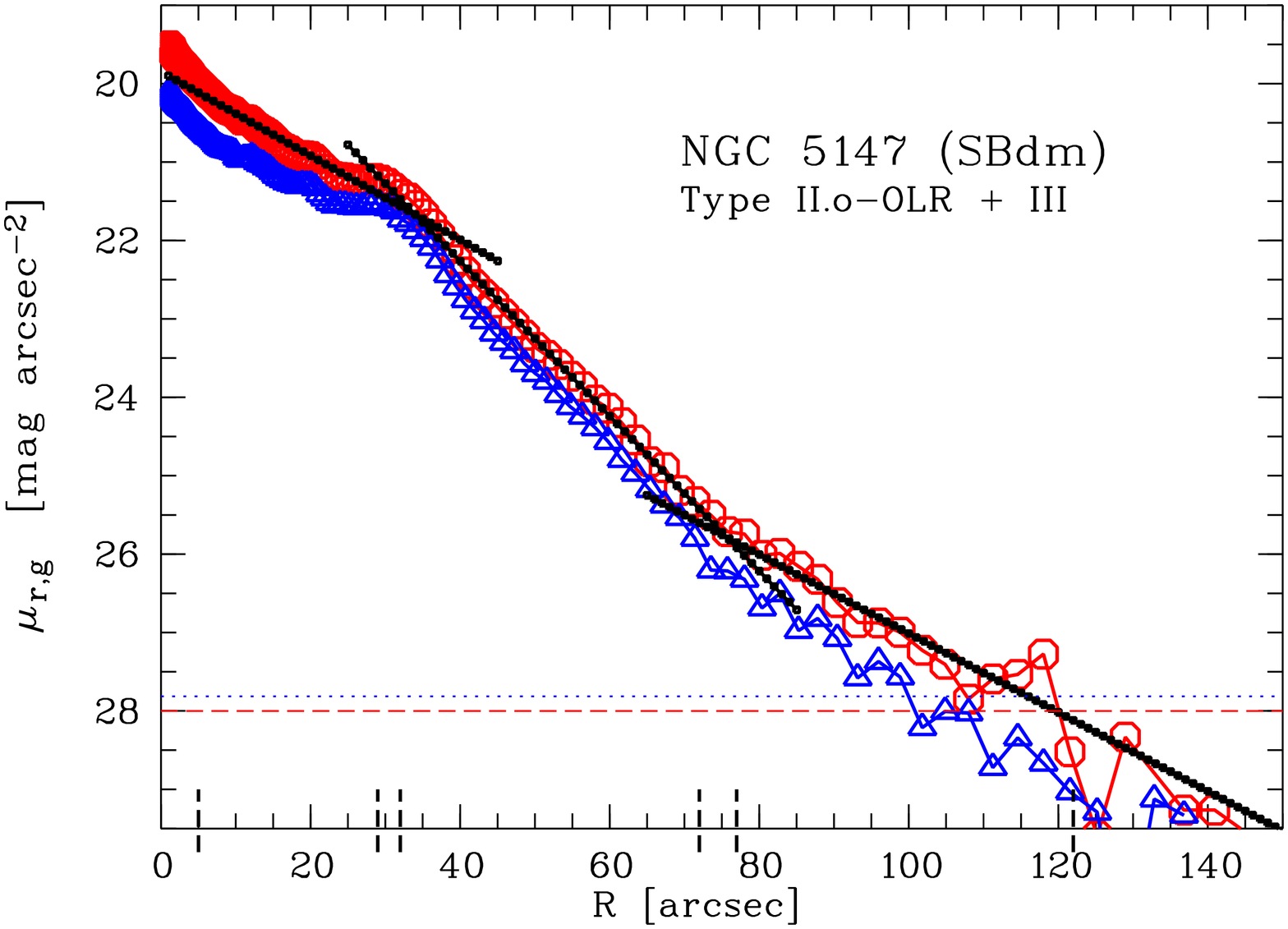}
}
\vfill
\parbox[b][0.5\height][t]{0.47\textwidth}
{
\noindent {\bf NGC\,5300     :}        \typetoct                \\          
\texttt{J134816.1+035703 .SXR5.  5.0 -18.93   3.5 1246}\\[0.25cm]
Galaxy close to border of SDSS field but almost complete. The size 
of the weak bar is about $R\sim\!12\arcsec$ (measured from the image).  
The final profile shows the prototypical \typetoct behaviour. A sharp 
break at $\sim\!85\arcsec$ followed by a downbending profile. The break 
is way to far out for a normal \typeolr break given the small bar 
size, and corresponds roughly to the end of the visible spiral 
arm structure. 

}
\hfill 
\parbox[b][0.5\height][b]{0.47\textwidth}
{
\includegraphics[width=5.7cm,angle=270]{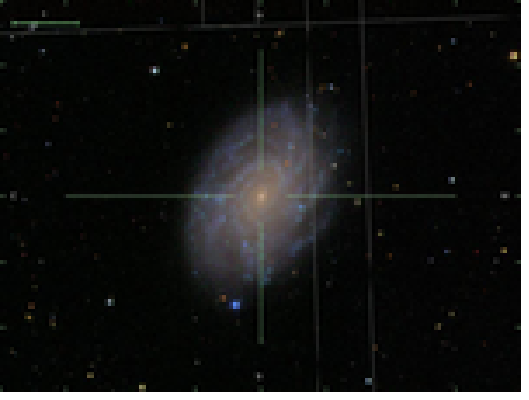}
\hspace*{-0.8cm}
\includegraphics[width=6.1cm,angle=270]{N5300_radn.ps}
}
\end{minipage}

\newpage
\onecolumn
\begin{minipage}[t][\textheight][t]{\textwidth}
\parbox[t][0.5\height][t]{0.47\textwidth}
{
\noindent {\bf NGC\,5334     :}        \typetoct                \\          
\texttt{J135254.4-010651 .SBT5*  5.0 -19.22   3.9 1433}\\[0.25cm]
Galaxy has a bar with a visible dust lane of size $R \ltsim 20$\arcsec
as measured from the image. 
The final profile is clearly downbending, but the exact position of 
the break is difficult to place. The extended break region resembles
in this case again a straight line, so that one could also define two 
break radii at $\sim\!85\arcsec$ and $\sim\!125$\arcsec. However, since
the first one is not even close to twice the bar radius and no other 
particular feature is responsible for the second break, the galaxy
is classified \typetoct with a single break at $\gtsim 90$\arcsec.  

}
\hfill 
\parbox[t][0.5\height][t]{0.47\textwidth}
{
\includegraphics[width=5.7cm,angle=270,]{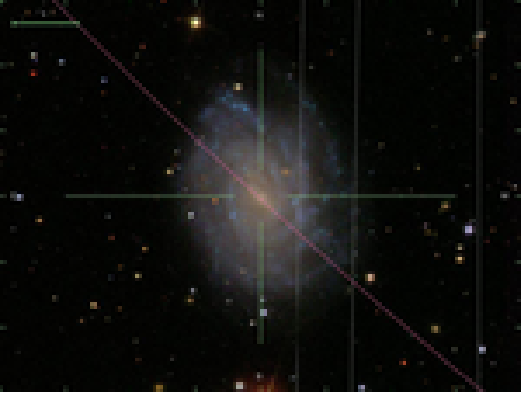}
\hspace*{-0.8cm}
\includegraphics[width=6.1cm,angle=270]{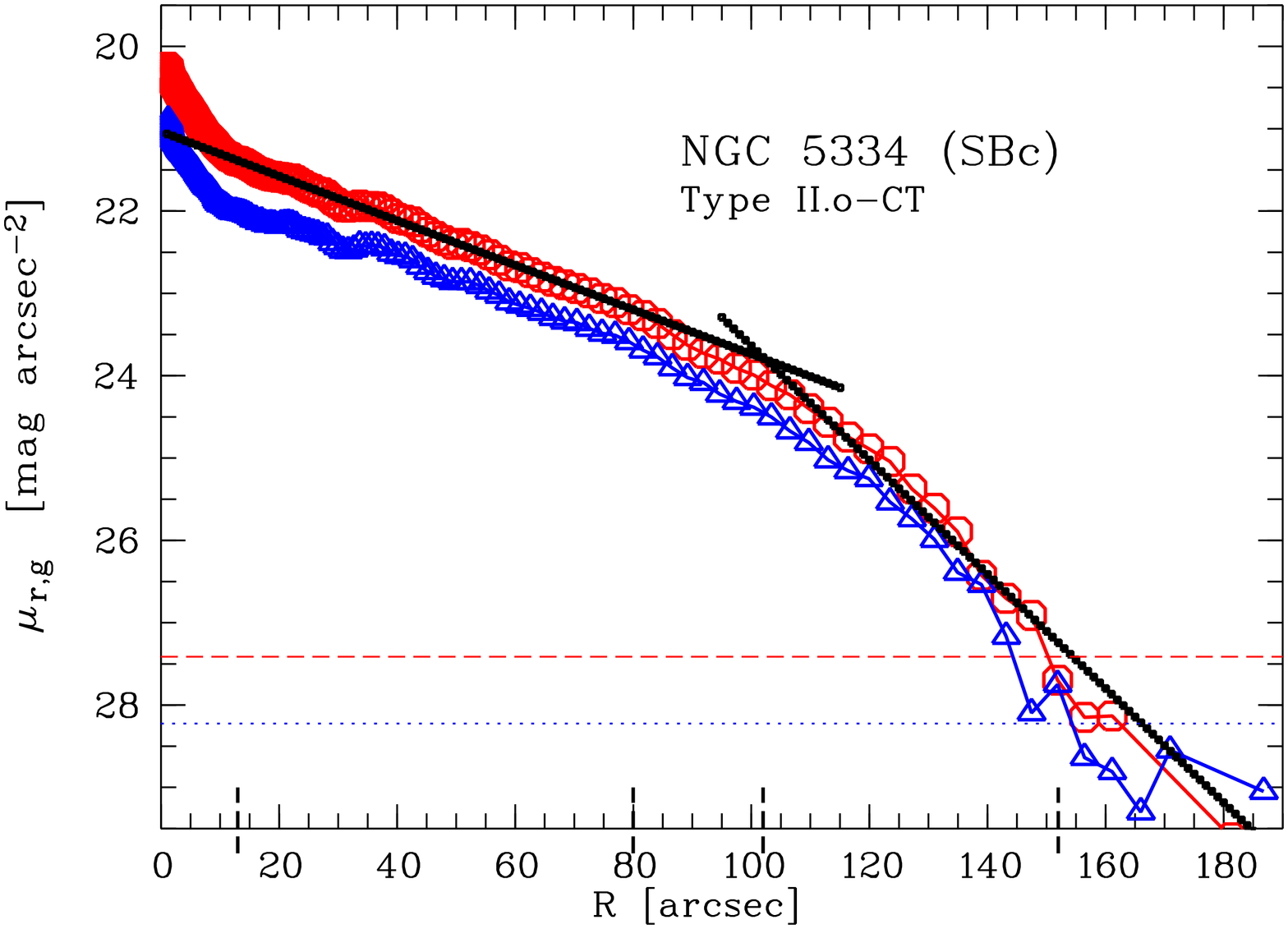}
}
\vfill
\parbox[b][0.5\height][t]{0.47\textwidth}
{
\noindent {\bf NGC\,5376     :}        \typeolr               \\           
\texttt{J135516.1+593024 .SXR3\$  3.0 -20.00   2.1 2293}\\[0.25cm]
Galaxy close to the border of the SDSS field but almost complete. 
It has a known, similar-sized companion (UGC\,08859, $v=1610$\kms) 
about $7.4\arcmin$ away and there is another possible, much 
fainter dwarf companion ($3.5\arcmin$ away towards north-west) 
visible on the image.  
The final profile shows a weak break at $\sim\!35\arcsec$ coinciding 
with the end of some very faint spiral arms. The dip at $\sim\!14$\arcsec
and the peak at $\sim\!18 \arcsec$ are related to a pseudoring structure 
(mentioned also in RC1) around a possible bar. Although the exact size of 
the bar is difficult to fix, assuming it is there, the break is close to 
twice the inner ring radius and therefore classified as \typeolrc. 

}
\hfill 
\parbox[b][0.5\height][b]{0.47\textwidth}
{
\includegraphics[width=5.7cm,angle=270]{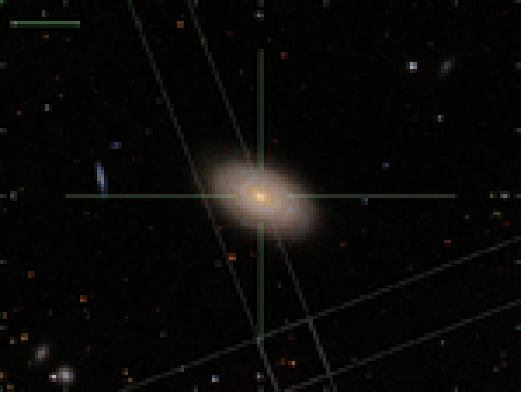}
\hspace*{-0.8cm}
\includegraphics[width=6.1cm,angle=270]{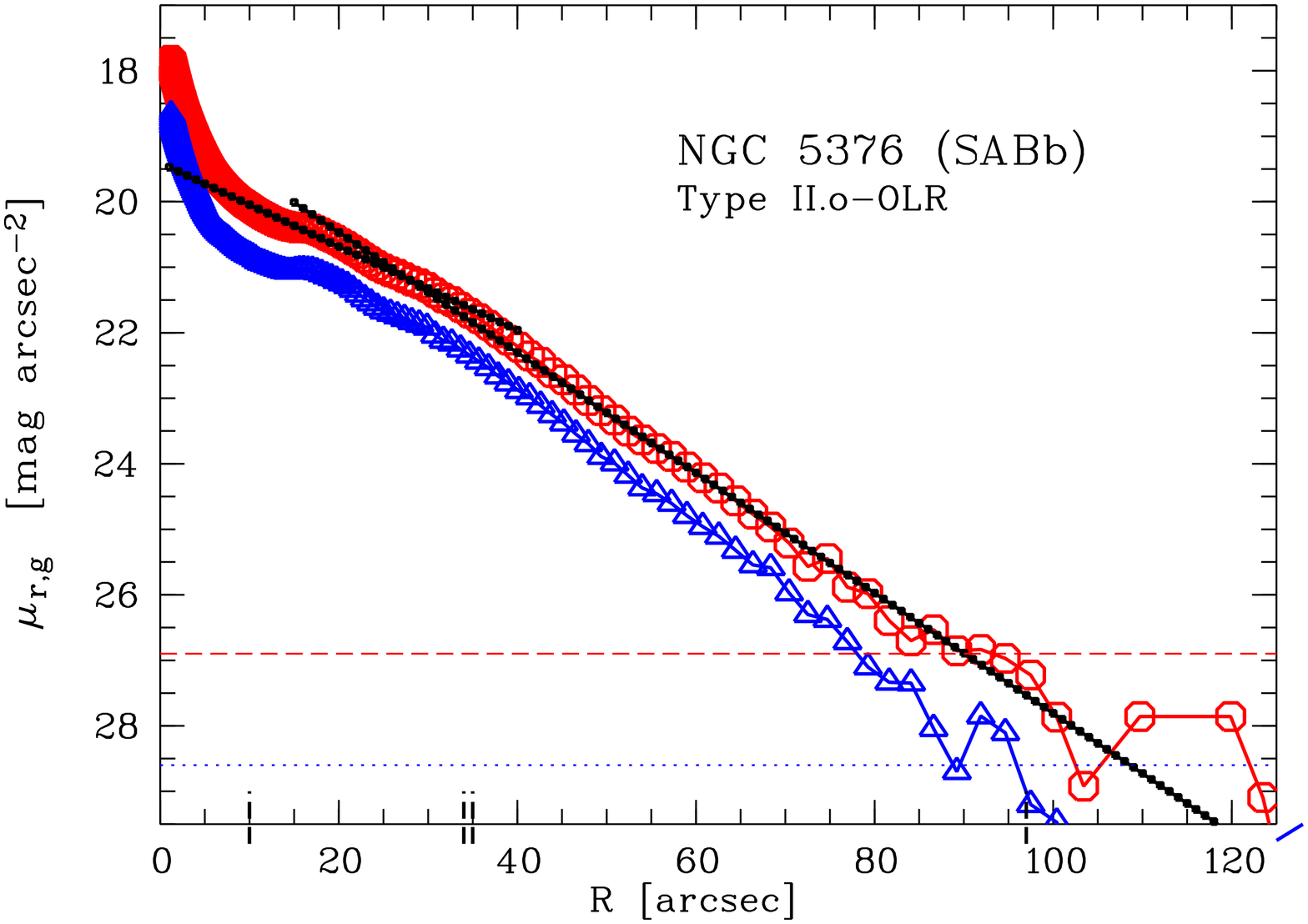}
}
\end{minipage}

\newpage
\onecolumn
\begin{minipage}[t][\textheight][t]{\textwidth}
\parbox[t][0.5\height][t]{0.47\textwidth}
{
\noindent {\bf NGC\,5430     :}        \typeolr     \\ 
\texttt{J140045.8+591943 .SBS3.  3.1 -20.86   2.2 3238}\\[0.25cm]
Galaxy is isolated according to \cite{prada2003}. The center is
dominated by a strong bar with dustlanes of size 
$R\sim\!35$\arcsec, followed by two wrapped spiral arms. 
The very center is slightly elongated, looking twofold (effect of 
dust lane?), thus the centering is uncertain.   
The extended bump at $\sim\!45\arcsec$ in the final profile 
is associated with the wrapped spiral arms. We have included 
the bump and bar for fitting the inner scalelength, because it 
does not change the result much.  
It is in principle possible to fit the full profile with a single 
exponential (\typeoc), allowing for this extended bump. However, we 
chose to fit a break at $\sim\!70\arcsec$. This reduces the 
deviation from the model and since it is about twice the bar radius 
the galaxy is classified as \typeolrc.   

}
\hfill 
\parbox[t][0.5\height][t]{0.47\textwidth}
{
\includegraphics[width=5.7cm,angle=270,]{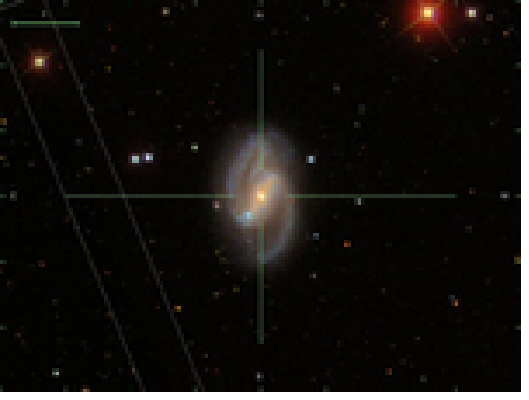}
\hspace*{-0.8cm}
\includegraphics[width=6.1cm,angle=270]{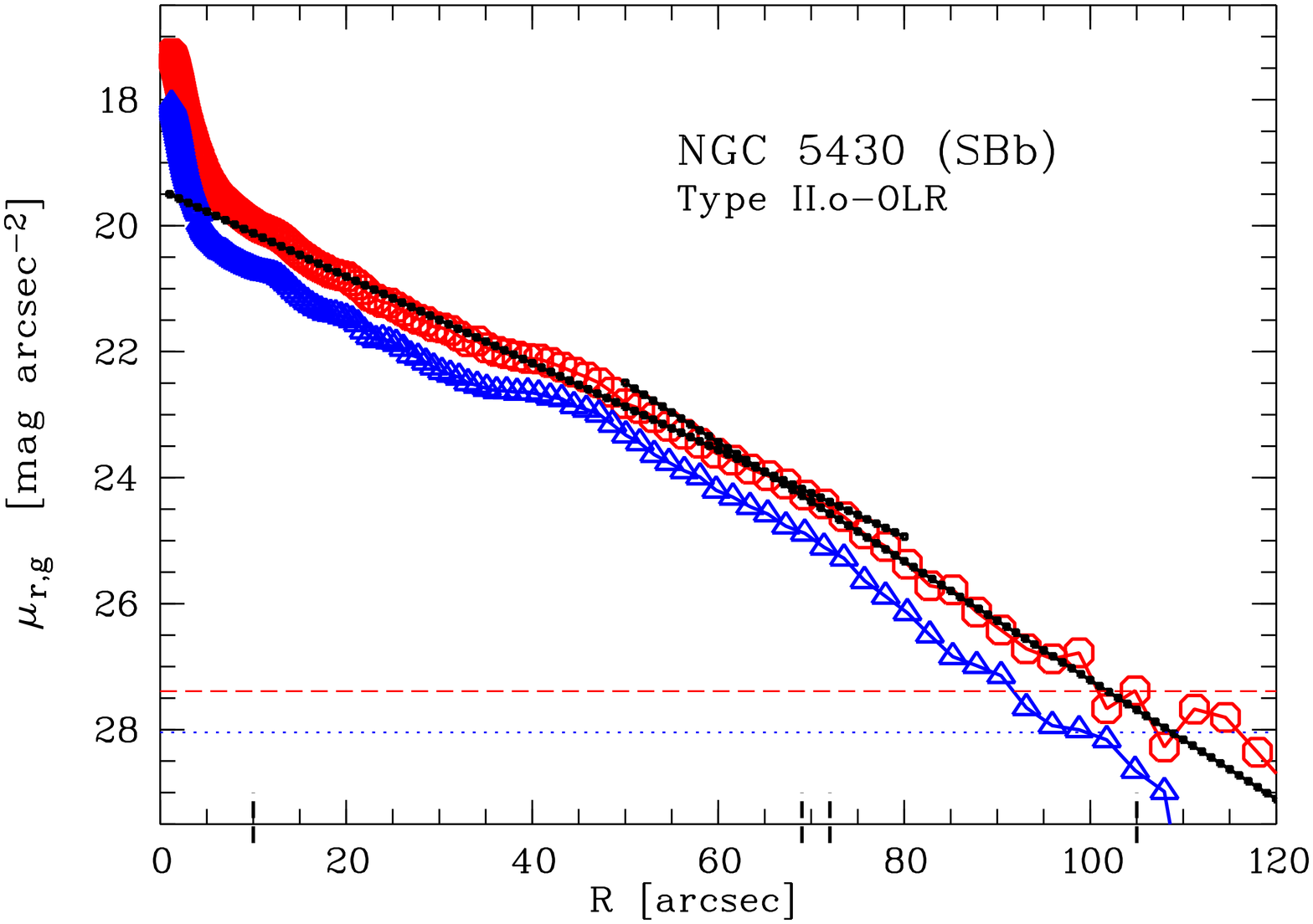}
}
\vfill
\parbox[b][0.5\height][t]{0.47\textwidth}
{
\noindent {\bf NGC\,5480     :}        \typeiiid                 \\        
\texttt{J140621.5+504332 .SAS5*  5.0 -19.72   1.7 2119}\\[0.25cm]
A large scale gradient (from top to bottom) in the background of 
the $r^{\prime}$ band image is removed with a linear fit.
Galaxy forms a close pair with NGC\,5481 ($v=2143$\kms) only 
$3.1\arcmin$ away with the very outer isophotes starting to 
overlap.
Galaxy appears to consist of two regions. An inner disk with 
the bright center and the spiral arms of size $R\sim\!40$\arcsec
with a common ellipticity and PA, sitting in a more diffuse outer
disk having a different ellipticity and a nearly orthogonal PA. 
The systems looks like an extended bar in an underlying disk, 
similar to NGC\,1068 and the photometric inclination (ellipticity) 
and PA is difficult to determine and therefore uncertain.  
The final profile exhibits a break with an upbending profile at 
$R\sim\!75$\arcsec, thus classified as \typeiiic, which is consistent
with the profile shown by \cite{courteau1996} (\cf UGC\,09026).
Between the inner disk and the break there is indication for two 
symmetric, thick spiral arms emerging from the inner bar-like disk.

}
\hfill 
\parbox[b][0.5\height][b]{0.47\textwidth}
{
\includegraphics[width=5.7cm,angle=270]{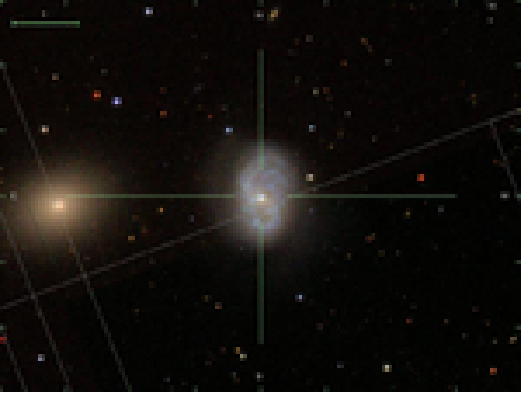}
\hspace*{-0.8cm}
\includegraphics[width=6.1cm,angle=270]{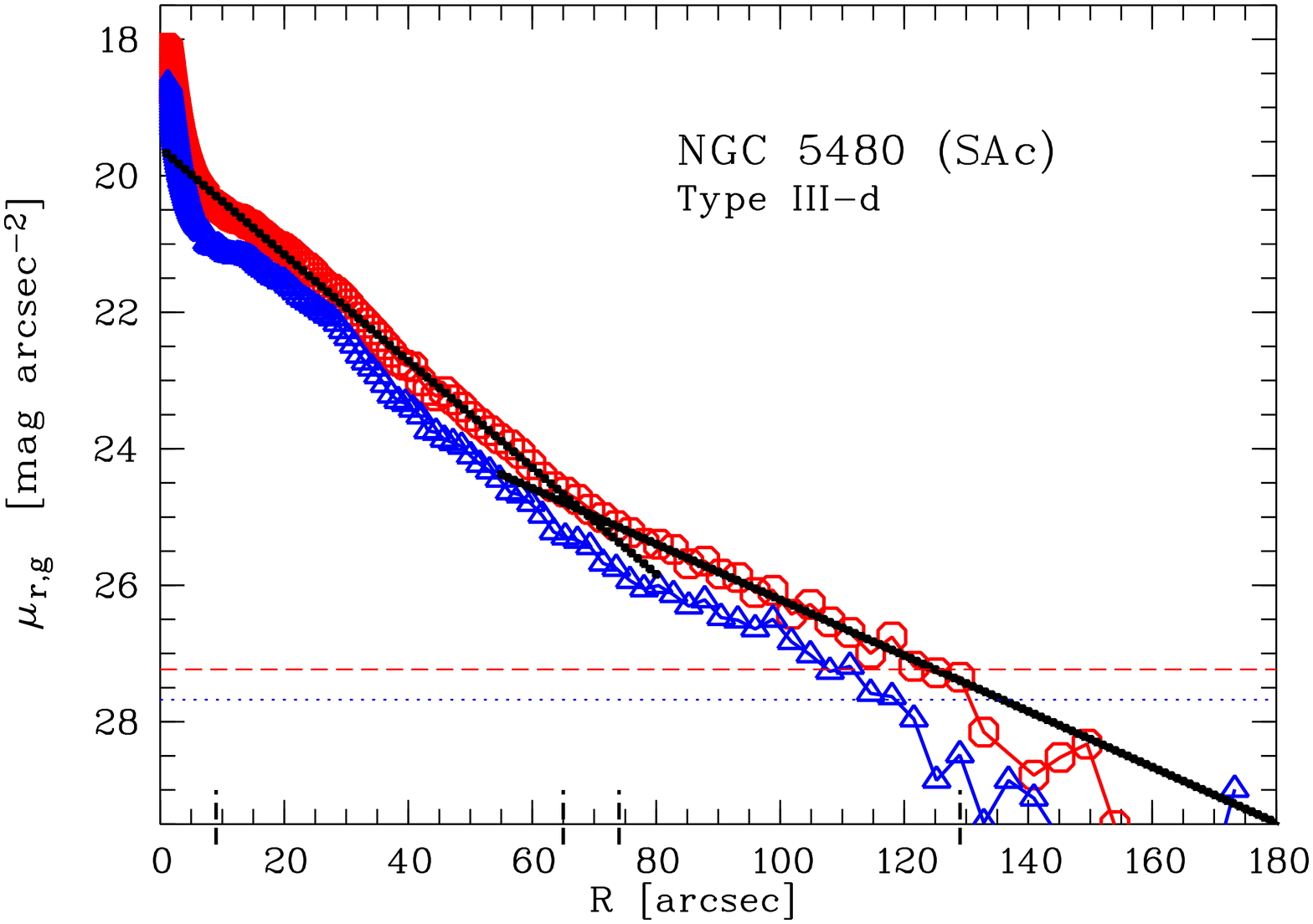}
}
\end{minipage}

\newpage
\onecolumn
\begin{minipage}[t][\textheight][t]{\textwidth}
\parbox[t][0.5\height][t]{0.47\textwidth}
{
\noindent {\bf NGC\,5584     :}        \typetoct                \\          
\texttt{J142223.8-002315 .SXT6.  6.1 -19.80   3.2 1702}\\[0.25cm]
Bright foreground star nearby but the influence is small. Two 
additional brighter stars in the outer disk force extended masking.  
The inner disk is dominated by two prominent spiral arms plus some 
more asymmetric ones with many starforming patches.  
The galaxy is classified as SAB, but the bar size ($R\sim\!10$\arcsec) 
is difficult to determine, since the spiral arms seem to continue 
towards the center.  
Similar to NGC\,3488 the profile shows a clear downbending, however, 
with uncertain break radius due to the extended break region which 
resembles in this case again a nearly straight line, so that one could 
also define two break radii at $\sim\!60\arcsec$ and $\sim\!110$\arcsec.
Since the bar is too small to be responsible for the inner break 
we use a single break at $\sim\!90\arcsec$ enclosing roughly the
inner spiral structure followed by a slightly asymmetric outer
disk (one side with a clearly more diffuse appearance).  

}
\hfill 
\parbox[t][0.5\height][t]{0.47\textwidth}
{
\includegraphics[width=5.7cm,angle=270,]{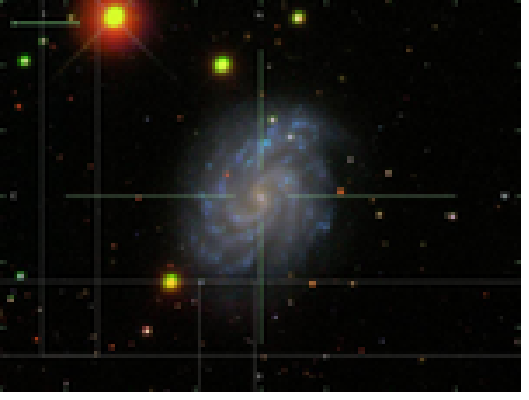}
\hspace*{-0.8cm}
\includegraphics[width=6.1cm,angle=270]{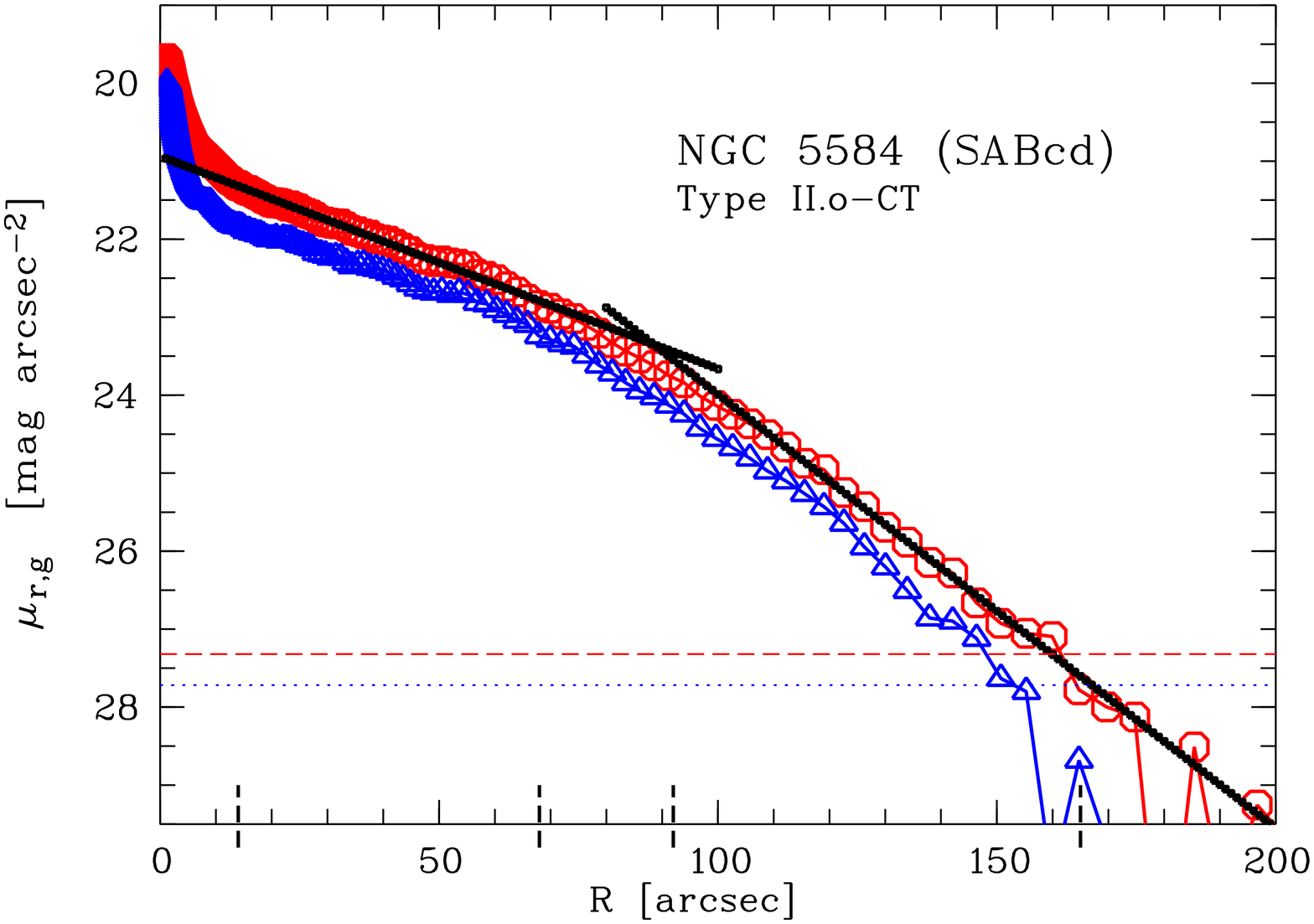}
}
\vfill
\parbox[b][0.5\height][t]{0.47\textwidth}
{
\noindent {\bf NGC\,5624     :}        \typeiii                 \\        
\texttt{J142635.3+513503 .S?...  5.0 -18.69   1.1 2186}\\[0.25cm]
Large scale gradient (from top to bottom) in the background of 
the $r^{\prime}$ band image is removed with a linear fit.
Very small galaxy which is most probably of noticeably later 
type than classified (Sc). 
Inner region dominated by an offcenterd (compared to the outer 
isophote used for centering) narrow bar, or rim-like structure of 
size $\sim\!10\arcsec$ which is responsible for the central dip 
in the profile. 
The PA of the outermost region is twisted compared to the inner, 
bar-like structure. 
The final profile exhibits a straight line starting from 
$\sim\!20\arcsec$ (enclosing the central region including the 
bar having the same PA) down to a break at $\sim\!45$ followed 
by an upbending profile (classified as \typeiiic) into the more 
symmetric, diffuse outer disk without visible substructure. 

}
\hfill 
\parbox[b][0.5\height][b]{0.47\textwidth}
{
\includegraphics[width=5.7cm,angle=270]{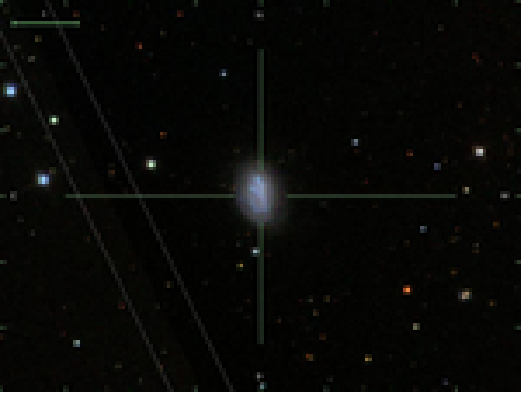}
\hspace*{-0.8cm}
\includegraphics[width=6.1cm,angle=270]{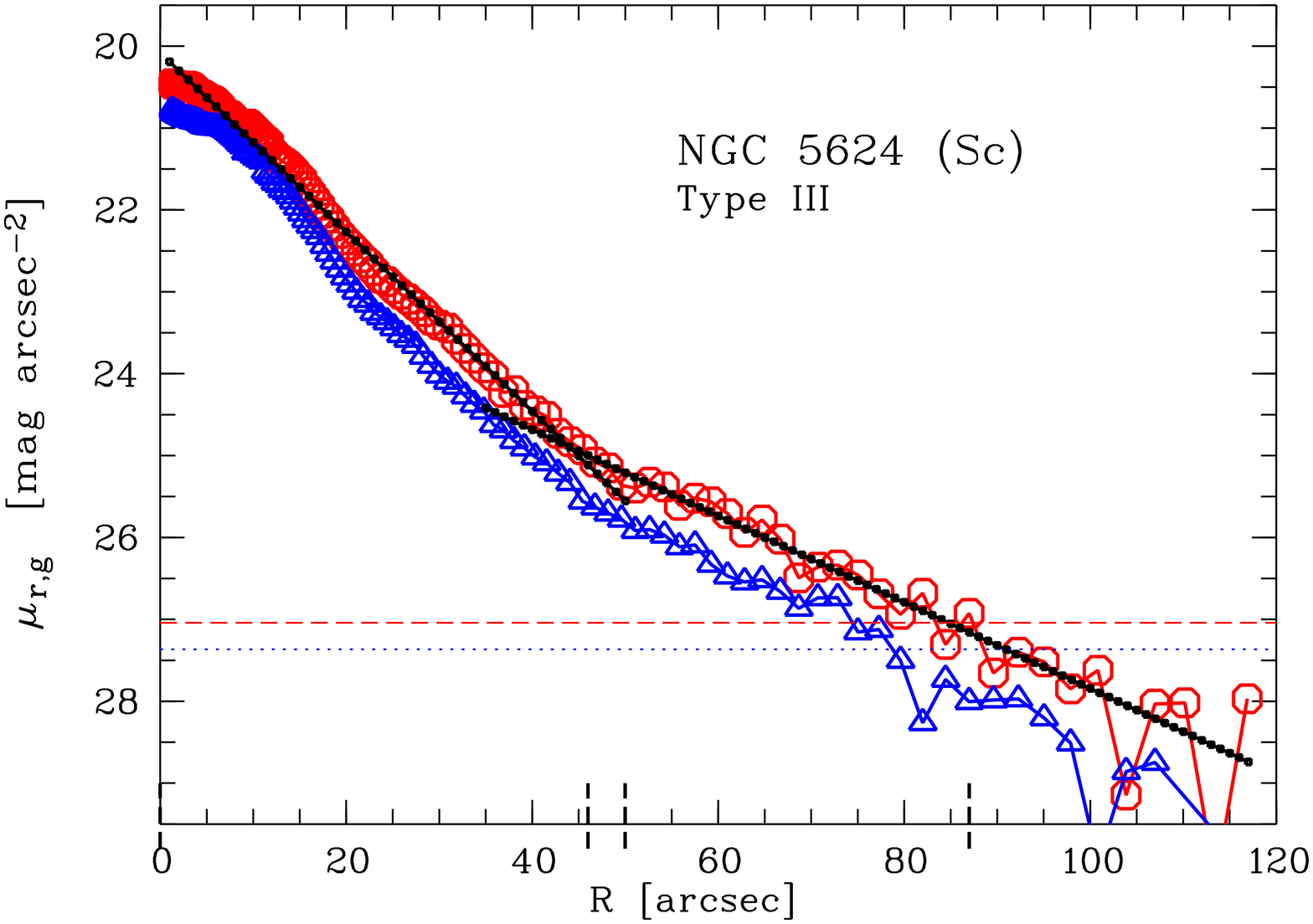}
}
\end{minipage}

\newpage
\onecolumn
\begin{minipage}[t][\textheight][t]{\textwidth}
\parbox[t][0.5\height][t]{0.47\textwidth}
{
\noindent {\bf NGC\,5660     :}        \typect                \\          
\texttt{J142949.8+493722 .SXT5.  5.1 -20.62   2.8 2585}\\[0.25cm]
Galaxy close to the edge of the SDSS field but almost complete 
sitting in a strip with higher background.  
According to \cite{sandage1994} it has a small companion nearby 
("a likely dwarf Im companion of unknown redshift"), however this 
could be associated with UGC\,09325 ($v=3730$\kms) being clearly
a background galaxy. 
Galaxy shows two, slightly asymmetric inner spiral arms plus 
an additional one extending towards the outer disk. Although 
classified as SAB there is no bar-like structure visible except 
for a possible ring-like structure of size $R\sim\!5$\arcsec.  
The final profile is not well described with a single exponential 
(\typeoc), already excluding the region beyond $R\sim\!90\arcsec$ 
which is most probably affected by an incorrect sky subtraction. 
Unaffected of this it is better described as having a break 
at $\sim\!70$\arcsec, corresponding to the end of the single extended 
spiral arm, with a following downbending profile, which is too far 
out to be related to the possible bar and therefore classified 
as \typectc.     
The bump at $\sim\!40\arcsec$ is related to the end of the inner 
spiral arms and beginning of the outer extended one. 
NIR imaging of the underlying old disk population should give 
clarification to the classification. 

}
\hfill 
\parbox[t][0.5\height][t]{0.47\textwidth}
{
\includegraphics[width=5.7cm,angle=270,]{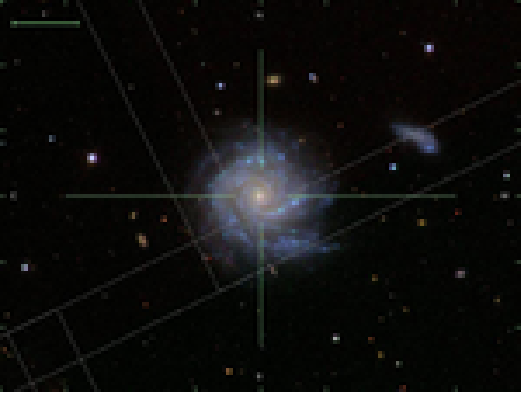}
\hspace*{-0.8cm}
\includegraphics[width=6.1cm,angle=270]{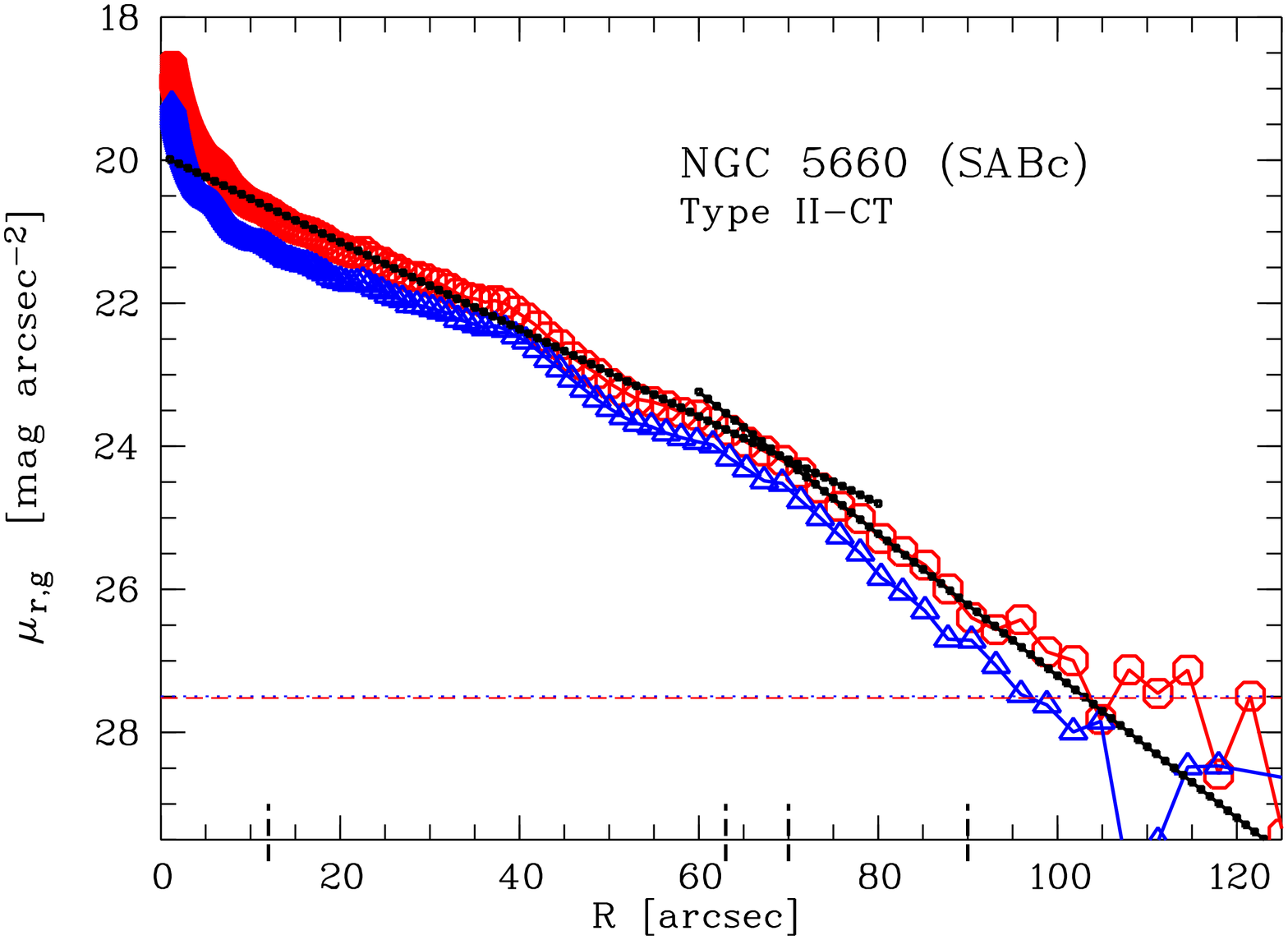}
}
\vfill
\parbox[b][0.5\height][t]{0.47\textwidth}
{
\noindent {\bf NGC\,5667     :}        \typetiiii      \\
\texttt{J143022.9+592811 .S..6*  5.8 -19.69   1.7 2222}\\[0.25cm]
Galaxy small and close to the edge of the SDSS field but almost 
complete with an untypical appearance. The inner region 
($R \ltsim 30$\arcsec) looks like an extended bar with dustlanes 
and a possible additional, tilted, secondary inner bar of size 
$R \ltsim 7$\arcsec. Outside this region there are two asymmetric 
narrow spiral arms visible towards north-west followed by an 
outer, slightly asymmetric, envelope, which makes the photometric 
inclination (ellipticity) and PA uncertain. 
The final profile shows a downbending break at $\sim\!25$\arcsec,
related to (and inside) the extended bar structure, therefore 
classified as \typetic, followed by an exponential down to another 
break at $\sim\!65\arcsec$ where an additional upbending starts
adding the \typeiii classification. 
The outer profile is uncertain due to a higher error in the sky.
A still justifiable sky value could create a continuous outer slope, 
but this is unlikely, since it will not create the asymmetric shape 
of the outer isophotes. 

}
\hfill 
\parbox[b][0.5\height][b]{0.47\textwidth}
{
\includegraphics[width=5.7cm,angle=270]{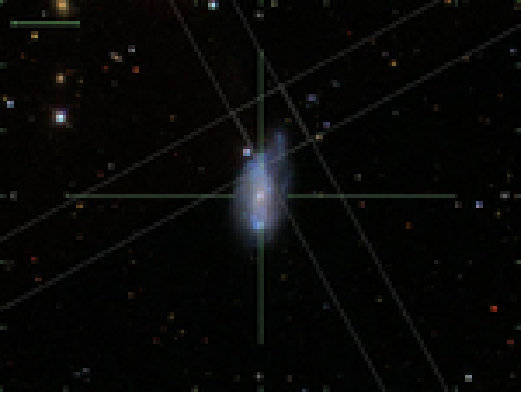}
\hspace*{-0.8cm}
\includegraphics[width=6.1cm,angle=270]{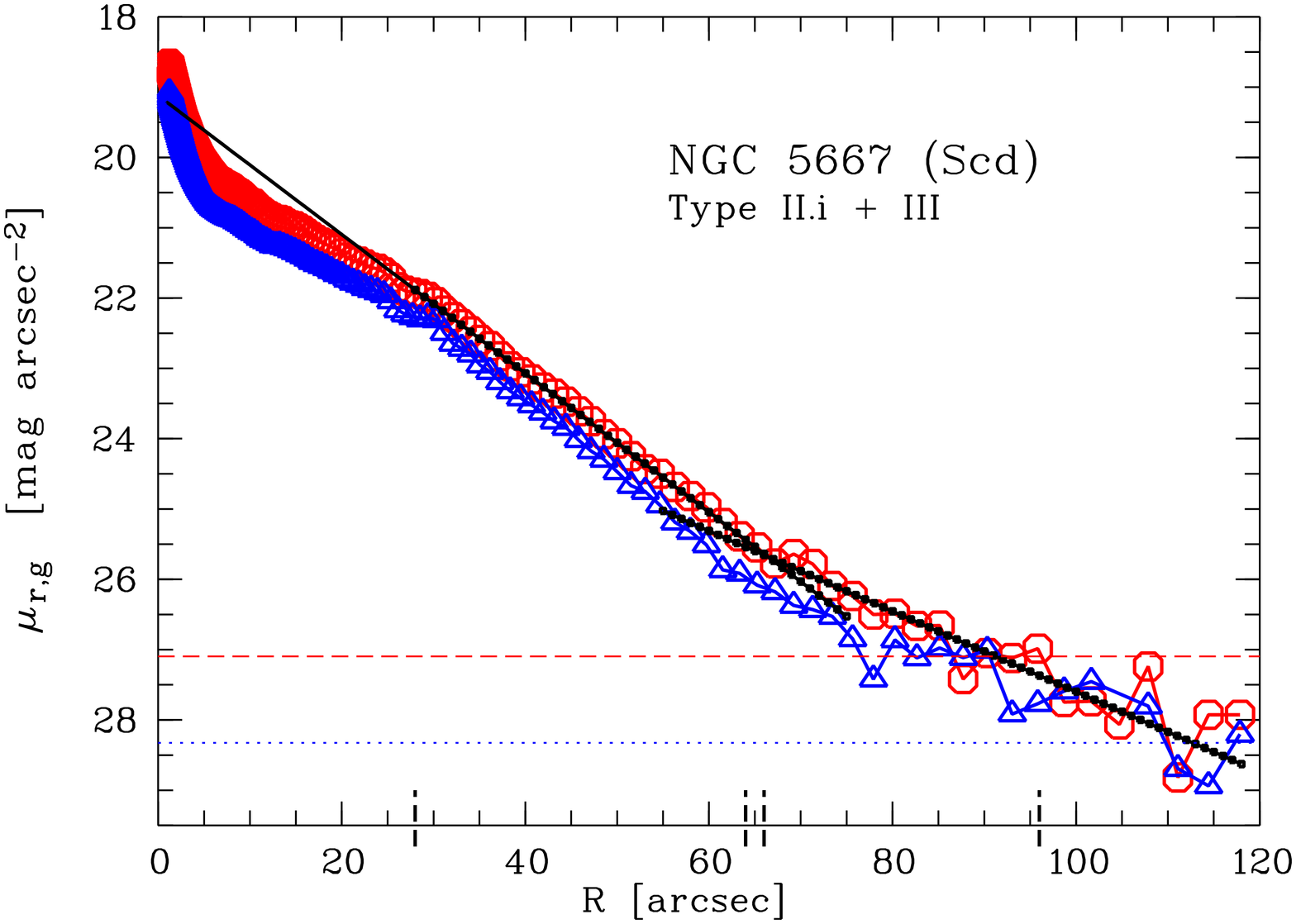}
}
\end{minipage}

\newpage
\onecolumn
\begin{minipage}[t][\textheight][t]{\textwidth}
\parbox[t][0.5\height][t]{0.47\textwidth}
{
\noindent {\bf NGC\,5668     :}        \typeo                 \\      
\texttt{J143324.4+042702 .SAS7.  6.7 -19.65   2.8 1672}\\[0.25cm]
Galaxy close to the edge of the SDSS field but almost complete. 
Although classified as SA there is a weak bar or oval inner 
structure of size $R\sim\!12\arcsec$ visible on the image, reflected 
by a small shoulder in the final profile.  
The outer disk (beyond $R\sim\!100$\arcsec) is slightly asymmetric 
and more extended towards the North. 
The bump at $\sim\!70\arcsec$ in the final profile is related 
to brighter spiral arms in this region. 
There is a possible upbending break visible at $\sim\!100$\arcsec
(associated to the asymmetric outer disk) or $\sim\!140\arcsec$ 
but the scalelength contrast is too low and it is too close to 
the sky error to argue for a \typeiii classification. 

}
\hfill 
\parbox[t][0.5\height][t]{0.47\textwidth}
{
\includegraphics[width=5.7cm,angle=270,]{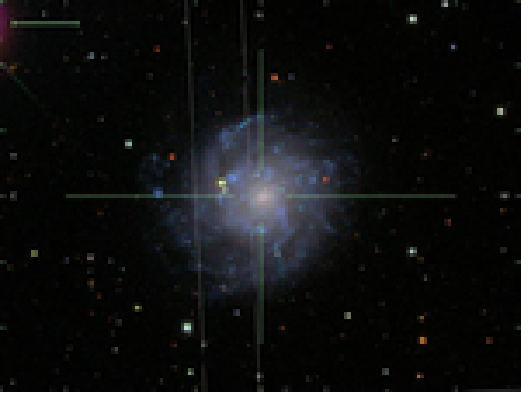}
\hspace*{-0.8cm}
\includegraphics[width=6.1cm,angle=270]{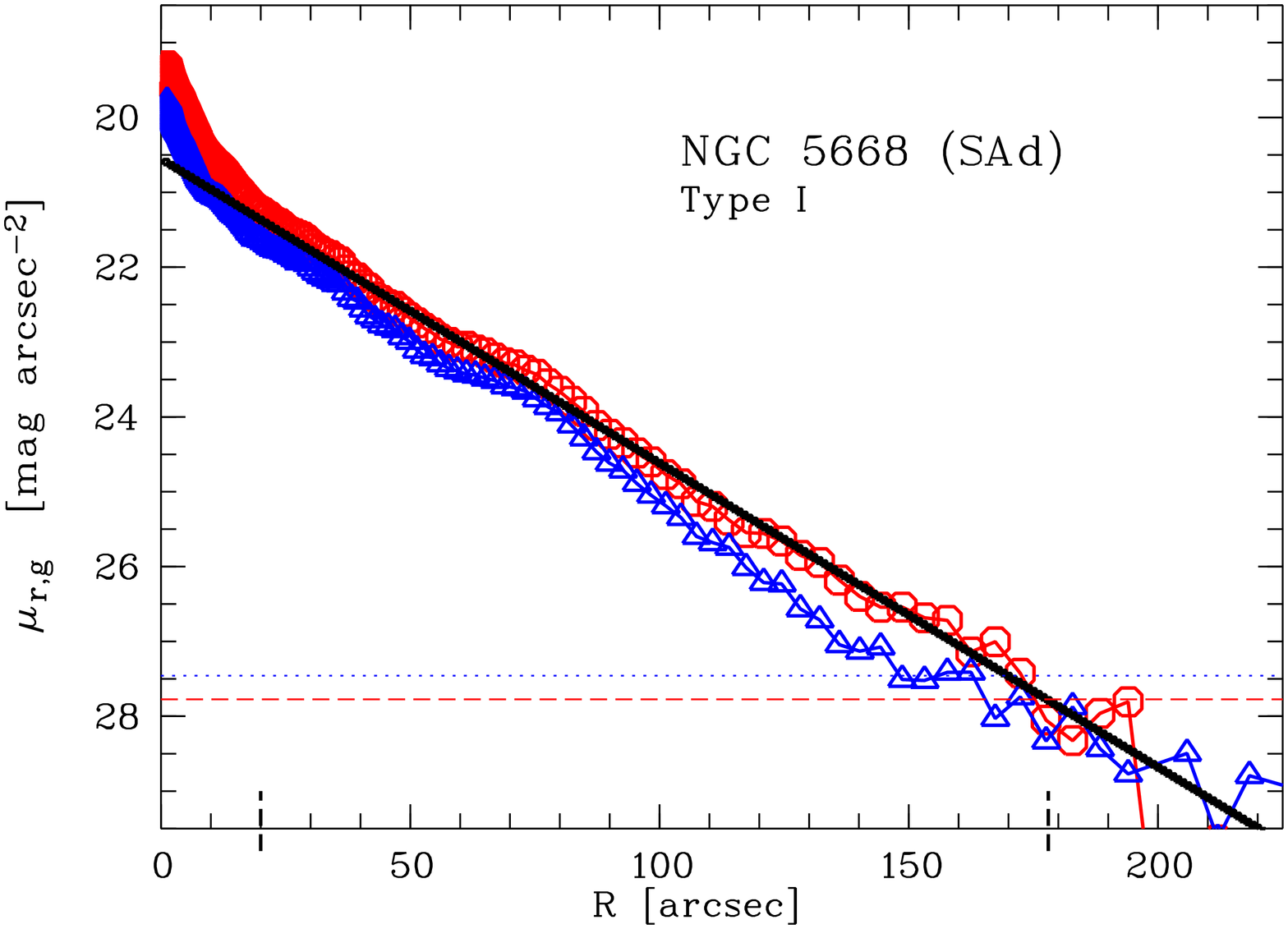}
}
\vfill
\parbox[b][0.5\height][t]{0.47\textwidth}
{
\noindent {\bf NGC\,5693     :}        \typeab                \\        
\texttt{J143611.2+483504 .SBT7.  6.8 -18.64   1.9 2537}\\[0.25cm]
Small galaxy with unusual appearance, showing an inner extended 
bar-like structure of size $R\sim\!15$\arcsec, apparently with two 
central peaks, probably caused by a superimposed foreground star. 
The brightest pixel in the bar structure does not correspond to 
the center obtained from the outer isophotes and is off by 
$\sim7\arcsec$ causing the inner dip in the final profile.   
The bar-like structure is followed by a single spiral arm extending 
towards the outer (quite roundish and symmetric) disk, which is used 
to obtain the ellipticity and PA.   
The final profile cannot be well fitted with a single exponential
(\typeoc) but rather shows a downbending behaviour albeit with an 
uncertain break radius due to the extended break region which 
resembles again a nearly straight line, so that one could also 
define two break radii at $\sim\!30\arcsec$ and $\sim\!50$\arcsec.
Although the first break is at twice the bar radius, we classify this
galaxy as \typeab since this whole region is clearly lopsided with 
an significantly offcentered bar.  

}
\hfill 
\parbox[b][0.5\height][b]{0.47\textwidth}
{
\includegraphics[width=5.7cm,angle=270]{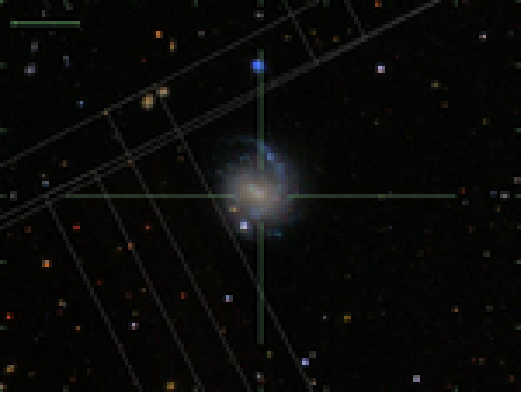}
\hspace*{-0.8cm}
\includegraphics[width=6.1cm,angle=270]{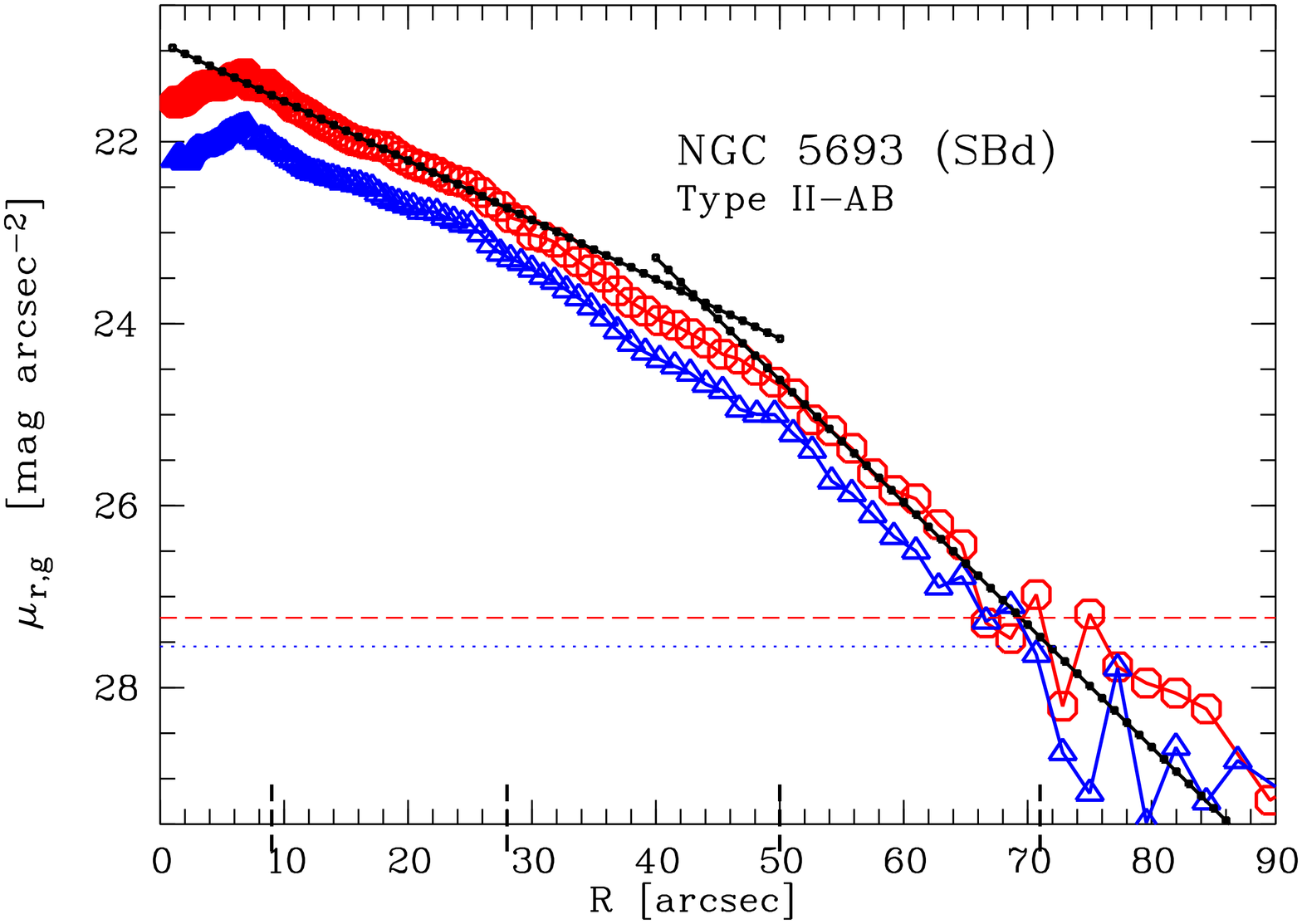}
}
\end{minipage}

\newpage
\onecolumn
\begin{minipage}[t][\textheight][t]{\textwidth}
\parbox[t][0.5\height][t]{0.47\textwidth}
{
\noindent {\bf NGC\,5713     :}        \typeiii                 \\        
\texttt{J144011.4-001725 .SXT4P  4.1 -20.50   2.8 1958}\\[0.25cm]
According to \cite{sandage1994} in a group of late-type spirals 
with four members in a region smaller than the Local Group.
No bright nucleus in the bar-like structure so the centering 
is done from the outer isophotes.  
In addition to a more chaotic inner region with spiral structure
(producing the bump between $R\sim\!10-35\arcsec$ in the final 
profile) there is clear substructure in the outer disk visible. 
Towards the South ($1.5\arcmin$ from the center) we see a very
diffuse additional patch of light (size $R\ltsim 20$\arcsec and 
masked), in addition to a more elongated feature ($\approx1\arcmin$ 
from the center, unmasked) towards north-east.
The profile shows a break at $\sim\!115\arcsec$ with an upbending 
profile associated with a rather symmetric featureless outer part.
The break is not caused by a possible sky error thus classified 
as \typeiiic.   

}
\hfill 
\parbox[t][0.5\height][t]{0.47\textwidth}
{
\includegraphics[width=5.7cm,angle=270,]{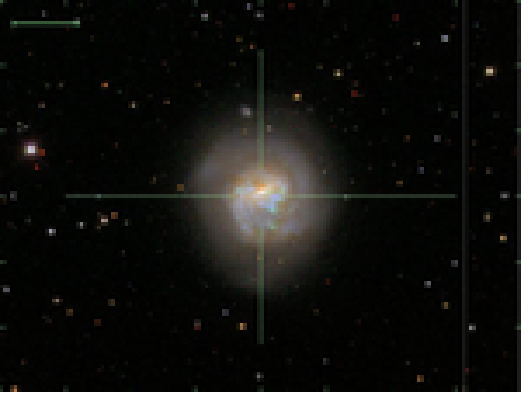}
\hspace*{-0.8cm}
\includegraphics[width=6.1cm,angle=270]{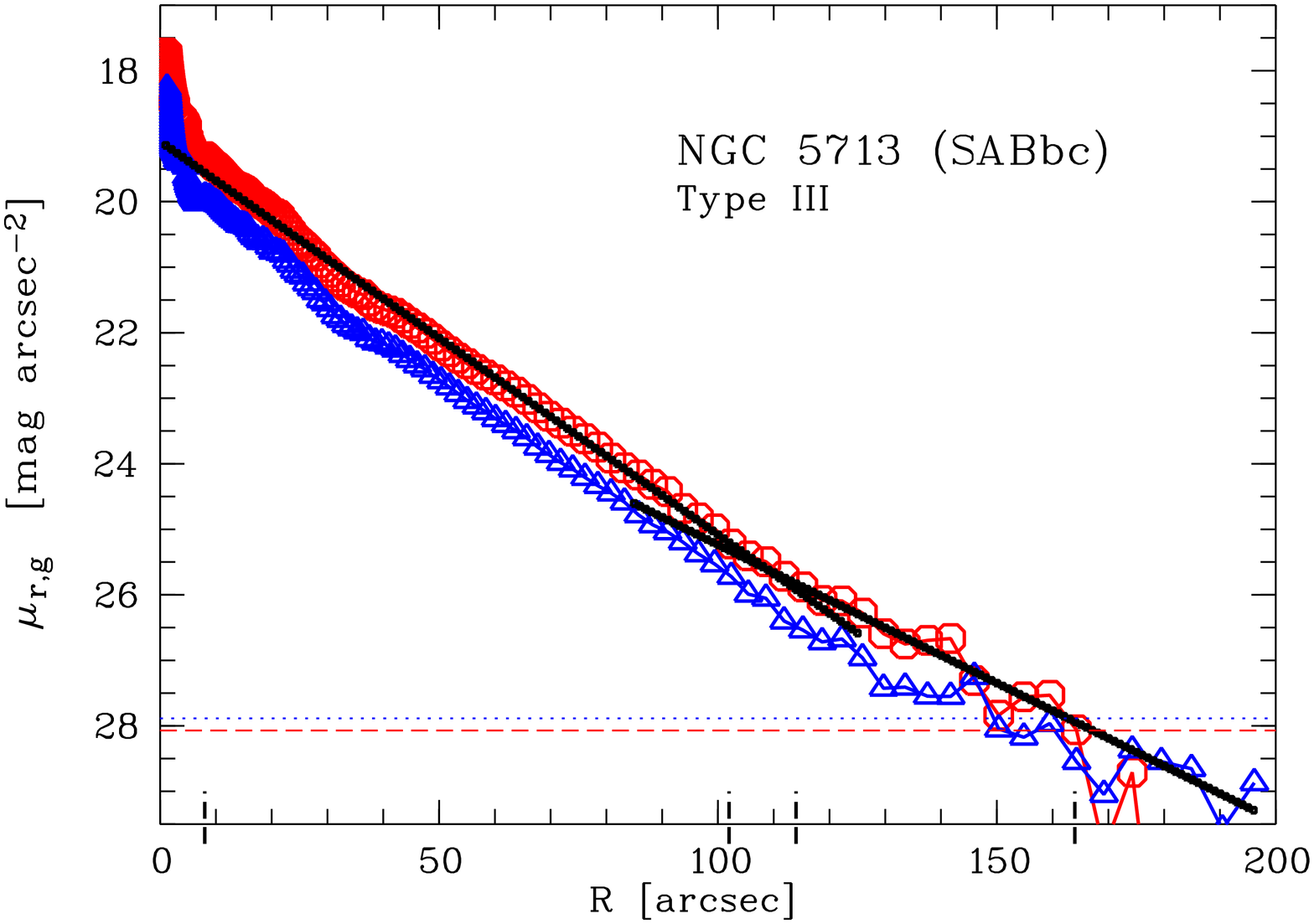}
}
\vfill
\parbox[b][0.5\height][t]{0.47\textwidth}
{
\noindent {\bf NGC\,5768     :}        \typeo                 \\          
\texttt{J145207.9-023147 .SAT5*  5.0 -19.24   1.6 2018}\\[0.25cm]
Bright star superimposed so using an extended mask. 
Photometric inclination (ellipticity) at $1\sigma$ in the free ellipse 
fit does not match the outer parts well, therefore we used the outermost 
ellipse which leaves the ellipticity and PA uncertain. 
Similar to NGC\,3888 the final profile exhibits an extended wiggle 
(with an unusual integral sign shape) at $\sim\!50\arcsec$ (without 
being a break) corresponding to a set of spiral arms in the outer 
disk which do not start at the center.     
The inner break at $\sim\!18\arcsec$ is associated with the inner 
spiral arms and the bar-like center. 
Although showing extended wiggles the galaxy is classified as 
\typeo since there is no better fitting alternative. 

}
\hfill 
\parbox[b][0.5\height][b]{0.47\textwidth}
{
\includegraphics[width=5.7cm,angle=270]{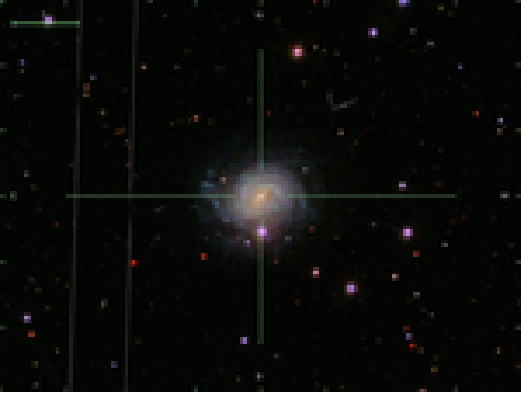}
\hspace*{-0.8cm}
\includegraphics[width=6.1cm,angle=270]{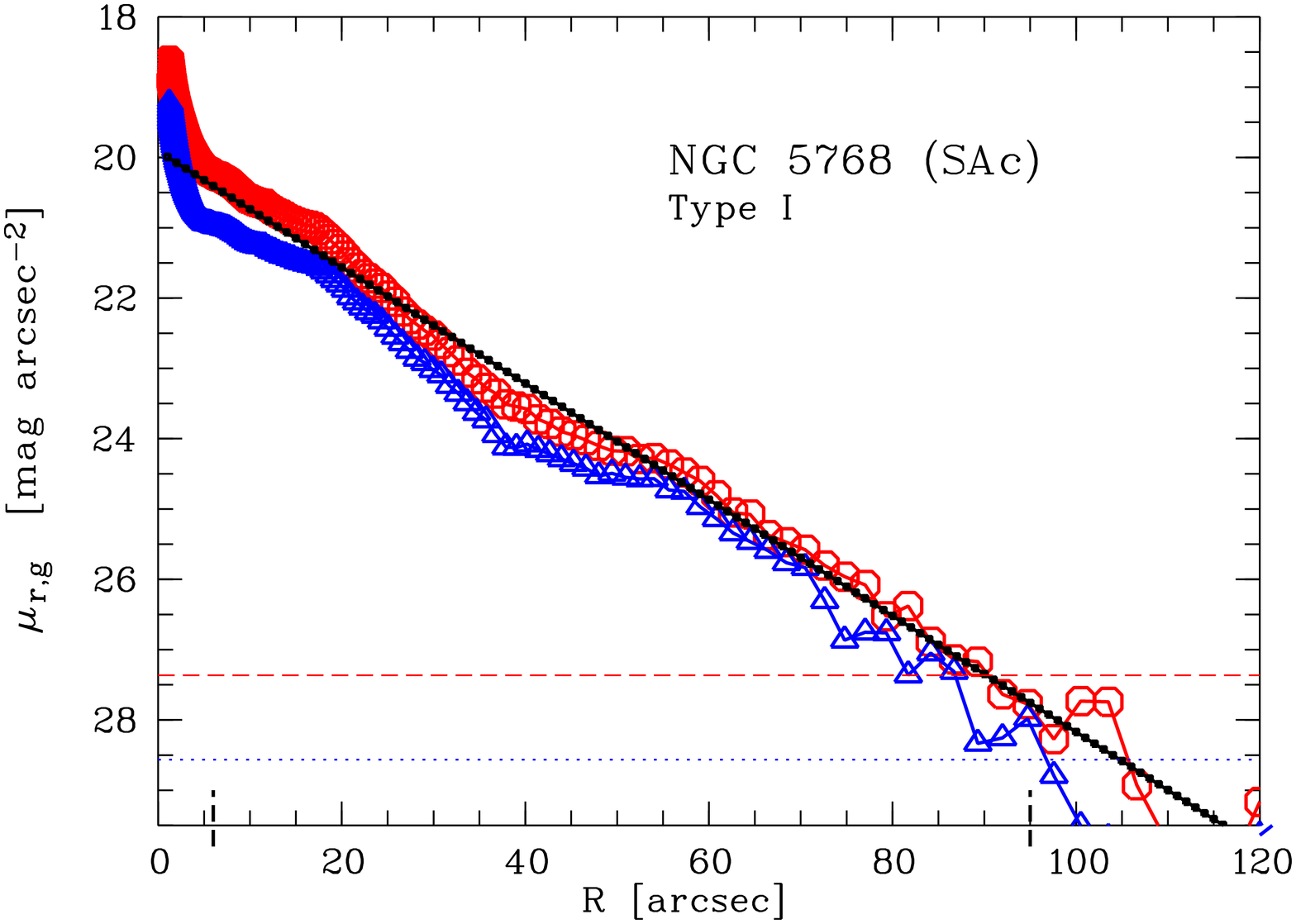}
}
\end{minipage}

\newpage
\onecolumn
\begin{minipage}[t][\textheight][t]{\textwidth}
\parbox[t][0.5\height][t]{0.47\textwidth}
{
\noindent {\bf NGC\,5774     :}        \typeolriii    \\
\texttt{J145342.6+033500 .SXT7.  6.9 -19.09   2.8 1655}\\[0.25cm]
Galaxy has a close physical, edge-on, companion (NGC\,5775, SBc?, 
$v=1681$\kms) and exhibits an outer, extended spiral arm pointing 
towards NGC\,5775 (similar to the M\,51/NGC\,5159 system but having 
a more similar sized mass) producing an asymmetric outer disk where 
the very outer isophotes already overlap.  
The center is obtained from outer ellipse fits (different from 
brightest pixel) since the inner elongated, narrow, bar-like 
region $R\sim\!10\arcsec$ is without an obvious nucleus. 
The shoulder in the final profile at $\gtsim 35\arcsec$ corresponds
to the end of the more extended bar-like structure enclosing the 
inner bar having a different PA.  
The break at $\sim\!80\arcsec$ with the downbending profile being 
at about twice the extended bar radius is therefore classified 
\typeolrc. 
The upbending profile (classified as \typeiiic) beyond a break at 
$\sim\!120\arcsec$ is due to the asymmetric outer disk with 
extended spiral arms. 
Fitting the profile with a single exponential (\typeoc) leaves 
three extended wiggles which are better explained by a 
\typeolriii break combination. 

}
\hfill 
\parbox[t][0.5\height][t]{0.47\textwidth}
{
\includegraphics[width=5.7cm,angle=270,]{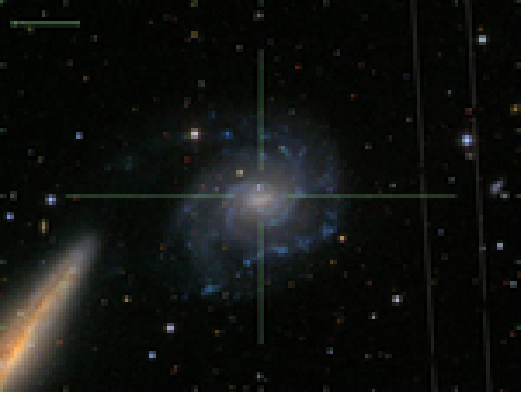}
\hspace*{-0.8cm}
\includegraphics[width=6.1cm,angle=270]{N5774_radn.ps}
}
\vfill
\parbox[b][0.5\height][t]{0.47\textwidth}
{
\noindent {\bf NGC\,5806     :}        \typeiii                 \\        
\texttt{J150000.4+015329 .SXS3.  3.3 -19.74   3.0 1440}\\[0.25cm]
Galaxy close to our high axis ratio limit (dust lane visible?) and 
close to the edge of the SDSS field but almost complete. 
From the inner disk there is an additional patch of diffuse light 
visible towards the south-east (not masked) extending into the 
asymmetric outer disk, which makes the photometric inclination 
(ellipticity) and PA uncertain since obtained further in. 
The final profile shows a shoulder at $\sim\!40\arcsec$ related to 
the inner bar-like center with the beginning of the spiral arms
and a break at $\sim\!120$\arcsec. The break is close to the 
position of the extra patch of light followed by an upbending 
profile (classified as \typeiiic) extending into the outer asymmetric 
disk without spiral structure. 
Our classification agrees well with \cite{erwin2006} and the 
\typeiii profile is also visible in \cite{courteau1996} 
(\cf UGC\,09645). 

}
\hfill 
\parbox[b][0.5\height][b]{0.47\textwidth}
{
\includegraphics[width=5.7cm,angle=270]{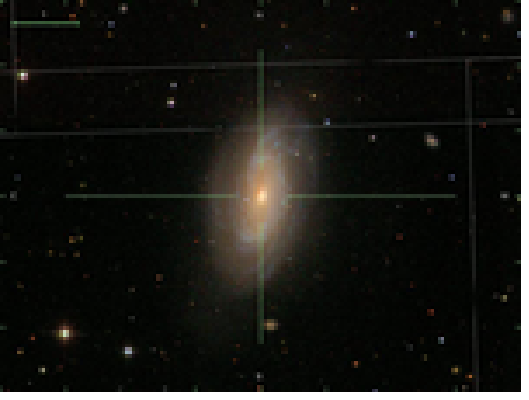}
\hspace*{-0.8cm}
\includegraphics[width=6.1cm,angle=270]{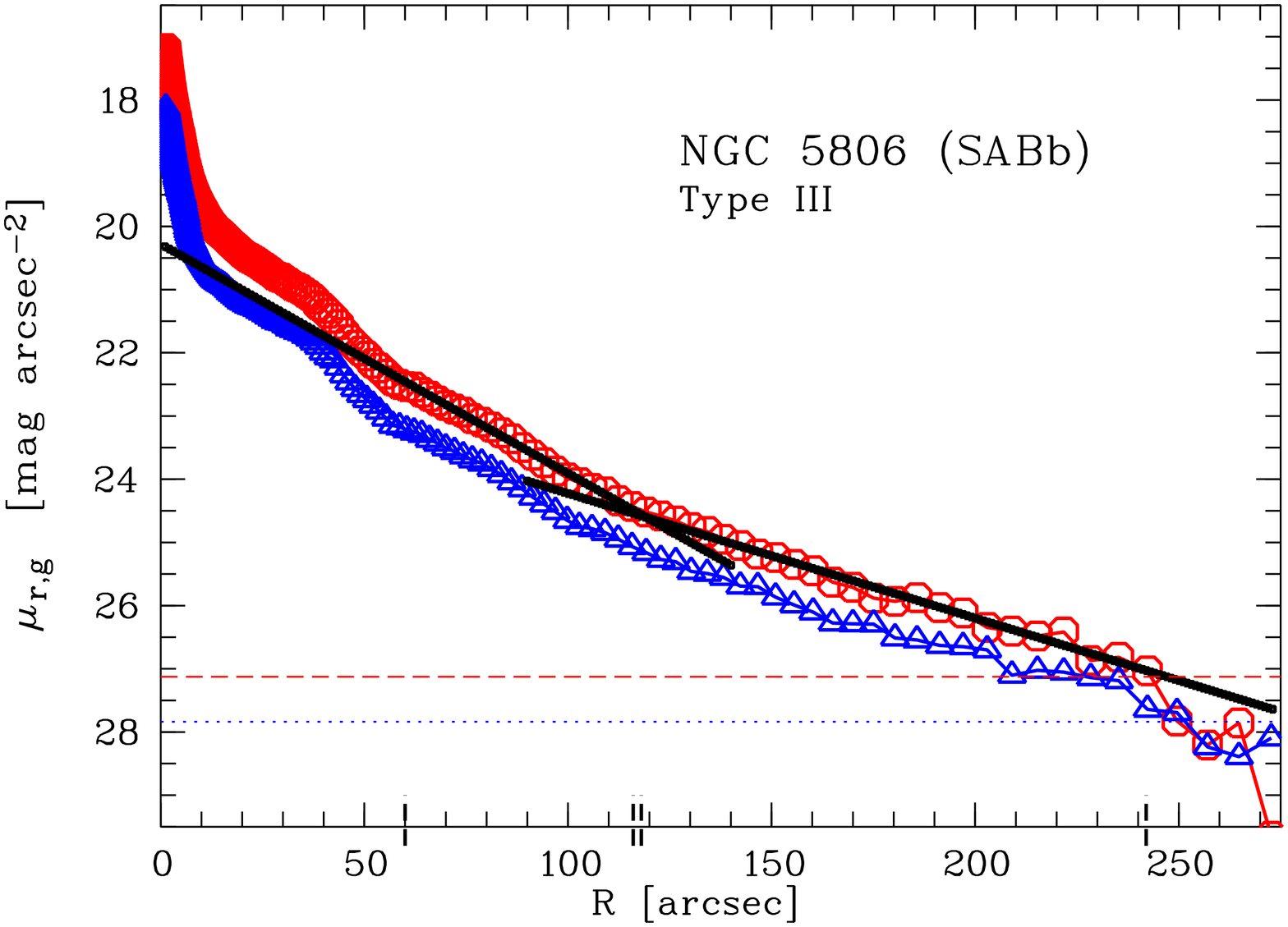}
}
\end{minipage}

\newpage
\onecolumn
\begin{minipage}[t][\textheight][t]{\textwidth}
\parbox[t][0.5\height][t]{0.47\textwidth}
{
\noindent {\bf NGC\,5850     :}        \typeolr               \\           
\texttt{J150707.7+013240 .SBR3.  3.0 -21.43   4.3 2637}\\[0.25cm]
Bright star nearby needs extended mask. The $r^{\prime}$ band background 
is slightly inhomogeneous. 
Galaxy is dominated by an extended bar of size $R\sim\!55\arcsec$ 
enclosed by an inner ring out to $R\sim\!75\arcsec$ having a 
secondary, inner bar of size $R\sim\!7\arcsec$ with clearly 
different ellipticity and PA. 
There is an additional, faint outer ring (or pseudo-ring, since 
looking similar to some wrapped spiral arms) visible on the image 
related to the break at $\sim\!130\arcsec$ in the final profile, 
thus obviously classified as \typeolrc. Since the outer ring is 
used to determine the photometric inclination (ellipticity) and 
PA they are rather uncertain. 

}
\hfill 
\parbox[t][0.5\height][t]{0.47\textwidth}
{
\includegraphics[width=5.7cm,angle=270,]{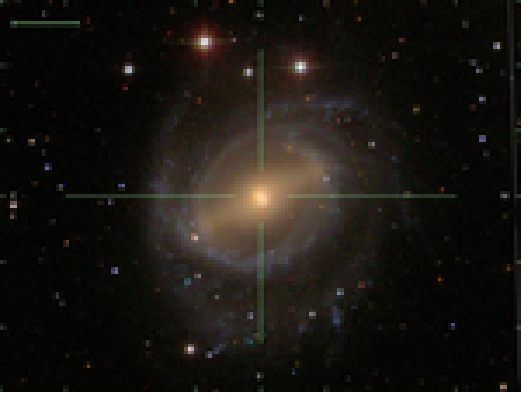}
\hspace*{-0.8cm}
\includegraphics[width=6.1cm,angle=270]{N5850_radn.ps}
}
\vfill
\parbox[b][0.5\height][t]{0.47\textwidth}
{
\noindent {\bf NGC\,5937     :}        \typeiii                 \\        
\texttt{J153046.1-024946 PSXT3P  3.2 -20.90   1.9 2870}\\[0.25cm]
Bright stars nearby need extended mask without really covering  
the outer disk. 
There is a possible, small dwarf companion (not in NED) visible 
about $1.5\arcmin$ away towards the north-west.
The outer disk is slightly asymmetric with continuously changing 
ellipticity and PA, so the values applied for the final 
profile are rather uncertain. 
The break at $\sim\!70\arcsec$ with an upbending profile (classified 
as \typeiiic) is most probably related to the changing ellipticity 
and should be taken with caution. 

}
\hfill 
\parbox[b][0.5\height][b]{0.47\textwidth}
{
\includegraphics[width=5.7cm,angle=270]{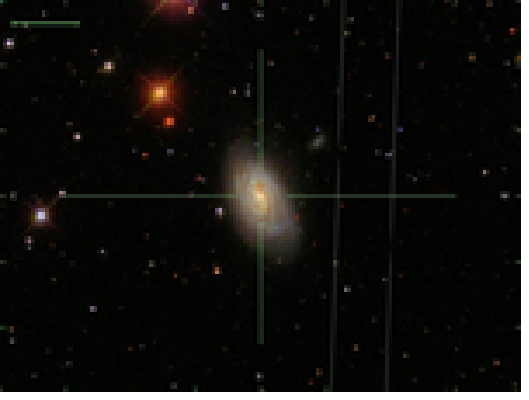}
\hspace*{-0.8cm}
\includegraphics[width=6.1cm,angle=270]{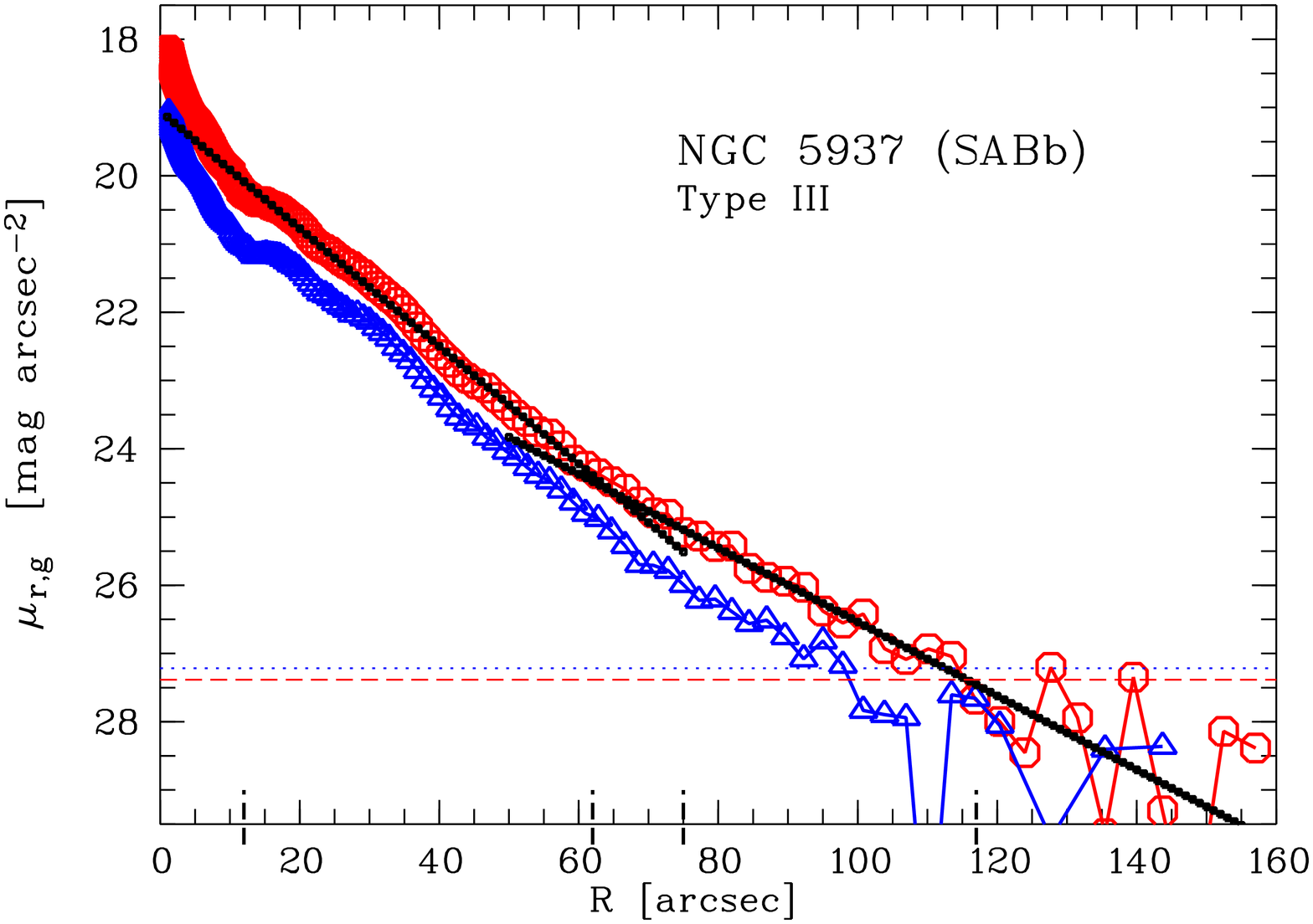}
}
\end{minipage}

\newpage
\onecolumn
\begin{minipage}[t][\textheight][t]{\textwidth}
\parbox[t][0.5\height][t]{0.47\textwidth}
{
\noindent {\bf NGC\,6070     :}        \typetoct                \\          
\texttt{J160958.9+004234 .SAS6.  6.0 -21.04   3.4 2085}\\[0.25cm]
Galaxy close to our high axis ratio limit having a very bright star 
(with an extended halo) in the FOV which is still small enough to 
avoid significant influence. Only partly fitted since \ltsim $1/5$ 
of galaxy is beyond SDSS field.
The outer isophotes are slightly asymmetric, with one sharp and one 
more fluffy side, so the ellipticity and PA are uncertain. 
The final profile exhibits a clear downbending with a more extended 
break region around $R\sim\!95\arcsec$ corresponding to the end of 
the spiral arm structure and classified as \typetoctc.  
Although the galaxy is classified as SA there is an elliptical central
bar-like structure of size $\ltsim 10\arcsec$ visible which would be 
however too small to argue for a \typeolr break. 
The small dip at $\sim\!30\arcsec$ is produced by spiral arm 
contrast. 

}
\hfill 
\parbox[t][0.5\height][t]{0.47\textwidth}
{
\includegraphics[width=5.7cm,angle=270,]{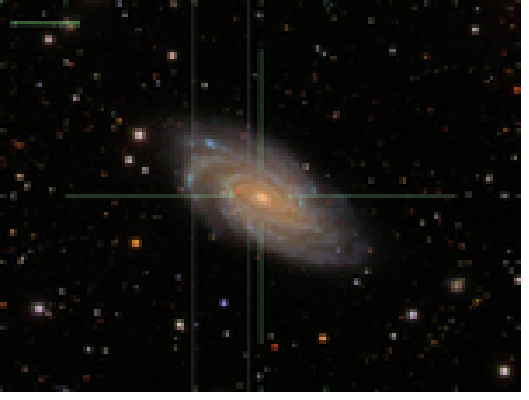}
\hspace*{-0.8cm}
\includegraphics[width=6.1cm,angle=270]{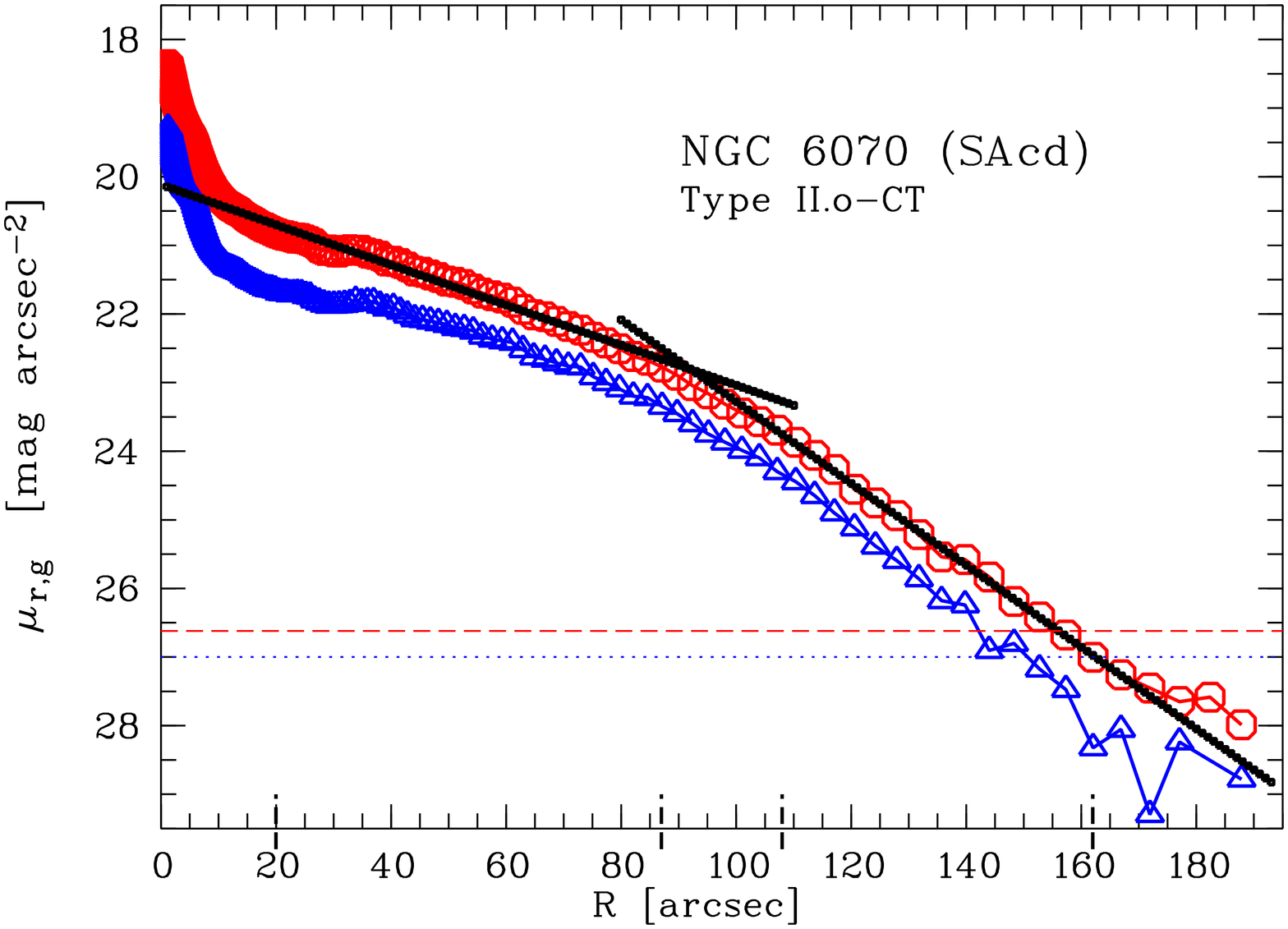}
}
\vfill
\parbox[b][0.5\height][t]{0.47\textwidth}
{
\noindent {\bf NGC\,6155     :}        \typeab                \\          
\texttt{J162608.3+482201 .S?...  7.8 -19.45   1.3 2684}\\[0.25cm]
Small galaxy with an extended mask reaching into the outer disk
to cover a very bright star close by.  
Lopsided disk with centering done from the outer isophotes. The 
center is $\sim\!3\arcsec$ off the brightest pixel associated to 
a small bar-like central structure of size $R\sim\!5\arcsec$ 
followed by a thin, star-forming, blue spiral arm feature 
towards north-east. 
The final profile shows a downbending break at $\sim\!35\arcsec$ 
corresponding to the inner lopsided disk thus classified as a 
\typeab break, although this is just at the radius of an outer 
ring (best visible in the colour image).

}
\hfill 
\parbox[b][0.5\height][b]{0.47\textwidth}
{
\includegraphics[width=5.7cm,angle=270]{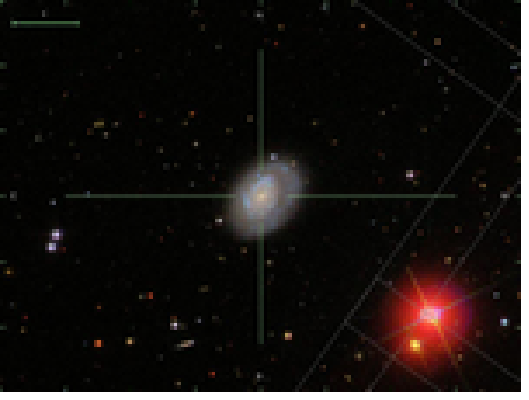}
\hspace*{-0.8cm}
\includegraphics[width=6.1cm,angle=270]{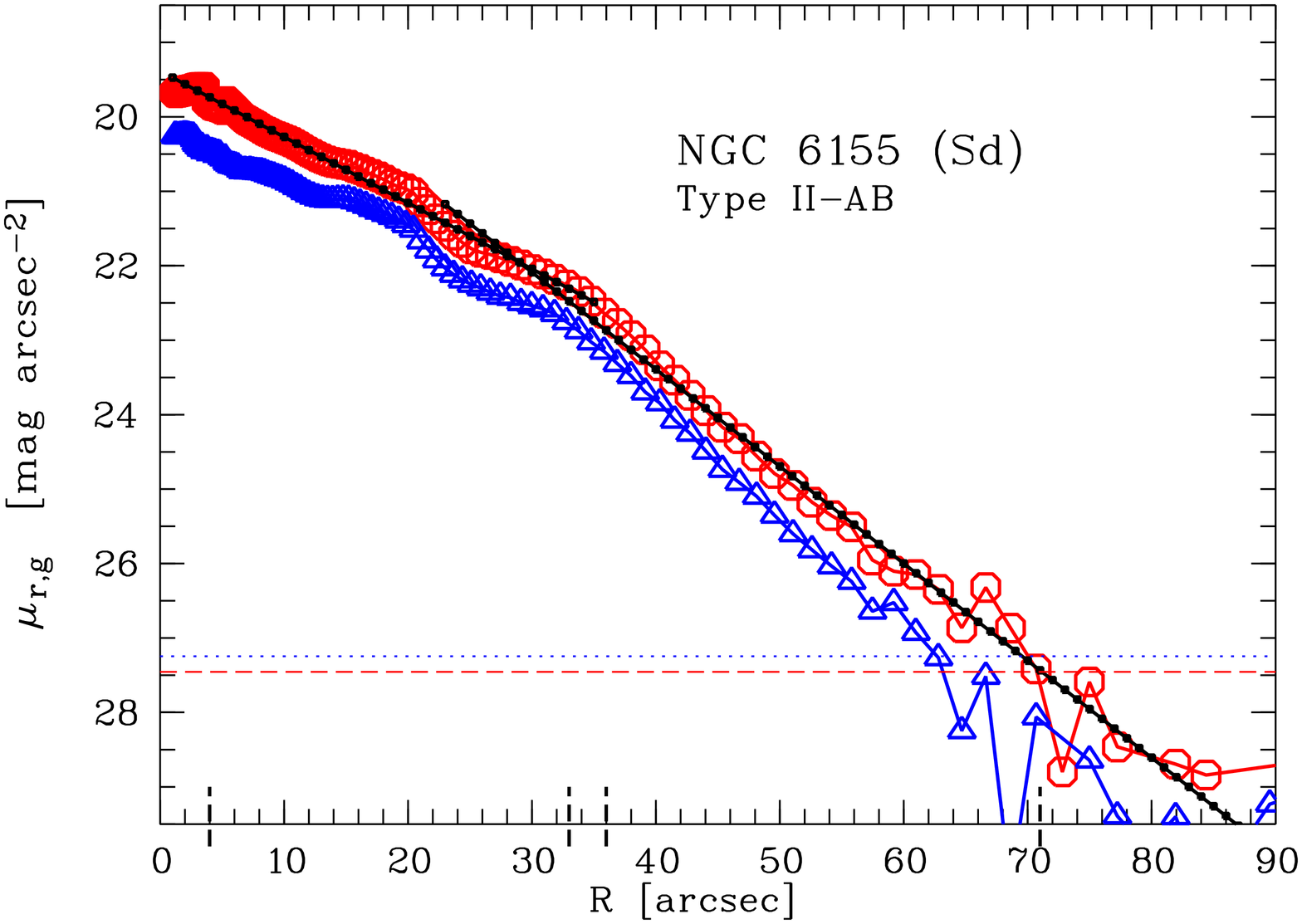}
}
\end{minipage}

\newpage
\onecolumn
\begin{minipage}[t][\textheight][t]{\textwidth}
\parbox[t][0.5\height][t]{0.47\textwidth}
{
\noindent {\bf NGC\,7437     :}        \typect                \\          
\texttt{J225810.0+141832 .SXT7.  7.1 -18.88   1.8 2190}\\[0.25cm]
Small galaxy close to the edge of the SDSS field but almost 
complete.
Similar to NGC\,5300 the final profile shows the prototypical 
\typetoct behaviour. A sharp break at $\sim\!40$\arcsec, corresponding roughly 
to the end of the apparent flocculent spiral arm structure, followed by 
a downbending profile. Although classified as SAB there is no obvious
structure visible which could be identified with a bar. Using the 
beginning spiral arms as an upper limit ($R\ltsim 8$\arcsec) the break 
is still too far out for a normal \typeolr break.

}
\hfill 
\parbox[t][0.5\height][t]{0.47\textwidth}
{
\includegraphics[width=5.7cm,angle=270,]{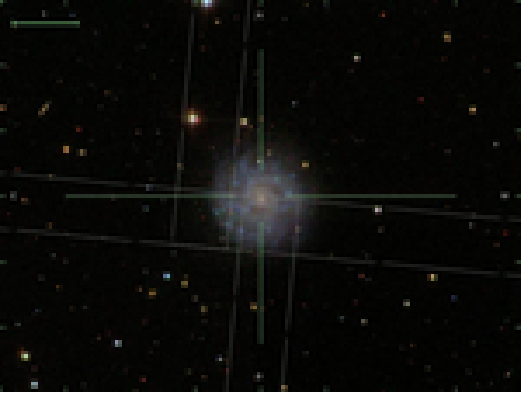}
\hspace*{-0.8cm}
\includegraphics[width=6.1cm,angle=270]{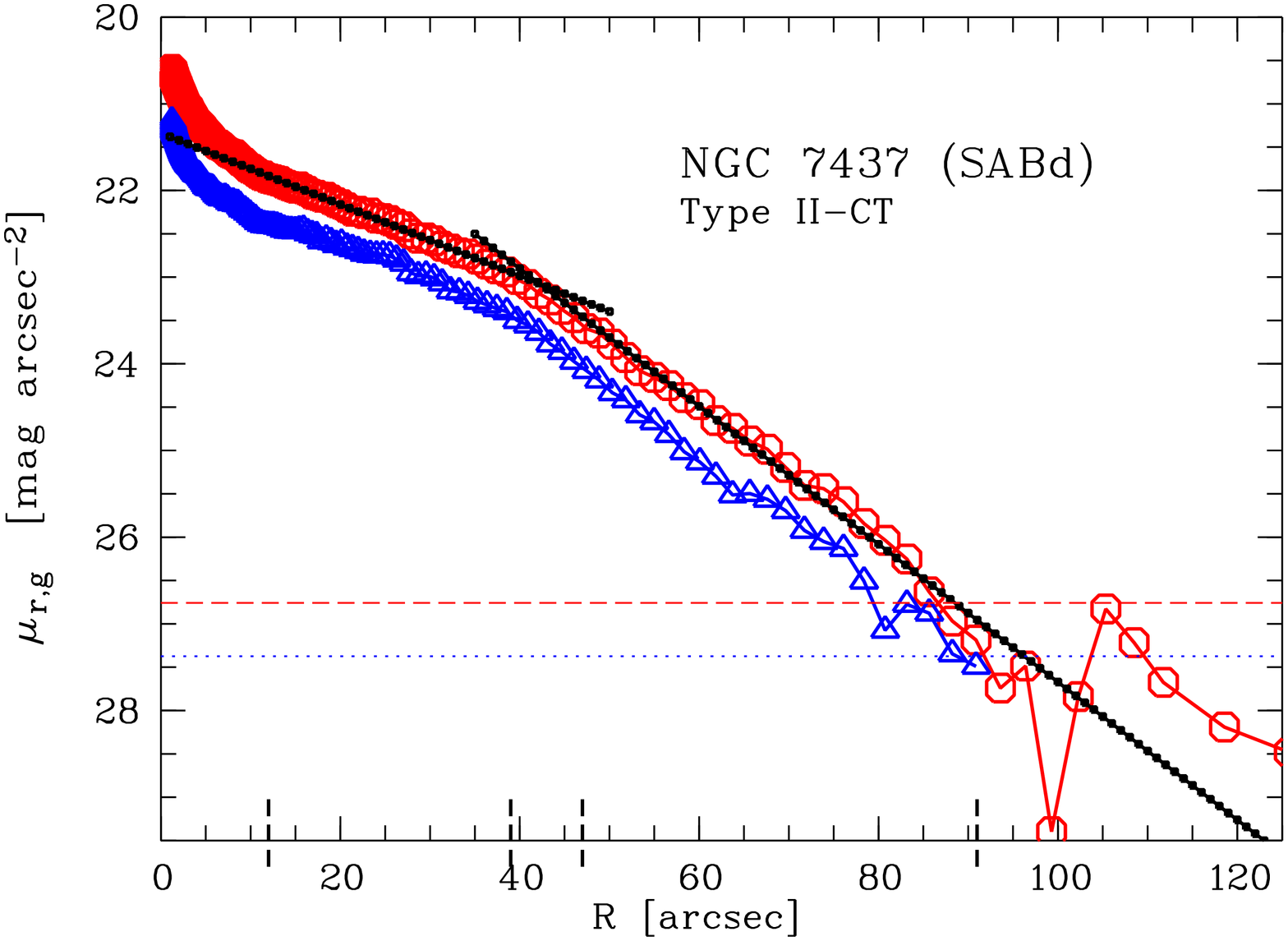}
}
\vfill
\parbox[b][0.5\height][t]{0.47\textwidth}
{
\noindent {\bf NGC\,7606     :}        \typect                \\          
\texttt{J231904.8-082906 .SAS3.  3.2 -21.38   4.3 2187}\\[0.25cm]
Galaxy with multiple thin spiral arms. 
The final profile exhibits clearly a break at $\sim\!110$\arcsec
corresponding roughly to the end of the visible spiral arm 
structure, following part of the ring enclosing an apparent 
hole (enhanced dust extinction?) in the disk towards the north-west.  
Since the galaxy is classified as SA and no bar structure visible 
the break is a \typectc.  
The apparent upbending of the profile beyond $\sim\!190$\arcsec
is not related to a sky error but due to a changing ellipticity in 
the very outer disk (maybe due to a warp?). 

}
\hfill 
\parbox[b][0.5\height][b]{0.47\textwidth}
{
\includegraphics[width=5.7cm,angle=270]{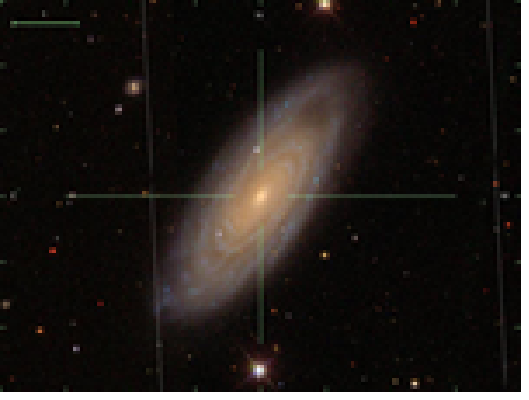}
\hspace*{-0.8cm}
\includegraphics[width=6.1cm,angle=270]{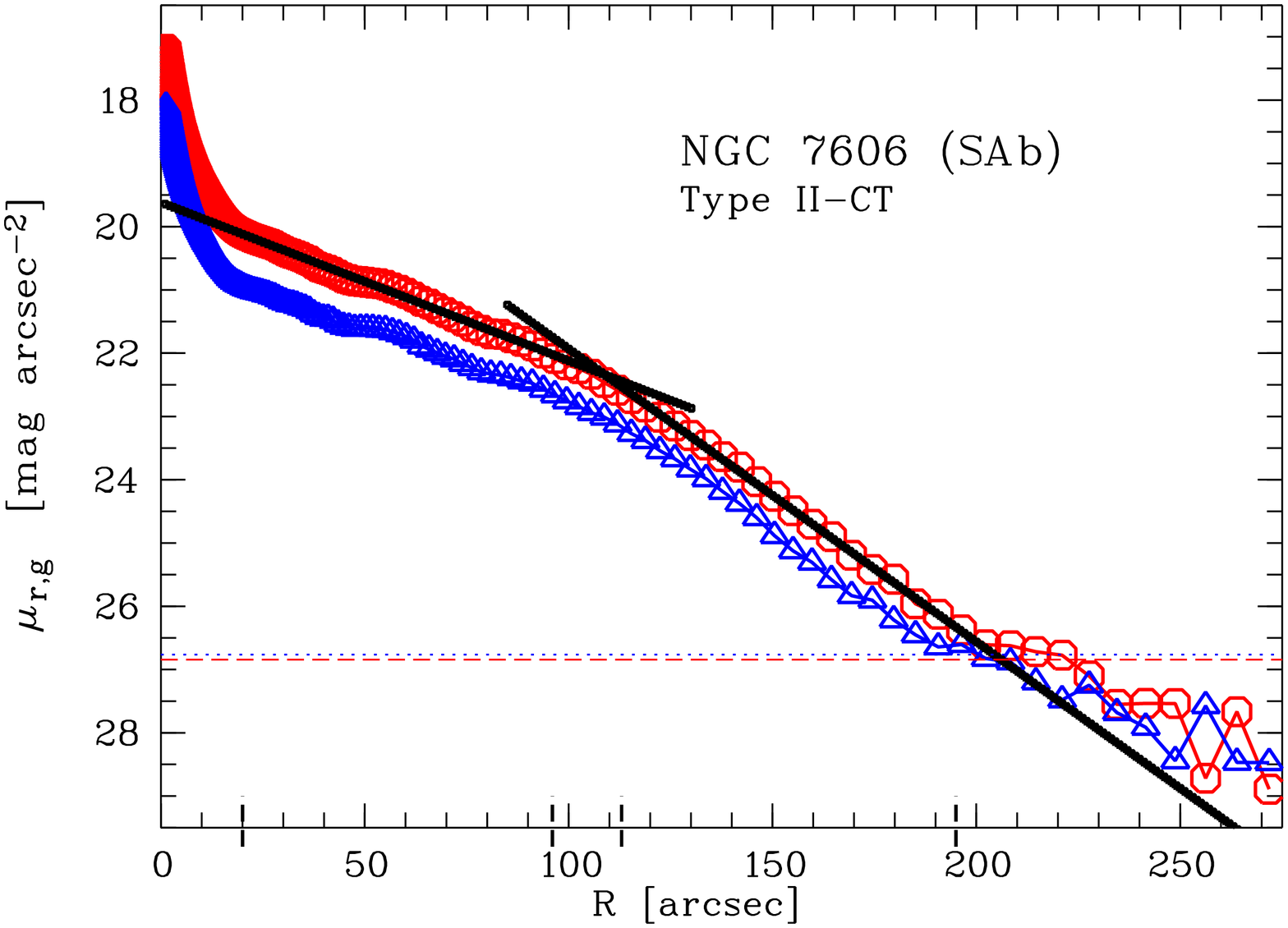}
}
\end{minipage}

\newpage
\onecolumn
\begin{minipage}[t][\textheight][t]{\textwidth}
\parbox[t][0.5\height][t]{0.47\textwidth}
{
\noindent {\bf PGC\,006667 $\equiv$ UGCA\,021  :}    \typeab      \\          
\texttt{J014910.2-100345 .SBS7.  7.1 -18.77   2.9 1887}\\[0.25cm]
A large scale gradient (from top to bottom) in the background of 
the $r^{\prime}$ band image is removed with linear fit.
Galaxy in NGC\,701 group (see above).
Final profile shows a bump at $\sim\!20\arcsec$ due to the spiral 
arms starting from the bar-like center. The clear downbending break 
at $\sim\!75\arcsec$ corresponds to a transition from the inner more 
symmetric to an asymmetric, extended outer disk region and is 
therefore classified as \typeabc.

}
\hfill 
\parbox[t][0.5\height][t]{0.47\textwidth}
{
\includegraphics[width=5.7cm,angle=270,]{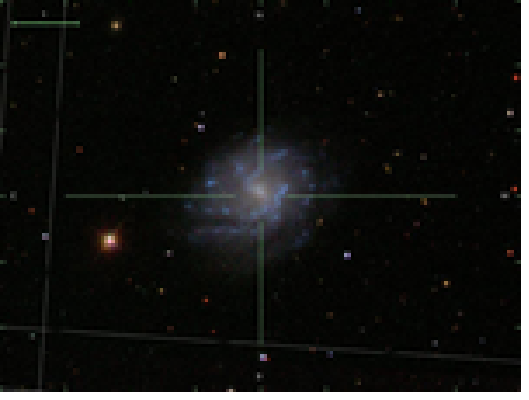}
\hspace*{-0.8cm}
\includegraphics[width=6.1cm,angle=270]{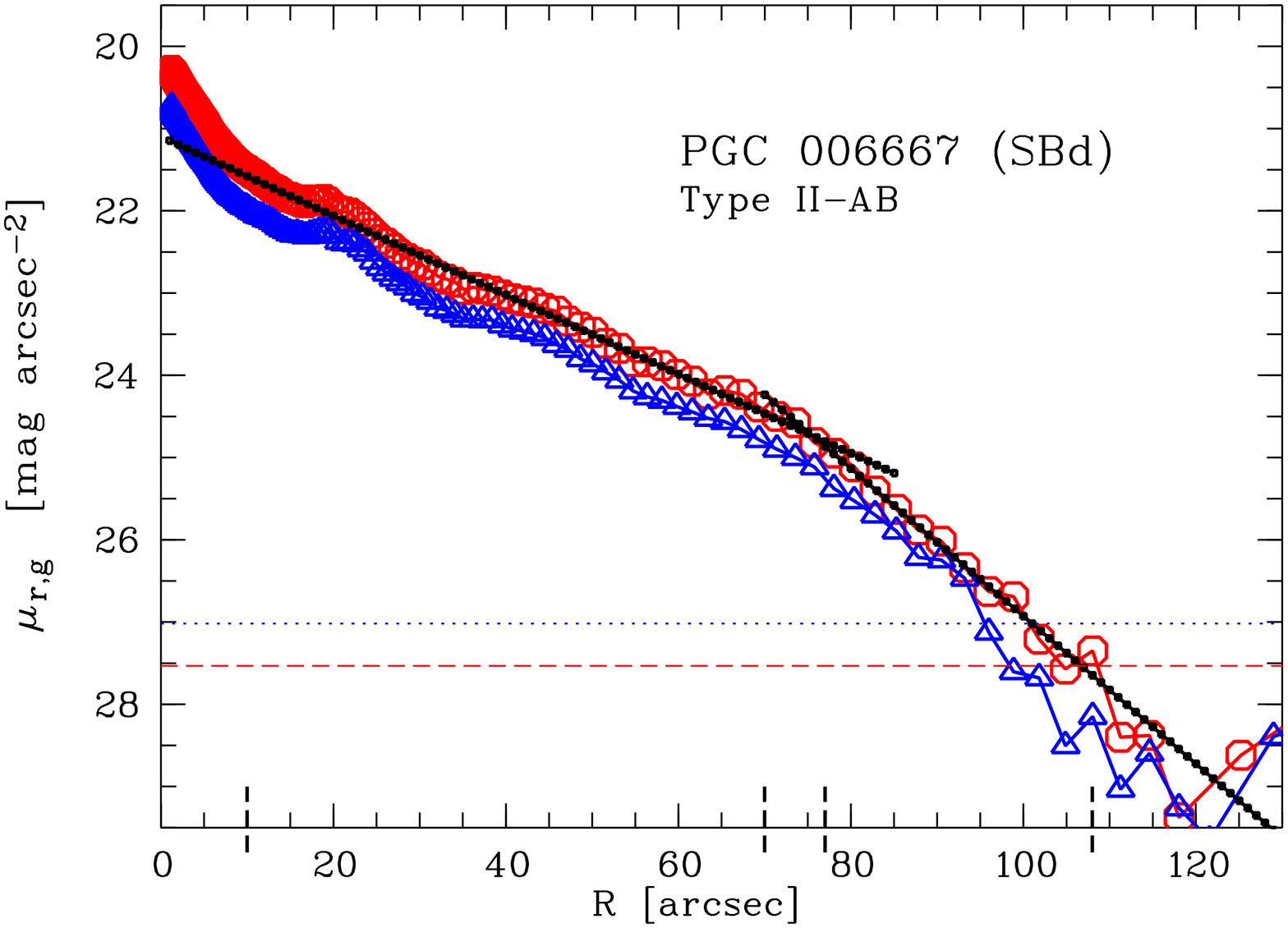}
}
\vfill
\parbox[b][0.5\height][t]{0.47\textwidth}
{
\noindent {\bf UGC\,02081    :}        \typect                \\          
\texttt{J023600.9+002512 .SXS6.  5.8 -18.56   1.8 2549}\\[0.25cm]
Galaxy close to the edge of the SDSS field but almost complete.
The final profile shows a downbending break starting around 
$\sim\!55\arcsec$ (clearly at $\sim\!70$\arcsec) with a very faint, 
but symmetric light distribution beyond. Although classified as SAB 
the bar is not obvious from the image but the spiral arms allow to 
estimate an upper limit of $R\ltsim 7$\arcsec. Therefore the break 
(which is not produced by a sky error) is classified \typectc.  
The profile published in \cite{dejong1994} is not deep enough to 
confirm the downbending. 

}
\hfill 
\parbox[b][0.5\height][b]{0.47\textwidth}
{
\includegraphics[width=5.7cm,angle=270]{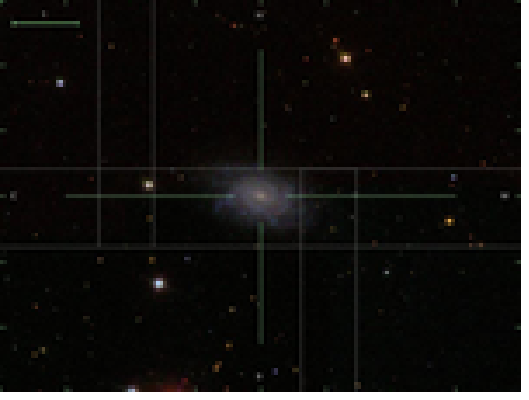}
\hspace*{-0.8cm}
\includegraphics[width=6.1cm,angle=270]{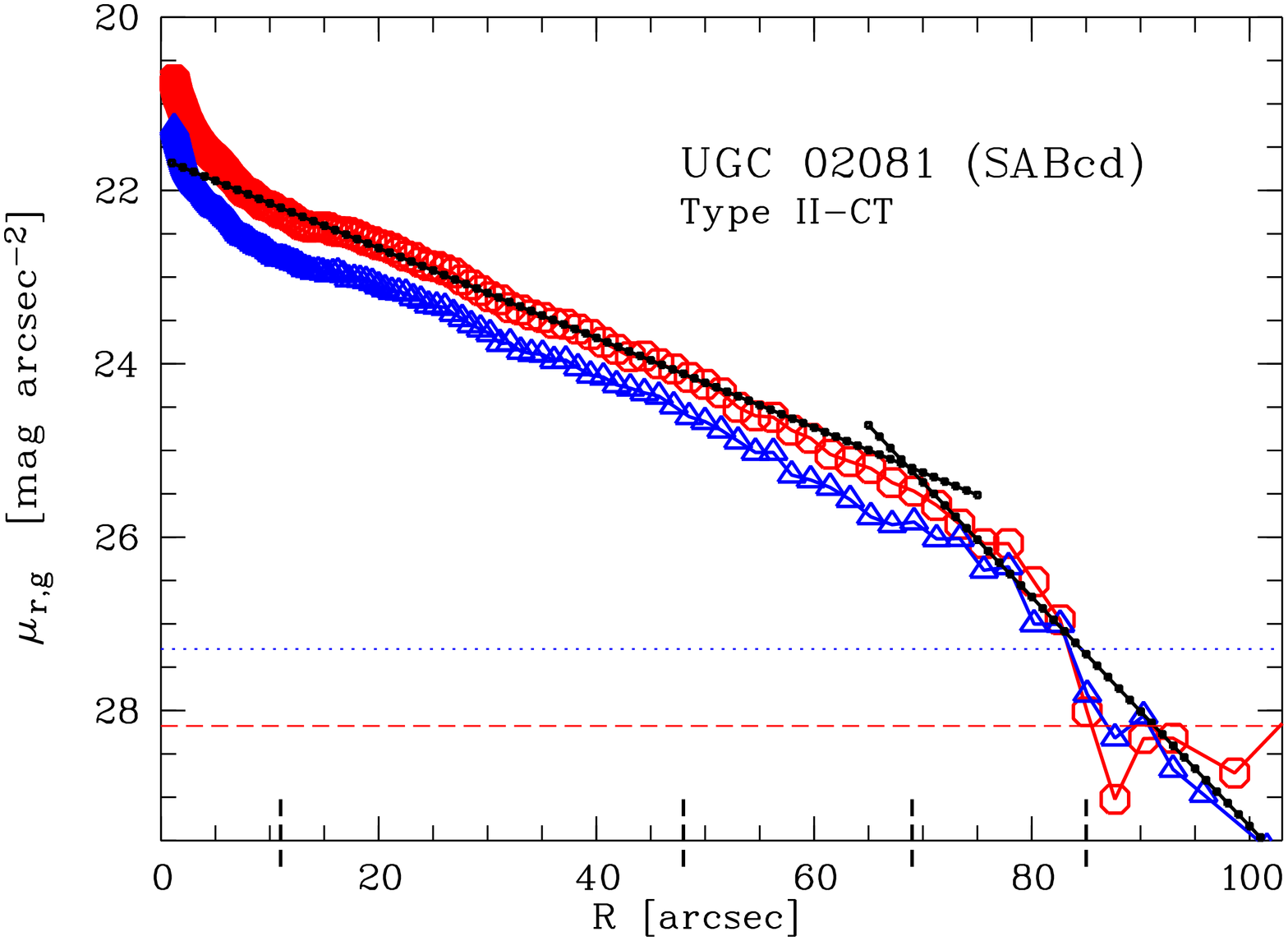}
}
\end{minipage}

\newpage
\onecolumn
\begin{minipage}[t][\textheight][t]{\textwidth}
\parbox[t][0.5\height][t]{0.47\textwidth}
{
\noindent {\bf UGC\,04393    :}        \typeo              \\        
\texttt{J082604.3+455803 .SB?..  5.5 -19.49   2.2 2290}\\[0.25cm]
Galaxy is classified as SB and the inner region is dominated by an 
elongated bar-like structure of size $R\sim\!30$\arcsec. Therefore 
this region is excluded from the fit. 
The extended bar, having a kind of double nucleus, is slightly 
offcenter (central dip in profile) compared to the center obtained 
from the outer isophotes and is followed by a single, narrow 
spiral arm like structure extending south-west.  
In the outer disk there is an additional, faint spiral arm visible 
towards the south-east. 
There is also a very faint shell or stream-like structure detected, 
detached from the galaxy about $2\arcmin$ away to the north-west. 
The inclination (ellipticity) and PA are rather uncertain since 
they are continuously changing towards the outer disk.
The final profile, beyond the inner bar region, is classified as \typeoc,
due to the uncertain ellipticity and the asymmetric outer disk with the 
spiral arm, although there are two possible breaks visible at  
$\sim\!60\arcsec$ and $\sim\!80$\arcsec, of which the first could 
correspond to a \typeolr break and the latter to a \typeiiic. 

}
\hfill 
\parbox[t][0.5\height][t]{0.47\textwidth}
{
\includegraphics[width=5.7cm,angle=270,]{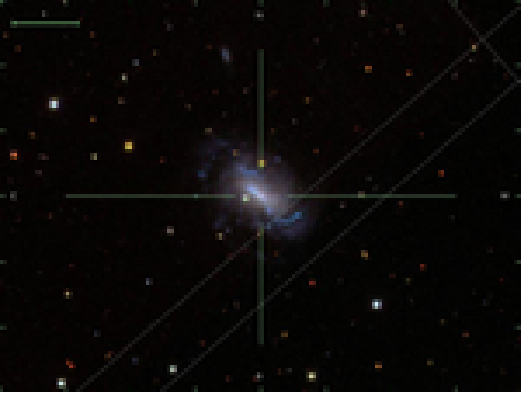}
\hspace*{-0.8cm}
\includegraphics[width=6.1cm,angle=270]{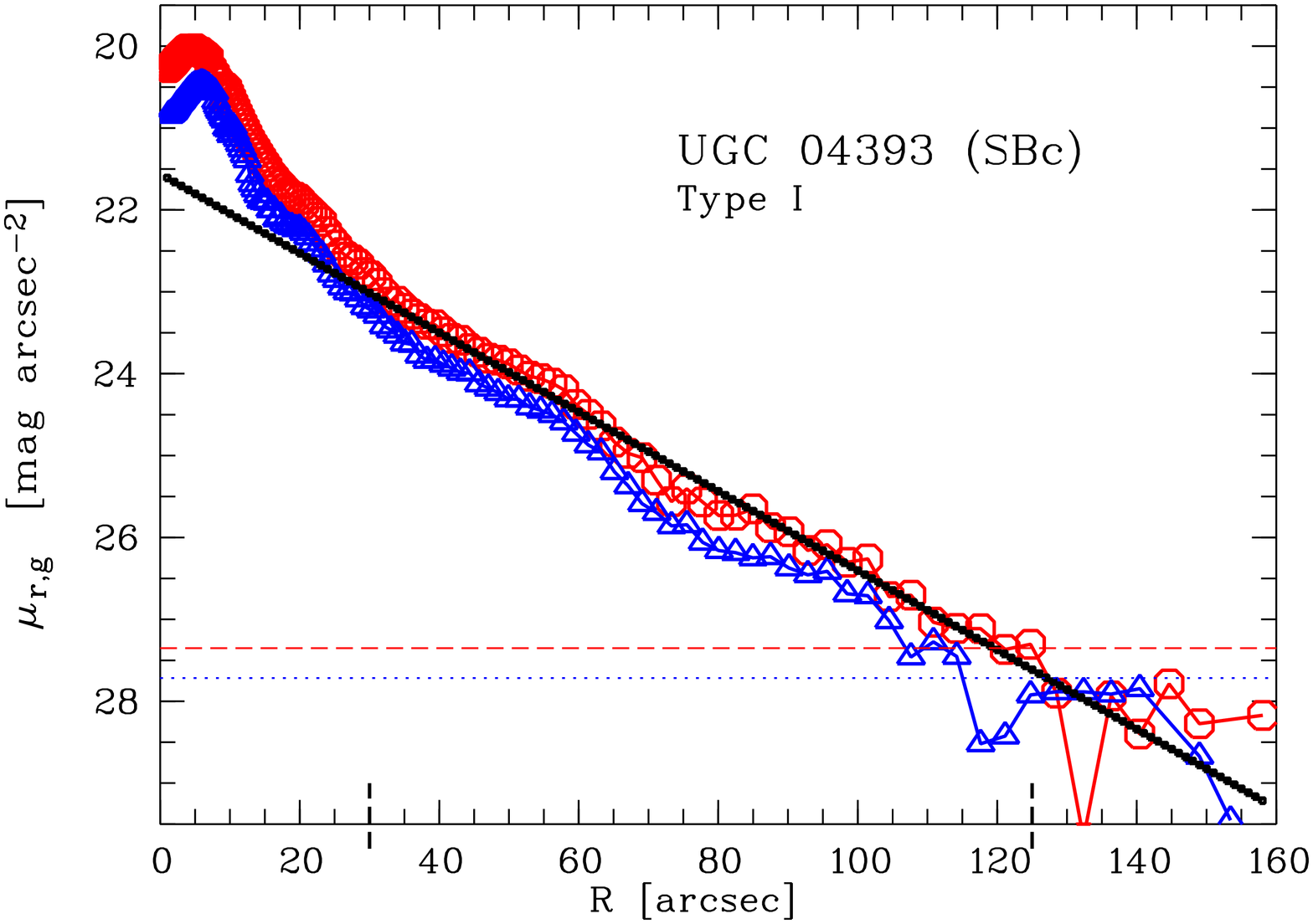}
}
\vfill
\parbox[b][0.5\height][t]{0.47\textwidth}
{
\noindent {\bf UGC\,06309    :}        \typeti                \\         
\texttt{J111746.4+512836 .SB?..  5.0 -19.63   1.3 3097}\\[0.25cm]
There is a small BCG (MRK\,1445) about $3.5\arcmin$ away 
at roughly the same distance ($v=2863$\kms).
Galaxy exhibits very unusual, disturbed shape. The inner disk is
dominated by a narrow bar of size $R\ltsim 20\arcsec$ followed 
by two asymmetric spiral arms (one highly wrapped, forming an inner 
ring, the other almost extending straight). The disk beyond the 
spiral arms is also highly asymmetric and more egg-shaped with a 
region of low emission (high extinction?) towards the south-east, 
which makes the bar appear to be offcentered. However, the center 
from the very outer isophotes coincides again with the nucleus 
inside the bar. 
The extended dip and peak structure around $15-20\arcsec$ in the 
final profile corresponds to the inter-arm region, respectively 
the pseudoring build by the arms, so the break around this radius 
is classified as \typetic.
This should be used with caution, since alternatively one can argue 
for a break at $\sim\!35$\arcsec, at about twice the bar radius thus  
a \typeolr break, while excluding the region between $15-30$\arcsec. 
   
}
\hfill 
\parbox[b][0.5\height][b]{0.47\textwidth}
{
\includegraphics[width=5.7cm,angle=270]{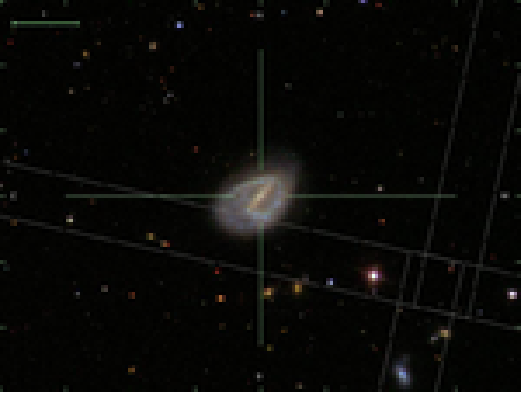}
\hspace*{-0.8cm}
\includegraphics[width=6.1cm,angle=270]{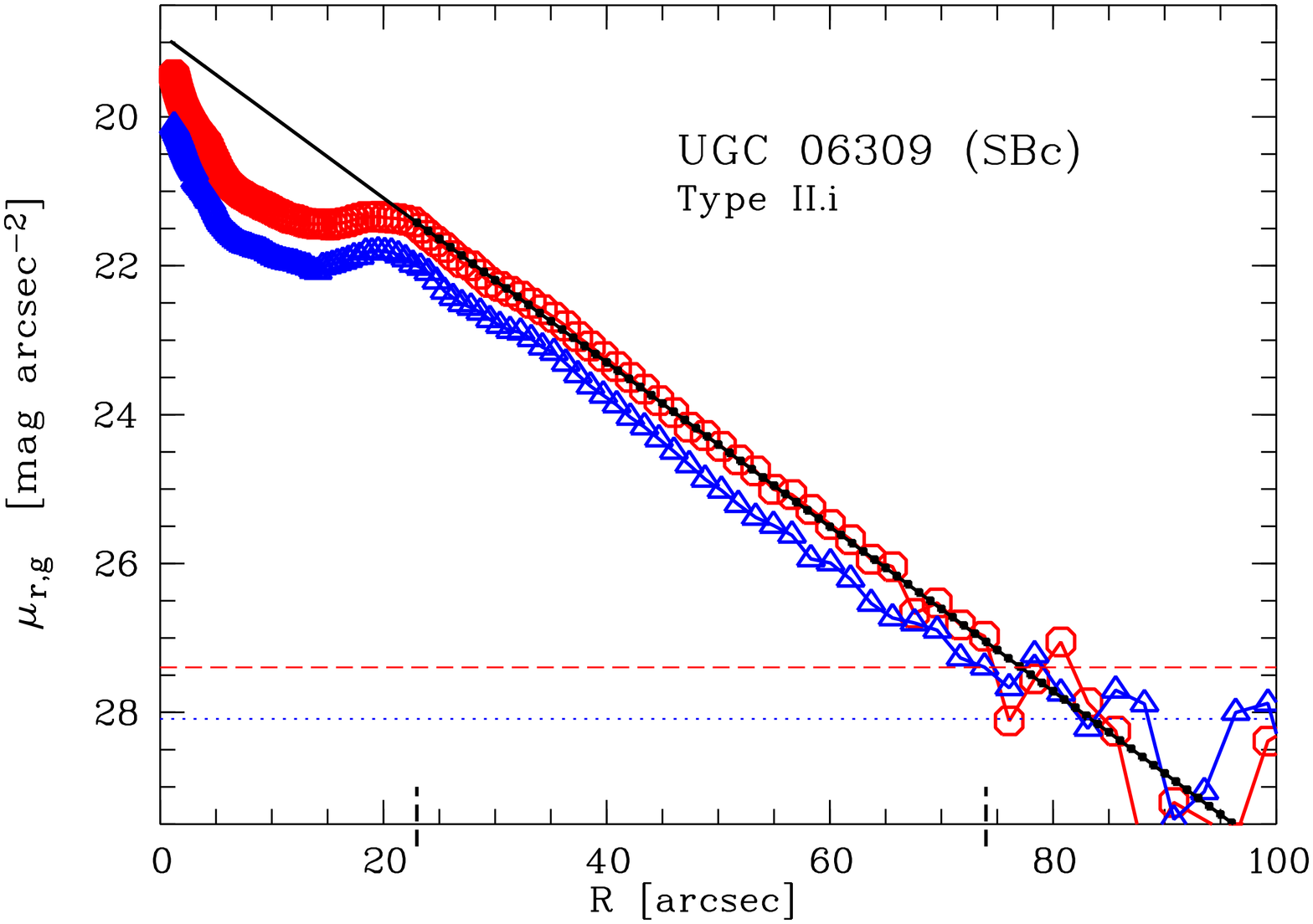}
}
\end{minipage}

\newpage
\onecolumn
\begin{minipage}[t][\textheight][t]{\textwidth}
\parbox[t][0.5\height][t]{0.47\textwidth}
{
\noindent {\bf UGC\,06518    :}        \typeti                 \\        
\texttt{J113220.4+535417 ......  4.3 -18.94   1.0 3044}\\[0.25cm]
Small galaxy where, although not classified as barred in NED or LEDA, 
there is clearly a bar of size $R\sim\!10\arcsec$ visible, with indication
for a ring-like structure around, already reported by \cite{takase1984}.
Despite the possibility to argue for a break at $\sim\!15\arcsec$ in
the final profile we classify the galaxy as \typeti excluding the 
region inside $R=10\arcsec$ from the fit. 

}
\hfill 
\parbox[t][0.5\height][t]{0.47\textwidth}
{
\includegraphics[width=5.7cm,angle=270,]{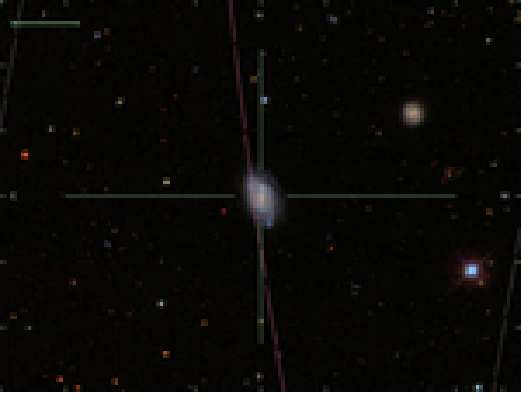}
\hspace*{-0.8cm}
\includegraphics[width=6.1cm,angle=270]{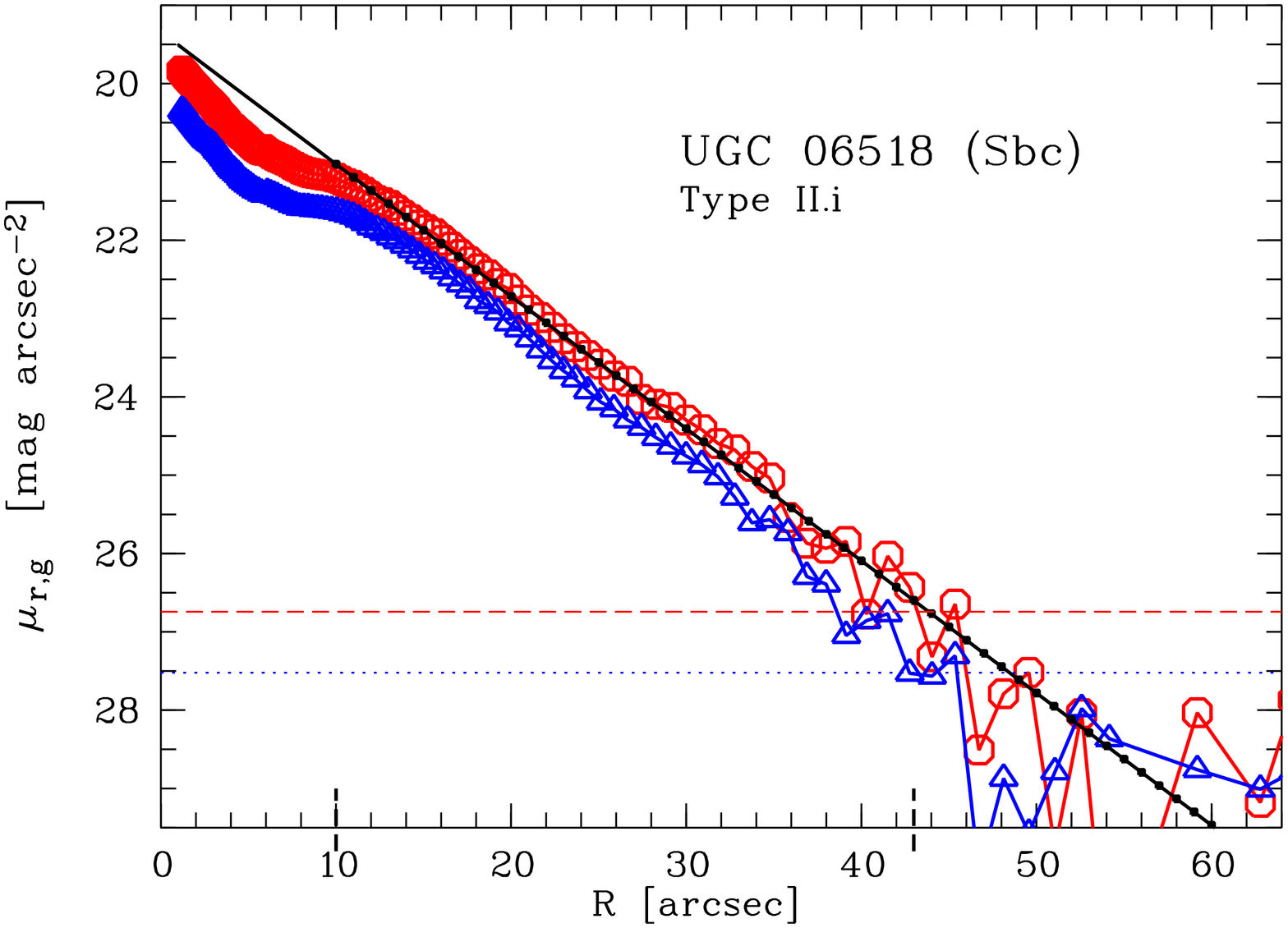}
}
\vfill
\parbox[b][0.5\height][t]{0.47\textwidth}
{
\noindent {\bf UGC\,06903    :}        \typetoct                \\          
\texttt{J115536.9+011415 .SBS6.  5.8 -18.80   2.5 1916}\\[0.25cm]
Very bright star in FOV but galaxy not influenced.
Final profile shows clearly a downbending break around 
$\sim\!65\arcsec$ corresponding roughly to the end of the visible 
spiral arm structure and classified \typetoctc, since the size of the 
bar with $R\ltsim 17$ is too small. 
The extended bump between $R\sim\!20-40\arcsec$ is related to the 
inner two, highly wrapped (pseudoring?) spiral arms. 
The downbending shape is consistent with the profile shown in  
\cite{jansen2000} (\cf ID\,99).

}
\hfill 
\parbox[b][0.5\height][b]{0.47\textwidth}
{
\includegraphics[width=5.7cm,angle=270]{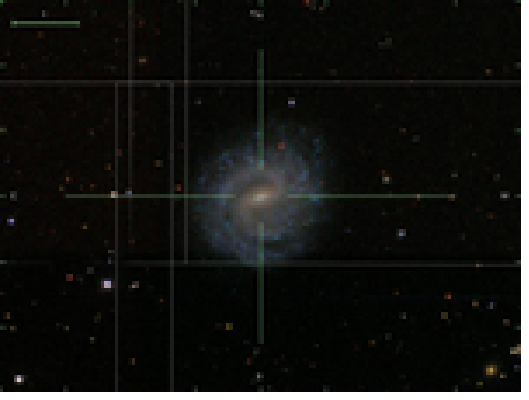}
\hspace*{-0.8cm}
\includegraphics[width=6.1cm,angle=270]{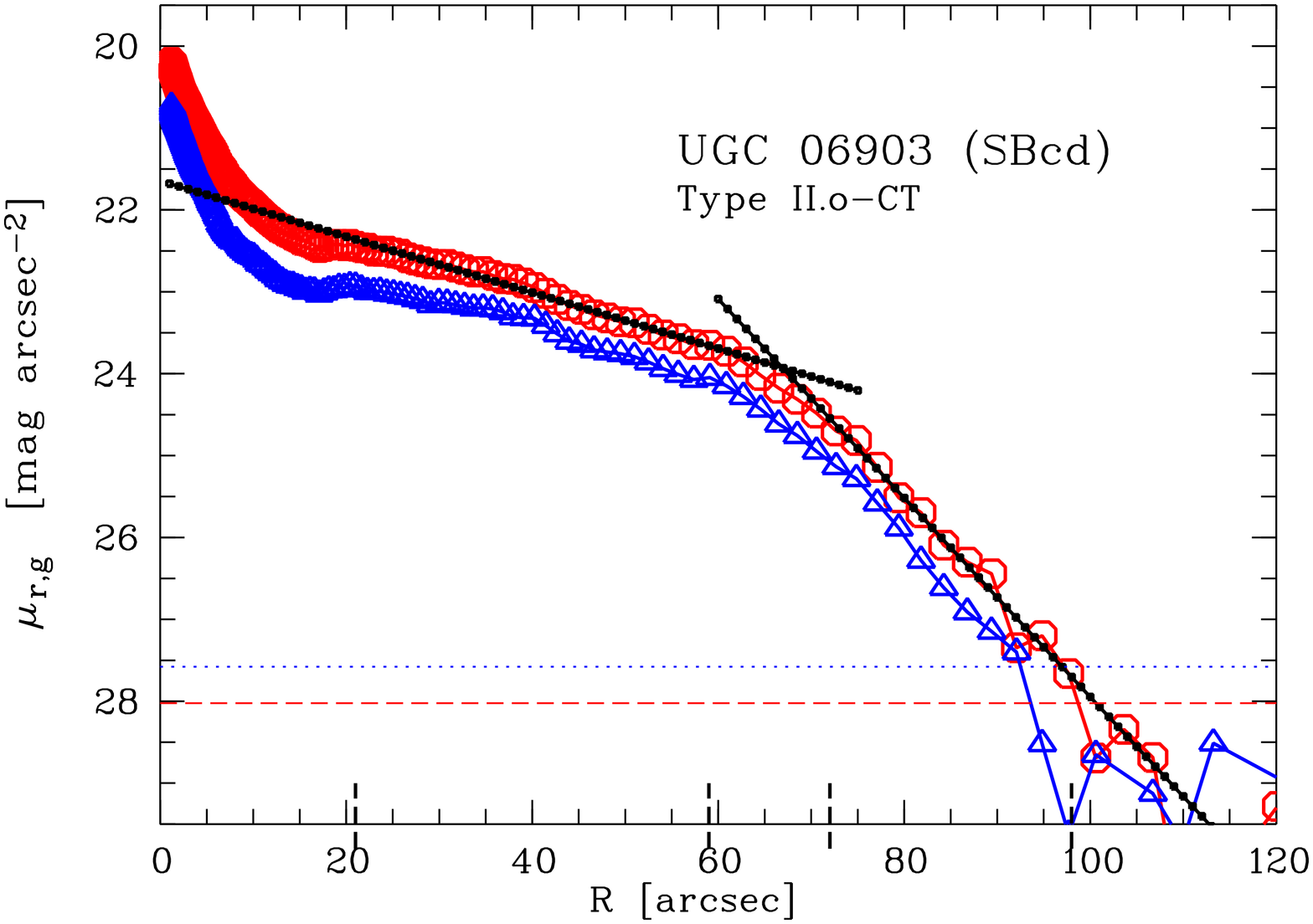}
}
\end{minipage}

\newpage
\onecolumn
\begin{minipage}[t][\textheight][t]{\textwidth}
\parbox[t][0.5\height][t]{0.47\textwidth}
{
\noindent {\bf UGC\,07700    :}        \typetoct                \\          
\texttt{J123232.8+635238 .SBS8.  7.8 -18.50   1.8 3239}\\[0.25cm]
There is a very bright star in the FOV but the galaxy is not influenced. 
On same image there is an edge-on companion (S0/a galaxy NGC\,4512) at 
similar distance ($v=2536$\kms) visible. 
Ellipticity obtained from the very outer isophotes and not 
by the usual $1\sigma$ criterion.  
The final profile shows clearly a downbending behaviour outside 
the bar region. The bar is about $R\sim\!17\arcsec$ in size and 
slightly offcentered (by $\sim\!5$\arcsec) compared to the center 
obtained from the outer, symmetric isophotes. This causes the 
dip at the center.   
The extended break region is not exponential but shows a 
curvature, but still one could define two break radii at 
$\sim\!40\arcsec$ and $\sim\!50$\arcsec. The region in between 
corresponds to a single spiral arm extending towards the 
north-east. 
Although the first one would roughly classify as a \typeolr break 
we used a single break slightly further out and classified the 
galaxy as \typetoctc. 

}
\hfill 
\parbox[t][0.5\height][t]{0.47\textwidth}
{
\includegraphics[width=5.7cm,angle=270,]{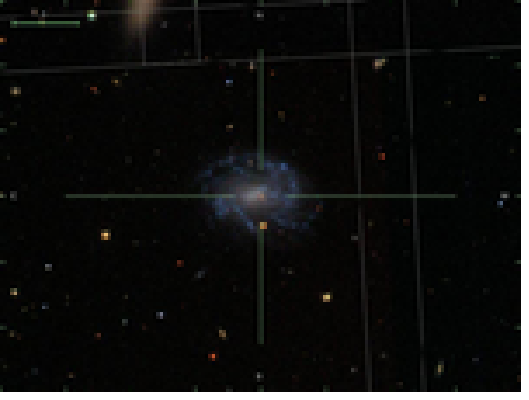}
\hspace*{-0.8cm}
\includegraphics[width=6.1cm,angle=270]{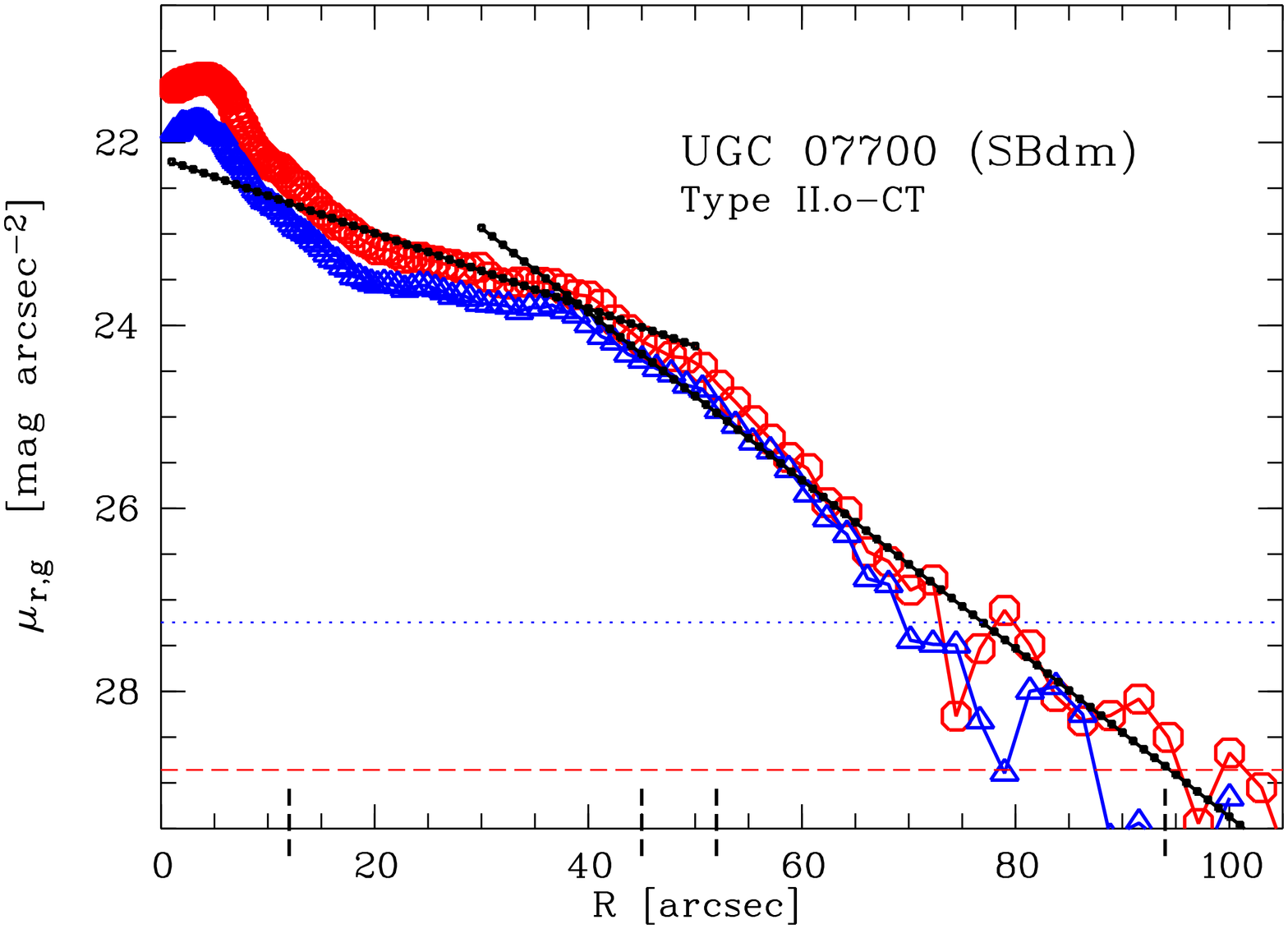}
}
\vfill
\parbox[b][0.5\height][t]{0.47\textwidth}
{
\noindent {\bf UGC\,08041    :}        \typetoct                \\          
\texttt{J125512.7+000700 .SBS7.  6.8 -18.62   3.2 1376}\\[0.25cm]
Galaxy close to our high axis ratio limit and not influenced by 
the very bright star in FOV.
The inner disk with a bright nucleus in the bar and weak spiral arm 
structure extends to about $\sim\!100$\arcsec, followed by a rather 
asymmetric, lopsided outer disk extending to the south-west. 
Centering on the outer isophotes ($R\gtsim 100$\arcsec) would be 
off by more than $\sim\!10$\arcsec.
The final profile shows a downbending break at $\sim\!75$\arcsec
roughly corresponding to the end of the spiral structure. 
Since the bar is only of size $R\ltsim 12\arcsec$ the break,
at significantly more than twice the bar radius, is classified as 
\typetoctc.
The fixed ellipse leaves the inner symmetric disk at about 
$\sim\!100\arcsec$ into the asymmetric outer part causing
the apparent upbending in the profile, which probably increases 
the fitted outer scalelength. 

}
\hfill 
\parbox[b][0.5\height][b]{0.47\textwidth}
{
\includegraphics[width=5.7cm,angle=270]{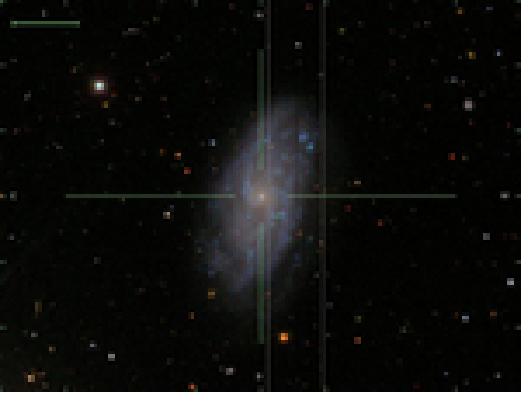}
\hspace*{-0.8cm}
\includegraphics[width=6.1cm,angle=270]{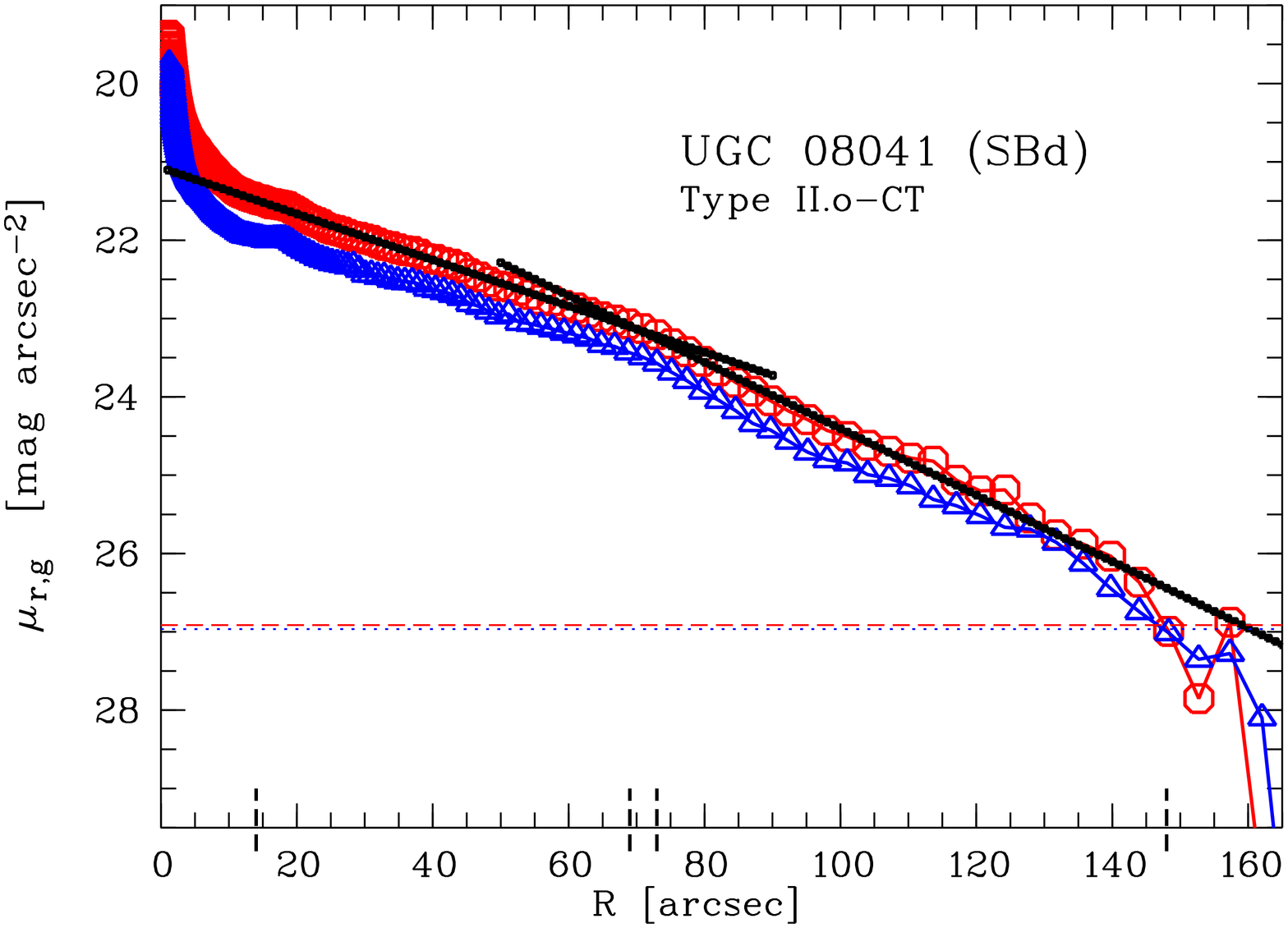}
}
\end{minipage}

\newpage
\onecolumn
\begin{minipage}[t][\textheight][t]{\textwidth}
\parbox[t][0.5\height][t]{0.47\textwidth}
{
\noindent {\bf UGC\,08084    :}        \typeolr                \\          
\texttt{J125822.4+024733 .SBS8.  8.1 -18.61   1.3 2824}\\[0.25cm]
Small galaxy dominated by an inner offcentered, asymmetric bar of 
size $R\sim\!17$\arcsec, which appears to be z-shaped together with 
the two spiral arms. The center is obtained from outer isophotes 
($\sim\!50 \arcsec$) and is about $\sim\!12\arcsec$ away from the 
brightest pixel in bar, which is responsible for the flat inner profile.  
The final profile shows clearly a downbending break at $\sim\!40$\arcsec
corresponding to the spiral arms forming a kind of pseudoring in the 
outer parts. Since the break is at about twice the bar size the galaxy 
is classified as \typeolrc. 

}
\hfill 
\parbox[t][0.5\height][t]{0.47\textwidth}
{
\includegraphics[width=5.7cm,angle=270,]{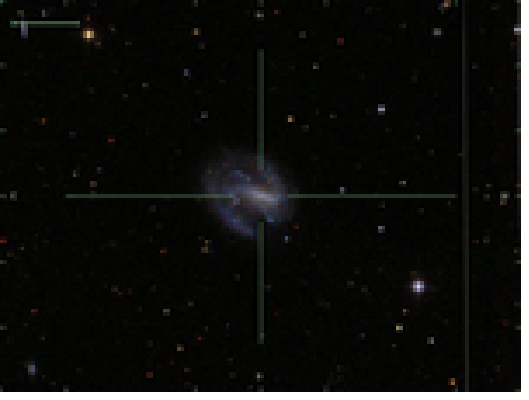}
\hspace*{-0.8cm}
\includegraphics[width=6.1cm,angle=270]{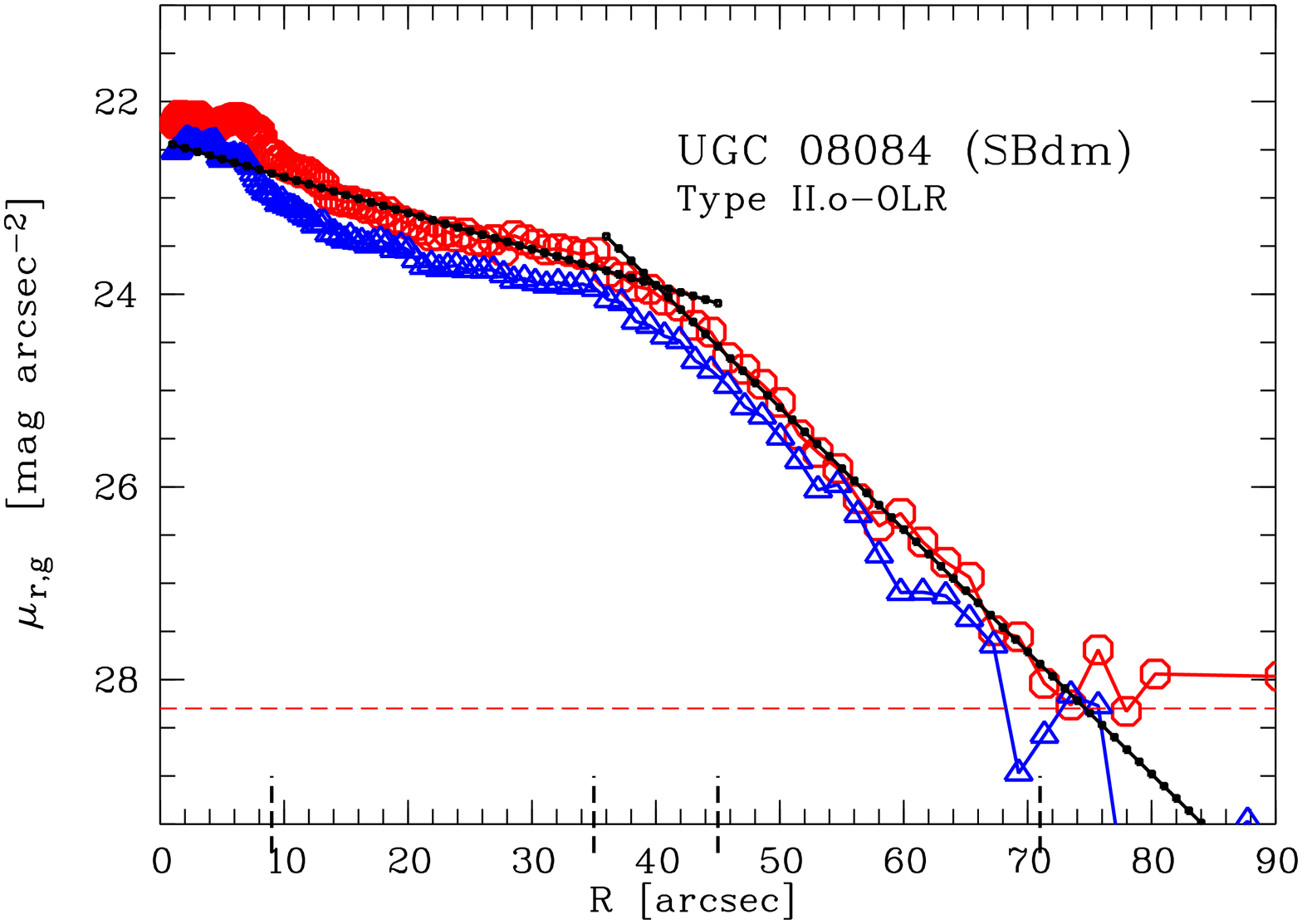}
}
\vfill
\parbox[b][0.5\height][t]{0.47\textwidth}
{
\noindent {\bf UGC\,08237    :}        \typeolr               \\           
\texttt{J130854.5+621823 PSB.3*  3.2 -19.61   0.9 3120}\\[0.25cm]
Galaxy is isolated according to \cite{prada2003}. The similar sized 
companion, only $2.3\arcmin$ away, is a confirmed background galaxy 
(UGC\,08234, $v=8100$\kms). 
Background needs extended mask due to straylight from a very bright 
star off the field. 
UGC\,08237 is a small galaxy dominated by a bright, thick bar of 
size $R\ltsim 12\arcsec$ (producing the shoulder in the final profile)
followed by two nearly symmetric, wrapped spiral arms forming a pseudoring
peaking at $\sim\!20$\arcsec.
Similar to NGC\,5701 (an early-type SB0/a galaxy from \cite{erwin2006})
one could define a break at $\sim\!25\arcsec$ just outside of the pseudoring
forming the region $R\sim\!12-25$ which is no longer well fitted by an 
exponential profile. \cite{erwin2006} call these systems 'extreme' OLR 
breaks, so we also classify the galaxy as \typeolr and apply a crude 
fit to obtain an inner scalelength.  

}
\hfill 
\parbox[b][0.5\height][b]{0.47\textwidth}
{
\includegraphics[width=5.7cm,angle=270]{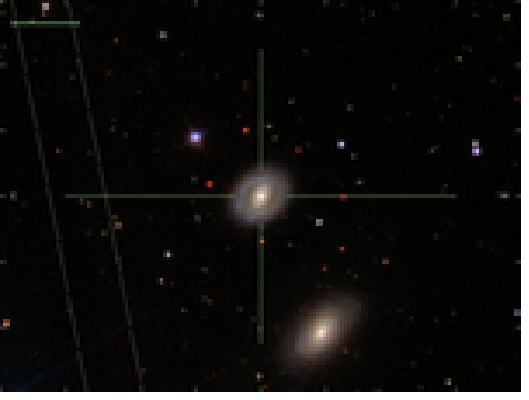}
\hspace*{-0.8cm}
\includegraphics[width=6.1cm,angle=270]{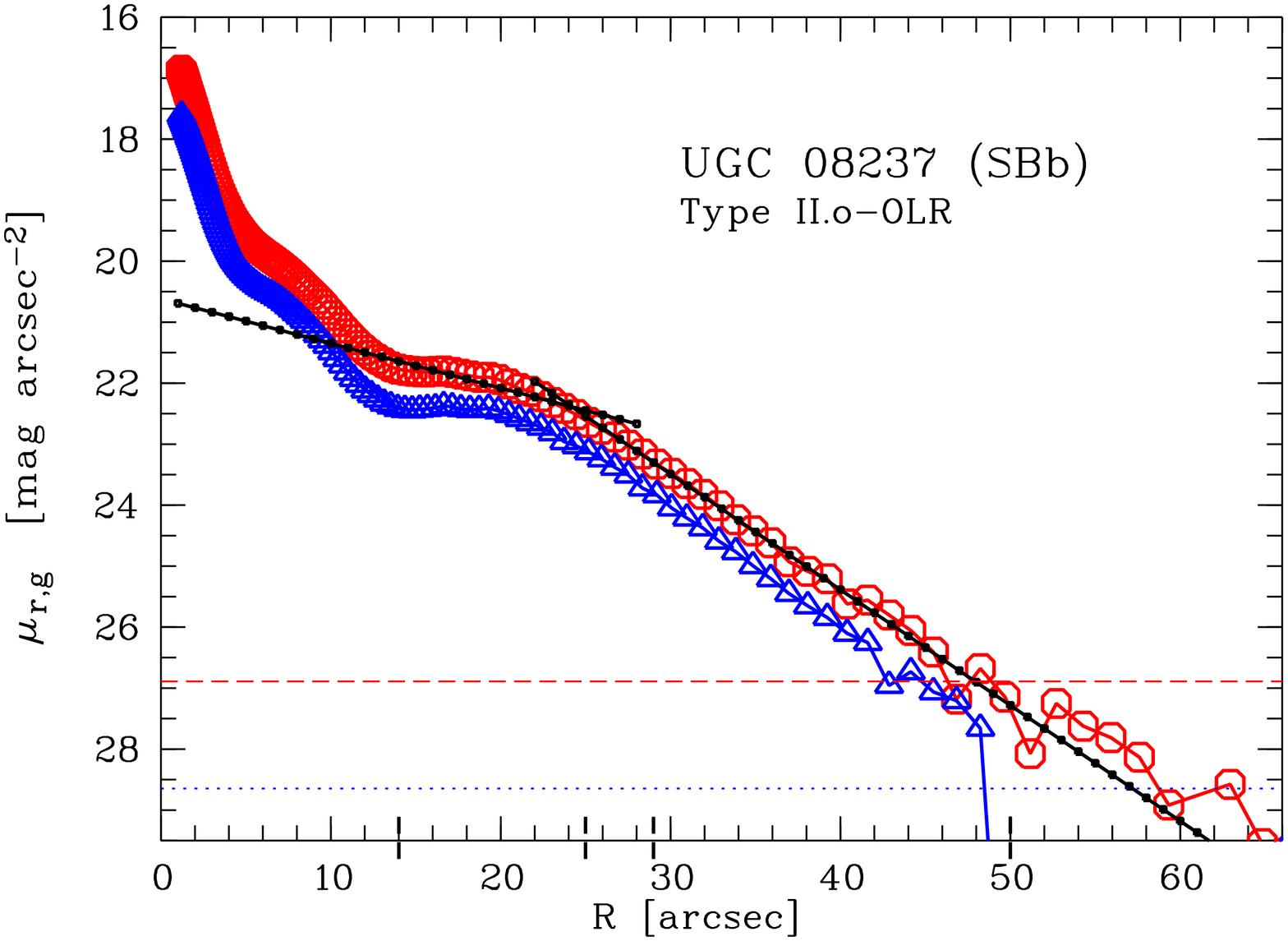}
}
\end{minipage}

\newpage
\onecolumn
\begin{minipage}[t][\textheight][t]{\textwidth}
\parbox[t][0.5\height][t]{0.47\textwidth}
{
\noindent {\bf UGC\,08658 $\equiv$ Holmberg\,V   :}        \typect     \\   
\texttt{J134039.9+541959 .SXT5.  5.1 -19.91   2.6 2285}\\[0.25cm]
Galaxy close to the border of the SDSS field but almost complete. Although 
classified as SAB, the bar size is not obvious from the image, but could be 
restricted by the spiral arms to be maximal $R\ltsim 10$\arcsec.  
The shoulder in the final profile at $\sim\!20\arcsec$ is due to the inner 
spiral arms being aligned with the ellipse.
There is a weak break at $\sim\!65\arcsec$ visible which roughly 
coincides with the end of the inner spiral arm structure, although 
there are maybe two faint extensions into the outer disk. 
Since being at significantly more than two bar radii we classify 
the break as \typectc.

}
\hfill 
\parbox[t][0.5\height][t]{0.47\textwidth}
{
\includegraphics[width=5.7cm,angle=270,]{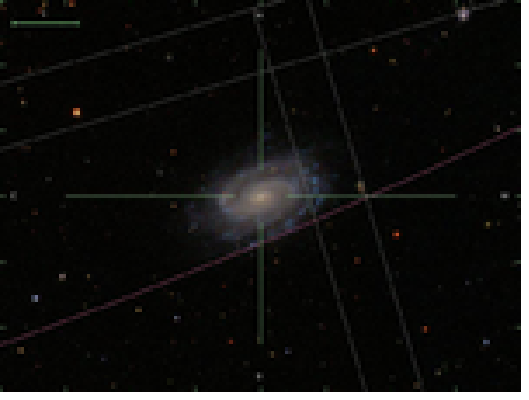}
\hspace*{-0.8cm}
\includegraphics[width=6.1cm,angle=270]{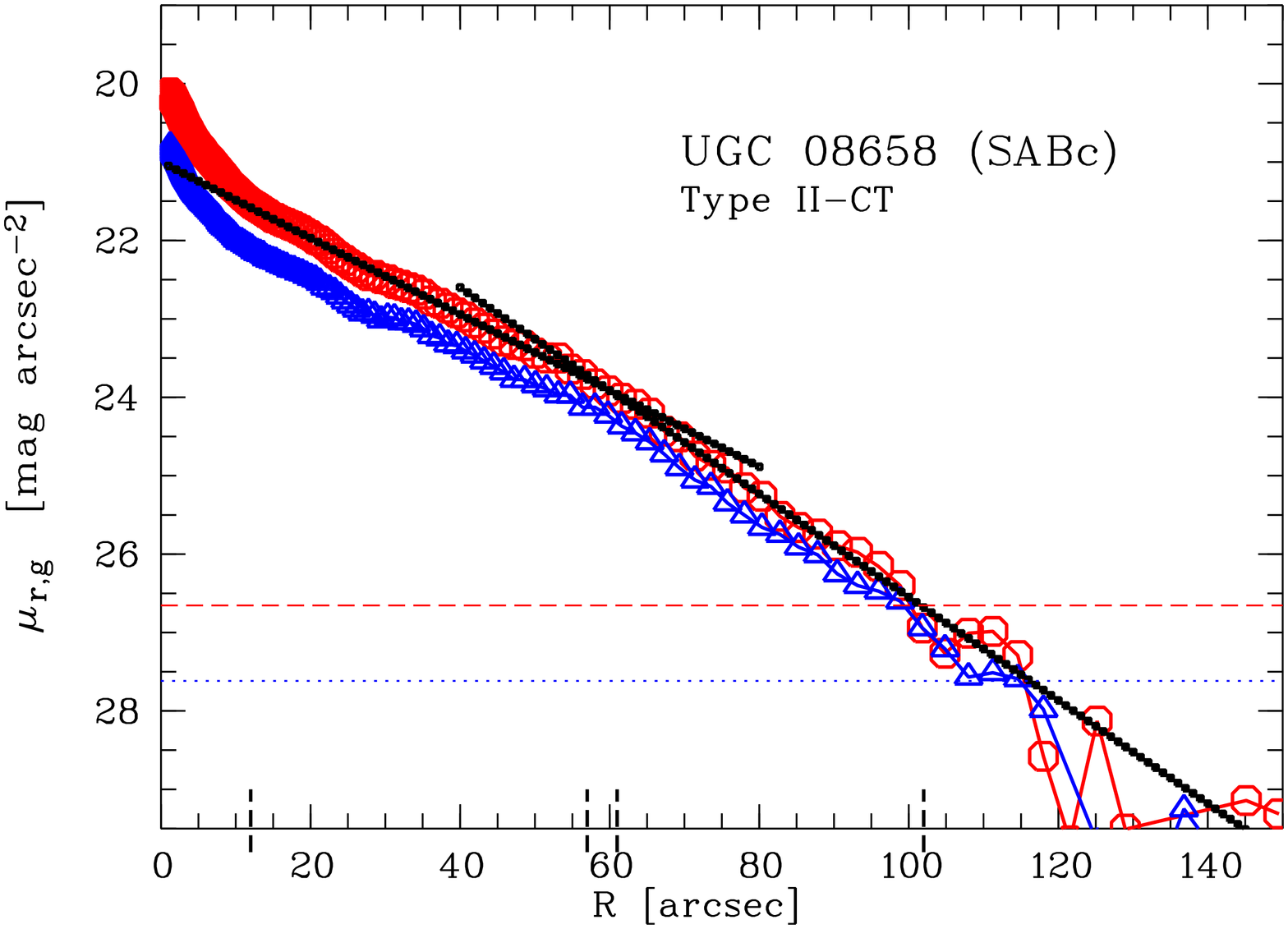}
}
\vfill
\parbox[b][0.5\height][t]{0.47\textwidth}
{
\noindent {\bf UGC\,09741 $\equiv$ NGC\,5875A  :}        \typeiii    \\
\texttt{J150833.5+521746 .S?...  6.0 -18.54   0.9 2735}\\[0.25cm]
The patch of faint light on the image towards the north-west 
is caused by straylight from a very bright star off field, but 
the galaxy is not affected. 
Very small galaxy dominated by a long, narrow bar of size 
$R\sim\!9$\arcsec, enclosed by a ring like structure building the 
inner $R\ltsim 17\arcsec$ disk.  
The photometric inclination (ellipticity) and PA is obtained from 
the outer disk at $\sim\!45\arcsec$ and not from the $1\sigma$ ellipse. 
Although it is possible to fit the profile beyond $\sim\!17$\arcsec
with a single exponential the offset inside would be too large to 
be assigned solely to a bulge component of a typical Sc galaxy. In 
addition, an extended bulge component is also not visible on the 
image inside the narrow bar. 
Thus we argue for a break, classified as \typeiiic, at $\sim\!25\arcsec$ 
followed by an upbending profile into the symmetric featureless  
outer disk. 
The shape of the profile is consistent with the one shown by  
\cite{jansen2000} (\cf ID\,166). 

}
\hfill 
\parbox[b][0.5\height][b]{0.47\textwidth}
{
\includegraphics[width=5.7cm,angle=270]{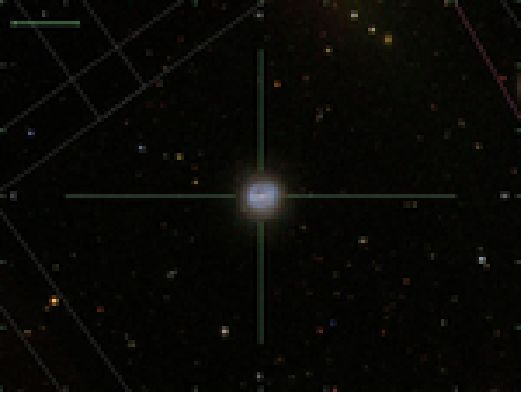}
\hspace*{-0.8cm}
\includegraphics[width=6.1cm,angle=270]{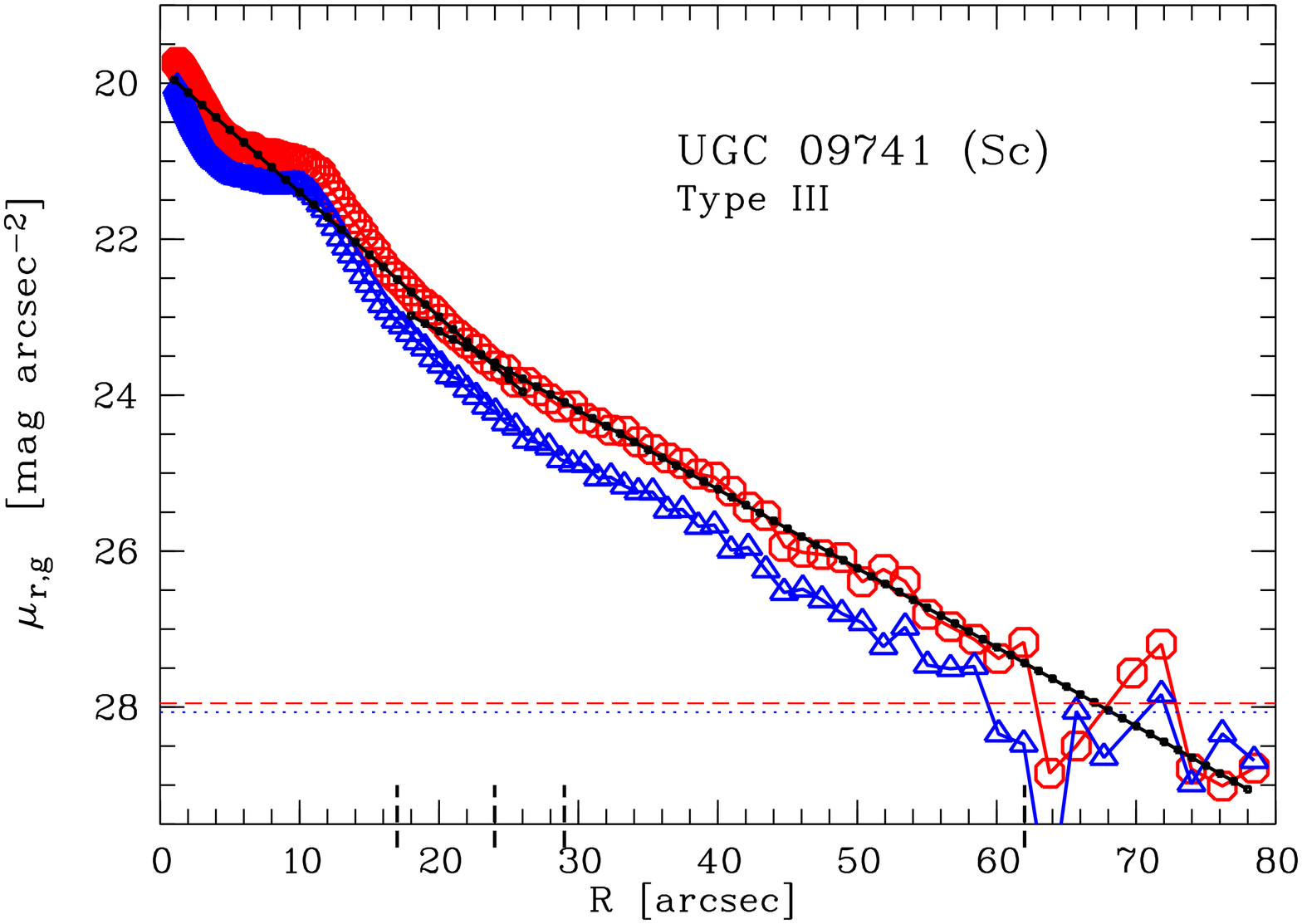}
}
\end{minipage}

\newpage
\onecolumn
\begin{minipage}[t][\textheight][t]{\textwidth}
\parbox[t][0.5\height][t]{0.47\textwidth}
{
\noindent {\bf UGC\,09837    :}        \typect                \\          
\texttt{J152351.7+580311 .SXS5.  5.1 -19.45   1.8 2938}\\[0.25cm]
Although classified as SAB, the bar size is not obvious from the 
image, but could be restricted to be maximal $R\ltsim 9$\arcsec
by the two symmetric spiral arms starting beyond the bulge/bar 
region out to $R\sim\!25$\arcsec.  
Further out, there is more asymmetric spiral arm structure visible,
with one arm almost forming a ring-like section, which makes the 
photometric inclination (ellipticity) and PA determination 
impossible at the $1\sigma$ ellipse.    
Since the ring like section extends to the end of our free ellipse
fit, we set the ellipticity to 0.05 (round).    
In the final profile there is a break visible at $\sim\!50\arcsec$ 
just beyond the bump produced by the ring-like spiral arm.  
Since this break is at significantly more than two times the upper
limit for the bar radius we classify the break as \typectc.
The deep surface photometry from \cite{pohlen2002} is consistent 
with the results here (\cf\fig\ref{SDSSvsCCD}) aside from this bump 
produced by the outer spiral arm, which seems to be suppressed in 
their averaged profile.  
The downbending shape with a break at $\sim\!50\arcsec$ is 
consistent with the profile shown by \cite{vdk1987}.
}
\hfill 
\parbox[t][0.5\height][t]{0.47\textwidth}
{
\includegraphics[width=5.7cm,angle=270,]{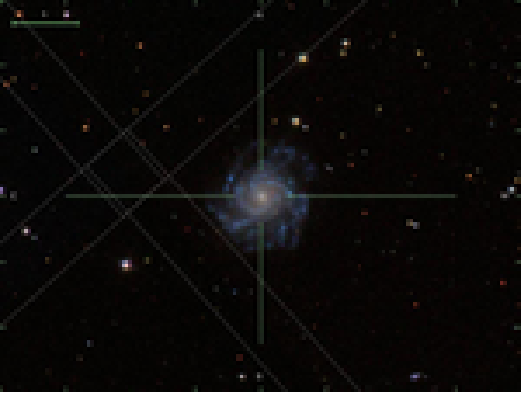}
\hspace*{-0.8cm}
\includegraphics[width=6.1cm,angle=270]{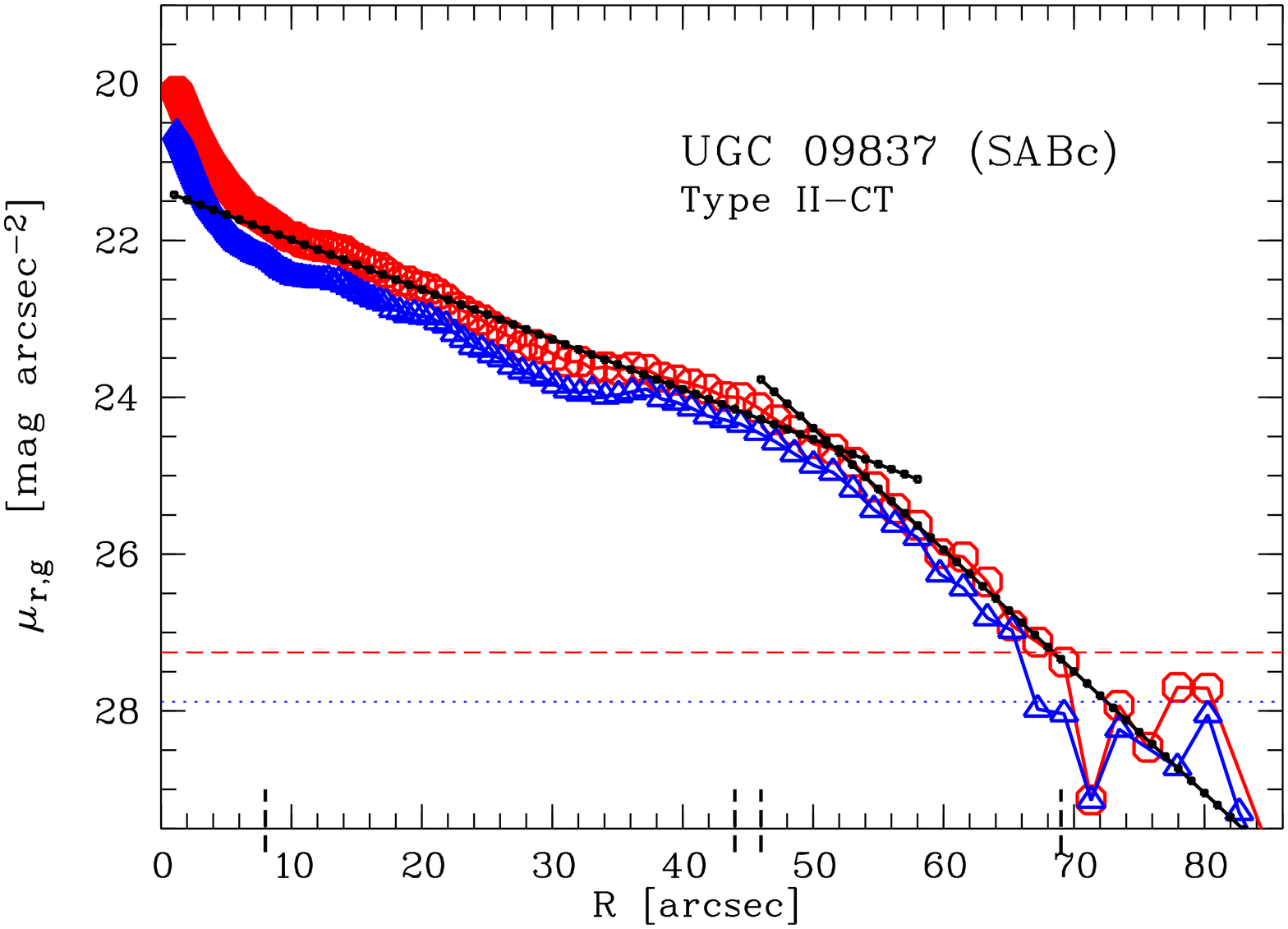}
}
\vfill
\parbox[b][0.5\height][t]{0.47\textwidth}
{
\noindent {\bf UGC\,10721    :}        \typeiii                 \\        
\texttt{J170825.6+253103 .S..6?  5.8 -19.68   1.2 3118}\\[0.25cm]
Small galaxy close to our high axis ratio limit so the dust 
may get important. Galaxy shows an inner apparently inclined 
disk with a nucleus and some thin spiral arms sitting in an almost
round outer isophote most probably related to a vertical structure
(halo or thick disk). Thus the photometric inclination (ellipticity) 
and PA are highly uncertain since continously changing with radius. 
The shoulder at $\sim\!12\arcsec$ in the final profile corresponds 
to an aligned spiral arm. 
There is a clear break visible at $\sim\!40\arcsec$ followed by an 
upbending profile which is classified as \typeiiic, associated to the 
transition between inner and outer disk (vertical structure?), so 
to be taken with some caution. The sharp transition in the profile 
argues against a thick disk.
The upbending behaviour is confirmed by the profile from 
\cite{courteau1996}.
}
\hfill 
\parbox[b][0.5\height][b]{0.47\textwidth}
{
\includegraphics[width=5.7cm,angle=270]{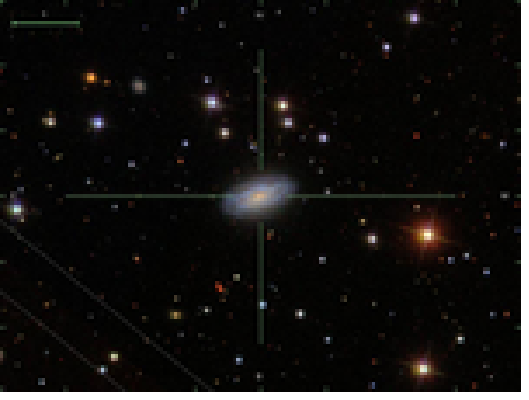}
\hspace*{-0.8cm}
\includegraphics[width=6.1cm,angle=270]{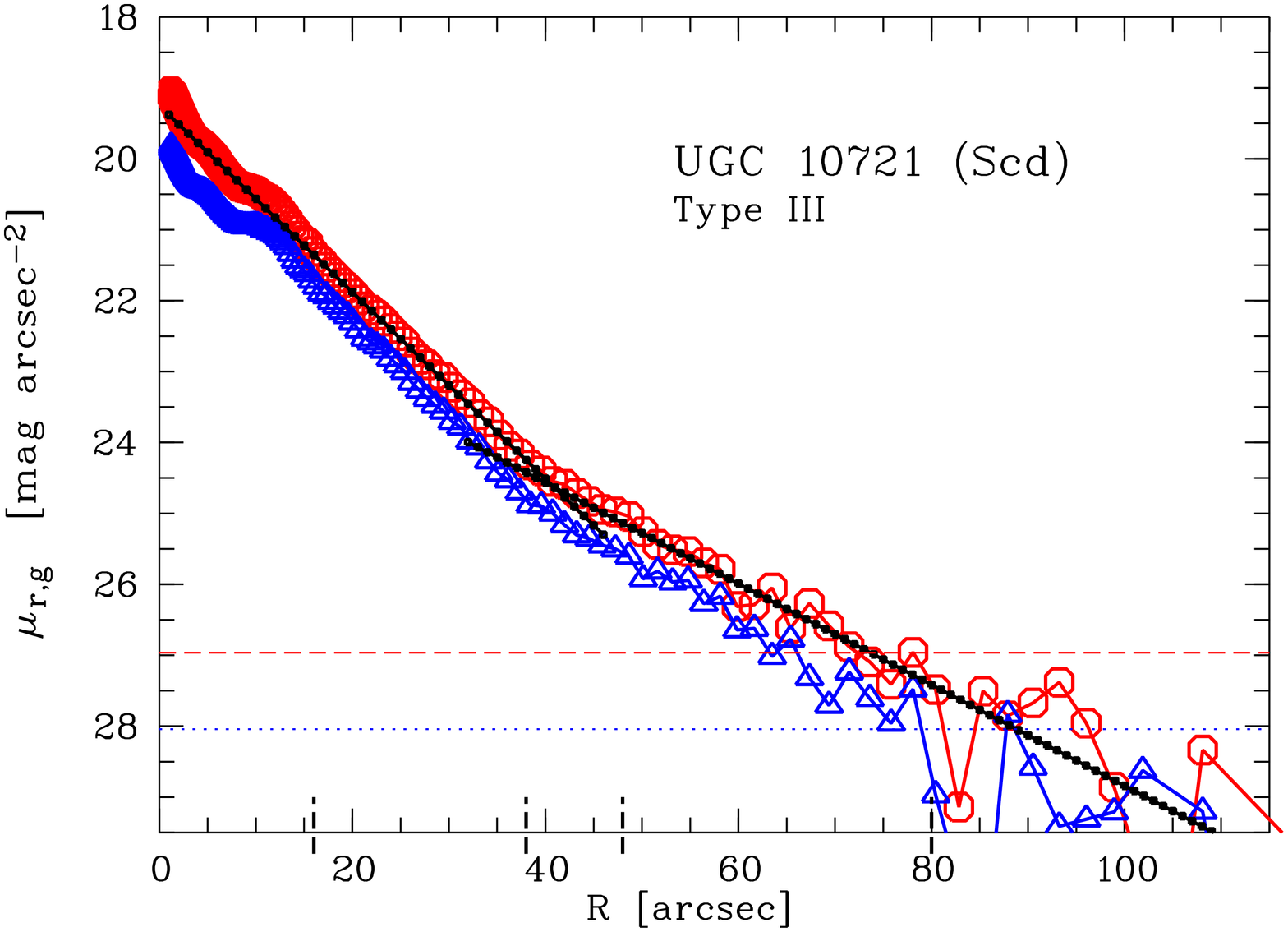}
}
\end{minipage}
\newpage
\onecolumn
\begin{minipage}[t][\textheight][t]{\textwidth}
\parbox[t][0.5\height][t]{0.47\textwidth}
{
\noindent {\bf UGC\,12709    :}        \typect                \\
\texttt{J233724.0+002327 .SXS9.  8.3 -19.05   3.0 2672}\\[0.25cm]
This is a faint, low surface brightness galaxy without a well 
defined center, so the centering (obtained from the $\sim\!80$\arcsec
ellipse) and the photometric inclination (ellipticity) and PA 
(obtained at $\sim\!70$\arcsec) are uncertain. 
The offcentered outer disk is reflected by the central dip in the 
final profile, which shows a clear downbending break at 
$\sim\!70\arcsec$ corresponding roughly to the end of the 
flocculent starforming patches.  
Although classified as SAB the central region ($\sim\!10\arcsec$) is 
too fuzzy to argue for or against a bar, but in both cases it is to small 
to be responsible for the break, which is therefore classified as \typectc.
}
\hfill 
\parbox[t][0.5\height][t]{0.47\textwidth}
{
\includegraphics[width=5.7cm,angle=270]{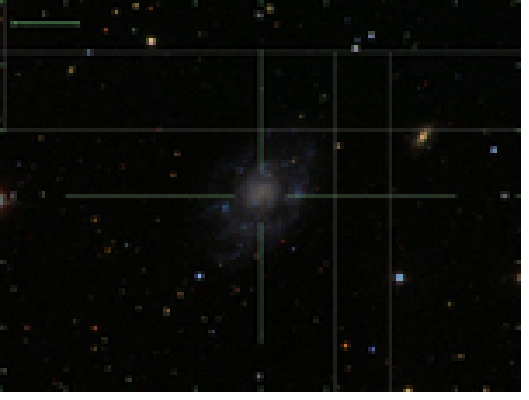}
\hspace*{-0.8cm}
\includegraphics[width=6.1cm,angle=270]{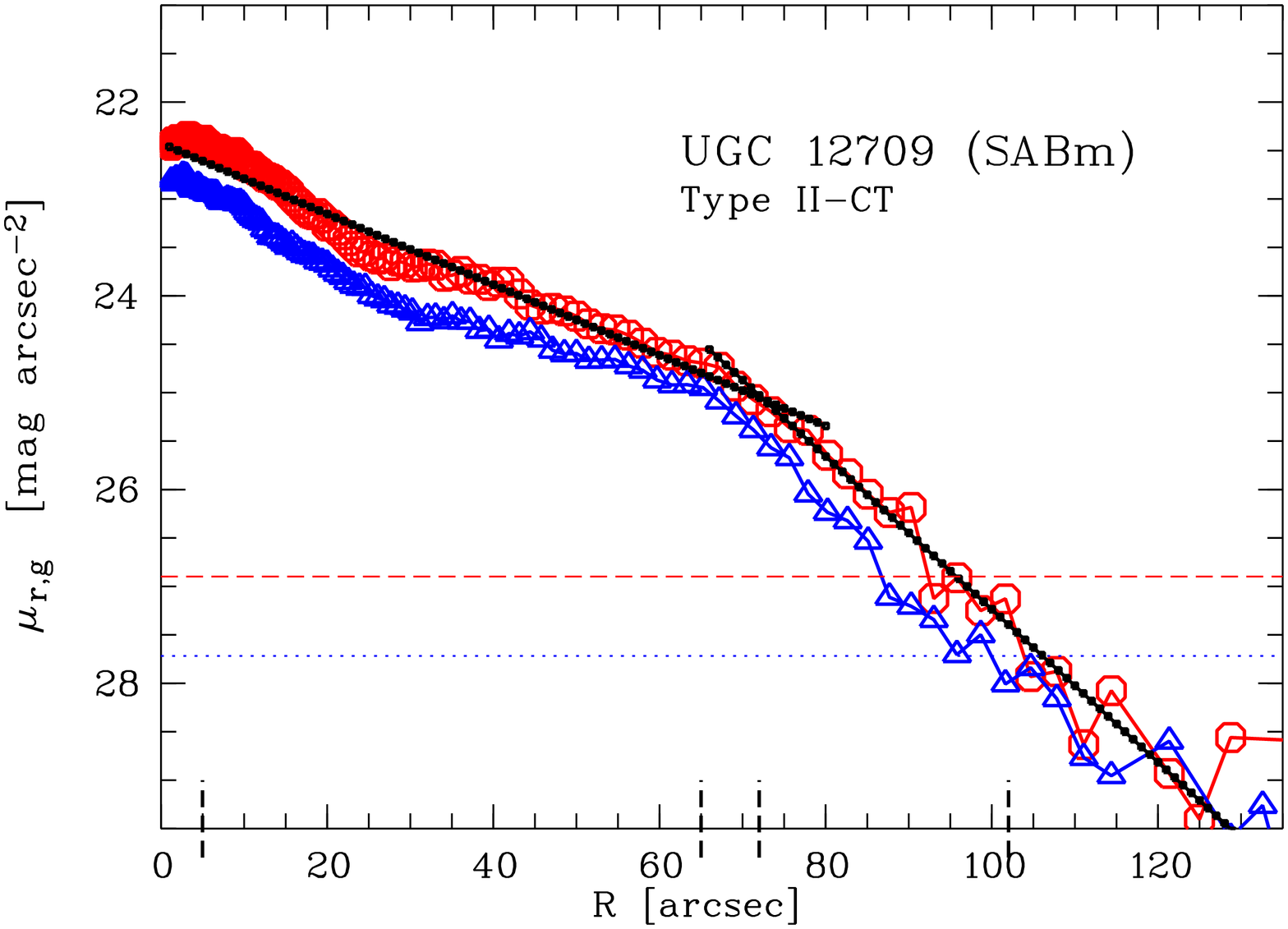}
}
\end{minipage}
\newpage
\onecolumn
\section{Rejected galaxies}
\label{rejected}
\begin{figure}[h]
\begin{center}
\includegraphics[width=5.7cm,angle=270]{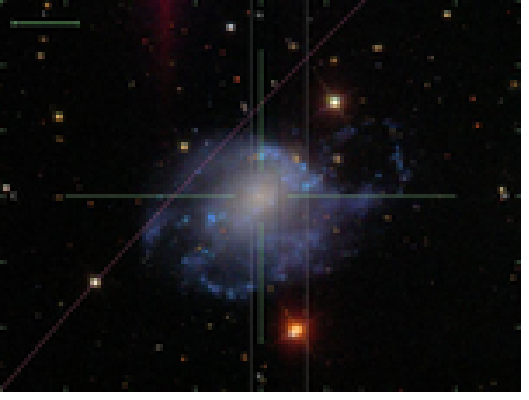}
\includegraphics[width=5.7cm,angle=270]{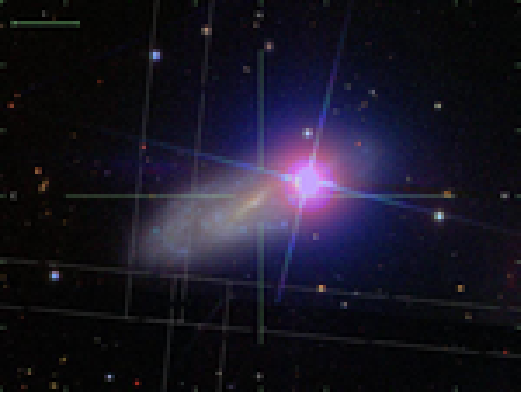}
\includegraphics[width=5.7cm,angle=270]{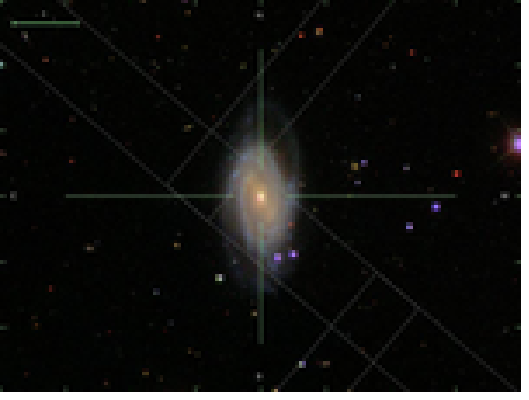}
\includegraphics[width=5.7cm,angle=270]{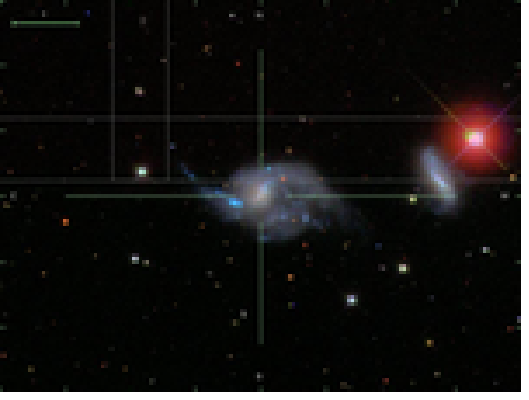}
\includegraphics[width=5.7cm,angle=270]{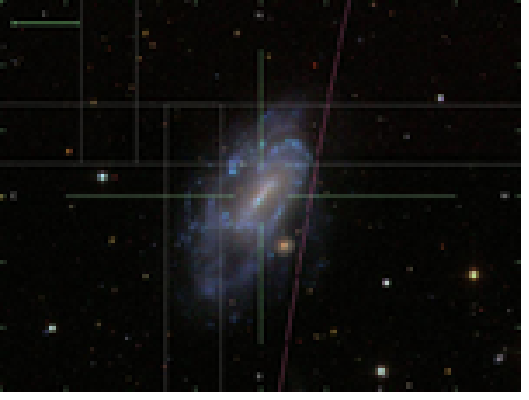}
\includegraphics[width=5.7cm,angle=270]{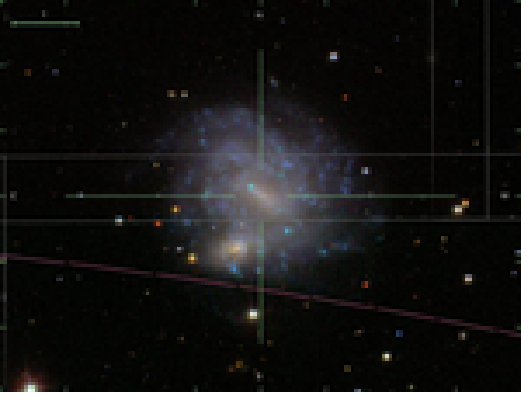}
\end{center}
\caption{Galaxies rejected from the original LEDA-SDSS(DR2) sample: 
SDSS colour plates of NGC\,0428,
NGC\,0988,
NGC\,2712,
NGC\,3023,
NGC\,4116, and
NGC\,4496A 
{\it (from left to right and top to bottom)}}
\end{figure}
\addtocounter{figure}{-1}
\begin{figure*}
\begin{center}
\includegraphics[width=5.7cm,angle=270]{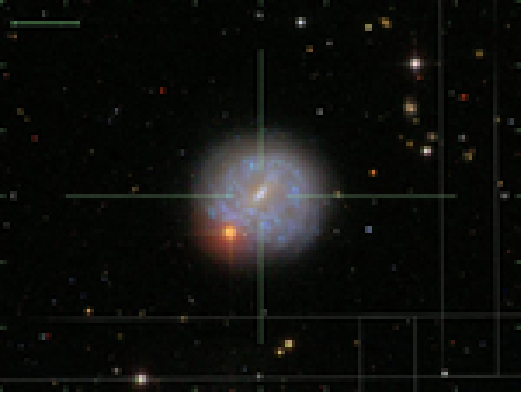}
\includegraphics[width=5.7cm,angle=270]{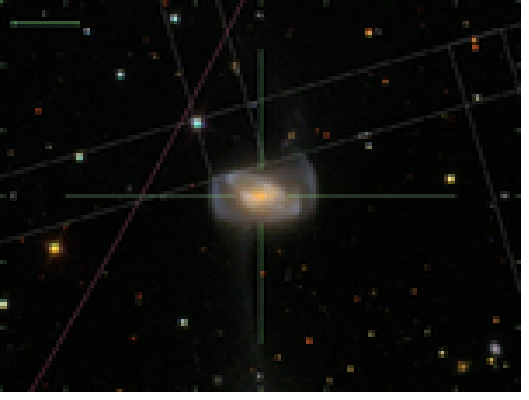}
\includegraphics[width=5.7cm,angle=270]{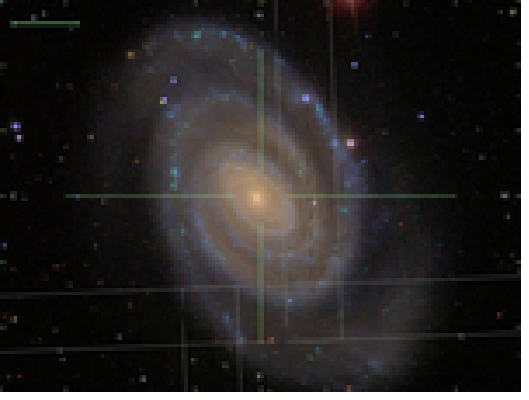}
\includegraphics[width=5.7cm,angle=270]{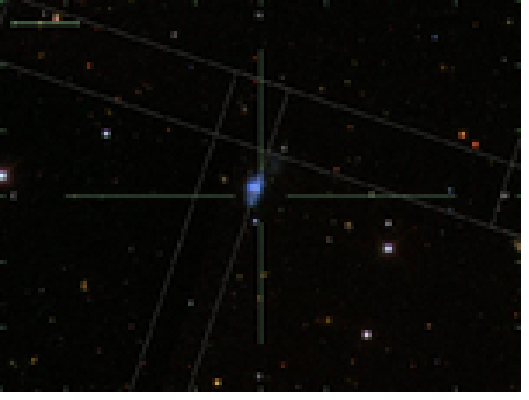}
\includegraphics[width=5.7cm,angle=270]{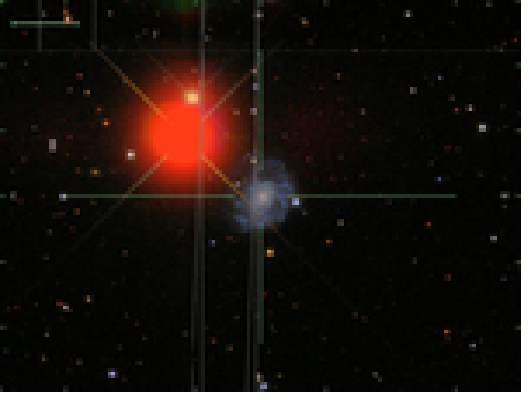}
\includegraphics[width=5.7cm,angle=270]{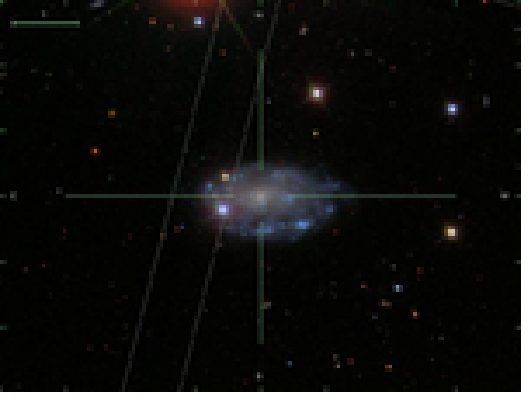}
\includegraphics[width=5.7cm,angle=270]{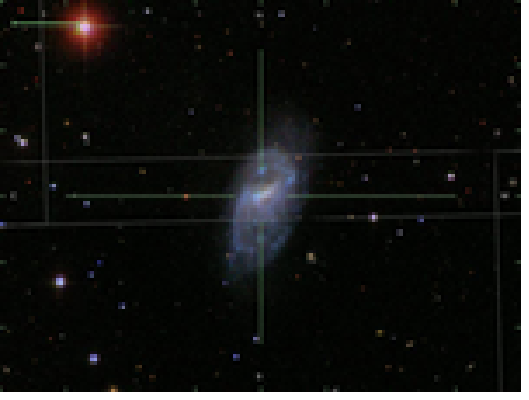}
\end{center}
\caption{(continued): Galaxies rejected from the original LEDA-SDSS(DR2) sample 
SDSS colour plates of NGC\,4900,
NGC\,5218,
NGC\,5364,
PGC\,032356,
UGC\,04684,
UGC\,06162, and
UGC\,09215
{\it (from left to right and top to bottom)}}
\end{figure*}
\end{document}